%\input harvmac
%\draftmode
%%%%%%%%%%%%%%%%%%%%%%%%%%%%%%%%%%%%%%%%%%%%%%%
%%%%%%%%%%%%%%%%%%  tex macros for preprints, cm version %%%%%%%%%%%%%%
%         (P. Ginsparg <ginsparg@lanl.gov>, last updated 7/94)
%                if confused, type `b' in response to query 
%           hypertex extensions (still provisional), 7/26/94
%
%---------------------------------------------------------------------%
%\input hyperbasics %comment out this line to restore non-hyper functionality
%
%% site dependent options:
%% \unredoffs and \redoffs define horizontal and vertical offsets
%% respectively for unreduced and reduced modes. \speclscape defines
%% the \special{} call that sets printer to landscape (sideways) mode.
%% from standard set below, leave uncommented as appropriate or redefine
%
%%% next 400dpi
\def\unredoffs{} \def\redoffs{\voffset=-.31truein\hoffset=-.48truein}
\def\speclscape{}
%\def\speclscape{\special{papersize=11in,8.5in}}
%
%%% apple lw
%\def\unredoffs{} \def\redoffs{\voffset=-.31truein\hoffset=-.59truein}
%\def\speclscape{\special{ps: landscape}}
%
%%% qms lasergrafix:
%\def\unredoffs{} \def\redoffs{\voffset=-.4truein\hoffset=.125truein}
%\def\speclscape{\special{qms: landscape}}
%
%%% saclay A4 paper:
%\def\unredoffs{\hoffset-.14truein\voffset-.2truein}
%\def\redoffs{\voffset=-.45truein\hoffset=-.21truein}
%\def\speclscape{\special{landscape}}
%
%---------------------------------------------------------------------%
%
\newbox\leftpage \newdimen\fullhsize \newdimen\hstitle \newdimen\hsbody
\tolerance=1000\hfuzz=2pt
\catcode`\@=11 % This allows us to modify PLAIN macros.
\ifx\hyperdef\UNd@FiNeD\def\hyperdef#1#2#3#4{#4}\def\hyperref#1#2#3#4{#4}\fi
\def\bigans{b }
\def\answ{b }
%\message{ big or little (b/l)? }\read-1 to\answ
%
\ifx\answ\bigans\message{(This will come out unreduced.}
\magnification=1200\unredoffs\baselineskip=16pt plus 2pt minus 1pt
\hsbody=\hsize \hstitle=\hsize %take default values for unreduced format
\else\message{(This will be reduced.} \let\l@r=L
\magnification=1000\baselineskip=16pt plus 2pt minus 1pt \vsize=7truein
\redoffs \hstitle=8truein\hsbody=4.75truein\fullhsize=10truein\hsize=\hsbody
\output={\ifnum\pageno=0 %%% This is the HUTP version
  \shipout\vbox{\speclscape{\hsize\fullhsize\makeheadline}
    \hbox to \fullhsize{\hfill\pagebody\hfill}}\advancepageno
  \else
  \almostshipout{\leftline{\vbox{\pagebody\makefootline}}}\advancepageno
  \fi}
\def\almostshipout#1{\if L\l@r \count1=1 \message{[\the\count0.\the\count1]}
      \global\setbox\leftpage=#1 \global\let\l@r=R
 \else \count1=2
  \shipout\vbox{\speclscape{\hsize\fullhsize\makeheadline}
      \hbox to\fullhsize{\box\leftpage\hfil#1}}  \global\let\l@r=L\fi}
\fi
%---------------------------------------------------------------------
%
\newcount\yearltd\yearltd=\year\advance\yearltd by -1900

\def\Title#1#2{\nopagenumbers\abstractfont\hsize=\hstitle\rightline{#1}%
\vskip 1in\centerline{\titlefont #2}\abstractfont\vskip .5in\pageno=0}
\def\Date#1{\vfill\leftline{#1}\tenpoint\supereject\global\hsize=\hsbody%
\footline={\hss\tenrm\hyperdef\hypernoname{page}\folio\folio\hss}}%
% (restores pagenumbers)
%
%       use following instead of \Date on the preliminary draft,
%       puts date/time on each page in big mode, writes labels in margins

\def\draftmode{\message{ DRAFTMODE }\def\draftdate{{\rm preliminary draft:
\number\month/\number\day/\number\yearltd\ \ \hourmin}}%
\headline={\hfil\draftdate}\writelabels\baselineskip=20pt plus 2pt minus 2pt
 {\count255=\time\divide\count255 by 60 \xdef\hourmin{\number\count255}
  \multiply\count255 by-60\advance\count255 by\time
  \xdef\hourmin{\hourmin:\ifnum\count255<10 0\fi\the\count255}}}
%       use \nolabels to get rid of eqn, ref, and fig labels in draft mode
\def\nolabels{\def\wrlabeL##1{}\def\eqlabeL##1{}\def\reflabeL##1{}}
\def\writelabels{\def\wrlabeL##1{\leavevmode\vadjust{\rlap{\smash%
{\line{{\escapechar=` \hfill\rlap{\sevenrm\hskip.03in\string##1}}}}}}}%
\def\eqlabeL##1{{\escapechar-1\rlap{\sevenrm\hskip.05in\string##1}}}%
\def\reflabeL##1{\noexpand\llap{\noexpand\sevenrm\string\string\string##1}}}
\nolabels
%
% tagged sec numbers
\global\newcount\secno \global\secno=0
\global\newcount\meqno \global\meqno=1
\def\s@csym{}
\def\newsec#1{\global\advance\secno by1%
{\toks0{#1}\message{(\the\secno. \the\toks0)}}%
%\ifx\answ\bigans \vfill\eject \else \bigbreak\bigskip \fi  %if desired
\global\subsecno=0\eqnres@t\let\s@csym\secsym\xdef\secn@m{\the\secno}\noindent
{\bf\hyperdef\hypernoname{section}{\the\secno}{\the\secno.} #1}%
\writetoca{{\string\hyperref{}{section}{\the\secno}{\the\secno.}} {#1}}%
\par\nobreak\medskip\nobreak}
\def\eqnres@t{\xdef\secsym{\the\secno.}\global\meqno=1\bigbreak\bigskip}
\def\sequentialequations{\def\eqnres@t{\bigbreak}}\xdef\secsym{}
\global\newcount\subsecno \global\subsecno=0
\def\subsec#1{\global\advance\subsecno by1%
{\toks0{#1}\message{(\s@csym\the\subsecno. \the\toks0)}}%
\ifnum\lastpenalty>9000\else\bigbreak\fi
\noindent{\it\hyperdef\hypernoname{subsection}{\secn@m.\the\subsecno}%
{\secn@m.\the\subsecno.} #1}\writetoca{\string\quad
{\string\hyperref{}{subsection}{\secn@m.\the\subsecno}{\secn@m.\the\subsecno.}}
{#1}}\par\nobreak\medskip\nobreak}
\def\appendix#1#2{\global\meqno=1\global\subsecno=0\xdef\secsym{\hbox{#1.}}%
\bigbreak\bigskip\noindent{\bf Appendix \hyperdef\hypernoname{appendix}{#1}%
{#1.} #2}{\toks0{(#1. #2)}\message{\the\toks0}}%
\xdef\s@csym{#1.}\xdef\secn@m{#1}%
\writetoca{\string\hyperref{}{appendix}{#1}{Appendix {#1.}} {#2}}%
\par\nobreak\medskip\nobreak}
%
%       \eqn\label{a+b=c}	gives displayed equation, numbered
%				consecutively within sections.
%     \eqnn and \eqna define labels in advance (of eqalign?)
%
\def\checkm@de#1#2{\ifmmode{\def\f@rst##1{##1}\hyperdef\hypernoname{equation}%
{#1}{#2}}\else\hyperref{}{equation}{#1}{#2}\fi}
\def\eqnn#1{\DefWarn#1\xdef #1{(\noexpand\relax\noexpand\checkm@de%
{\s@csym\the\meqno}{\secsym\the\meqno})}%
\wrlabeL#1\writedef{#1\leftbracket#1}\global\advance\meqno by1}
\def\f@rst#1{\c@t#1a\em@ark}\def\c@t#1#2\em@ark{#1}
\def\eqna#1{\DefWarn#1\wrlabeL{#1$\{\}$}%
\xdef #1##1{(\noexpand\relax\noexpand\checkm@de%
{\s@csym\the\meqno\noexpand\f@rst{##1}}{\hbox{$\secsym\the\meqno##1$}})}
\writedef{#1\numbersign1\leftbracket#1{\numbersign1}}\global\advance\meqno by1}
\def\eqn#1#2{\DefWarn#1%
\xdef #1{(\noexpand\hyperref{}{equation}{\s@csym\the\meqno}%
{\secsym\the\meqno})}$$#2\eqno(\hyperdef\hypernoname{equation}%
{\s@csym\the\meqno}{\secsym\the\meqno})\eqlabeL#1$$%
\writedef{#1\leftbracket#1}\global\advance\meqno by1}
\def\xeqn{\expandafter\xe@n}\def\xe@n(#1){#1}
\def\xeqna#1{\expandafter\xe@n#1}
\def\eqns#1{(\e@ns #1{\hbox{}})}
\def\e@ns#1{\ifx\UNd@FiNeD#1\message{eqnlabel \string#1 is undefined.}%
\xdef#1{(?.?)}\fi{\let\hyperref=\relax\xdef\next{#1}}%
\ifx\next\em@rk\def\next{}\else%
\ifx\next#1\xeqn#1\else\def\n@xt{#1}\ifx\n@xt\next#1\else\xeqna#1\fi
\fi\let\next=\e@ns\fi\next}

\def\DefWarn#1{\ifx\UNd@FiNeD#1\else
\immediate\write16{*** WARNING: the label \string#1 is already defined ***}\fi}
%
%			 footnotes
\newskip\footskip\footskip14pt plus 1pt minus 1pt %sets footnote baselineskip
\def\footnotefont{\ninepoint}\def\f@t#1{\footnotefont #1\@foot}
\def\f@@t{\baselineskip\footskip\bgroup\footnotefont\aftergroup\@foot\let\next}
\setbox\strutbox=\hbox{\vrule height9.5pt depth4.5pt width0pt}
\global\newcount\ftno \global\ftno=0
\def\foot{\global\advance\ftno by1\def\foot@rg{\hyperref{}{footnote}%
{\the\ftno}{\the\ftno}\xdef\foot@rg{\noexpand\hyperdef\noexpand\hypernoname%
{footnote}{\the\ftno}{\the\ftno}}}\footnote{$^{\foot@rg}$}}
%
%say \footend to put footnotes at end
%will cause problems if \ref used inside \foot, instead use \nref before
\newwrite\ftfile
\def\footend{\def\foot{\global\advance\ftno by1\chardef\wfile=\ftfile
%%$^{\the\ftno}$\ifnum\ftno=1\immediate\openout\ftfile=\jobname.fts\fi%
\hyperref{}{footnote}{\the\ftno}{$^{\the\ftno}$}%
\ifnum\ftno=1\immediate\openout\ftfile=\jobname.fts\fi%
\immediate\write\ftfile{\noexpand\smallskip%
%%\noexpand\item{f\the\ftno:\ }\pctsign}\findarg}%
\noexpand\item{\noexpand\hyperdef\noexpand\hypernoname{footnote}
{\the\ftno}{f\the\ftno}:\ }\pctsign}\findarg}%
\def\footatend{\vfill\eject\immediate\closeout\ftfile{\parindent=20pt
\centerline{\bf Footnotes}\nobreak\bigskip\input \jobname.fts }}}
\def\footatend{}
%
%     \ref\label{text}
% generates a number, assigns it to \label, generates an entry.
% To list the refs on a separate page,  \listrefs
%
\global\newcount\refno \global\refno=1
\newwrite\rfile
\def\ref{[\hyperref{}{reference}{\the\refno}{\the\refno}]\nref}
\def\nref#1{\DefWarn#1%
\xdef#1{[\noexpand\hyperref{}{reference}{\the\refno}{\the\refno}]}%
\writedef{#1\leftbracket#1}%
\ifnum\refno=1\immediate\openout\rfile=\jobname.refs\fi
\chardef\wfile=\rfile\immediate\write\rfile{\noexpand\item{[\noexpand\hyperdef%
\noexpand\hypernoname{reference}{\the\refno}{\the\refno}]\ }%
\reflabeL{#1\hskip.31in}\pctsign}\global\advance\refno by1\findarg}
%	horrible hack to sidestep tex \write limitation
\def\findarg#1#{\begingroup\obeylines\newlinechar=`\^^M\pass@rg}
{\obeylines\gdef\pass@rg#1{\writ@line\relax #1^^M\hbox{}^^M}%
\gdef\writ@line#1^^M{\expandafter\toks0\expandafter{\striprel@x #1}%
\edef\next{\the\toks0}\ifx\next\em@rk\let\next=\endgroup\else\ifx\next\empty%
\else\immediate\write\wfile{\the\toks0}\fi\let\next=\writ@line\fi\next\relax}}
\def\striprel@x#1{} \def\em@rk{\hbox{}}
\def\lref{\begingroup\obeylines\lr@f}
\def\lr@f#1#2{\DefWarn#1\gdef#1{\let#1=\UNd@FiNeD\ref#1{#2}}\endgroup\unskip}

\def\addref#1{\immediate\write\rfile{\noexpand\item{}#1}} %now unnecessary
\def\listrefs{\footatend\vfill\supereject\immediate\closeout\rfile\writestoppt
\baselineskip=\footskip\centerline{{\bf References}}\bigskip{\parindent=20pt%
\frenchspacing\escapechar=` \input \jobname.refs\vfill\eject}\nonfrenchspacing}
\def\startrefs#1{\immediate\openout\rfile=\jobname.refs\refno=#1}
\def\xref{\expandafter\xr@f}\def\xr@f[#1]{#1}
\def\refs#1{\count255=1[\r@fs #1{\hbox{}}]}
\def\r@fs#1{\ifx\UNd@FiNeD#1\message{reflabel \string#1 is undefined.}%
\nref#1{need to supply reference \string#1.}\fi%
\vphantom{\hphantom{#1}}{\let\hyperref=\relax\xdef\next{#1}}%
\ifx\next\em@rk\def\next{}%
\else\ifx\next#1\ifodd\count255\relax\xref#1\count255=0\fi%
\else#1\count255=1\fi\let\next=\r@fs\fi\next}
%

%
% this is ugly, but moore insists
\newwrite\ffile\global\newcount\figno \global\figno=1
\def\fig{fig.~\hyperref{}{figure}{\the\figno}{\the\figno}\nfig}
\def\nfig#1{\DefWarn#1%
\xdef#1{fig.~\noexpand\hyperref{}{figure}{\the\figno}{\the\figno}}%
\writedef{#1\leftbracket fig.\noexpand~\xfig#1}%
\ifnum\figno=1\immediate\openout\ffile=\jobname.figs\fi\chardef\wfile=\ffile%
{\let\hyperref=\relax
\immediate\write\ffile{\noexpand\medskip\noexpand\item{Fig.\ %
\noexpand\hyperdef\noexpand\hypernoname{figure}{\the\figno}{\the\figno}. }
\reflabeL{#1\hskip.55in}\pctsign}}\global\advance\figno by1\findarg}
\def\listfigs{\vfill\eject\immediate\closeout\ffile{\parindent40pt
\baselineskip14pt\centerline{{\bf Figure Captions}}\nobreak\medskip
\escapechar=` \input \jobname.figs\vfill\eject}}
\def\xfig{\expandafter\xf@g}\def\xf@g fig.\penalty\@M\ {}
\def\figs#1{figs.~\f@gs #1{\hbox{}}}
\def\f@gs#1{{\let\hyperref=\relax\xdef\next{#1}}\ifx\next\em@rk\def\next{}\else
\ifx\next#1\xfig #1\else#1\fi\let\next=\f@gs\fi\next}
\def\figin{\epsfcheck\figin}\def\figins{\epsfcheck\figins}
\def\epsfcheck{\ifx\epsfbox\UNd@FiNeD
\message{(NO epsf.tex, FIGURES WILL BE IGNORED)}
\gdef\figin##1{\vskip2in}\gdef\figins##1{\hskip.5in}% blank space instead
\else\message{(FIGURES WILL BE INCLUDED)}%
\gdef\figin##1{##1}\gdef\figins##1{##1}\fi}
\def\DefWarn#1{}
\def\figinsert{\goodbreak\midinsert}
\def\ifig#1#2#3{\DefWarn#1\xdef#1{fig.~\noexpand\hyperref{}{figure}%
{\the\figno}{\the\figno}}\writedef{#1\leftbracket fig.\noexpand~\xfig#1}%
\figinsert\figin{\centerline{#3}}\medskip\centerline{\vbox{\baselineskip12pt
\advance\hsize by -1truein\noindent\wrlabeL{#1=#1}\footnotefont%
{\bf Fig.~\hyperdef\hypernoname{figure}{\the\figno}{\the\figno}:} #2}}
\bigskip\endinsert\global\advance\figno by1}
\newwrite\lfile
{\escapechar-1\xdef\pctsign{\string\%}\xdef\leftbracket{\string\{}
\xdef\rightbracket{\string\}}\xdef\numbersign{\string\#}}
\def\writedefs{\immediate\openout\lfile=\jobname.defs \def\writedef##1{%
{\let\hyperref=\relax\let\hyperdef=\relax\let\hypernoname=\relax
 \immediate\write\lfile{\string\def\string##1\rightbracket}}}}%
\def\writestop{\def\writestoppt{\immediate\write\lfile{\string\pageno
 \the\pageno\string\startrefs\leftbracket\the\refno\rightbracket
 \string\def\string\secsym\leftbracket\secsym\rightbracket
 \string\secno\the\secno\string\meqno\the\meqno}\immediate\closeout\lfile}}
\def\writestoppt{}\def\writedef#1{}
\def\seclab#1{\DefWarn#1%
\xdef #1{\noexpand\hyperref{}{section}{\the\secno}{\the\secno}}%
\writedef{#1\leftbracket#1}\wrlabeL{#1=#1}}
\def\subseclab#1{\DefWarn#1%
\xdef #1{\noexpand\hyperref{}{subsection}{\secn@m.\the\subsecno}%
{\secn@m.\the\subsecno}}\writedef{#1\leftbracket#1}\wrlabeL{#1=#1}}
\def\applab#1{\DefWarn#1%
\xdef #1{\noexpand\hyperref{}{appendix}{\secn@m}{\secn@m}}%
\writedef{#1\leftbracket#1}\wrlabeL{#1=#1}}
\newwrite\tfile \def\writetoca#1{}
\def\leaderfill{\leaders\hbox to 1em{\hss.\hss}\hfill}
%	use this to write file with table of contents
\def\writetoc{\immediate\openout\tfile=\jobname.toc
   \def\writetoca##1{{\edef\next{\write\tfile{\noindent ##1
   \string\leaderfill {\string\hyperref{}{page}{\noexpand\number\pageno}%
                       {\noexpand\number\pageno}} \par}}\next}}}
%       and this lists table of contents on second pass
\newread\ch@ckfile
\def\listtoc{\immediate\closeout\tfile\immediate\openin\ch@ckfile=\jobname.toc
\ifeof\ch@ckfile\message{no file \jobname.toc, no table of contents this pass}%
\else\closein\ch@ckfile\centerline{\bf Contents}\nobreak\medskip%
{\baselineskip=18pt
%   \footnotefont
\parskip=2pt\catcode`\@=11\input\jobname.toc
\catcode`\@=12\bigbreak\bigskip}\fi}
\catcode`\@=12 % at signs are no longer letters
%
%	Unpleasantness in calling in abstract and title fonts
\edef\tfontsize{\ifx\answ\bigans scaled\magstep3\else scaled\magstep4\fi}
\font\titlerm=cmr10 \tfontsize \font\titlerms=cmr7 \tfontsize
\font\titlermss=cmr5 \tfontsize \font\titlei=cmmi10 \tfontsize
\font\titleis=cmmi7 \tfontsize \font\titleiss=cmmi5 \tfontsize
\font\titlesy=cmsy10 \tfontsize \font\titlesys=cmsy7 \tfontsize
\font\titlesyss=cmsy5 \tfontsize \font\titleit=cmti10 \tfontsize
\skewchar\titlei='177 \skewchar\titleis='177 \skewchar\titleiss='177
\skewchar\titlesy='60 \skewchar\titlesys='60 \skewchar\titlesyss='60
\def\titlefont{\def\rm{\fam0\titlerm}% switch to title font
\textfont0=\titlerm \scriptfont0=\titlerms \scriptscriptfont0=\titlermss
\textfont1=\titlei \scriptfont1=\titleis \scriptscriptfont1=\titleiss
\textfont2=\titlesy \scriptfont2=\titlesys \scriptscriptfont2=\titlesyss
\textfont\itfam=\titleit \def\it{\fam\itfam\titleit}\rm}
 \ifx\answ\bigans\else scaled\magstep1\fi
\ifx\answ\bigans\def\abstractfont{\tenpoint}\else
\font\absit=cmti10 scaled \magstep1
\font\abssl=cmsl10 scaled \magstep1
\font\absrm=cmr10 scaled\magstep1 \font\absrms=cmr7 scaled\magstep1
\font\absrmss=cmr5 scaled\magstep1 \font\absi=cmmi10 scaled\magstep1
\font\absis=cmmi7 scaled\magstep1 \font\absiss=cmmi5 scaled\magstep1
\font\abssy=cmsy10 scaled\magstep1 \font\abssys=cmsy7 scaled\magstep1
\font\abssyss=cmsy5 scaled\magstep1 \font\absbf=cmbx10 scaled\magstep1
\skewchar\absi='177 \skewchar\absis='177 \skewchar\absiss='177
\skewchar\abssy='60 \skewchar\abssys='60 \skewchar\abssyss='60
\def\abstractfont{\def\rm{\fam0\absrm}% switch to abstract font
\textfont0=\absrm \scriptfont0=\absrms \scriptscriptfont0=\absrmss
\textfont1=\absi \scriptfont1=\absis \scriptscriptfont1=\absiss
\textfont2=\abssy \scriptfont2=\abssys \scriptscriptfont2=\abssyss
\textfont\itfam=\absit \def\it{\fam\itfam\absit}\def\footnotefont{\tenpoint}%
\textfont\slfam=\abssl \def\sl{\fam\slfam\abssl}%
\textfont\bffam=\absbf \def\bf{\fam\bffam\absbf}\rm}\fi
\def\tenpoint{\def\rm{\fam0\tenrm}% switch back to 10-point type
\textfont0=\tenrm \scriptfont0=\sevenrm \scriptscriptfont0=\fiverm
\textfont1=\teni  \scriptfont1=\seveni  \scriptscriptfont1=\fivei
\textfont2=\tensy \scriptfont2=\sevensy \scriptscriptfont2=\fivesy
\textfont\itfam=\tenit \def\it{\fam\itfam\tenit}\def\footnotefont{\ninepoint}%
\textfont\bffam=\tenbf \def\bf{\fam\bffam\tenbf}\def\sl{\fam\slfam\tensl}\rm}
\font\ninerm=cmr9 \font\sixrm=cmr6 \font\ninei=cmmi9 \font\sixi=cmmi6
\font\ninesy=cmsy9 \font\sixsy=cmsy6 \font\ninebf=cmbx9
\font\nineit=cmti9 \font\ninesl=cmsl9 \skewchar\ninei='177
\skewchar\sixi='177 \skewchar\ninesy='60 \skewchar\sixsy='60
\def\ninepoint{\def\rm{\fam0\ninerm}% switch to footnote font
\textfont0=\ninerm \scriptfont0=\sixrm \scriptscriptfont0=\fiverm
\textfont1=\ninei \scriptfont1=\sixi \scriptscriptfont1=\fivei
\textfont2=\ninesy \scriptfont2=\sixsy \scriptscriptfont2=\fivesy
\textfont\itfam=\ninei \def\it{\fam\itfam\nineit}\def\sl{\fam\slfam\ninesl}%
\textfont\bffam=\ninebf \def\bf{\fam\bffam\ninebf}\rm}
%
%---------------------------------------------------------------------
%
\def\noblackbox{\overfullrule=0pt}
\hyphenation{anom-aly anom-alies coun-ter-term coun-ter-terms}
\def\inv{^{\raise.15ex\hbox{${\scriptscriptstyle -}$}\kern-.05em 1}}

\def\Dsl{\,\raise.15ex\hbox{/}\mkern-13.5mu D} %this one can be subscripted
\def\dsl{\raise.15ex\hbox{/}\kern-.57em\partial}

\def\tr{{\rm tr}} \def\Tr{{\rm Tr}}
 %pound sterling
\def\lspace{\ifx\answ\bigans{}\else\qquad\fi}
\def\lbspace{\ifx\answ\bigans{}\else\hskip-.2in\fi} % $$\lbspace...$$
\def\boxeqn#1{\vcenter{\vbox{\hrule\hbox{\vrule\kern3pt\vbox{\kern3pt
	\hbox{${\displaystyle #1}$}\kern3pt}\kern3pt\vrule}\hrule}}}
\def\mbox#1#2{\vcenter{\hrule \hbox{\vrule height#2in
		\kern#1in \vrule} \hrule}}  %e.g. \mbox{.1}{.1}
%	matters of taste
%\def\tilde{\widetilde} \def\bar{\overline} \def\hat{\widehat}
%
% some sample definitions
  %     curly letters

\def\vev#1{\langle #1 \rangle}

\def\darr#1{\raise1.5ex\hbox{$\leftrightarrow$}\mkern-16.5mu #1}
 %pound sterling

 %puts a small half in a displayed eqn
\def\roughly#1{\raise.3ex\hbox{$#1$\kern-.75em\lower1ex\hbox{$\sim$}}}

\noblackbox
%%%%%%%%%%%%%%%%%%%%%%%%%%%%%%%%%%%%%%%%%%%%%%%%%%%%%%%%%%%%%%%%%%%%%
%%%%%%  TeX twice to generate table of contents 
%%%%%%%%%%%%%%%%%%%%%%%%%%%%%%%%%%%%%%%%%%%%%%%%%%%%%%%%%%%%%%%%%%%%%
%  Local Definitions
%%%%%%%%%%%%%%%%%%%%%%%%%%%%%%%%%%%%%%%%%%%%%%%%%%%%%%%%%%%%%%%%%%%%%

\def\F#1#2{{   _{#1}F_{#2}  }}
\def\Ga{\Gamma}
\def\ie{{\it i.e.\ }}
\def\cf{{\it c.f.\ }}
\def\frac#1#2{{#1 \over #2}}
\def\al{{\alpha}}
\def\xxi#1#2{{(\xi_#1\xi_#2)}}

\def\cf{{\it cf.\ }}
\def\ie{{\it i.e.\ }}
\def\eg{{\it e.g.\ }}
\def\eqq{{\it Eq.\ }}
\def\eqqs{{\it Eqs.\ }}
\def\Ec{{\cal E}}
\def\ap{\alpha'}
\def\Vc{{\cal V}}
\def\STr{{\rm STr}}
\def\si{\sigma}
\def\ep{\epsilon}
\def\be{\beta}
%%%%%%%%%%%%%%%%%%%%%%%%%%%%%%%%%%%%%%%%%%%%%%%%%%%%%%%%%%%%%%%%%
%%%%% Referencing  %%%%%%%%%%%%%%%%%%%%%%%%%%%%%%%%%%%%%%%%%%%%%%
%%%%%%%%%%%%%%%%%%%%%%%%%%%%%%%%%%%%%%%%%%%%%%%%%%%%%%%%%%%%%%%%%
\newif\ifnref
\def\rrr#1#2{\relax\ifnref\nref#1{#2}\else\ref#1{#2}\fi}
\def\ldf#1#2{\begingroup\obeylines
\gdef#1{\rrr{#1}{#2}}\endgroup\unskip}

\def\doubref#1#2{\refs{{#1},{#2} }}
\def\threeref#1#2#3{\refs{{#1},{#2},{#3} }}

\nreffalse

\def\lref{\ldf}
%%%%%%%%%%%%%%%%%%%%%%%%%%%%%%%%%%%%%%%%%%%%%%%%%%%%%%%%%%%%%%%%%%
%%%%%%%%%%%%%%%%%   Stuff for Figures  %%%%%%%%%%%%%%%%%%%%%%%%%%%
%%%%%%%%%%%%%%%%%%%%%%%%%%%%%%%%%%%%%%%%%%%%%%%%%%%%%%%%%%%%%%%%%%
\newcount\figno
\figno=0
\def\fig#1#2#3{
\par\begingroup\parindent=0pt\leftskip=1cm\rightskip=1cm\parindent=0pt
\baselineskip=11pt \global\advance\figno by 1 \midinsert
\epsfxsize=#3 \centerline{\epsfbox{#2}} \vskip 12pt
\centerline{{\bf Figure \the\figno :}{\it ~~ #1}}\par
\endinsert\endgroup\par}
\def\figlabel#1{\xdef#1{\the\figno}}
%%%%%%%%%%%%%%%%%%%%%%%%%%%%%%%%%%%%%%%%%%%%%%%%%%%%%%%%%%%%%%%%%%

\input epsf
\input psfig
\def\figin{\epsfcheck\figin}\def\figins{\epsfcheck\figins}
\def\epsfcheck{\ifx\epsfbox\UnDeFiNeD
\message{(NO epsf.tex, FIGURES WILL BE IGNORED)}
\gdef\figin##1{\vskip2in}\gdef\figins##1{\hskip.5in}% blank space instead
\else\message{(FIGURES WILL BE INCLUDED)}%
\gdef\figin##1{##1}\gdef\figins##1{##1}\fi}
\def\DefWarn#1{}
\def\figinsert{\goodbreak\midinsert}
\def\ifig#1#2#3{\DefWarn#1\xdef#1{fig.~\the\figno}
\writedef{#1\leftbracket fig.\noexpand~\the\figno}%
\figinsert\figin{\centerline{#3}}\medskip\centerline{\vbox{\baselineskip12pt
\advance\hsize by -1truein\noindent\footnotefont{\bf Fig.~\the\figno } #2}}
\bigskip\endinsert\global\advance\figno by1}

%%%%%%%%%%%%%%%   Standard alltime definitions   %%%%%%%%%%%%%%%%%%%%
%%%%%%%%%%%%%%%%%%%%%%%%%%%%%%%%%%%%%%%%%%%%%%%%%%%%%%%%%%%%%%%%%%%%%

\def\appA{A}
\def\appB{B}
\def\appC{C}
\def\appD{D}
\def\tilde{\widetilde}

\def\h {{1\over 2}}

\def\ov {\overline}
\def\o {\over}
\def\fc#1#2{{#1 \o #2}}

\def\IZ{ {\bf Z}}

\def\IR{ {\bf R}}

      % For Eisenstein E2
\def\Li {{\cal L}i}  % For Polylogarithm

\def\br{\hfill\break}
\def\tr {{\rm tr}}
\def\det {{\rm det}}

\def\lf {\left}
\def\ri {\right}
\def\ra {\rightarrow}

\def\re {{\rm Re}}
\def\im {{\rm Im}}
\def\p {\partial}

\def\Bc {{\cal B}}

\def\Fc {{\cal F}} 
\def\Cc {{\cal C}} \def\Oc {{\cal O}}
\def\Lc {{\cal L}} 
\def\Mc {{\cal M}}  \def\Ac {{\cal A}}
\def\Pc {{\cal P}}

\def\Ic {{\cal I}}

%%%%%%%%%%%%%%%%%%%%%%%%%%%%%%%%%%%%%%%%%%%%%%%%%%%%%%%%%%%%%%%%%%%

\lref\BRAZIL{L.A.~Barreiro and R.~Medina,
``5-field terms in the open superstring effective action,''
  JHEP {\bf 0503}, 055 (2005)
  [arXiv:hep-th/0503182];\br
  %%CITATION = HEP-TH 0503182;%%
F.~Machado and R.~Medina,
``The open superstring and the non-abelian Born-Infeld theory,''
  Nucl.\ Phys.\ Proc.\ Suppl.\  {\bf 127}, 166 (2004);\br
  %%CITATION = NUPHZ,127,166;%%
O.~Chandia and R.~Medina,
``4-point effective actions in open and closed superstring theory,''
  JHEP {\bf 0311}, 003 (2003)
  [arXiv:hep-th/0310015];\br
  %%CITATION = HEP-TH 0310015;%%
F.~Brandt, F.~Machado and R.~Medina,
``Open and closed superstring five-point amplitudes at tree-level,''
  Int.\ J.\ Mod.\ Phys.\ A {\bf 18}, 2127 (2003).
  %%CITATION = IMPAE,A18,2127;%%
}

\lref\BBBL{J.M. Borwein, D.M. Bradley, D.J. Broadhurst, and P. Lisonek, 
{\it Special values of multidimensional polylogarithms,} 
Trans. Amer. Math. Soc. 353 (2001), 907-941;\br
D.J. Broadhurst, J. M. Borwein, and D. M. Bradley, {\it 
Evaluation of k-fold Euler/Zagier sums: a compendium of results for arbitrary k,} 
Electronic J. Combinatorics 4(2) (1997), R5.}

\lref\ZAGIER{D. Zagier, 
{\it Values of zeta functions and their applications,} 
First European Congress of Mathematics, Paris, 1992, Vol. II, Ed. A. Joseph et. al, 
Birkh\"auser, Basel, 1994, pp. 497--512.}

\lref\Tornheim{L. Tornheim, {\it Harmonic double series,} Amer. J. Math. 72 (1950), 303-314;\br
O. Espinosa and V. H. Moll,
{\it The evaluation of Tornheim double sums. Part 1}, arXiv: math-CA/0505647.}

\lref\Crandall{R.E. Crandall and J.P. Buhler
{\it On the evaluation of Euler sums}, Experiment. Math.  3 (1994) 275;\br
K. Boyadzhiev, 
{\it Evaluation of Euler-Zagier sums}, 
Internat. J. Math. Math. Sci. 27 (2001), 407-412.}

\lref\Witten{E.~Witten,
``Perturbative gauge theory as a string theory in twistor space,''
Commun.\ Math.\ Phys.\  {\bf 252}, 189 (2004)
[arXiv:hep-th/0312171].
%%CITATION = HEP-TH 0312171;%%
}

\lref\twist{
F.~Cachazo, P.~Svrcek and E.~Witten,
``MHV vertices and tree amplitudes in gauge theory,''
JHEP {\bf 0409}, 006 (2004)
[arXiv:hep-th/0403047];\br
%%CITATION = HEP-TH 0403047;%%
R.~Britto, F.~Cachazo, B.~Feng and E.~Witten,
``Direct proof of tree-level recursion relation in Yang-Mills theory,''
arXiv:hep-th/0501052;\br
%%CITATION = HEP-TH 0501052;%%
R.~Britto, F.~Cachazo and B.~Feng,
``New recursion relations for tree amplitudes of gluons,''
arXiv:hep-th/0412308;\br
%%CITATION = HEP-TH 0412308;%%
F.~Cachazo and P.~Svrcek,
 ``Lectures on twistor strings and perturbative Yang-Mills theory,''
  arXiv:hep-th/0504194.
  %%CITATION = HEP-TH 0504194;%%
}

\lref\Alex{A.~Sevrin and A.~Wijns,
``Higher order terms in the non-Abelian D-brane effective action and  magnetic
%background fields,''
JHEP {\bf 0308}, 059 (2003)
[arXiv:hep-th/0306260];\br
%%CITATION = HEP-TH 0306260;%%
P.~Koerber and A.~Sevrin,
``The non-abelian D-brane effective action through order $\alpha'^4$,''
JHEP {\bf 0210}, 046 (2002)
[arXiv:hep-th/0208044].
%%CITATION = HEP-TH 0208044;%%
% P.~Koerber and A.~Sevrin,
``Testing the $\alpha'^3$ term in the non-abelian open superstring  effective
%action,''
JHEP {\bf 0109}, 009 (2001)
[arXiv:hep-th/0109030].
%%CITATION = HEP-TH 0109030;%%
%P.~Koerber and A.~Sevrin,
``The non-Abelian Born-Infeld action through order $\alpha'^3$,''
JHEP {\bf 0110}, 003 (2001)
[arXiv:hep-th/0108169];\br
%%CITATION = HEP-TH 0108169;%%
A.~Sevrin, J.~Troost and W.~Troost,
``The non-abelian Born-Infeld action at order $F^6$,''
Nucl.\ Phys.\ B {\bf 603}, 389 (2001)
[arXiv:hep-th/0101192];\br
%%CITATION = HEP-TH 0101192;%%
L.~De Fosse, P.~Koerber and A.~Sevrin,
``The uniqueness of the Abelian Born-Infeld action,''
Nucl.\ Phys.\ B {\bf 603}, 413 (2001)
[arXiv:hep-th/0103015];\br
%%CITATION = HEP-TH 0103015;%%
P.~Koerber,
``Abelian and non-Abelian D-brane effective actions,''
Fortsch.\ Phys.\  {\bf 52}, 871 (2004)
[arXiv:hep-th/0405227].
%%CITATION = HEP-TH 0405227;%%
}

\lref\sevrin{P.~Koerber and A.~Sevrin,
``The non-abelian D-brane effective action through order $\ap^4$,''
JHEP {\bf 0210}, 046 (2002)
[arXiv:hep-th/0208044];\br
%%CITATION = HEP-TH 0208044;%%
A.~Sevrin and A.~Wijns,
``Higher order terms in the non-Abelian D-brane effective action and  magnetic
background fields,''
JHEP {\bf 0308}, 059 (2003)
[arXiv:hep-th/0306260].
%%CITATION = HEP-TH 0306260;%%
}

\lref\STi{S.~Stieberger and T.R.~Taylor,
``Non-Abelian Born-Infeld action and type I - heterotic duality  (I): 
Heterotic $F^6$ terms at two loops,''
Nucl.\ Phys.\ B {\bf 647}, 49 (2002);
[arXiv:hep-th/0207026].
%%CITATION = HEP-TH 0207026;%%
}
\lref\STii{S.~Stieberger and T.R.~Taylor,
``Non-Abelian Born-Infeld action and type I - heterotic duality (II):
Nonrenormalization theorems,''
Nucl.\ Phys.\ B {\bf 648}, 3 (2003)
[arXiv:hep-th/0209064].
%%CITATION = HEP-TH 0209064;%%
}

\lref\STiii{S.~Stieberger and T.R.~Taylor, work to appear.}

\lref\Myers{R.C.~Myers,
``Dielectric-branes,''
JHEP {\bf 9912}, 022 (1999)
[arXiv:hep-th/9910053].
%%CITATION = HEP-TH 9910053;%%
}

\lref\GREEN{
M.B.~Green and J.H.~Schwarz,
``Supersymmetrical Dual String Theory. 2. Vertices And Trees,''
Nucl.\ Phys.\ B {\bf 198}, 252 (1982);\br
%%CITATION = NUPHA,B198,252;%%}
J.H.~Schwarz,
``Superstring Theory,''
Phys.\ Rept.\  {\bf 89}, 223 (1982).
%%CITATION = PRPLC,89,223;%%
}

\lref\Bilal{A.~Bilal,
``Higher-derivative corrections to the non-abelian Born-Infeld action,''
Nucl.\ Phys.\ B {\bf 618}, 21 (2001)
[arXiv:hep-th/0106062].
%%CITATION = HEP-TH 0106062;%%
}

\lref\witten{
E.~Witten,
``String theory dynamics in various dimensions,''
Nucl.\ Phys.\ B {\bf 443} (1995) 85
[arXiv:hep-th/9503124];\br
J.~Polchinski and E.~Witten,
``Evidence for Heterotic - Type I String Duality,''
Nucl.\ Phys.\ B {\bf 460} (1996) 525
[arXiv:hep-th/9510169].
%%CITATION = HEP-TH 9510169;%%
}
\lref\review{For a review, see e.g.:
I.~Antoniadis, H.~Partouche and T.R.~Taylor,
``Lectures on heterotic-type I duality,''
Nucl.\ Phys.\ Proc.\ Suppl.\  {\bf 61A} (1998) 58
[arXiv:hep-th/9706211];\br
E.~Kiritsis,
``Introduction to non-perturbative string theory,''
arXiv:hep-th/9708130;\br
%%CITATION = HEP-TH 9708130;%%
A.~Sen,
``An introduction to non-perturbative string theory,''
arXiv:hep-th/9802051.
%%CITATION = HEP-TH 9802051;%%
}

\lref\AA{A.A. Tseytlin,
``On SO(32) heterotic - type I superstring duality in ten dimensions,''
Phys.\ Lett.\ B {\bf 367} (1996) 84
[arXiv:hep-th/9510173];
% A.~A.~Tseytlin,
``Heterotic - type I superstring duality and low-energy effective actions,''
Nucl.\ Phys.\ B {\bf 467} (1996) 383
[arXiv:hep-th/9512081].
%%CITATION = HEP-TH 9512081;%%
}

\lref\seealso{M.~de Roo and M.G.C.~Eenink,
``The effective action for the 4-point functions in abelian open  superstring theory,''
  JHEP {\bf 0308}, 036 (2003)
  [arXiv:hep-th/0307211];\br
  %%CITATION = HEP-TH 0307211;%%
A.~Collinucci, M.~De Roo and M.G.C.~Eenink,
  ``Supersymmetric Yang-Mills theory at $\ap^3$,''
  JHEP {\bf 0206}, 024 (2002)
  [arXiv:hep-th/0205150];
  %%CITATION = HEP-TH 0205150;%%
%A.~Collinucci, M.~de Roo and M.~G.~C.~Eenink,
``Derivative corrections in 10-dimensional super-Maxwell theory,''
  JHEP {\bf 0301}, 039 (2003)
  [arXiv:hep-th/0212012].
  %%CITATION = HEP-TH 0212012;%%
}

\lref\paris{C.~Bachas and E.~Kiritsis,
``$F^4$ terms in N = 4 string vacua,''
Nucl.\ Phys.\ Proc.\ Suppl.\  {\bf 55B}, 194 (1997)
[arXiv:hep-th/9611205];\br
%%CITATION = HEP-TH 9611205;%%
C. Bachas, C. Fabre, E.~Kiritsis, N.A. Obers and P. Vanhove,
``Heterotic/type-I duality and D-brane instantons,''
Nucl.\ Phys.\ B {\bf 509}, 33 (1998)
[arXiv:hep-th/9707126];\br
%%CITATION = HEP-TH 9707126;%%
E. Kiritsis and N.A. Obers,
``Heterotic/type-I duality in $D<10$ dimensions, threshold corrections  and 
D-instantons,''
JHEP {\bf 9710}, 004 (1997)
[arXiv:hep-th/9709058].
%%CITATION = HEP-TH 9709058;%%
}

\lref\LSWF{W. Lerche and S. Stieberger,
``Prepotential, mirror map and F-theory on K3,''
Adv.\ Theor.\ Math.\ Phys.\  {\bf 2}, 1105 (1998)
[Erratum-ibid.\  {\bf 3}, 1199 (1999)]
[arXiv:hep-th/9804176];\br
%%CITATION = HEP-TH 9804176;%%
W. Lerche, S. Stieberger and N.P. Warner,
``Quartic gauge couplings from K3 geometry,''
Adv.\ Theor.\ Math.\ Phys.\  {\bf 3}, 1575 (1999)
[arXiv:hep-th/9811228];
%%CITATION = HEP-TH 9811228;%%
``Prepotentials from symmetric products,''
Adv.\ Theor.\ Math.\ Phys.\  {\bf 3}, 1613 (1999)
[arXiv:hep-th/9901162];\br
%%CITATION = HEP-TH 9901162;%%
K. Foerger and S. Stieberger,
``Higher derivative couplings and heterotic-type I duality in eight  
dimensions,''
Nucl.\ Phys.\ B {\bf 559}, 277 (1999)
[arXiv:hep-th/9901020].
%%CITATION = HEP-TH 9901020;%%
}

\lref\LS{W.~Lerche and S.~Stieberger,
``1/4 BPS states and non-perturbative couplings in N = 4 string theories,''
Adv.\ Theor.\ Math.\ Phys.\  {\bf 3}, 1539 (1999)
[arXiv:hep-th/9907133].
%%CITATION = HEP-TH 9907133;%%
}

\lref\FT{E.S.~Fradkin and A.A.~Tseytlin,
``Effective Action Approach To Superstring Theory,''
Phys.\ Lett.\ B {\bf 160}, 69 (1985);
%%CITATION = PHLTA,B160,69;%%
% E.~S.~Fradkin and A.~A.~Tseytlin,
``Nonlinear Electrodynamics From Quantized Strings,''
Phys.\ Lett.\ B {\bf 163}, 123 (1985).
%%CITATION = PHLTA,B163,123;%%
}

\lref\abder{O.D.~Andreev and A.A.~Tseytlin,
 ``Partition Function Representation For The Open Superstring Effective Action:
Cancellation Of M\"obius Infinities And Derivative Corrections To Born-Infeld
Lagrangian,''
Nucl.\ Phys.\ B {\bf 311}, 205 (1988);\br
%%CITATION = NUPHA,B311,205;%%
A.A.~Tseytlin,
``Born-Infeld action, supersymmetry and string theory,''
arXiv:hep-th/9908105;\br
%%CITATION = HEP-TH 9908105;%%
N. Wyllard,
``Derivative corrections to D-brane actions with constant background  fields,''
Nucl.\ Phys.\ B {\bf 598}, 247 (2001)
[arXiv:hep-th/0008125].
%%CITATION = HEP-TH 0008125;%%
}

\lref\Kitz{Y.~Kitazawa,
``Effective Lagrangian For Open Superstring From Five Point Function,''
Nucl.\ Phys.\ B {\bf 289}, 599 (1987).
%%CITATION = NUPHA,B289,599;%%
}

\lref\Fisk{
M.D.~Fisk,
``Five Point Tree Amplitude For The Heterotic String,''
Nucl.\ Phys.\ B {\bf 320}, 377 (1989).
%%CITATION = NUPHA,B320,377;%%
}

\lref\GM{M.R.~Garousi and R.C.~Myers,
``World-volume potentials on D-branes,''
JHEP {\bf 0011}, 032 (2000)
[arXiv:hep-th/0010122].
%%CITATION = HEP-TH 0010122;%%
}

\lref\LMRS{D. L\"ust, P. Mayr, R. Richter and S. Stieberger,
``Scattering of gauge, matter, and moduli fields from intersecting branes,''
Nucl.\ Phys.\ B {\bf 696}, 205 (2004)
[arXiv:hep-th/0404134].
%%CITATION = HEP-TH 0404134;%%
}

\lref\TSE{
A.A.~Tseytlin,
``Vector Field Effective Action In The Open Superstring Theory,''
Nucl.\ Phys.\ B {\bf 276}, 391 (1986)
[Erratum-ibid.\ B {\bf 291}, 876 (1987)].
%%CITATION = NUPHA,B276,391;%%
}

\lref\TSEBI{
A.A.~Tseytlin,
``On non-abelian generalisation of the Born-Infeld action in string  theory,''
Nucl.\ Phys.\ B {\bf 501}, 41 (1997)
[arXiv:hep-th/9701125].
%%CITATION = HEP-TH 9701125;%%
}

\lref\nielsen{Z.~Koba and H.B.~Nielsen,
``Reaction Amplitude For N Mesons: A Generalization Of The
Veneziano-Bardakci-Ruegg-Virasora Model,''
Nucl.\ Phys.\ B {\bf 10}, 633 (1969);
%%CITATION = NUPHA,B10,633;%%
``Manifestly Crossing Invariant Parametrization Of N Meson Amplitude,''
Nucl.\ Phys.\ B {\bf 12}, 517 (1969).
%%CITATION = NUPHA,B12,517;%%
}

\lref\DP{E. D'Hoker and D.H.~Phong,
``Two-loop superstrings. VI: Non-renormalization theorems and the 4-point
function,''
arXiv:hep-th/0501197;
%%CITATION = HEP-TH 0501197;%%
``Two-loop superstrings. V: Gauge slice independence of the N-point
function,''
arXiv:hep-th/0501196.
%%CITATION = HEP-TH 0501196;%%
}

\lref\Sloan{D.J.~Gross and J.H.~Sloan,
``The Quartic Effective Action For The Heterotic String,''
Nucl.\ Phys.\ B {\bf 291}, 41 (1987).
%%CITATION = NUPHA,B291,41;%%
}

\lref\GW{D.J.~Gross and E.~Witten,
``Superstring Modifications Of Einstein's Equations,''
Nucl.\ Phys.\ B {\bf 277}, 1 (1986);\br
%%CITATION = NUPHA,B277,1;%%
H.~Kawai, D.C.~Lewellen and S.H.H.~Tye,
``A Relation Between Tree Amplitudes Of Closed And Open Strings,''
Nucl.\ Phys.\ B {\bf 269}, 1 (1986).
%%CITATION = NUPHA,B269,1;%%
}

\lref\Mangano{
M.L.~Mangano and S.J.~Parke,
``Multiparton Amplitudes In Gauge Theories,''
Phys.\ Rept.\  {\bf 200}, 301 (1991);\br
%%CITATION = PRPLC,200,301;%%
M.L.~Mangano, S.J.~Parke and Z.~Xu,
``Duality And Multi - Gluon Scattering,''
Nucl.\ Phys.\ B {\bf 298}, 653 (1988).
%%CITATION = NUPHA,B298,653;%%
}

\lref\PT{S.J.~Parke and T.R.~Taylor,
``An Amplitude For N Gluon Scattering,''
Phys.\ Rev.\ Lett.\  {\bf 56}, 2459 (1986).
%%CITATION = PRLTA,56,2459;%%
}

\lref\progress{Work in progress.}
\lref\MHV{S. Stieberger and T.R. Taylor, work to appear.}

\lref\berko{N.~Berkovits,
``Multiloop amplitudes and vanishing theorems using the pure spinor formalism
for the superstring,''
JHEP {\bf 0409}, 047 (2004)
[arXiv:hep-th/0406055];
%%CITATION = HEP-TH 0406055;%%
``Covariant multiloop superstring amplitudes,''
arXiv:hep-th/0410079.
%%CITATION = HEP-TH 0410079;%%
}

\lref\HT{A.~Hashimoto and W.I.~Taylor,
``Fluctuation spectra of tilted and intersecting D-branes from the  Born-Infeld
action,''
Nucl.\ Phys.\ B {\bf 503}, 193 (1997)
[arXiv:hep-th/9703217].
%%CITATION = HEP-TH 9703217;%%
}

\lref\TZ{W.~Taylor,
``Lectures on D-branes, tachyon condensation, and string field theory,''
arXiv:hep-th/0301094;\br
%%CITATION = HEP-TH 0301094;%%
W.~Taylor and B.~Zwiebach,
``D-branes, tachyons, and string field theory,''
arXiv:hep-th/0311017.
%%CITATION = HEP-TH 0311017;%%
}

\lref\Brazil{
R.~Medina, F.T.~Brandt and F.R.~Machado,
``The open superstring 5-point amplitude revisited,''
JHEP {\bf 0207}, 071 (2002)
[arXiv:hep-th/0208121].
%%CITATION = HEP-TH 0208121;%%
}

\lref\bailey{W.N. Bailey, 
{\it Generalised Hypergeometric Series}, 
Cambridge, England: Cambridge University Press, 1935.}

\lref\Grad{
I.S. Gradshteyn and I.M. Ryzhik,
{\it Table of Integrals, Series and Products}, Academic Press 1994.}

\lref\mathworld{E.W. Weisstein, {\it Harmonic Number}, From MathWorld--A Wolfram Web Resource. 
http://mathworld.wolfram.com/HarmonicNumber.html.}

\lref\borwein{D. Borwein, and J.M. Borwein, 
{\it On an Intriguing Integral and Some Series Related to $\zeta(4)$}, 
Proc. Amer. Math. Soc. 123, 1191 (1995).}

\lref\BaileyBorwein{D.H. Bailey, J.M. Borwein, and R. Girgensohn, 
{\it Experimental evaluation of Euler sums,} Experiment. Math. 3 (1994), 17-30.}

\lref\Girgensohn{J.M. Borwein and R. Girgensohn, 
{\it Evaluation of triple Euler sums}, with appendix {\it Euler sums in quantum field theory} 
by D. J. Broadhurst, Electronic J. Combinatorics 3 (1996), R23.}

\lref\BBG{D. Borwein, J.M. Borwein, and R. Girgensohn, {\it Explicit evaluation of Euler sums,}
Proc. Edinburgh Math. Soc. 38 (1995), 277-294.}

\lref\appell{P. Appell, and J. Kamp\'e de F\'eriet, {\it Fonctions
hyperg\'eom\'etriques et hypersph\'eriques: polynomes d'Hermite}, Paris: Gauthier-Villars, 1926.}

\lref\PARIS{P. Appell, {\it Sur le fonctions hyperg\'eom\'etriques de
plusieurs variables}, Paris: Gauthier-Villars, 1925;\br
J. Kamp\'e de F\'eriet, {\it La fonction hyperg\'eom\'etrique},  Paris: Gauthier-Villars, 1937;\br
H. Exton, {\it Multiple Hypergeometric Functions and Applications},
Chichester, England: Ellis Horwood, 1976;\br
H.M. Srivastava, and P.W. Karlsson,  {\it Multiple Gaussian Hypergeometric
Series}, Chichester, England: Ellis Horwood, 1985.}

\lref\slater{L.J. Slater, {\it 
Generalized Hypergeometric Functions,} Cambridge University Press, 1966.}

\lref\noerlund{N.E. N\"orlund, {\it Vorlesungen \"uber Differenzenrechnung}, Berlin: 
Springer, 1924.}

\lref\siegel{C.L. Siegel, {\it Symplectic Geometry}, Am. Jour. Math.  65 (1943), 1-86.}

\Title{
\vbox{
\rightline{LMU--ASC 07/05,\ \  MPP--2005--12}
\rightline{\tt hep-th/0509042}  }} 
{\vbox{\vskip -3cm\centerline{Six Gluon Open Superstring  Disk Amplitude,}
\bigskip\centerline{Multiple Hypergeometric Series and Euler--Zagier Sums}}}
%\smallskip
\centerline{D. Oprisa $^{a}$\ \ and\ \ S. Stieberger $^b$}
\medskip
\centerline{\it $^a$ Max--Planck--Institut f\"ur Physik,}
\centerline{\it F\"ohringer Ring 6, 80805 M\"unchen, Germany}
\vskip5pt
\centerline{\it $^b$ Arnold--Sommerfeld--Center for Theoretical Physics,}
\centerline{\it  Department f\"ur Physik, Ludwig--Maximilians--Universit\"at M\"unchen,}
\centerline{\it Theresienstra\ss e 37, 80333 M\"unchen, Germany}
% \medskip
\bigskip
%\bigskip
\centerline{\bf Abstract}
%\vskip .1in
\noindent

The six gluon disk amplitude is calculated in superstring theory. This 
amplitude probes the gauge interactions with six external legs 
on $Dp$--branes, in particular including \eg $F^6$--terms.
The full string $S$--matrix can be expressed by six generalized multiple hypergeometric
functions (triple hypergeometric functions), 
which in the effective action play an important r\^ole in arranging 
the higher order $\ap$ gauge interaction terms 
with six external legs (like $F^6, D^4F^4, D^2 F^5, D^6F^4, D^2F^6,\ldots$).

A systematic and efficient method is found to calculate tree--level string amplitudes
by equating seemingly different expressions for one and the same string $S$--matrix:
Comparable to Riemann identities appearing in string--loop 
calculations, we find an intriguing way of using world--sheet supersymmetry
to generate a system of non--trivial equations for string tree--level amplitudes.
These equations result in algebraic identities between different multiple  
hypergeometric functions. Their (six--dimensional) 
solution gives the ingredients of the string $S$--matrix.
We derive material relevant for any open string six--point
scattering process: relations between triple hypergeometric functions, their 
integral representations and their $\ap$--
(momentum)--expansions given by (generalized) Euler--Zagier sums or (related) 
Witten zeta--functions.

\Date{}
\noindent

\goodbreak
\listtoc 
\writetoc

\break
\newsec{Introduction}
\def\ss#1{{\scriptstyle{#1}}}

Due to the non--locality of string theory or the presence of infinite many massive
string states the world--volume effective action describing the dynamics of massless fields of a 
$Dp$--brane is organized as an infinite power series in the string tension $\ap$.
A famous example for such a series is represented by 
the tree--level gauge sector of a $Dp$--brane.
For $F$ the gauge field strength and $D$ the gauge covariant 
derivative\foot{$F$ and $D$ are in the adjoint representation of the gauge group.
Some group theoretical facts are summarized in
appendix \appD.}, 
this series may be formally written:
\eqn\Formally{
\Lc^{Dp}_{\rm effective}=\Tr\ \sum_{m\geq 2\atop n\geq 0}\hskip-0.6mm ^{'} 
\ap^{\h n+m-2}\ D^n F^m\ .}
Of course, this series contains a lot of redundant terms, which may be eventually eliminated
through applying Bianchi identities, equations of motions, partial integrations and
field redefinitions. Moreover, due to the relation
\eqn\relation{
[D_\mu,D_\nu]\ F_{\rho\sigma}=-i\ [F_{\mu\nu},\ F_{\rho\sigma}]}
certain combinations of derivative terms may be converted into non--derivative terms
and vice versa. 
To this end, the prime at the sum indicates, that the sum \Formally\  
should be evaluated over only all independent invariants.
In the supersymmetric case the latter are supposed to represent 
supersymmetric invariants. Furthermore, according to the definition of the effective
action, each term in the series \Formally\ accounts for a gluon contact interaction represented by 
an irreducible Feynman diagram of the underlying (higher derivative) gauge theory.
Whether the series converges or may be even written as an analytic function in $F$ and $DF$
is of burning interest. 
To determine the expansion \Formally\ for a given string background, in that background 
one calculates gauge boson string tree--level scattering amplitudes (their string $S$--matrices)
and their $\ap$--expansion 
matches certain terms of \eqq \Formally\ (after proper subtraction of reducible diagrams).
The fact, that \eqq \relation\ allows us to write terms in 
the effective action in various different ways, which are all equivalent from the 
string $S$--matrix point of view, is related to the fact, that a string $S$--matrix
does not feel those redefinitions (on--shell).
The lowest order terms of \Formally, \ie the $\ap^0$--order, describes just the ordinary 
Non--Abelian Yang--Mills
theory, \ie the interaction term $\Tr F^2$. While the next leading order $\ap$ is absent
in the superstring, the $\ap^2$--order has been completely matched to pieces of a four--gluon 
open superstring scattering $S$--matrix.
To this end, up to the order $\ap^2$, the series \Formally\ looks as follows 
\threeref\GREEN\GW\TSE
\eqn\super{\eqalign{
\Lc_{\rm effective}^{Dp}&
=\Tr\lf\{\ F_{mn}^2-\fc{1}{3}\ (2\pi\ap)^2\ \lf(\ F_{ab} F_{bd} F_{ca} F_{dc}+\h\ 
F_{ab} F_{bc} F_{cd} F_{da}\ri.\ri.\cr
&\lf.\lf.-\fc{1}{4}\ F_{ab}  F_{ba} F_{cd} F_{dc}-
\fc{1}{8}\ F_{ab}F_{cd} F_{ba} F_{dc}\ \ri)+\Oc(\ap^3)\ \ri\}\ .}}
Up to this $\ap$--order derivative terms ($n\neq 0$) 
may be eliminated as a result of field redefinitions \TSE.
The series \Formally\ is different for the bosonic and superstring. In this 
article we shall only
discuss the more interesting superstring case. Furthermore, 
it drastically simplifies in the case of an Abelian gauge group, 
\ie a single $Dp$--brane, due to the absence of a huge set of gluon interactions. 
Up to the order $\ap^2$, displayed \eqq \super, the terms may be derived from 
the so--called (generalized) Born--Infeld action \TSEBI
\eqn\BORN{
\Lc=\STr \sqrt{\det(\delta_{mn}+2\pi\ap\ F_{mn})}
=\STr\sum_{\nu_i\geq 0} c_{\nu_1\nu_2\ldots\nu_n}\ 
(\tr F^2)^{\nu_1}\ (\tr F^4)^{\nu_2}\ldots(\tr F^{2n})^{\nu_n}\ ,}
with the coefficients
\eqn\coeffs{
c_{\nu_1\nu_2\ldots\nu_n}=(2\pi\ap)^{2\nu_1+4\nu_2+\ldots+2n\nu_n}\ 
\lf(-\fc{1}{4}\ri)^{\sum_{i}\nu_i}\ 
\fc{1}{\nu_1!\cdot\ldots\cdot\nu_n!}\ 
\fc{1}{2^{\nu_2}3^{\nu_3}\cdot\ldots\cdot n^{\nu_n}}}
after expansion to fourth order in $F$ and applying the rule of the symmetrized gauge 
trace $\rm STr$.
The latter ``averages'' over all group structures
\eqn\democ{
\STr(A_1\ldots A_n):=\fc{1}{n!}\ \Tr(A_{(1}\ldots A_{n)})\ ,} 
with all permutations of gauge fields taken into account.

However, the series \Formally\ departs already at the order $\ap^3$ from this nice 
and closed form \BORN, due to the presence of $F^5$ and $D^2 F^4$--terms. 
Those terms appear with a $\zeta(3)$ factor in the effective action.
While the $D^2 F^4$ can be fixed through expanding the four gluon string 
$S$--matrix up to order $\ap^3$ \Bilal, the pure $F^5$ terms can only be 
fixed by computing a five gluon string $S$--matrix \Brazil.
In fact, there are several important works, which investigate 
the structure of the higher order $\ap$ gauge interaction terms in the string effective 
action.
However, only $F^4, D^{2n} F^4$ and $F^5, D^{2n} F^5$ terms have been thoroughly 
investigated on the basis of the four-- and five--gluon string $S$--matrix results, respectively
\doubref\BRAZIL\seealso, since only those two amplitudes have been calculated \doubref\GREEN\Brazil. 
However, to probe the full $\ap^4$ order of \Formally, one has to compute six--point 
gluon scattering amplitudes. This amplitude comprises the necessary 
information to obtain the structure of the $F^6$ and also $D^{2n} F^6$--terms.
\vskip0.5cm
{\vbox{\ninepoint{
$$
\vbox{\offinterlineskip\tabskip=0pt
\halign{\strut\vrule#
%%%%%%%%%%%%%%%%%%
&~$#$~\hfil
&\vrule#
&~$#$~\hfil
&~$#$~\hfil
&~$#$~\hfil
&~$#$~\hfil
&~$#$~\hfil
&~$#$~\hfil
&~$#$~\hfil
&~$#$~\hfil
&\vrule#&\vrule#
\cr
%%%%%%%%%%%%%%%%%%
\noalign{\hrule}
&
\ap^0\ \ 1
&&&
{\bf F^2}
&&
\ 
&&
\ 
&&
\ 
&
\cr
%%%%%%%%%%%%%%%%%%
% \noalign{\hrule}
&
\ap^1\ \ 0
&&&
F^3
&&
D^2 F^2
&&
\ 
&&
\ 
&
\cr
%%%%%%%%%%%%%%%%%%
% \noalign{\hrule}
&
\ap^2\ \zeta(2)
&&&
{\bf F^4}
&&
D^2 F^3
&&
D^4 F^2 
&&
\ 
&
\cr
%%%%%%%%%%%%%%%%%%
% \noalign{\hrule}
&
\ap^3\ \zeta(3)
&&&
{\bf F^5}
&&
{\bf D^2 F^4}
&&
D^6 F^2
&&
\ 
&
\cr
%%%%%%%%%%%%%%%%%%
%\noalign{\hrule}
&
\ap^4\ \zeta(4)
&&&
{\bf F^6}
&&
{\bf D^4 F^4} 
&&
{\bf D^2 F^5}
&&
\ 
&
\cr
%%%%%%%%%%%%%%%%%%
%%%%%%%%%%%%%%%%%%
%\noalign{\hrule}
&
\ap^5\ \zeta(5)
&&&
{\bf F^7}
&&
{\bf D^6 F^4} 
&&
{\bf D^4 F^5}
&&
{\bf D^2 F^6} 
&
\cr
%%%%%%%%%%%%%%%%%%
%\noalign{\hrule}
&
\ \ \ \ \ \vdots
&&&
\ldots
&&
\ldots 
&&
\ldots
&&
\ldots\ldots
&
\cr
%%%%%%%%%%%%%%%%%%
\noalign{\hrule}}}$$
\vskip-10pt
\centerline{\noindent{\bf Table 1:}
{\sl Higher order $F$--terms appearing at a}}
\centerline{\sl  given $\ap$--order in the supersymmetric $D$--brane action \eqq \Formally.}
\vskip10pt}}}\ \br
The expected higher order $\ap$ gauge interaction terms are
displayed in Table 1 up to $\ap^5$. The terms in bold are the interactions, which contribute
to the series \Formally, while the others may be eliminated (\cf the comment below \Formally).
Furthermore, we show the respective zeta--function coefficients of those terms in the effective
action.

Computing the six gluon string $S$--matrix in superstring theory 
is the main part of the present article. More concretely, we calculate
the superstring tree--level $S$--matrix with six external gauge bosons. The latter are 
represented by open strings, hence the string world--sheet topology of this scattering
process is a disk.
Furthermore, we present the relevant material, which is needed for any open string six--point
scattering process involving six open strings: the 
integral representations of the triple hypergeometric function, various identities
between multiple hypergeometric functions  and Euler--Zagier sums.
Calculating tree--level string $S$--matrices with more than five external states is certainly quite
challenging from the technical point of view.
In fact, so far only string tree--level amplitudes involving five external states 
have been elaborated. More precisely\foot{From the technical point of view, five open string 
tree--level amplitudes are related to \eg  three open and one closed string disk amplitudes \GM.
Note, that in {\it Ref.} \LMRS\ a specific tree--level string amplitude with two open 
and two closed strings has been calculated. On the disk, this amplitude is related to
an open string six--point function of a very special kind, which has made tractable the problem.}, 
five gluon tree--level amplitudes have been determined
in the heterotic string in {\it Ref.} \Fisk, while in the type $I$ superstring in {\it Refs.}
\doubref\Kitz\Brazil. 
Hence, it is appropriate to state, that this is the first work, where the problem of 
tree--level string amplitudes involving six external states has been addressed.

We find a very powerful trick, to tackle the problem of calculating $N$--point tree--level 
string $S$--matrices in the superstring. 
More precisely, in order to guarantee a total ghost charge of $-2$ on the disk, two vertex 
operators of the $N$ gluons have to be chosen in the $(-1)$--ghost picture, while the 
remaining $N-2$ vertices are put into the zero--ghost picture. 
However, we are completely free, which pair of vertex operators we choose to put into the 
$(-1)$--ghost picture. There are $\lf(N\atop 2\ri)=\h(N-1)N$
possibilities, which all have to lead to the same result.
For a given space--time kinematics each of these choices leads to a seemingly 
different expression. Quite generically, such an expression is given by some multiple 
hypergeometric functions, multiplied with some polynomials in the kinematic invariants 
and some $\Gamma$--functions depending on the latter. 
All these expressions have to be equated. This gives a set of non--trivial relations
between different hypergeometric functions, which allows us to express many of them
in terms of just a few. These identities represent algebraic equations between different multiple  
hypergeometric functions. Their solution gives the ingredients of the string $S$--matrix.
Many of the identities between hypergeometric functions we obtain by the above procedure
cannot be found in the literature, see \eg \bailey. 
Hence, from the mathematical point of view identities between different
hypergeometric functions may be inferred from higher point string tree--level amplitudes
as a result of supersymmetry.
We shall point out, that the above described symmetry, following from world--sheet supersymmetry
should not be confused with the cyclic invariance of a gluon string $S$--matrix.
Applying this symmetry on a given amplitude would also lead to 
seemingly different expressions for one and the same kinematic structure. 
Equating the latter would also give a system of equations.
However, this system is much smaller, in fact too small, to write down the full
string $S$--matrix in terms of a few functions, which in contrast is possible using the above
described trick using world--sheet supersymmetry.

There is a striking relation between multiple Gaussian hypergeometric functions or
their integral representations, which
are the relevant objects for tree--level string calculations and zeta--functions
of various kinds, which play a key role in modern number theory.
A simple example for this relation may be given by the following integral ($\ap s>-1$):
\eqn\maylead{
\int_0^1 dx\int_0^1 dy\ (1-xy)^{\ap s-1}=\fc{H_{\ap s}}{\ap s}=\zeta(2)-\ap s\ \zeta(3)+
(\ap s)^2\ \zeta(4)-(\ap s)^3\ \zeta(5)+\ldots\ .}
This integral may calculate the $\ap$--expansion for a certain kinematics of a 
five--point string amplitude with $s$ being some kinematic invariant.
Besides, $H_n$ is the harmonic number (\cf section 4 for more details).
Similarly, one may derive
\eqn\Maylead{\eqalign{
\int_0^1 dx\int_0^1 dy\int_0^1 dz\ (1-xyz)^{\ap s-2}&=\zeta(2)+
\ap s\ [\ \zeta(2)-2\ \zeta(3)\ ]\cr
&+(\ap s)^2\ [\ \zeta(2)-2\ \zeta(3)+\fc{5}{4}\ \zeta(4)\ ]+\ldots}}
accounting for a certain (simplified) six--point scattering process\foot{
Due to the nine kinematic invariants the integrands, 
which appear in any six--point scattering process on the disk, are combinations of the 
nine polynomials: 
\eqn\ninepolynomials{
x\ \ ,\ \ (1-x)\ \ ,\ \ y\ \ ,\ \ (1-y)\ \ ,\ \ z\ \ ,\ \ (1-z)\ \ ,\ \ (1-xy)\ \ ,\ \ (1-yz)\ \ ,
\ \ (1-xyz)\ .}
Their integrals may be expressed through the triple hypergeometric function $F^{(3)}$ (\cf section 4).}.
In the above two examples the basic Riemann zeta--functions appear. The latter are given
by the following integer sum:
\eqn\basicRiemann{
\zeta(s)=\sum_{n=1}^\infty n^{-s}\ \ \ ,\ \ \ s\geq 2  .}
Generalization of them are the multiple zeta values, {\it e.g.:}
\eqn\multipleZETA{
\zeta(s_1,s_2,s_3)=\sum_{m_i=1 \atop m_3<m_2<m_1}^\infty \fc{1}{m_1^{s_1}m_2^{s_2}m_3^{s_3}}\ \ \ 
,\ \ \ s_1\geq 2\ ,\ s_2,s_3\geq 1\ ,}
which appear in the integral:
\eqn\Mayleadd{
\int_0^1 dx\int_0^1 dy\int_0^1 dz\ \fc{(1-xy)^{\ap s}}{(1-xyz)^2}=\zeta(2)-\ap s\ \zeta(3)+
(s\ap)^2\ \zeta(2,1,1)+\ldots\ ,}
with $\zeta(2,1,1)=\zeta(4)$.
A class of integrals, similar to \Maylead\ and \Mayleadd\ may be evaluated by
calculating the relevant multiple zeta values or related Euler sums
(\cf section 4 for more details). However, the $\ap$--expansion of 
the full disk six--point string $S$--matrix, which is 
eventually expressed through some triple hypergeometric functions, generically 
leads to new types of zeta--functions. The latter are multiple sums over integers, 
which in most of the cases we have to evaluate case by case. One simplified example is the 
following type:
\eqn\simplify{\eqalign{
\int_0^1 dx\int_0^1 dy\int_0^1 dz\ \fc{(1-xy)^{\ap s}\ (1-yz)^{\ap t}}{(1-xyz)^{2}}&=\zeta(2)-
\ap\ (s+t)\ \zeta(3)\cr
&+\ap^2\ [\ s^2+t^2+\fc{7}{4}\ st\ ]\ \zeta(4)+\ldots\ .}}
Extracting from the above integral the second order in $\ap$ requires to evaluate 
the following triple sum
\eqn\followtriple{
\sum_{m_i=1}^\infty\fc{m_3}{m_1\ m_2\ (m_1+m_3)\ (m_2+m_3)\ (m_1+m_2+m_3)}=\fc{7}{4}\ \zeta(4)}
among others. The evaluation of those sums, which is the topic of subsection 4.5 is closely
related to the subject of Tornheim double series, Witten zeta--functions 
and multiple Euler/Zagier sums. 
Those sums generically may not be simply expressed by the Riemann zeta--function.
An example of a Witten zeta--function is
\eqn\WittenZETA{
W(a,b,c)=\sum_{m,n=1}^\infty\fc{1}{m^a\ n^b\ (m+n)^c}\ ,}
which also occurs after expanding the triple hypergeometric function w.r.t. $\ap$. 

Just as special values of the Riemann zeta--functions \basicRiemann\ 
are related to the Euler characteristics of moduli spaces $\Mc_g$ of Riemann surfaces of 
genus $g$, \ie $\chi(\Mc_g)\sim\zeta(2g),\ g\geq 2$,
special values of the encountered new types of zeta--functions calculate the volume
of certain moduli spaces of vector bundles of curves. Hence the $\ap$--expansion of 
our six gluon disk amplitude may perhaps be related to some geometric setup 
described  by the relation between the  triple hypergeometric function and 
new types of zeta--functions.

While computing string amplitudes with external gluons allows to determine
\Formally\ directly by means of extracting the relevant parts of the effective action
from the string $S$--matrix amplitude there have been presented some indirect approaches 
to tackle this problem: By
matching the open string spectrum with a spectrum following from a higher order Yang--Mills
action one obtains information about the higher order terms in \Formally, see {\it Refs.}
\doubref\HT\Alex\ and {\it Refs.} \TZ\ for a review.
More recently, an other approach has been suggested \sevrin: Requiring the existence of certain 
$BPS$ solutions to the equations of motions allowed 
to propose the $\ap^4$--part of \Formally.
Since all those approaches are indirect, in particular based on field--theory, 
it is certainly important 
to check these findings against the results derived from the string amplitudes \progress.
Moreover, the various transcendental zeta values $\zeta(3), \zeta(5),\zeta(3)^2,\ldots,$
which appear in the $\ap$--expansion
of the six gluon string $S$--matrix, are certainly hard to obtain in any field--theory approach.

Our results may be summarized under the following three points:
$${\eqalign{
&\bullet\ \  {\rm An\ efficient\ method\ to\ calculate\ supersymmetric\ }N
{\rm -point\ tree-level\ string}\cr
&\hskip0.65cm{\rm amplitudes\ is\ presented.}\cr
&\bullet\ \  {\rm The\ six\ gluon\ open\ superstring\ disk\ amplitude\ can\ be\ expressed\ 
through\ a\ basis}\cr
&\hskip0.65cm{\rm of\ six\ triple\ hypergeometic\ 
functions,\ which\ encode\ the\ full}\ \ap- {\rm dependence.}\cr
&\bullet\ \  {\rm Material\ to\ obtain\ the\ } \ap- {\rm expansion\ of\ these\ functions\ is\
derived:\ We\ calculate}\cr
&\hskip0.65cm{\rm many\ multiple\ Euler-Zagier\ sums\ including\  multiple\ harmonic\ series.}}}$$

%\break

\newsec{Tree--level scattering of open superstrings on $Dp$--branes}

\subsec{Open superstring gluon $S$--matrix}

In the following we study the gauge dynamics on the $Dp$--brane world volume
with sixteen supercharges.
Our discussion holds for both type $I$ or type $IIA/IIB$ superstring theory.
The gauge interactions are described by an effective $D$--brane action of the sort \Formally, 
which is derived from open string scattering amplitudes. 
The $\ap$--expansion of the latter reproduces
certain interaction terms of the effective action.
The massless open string states of a $Dp$--brane are the gauge vectors $A^\mu$, which are
the longitudinal massless excitations of the $NS$--ground state and the gauge fermions $\psi^\mu$
originating from the $R$--ground state, with $\mu=0,\ldots,p$. 
The $Dp$--brane fields $\psi^\mu$ and $A^\mu$ obey Neumann boundary conditions. In addition,
there are the fields with transverse index $\mu=p+1,\ldots,9$, 
which obey Dirichlet boundary conditions.
Here, we shall not discuss their contribution to the $D$--brane action. Their effects may
be deduced from $T$--duality \Myers. 
Hence, without loss of generality, we may focus on an unwrapped
$D9$--brane, \ie $p=9$. The latter carries $16$ supercharges.
Hence, for concreteness we perform our calculation in an uncompactified
$D=10$ type $I$ superstring background with $SO(32)$ gauge group.
This is no restriction on the dimension nor on the gauge group of the $Dp$--brane. It only means, 
that no transversal oscillations of the branes are allowed and the fields $A^\mu,\psi^\mu$
with  $\mu=0,\ldots,9$, live on the $D9$--brane world--volume.
Qualitative different results are only to be expected for $D3$--branes due to the reduced number of
kinematic invariants in four space--time dimensions (\cf comments following later).

A string $S$--matrix $\Ac_N$ of $N$ external gluons with momenta $k_i$, 
polarizations $\xi_i$ and color indices $a_i$ may be formally arranged 
as the sum:
\eqn\formally{
\Ac_N(k_1,\xi_1,a_1;\ldots;k_N,\xi_N,a_N)
=\sum_{\pi\ \in\ \overline{(1,2,\ldots,N)}\atop (N-1)!\ \ {\rm permutations}\ \pi}   
\Tr(\lambda^{\pi(1)}\lambda^{\pi(2)}\ldots\lambda^{\pi(N)})\ A^\pi\ .}
The sum runs over all $(N-1)!$ cyclically inequivalent permutations 
$(\pi(1),\pi(2),\ldots,\pi(N))$ of the $N$ gluons, 
labeled by  the $N$ integers $i$, with $i=1,\ldots,N$, \ie $\pi\in S_N/\IZ_N$.
The matrices $\lambda^a$ are the Chan--Paton factors describing
the gauge degrees of freedom of the open string end points. Hence, they are in
the adjoint representation of the gauge group under consideration.
Throughout the whole article $\Tr$ denotes the group trace over the
Chan--Paton factors $\lambda^a$ in the adjoint representation. On the other
hand, the trace $\tr$ accounts for the trace over the Lorentz--indices ($\mu,\nu$)
of the gauge field strength $F_{\mu\nu}$.

In the string $S$--matrix the external gauge bosons are asymptotic states
incoming and outgoing at $t=\pm \infty$. For any open string scattering the string 
world--sheet may be conformally mapped to a Riemann surface with boundaries.
The external states are created through vertex operators inserted at the boundary. 
For open string tree--level scattering the world--sheet topology becomes 
a disk. The latter may be conformally mapped to the upper half plane $\im z\geq 0$ 
with the real axis as its boundary. Hence, all vertex positions $z_i$ are located
along the real axis, \ie $z_i\in \IR$. 
The gauge boson vertex operator is given by
\eqn\vertex{
V_{A^a}^{(-1)}(z,k)=\lambda^a\ \xi_\mu\ e^{-\phi(z)}\ \psi^\mu(z)\ e^{ik_\rho X^\rho(z)}}
in the $(-1)$--ghost picture, and by
\eqn\vertexi{
V_{A^a}^{(0)}(z,k)=\lambda^a\ \xi_\mu\ [\ \p X^\mu(z)+i\ (k\psi)\ \psi^\mu(z)\ ]\
e^{ik_\rho X^\rho(z)}}
in the zero--ghost picture.
Here, $\xi^\mu$ and $k^\mu$ are the polarization vector and space--time
momentum of the gauge boson, respectively. The latter obey
the on--shell constraints $\sum\limits_{\mu=0}^p\xi^\mu k^\mu=0$ and $k^2=0$. 

The string $S$--matrix \formally\ becomes ($N\geq 3$):
\eqn\START{\eqalign{
&\Ac_N(k_1,\xi_1,a_1;\ldots;k_N,\xi_N,a_N)\cr
&\hskip1cm=V_{\rm CKG}^{-1}\ \prod_{r=1}^N \int d^2 z_r\
\vev{V_{A^{a_1}}^{(-1)}(z_1)\ V_{A^{a_2}}^{(-1)}(z_2)\ V_{A^{a_3}}^{(0)}(z_3)\ldots\
V_{A^{a_N}}^{(0)}(z_N)}\ .}}
Here, $V_{\rm CKG}$ is the volume of the conformal Killing group $PSL(2,\IR)$, 
which leaves the boundary ($\im(z)=0$) of the disk fixed.
Two vertex operators have to be chosen in the $(-1)$--ghost picture \vertex\ 
and the remaining $N-2$ in the zero--ghost picture, in order to guarantee a total ghost 
charge of $-2$ on the disk. This requirement is a consequence of the superdiffeomorphism 
invariance on the string world sheet.
On the disk the conformal Killing volume $V_{\rm CKG}^{-1}$ is cancelled by fixing three
positions and introducing the respective $c$--ghost correlator.
We have total momentum conservation $\sum\limits_{i=1}^N k_i=0$ for the $N$ 
external momenta $k_i$.

Essentially to calculate the string $S$--matrix $\Ac_N$, the main work consists 
of determining, for a given group ordering $\pi$, the function $A^\pi$ contributing 
to \formally.
The latter captures the full space--time kinematics to all orders in $\ap$.
In \eqq \START\ the contribution of a given group contraction 
$\Tr(\lambda^{\pi(1)}\lambda^{\pi(2)}\ldots\lambda^{\pi(N)})$,
described by  a permutation $\pi$ of the  $N$ indices $i$, 
is given by specifying the  integration region $\Ic_\pi$ 
\eqn\region{
\Ic_\pi=\{\ \im(z_j)=0\ |\  z_{\pi(1)}<z_{\pi(2)}<\ldots<z_{\pi(N)}\ \}}
along the boundary of the disk. Hence we may write:
\eqn\specific{\eqalign{
\Tr(\lambda^{\pi(1)}\lambda^{\pi(2)}\ldots\lambda^{\pi(N)})&\ A^\pi\cr
&\hskip-1cm=V_{\rm CKG}^{-1}\ \int_{\Ic_\pi} \ \prod_{j=1}^N d^2 z_j\
\vev{V_{A^{a_1}}^{(-1)}(z_1)\ V_{A^{a_2}}^{(-1)}(z_2)\ V_{A^{a_3}}^{(0)}(z_3)\ldots\
V_{A^{a_N}}^{(0)}(z_N)}\ .}}
For the moment, in \eqq \START\  we have
chosen the first and second gauge boson vertex operator in the $(-1)$--ghost picture, while
the others are in the zero--ghost picture. 
To clearly mark this, in the following we shall write for \specific:
\eqn\express{
A^\pi\equiv A^\pi(1,2,3,\ldots,N)\ .}
Later we shall relax this special choice.
Hence, in the following, the argument of  $A^\pi(a,b,i_1,i_2,\ldots,i_{N-2})$ should indicate, 
that the pair $(a,b)$ of gauge vertex operators has been chosen to be in the 
$(-1)$--ghost picture, while the others $i_k$ are taken in the zero--ghost picture.
We should point out, that this ordering is independent of the group structure and we could 
equally calculate all group contributions $A^\pi$ by one and the same choice, \eg with
the first and second gauge vertex in the $(-1)$--ghost picture \express.
However, as we shall see in a moment, relaxing this choice will lead to non--trivial relations
between different expressions for one and the same group structure $\pi$.

The correlator in \START\ or \specific\ may be evaluated by performing Wick contractions. 
In effect the correlator decomposes into products of various two--point
functions. The latter are basic on the disk and given by:
\eqn\First{\eqalign{
\vev{X^\mu(z_1) X^\nu(z_2) }        &=-g^{\mu\nu}\ \ln(z_1-z_2)\ ,\cr
\vev{ \psi^\mu(z_1) \psi^\nu(z_2) } &=-\frac{g^{\mu\nu}}{z_1-z_2}\ ,\cr
\vev{ e^{ik_1 X(z_1)}e^{ik_2 X(z_2)}}   &= |\ z_1-z_2\ |^{k_1 k_2}\ ,\cr
\vev{ e^{-\phi(z_1)}e^{-\phi(z_2)}}   &= \frac{1}{z_1-z_2}\ .}}
Throughout the whole article we shall work with the convention $\ap=\h$.

\subsec{Four gluon open superstring $S$--matrix revisted}

The open string disk scattering of four gauge bosons is summarized in the string $S$--matrix:
\eqn\studyz{
\Ac_4(k_1,\xi_1,a_1;k_2,\xi_2,a_2;k_3,\xi_3,a_3;k_4,\xi_4,a_4)=
\sum_{\pi\ \in\ \overline{(1,2,3,4)} \atop 6\ \ {\rm permutations}\ \pi} 
\Tr(\lambda^{\pi(1)}\lambda^{\pi(2)}\lambda^{\pi(3)}\lambda^{\pi(4)})\ A^\pi\ .}
The sum runs over all six cyclically inequivalent permutations $\pi$ of the four 
external gluons labeled by $1,2,3$ and $4$.
For the group structure $\Tr(\lambda^1\lambda^2\lambda^4\lambda^3)$, \ie $\pi=(1,2,4,3)$,
the function $A^\pi$ is given by \doubref\GREEN\GW
\eqn\wehave{
A^{(1,2,4,3)}=\fc{\Ga(s)\ \Ga(t)}{\Ga(1+s+t)}\ \lf[\ tu\ (\xi_1 \xi_2)\ (\xi_3 \xi_4)+
su\ (\xi_1 \xi_3)\ (\xi_2 \xi_4)+st\ (\xi_1 \xi_4)\ (\xi_2 \xi_3)+\ldots\ \ri]\ ,}
with the three Mandelstam variables $s=k_1k_2,\ t=k_1k_3$ and $u=k_1k_4$, with
$s+t+u=0$. The dots stand for other kinematical contractions of the form $(\xi\xi)(\xi k)(\xi k)$.
With \wehave\ the full four gluon string $S$--matrix \studyz\ may be obtained by relabeling
gluon indices.
Furthermore, its field--theory interpretation, which leads to the terms 
displayed in \super, is fully understood \TSE. However, in this subsection we shall  
discuss the four gauge boson string $S$--matrix \wehave\ from a different perspective.

In fact, rather than studying the amplitude \studyz\ for a given (fixed) pair of
gluon vertices in the $(-1)$--ghost picture, we allow for any pairings. Hence 
we shall focus our attention on the more general expression (\cf also \eqq \specific):
\eqn\Studyz{\eqalign{
\Tr(\lambda^{\pi(1)}\lambda^{\pi(2)}\ldots\lambda^{\pi(N)})&\  
A^\pi(a,b,i,j)\cr
&=V_{\rm CKG}^{-1}\  \int_{\Ic_\pi} \prod_{r=1}^4 d^2 z_r\
\vev{V_{A^{a_a}}^{(-1)}(z_a)\ V_{A^{a_b}}^{(-1)}(z_b)\  V_{A^{a_i}}^{(0)}(z_i)\
V_{A^{a_j}}^{(0)}(z_j)}\ ,}}
with some permutation $(a,b,i,j)\in \overline{(1,2,3,4)}$ of vertex operators.
According to the definition \Studyz, any choice, \eg $(a,b,i,j)=(1,4,2,3)$ would be as good
to determine the full $S$--matrix \wehave.

After performing the Wick contractions in \Studyz, with \First\ we obtain the following 
expression for \Studyz:
\eqn\Gives{\eqalign{
A^\pi(a,b,i,j)&=A^\pi_2(a,b,i,j)\ (\xi_a\xi_b)\ (\xi_i\xi_j)\cr
&+A^\pi_1(a,b,i,j)\ (\xi_a\xi_i)\ (\xi_b\xi_j)+A^\pi_1(a,b,j,i)\ (\xi_a\xi_j)\ (\xi_b\xi_i)\cr
&+B^\pi_2(a,b,i,j)\ (\xi_a\xi_b)+B^\pi_3(i,j,a,b)\ (\xi_i\xi_j)+B^\pi_1(a,i,b,j)\ (\xi_a\xi_i)\cr
&+B^\pi_1(a,j,b,i)\ (\xi_a\xi_j)+B^\pi_1(b,i,a,j)\ (\xi_b\xi_i)+B^\pi_1(b,j,a,i)\ (\xi_b\xi_j)\ ,}}
with the following integrals:
\eqn\function{\eqalign{
A^\pi_1(a,b,i,j)&=\int_{\Ic_\pi} dz_4\ \vev{c(z_1)c(z_2)c(z_3)}\ \Ec\  
\fc{k_ik_j}{z_{ai}\ z_{jb}\ z_{ij}}\ \fc{(-1)}{z_{ab}}\ ,\cr
A^\pi_2(a,b,i,j)&=\int_{\Ic_\pi} dz_4\ \vev{c(z_1)c(z_2)c(z_3)}\ \Ec\ 
\fc{1}{z_{ab}^2\ z_{ij}^2}\ (1-k_ik_j)\ ,\cr
B^\pi_1(a,i,b,j)&=\int_{\Ic_\pi} dz_4\ \vev{c(z_1)c(z_2)c(z_3)}\ \Ec\  \fc{1}{z_{ab}\ z_{ij}}\cr
&\times\lf\{\ \fc{(\xi_bk_i)(\xi_jk_a)}{z_{aj}z_{bi}}+\fc{(\xi_bk_i)(\xi_jk_b)}{z_{ai}z_{bj}}
-\fc{(\xi_bk_j)(\xi_jk_i)}{z_{ai}z_{bj}}\ \ri\}\ ,\cr
B^\pi_2(a,b,i,j)&=\int_{\Ic_\pi} dz_4\ \vev{c(z_1)c(z_2)c(z_3)}\ \Ec\  \fc{(-1)}{z_{ab}^2\ z_{ij}^2}\ 
\lf\{\ (\xi_ik_b)(\xi_jk_b)-(\xi_ik_b)(\xi_jk_a)\ \fc{z_{bj}z_{ia}}{z_{aj}z_{ib}}\ri.\cr
&\lf.+(\xi_ik_a)(\xi_jk_a)+(\xi_ik_a)(\xi_jk_b)\ \fc{z_{aj}z_{ib}}{z_{bj}z_{ia}}
-(\xi_ik_j)(\xi_jk_i)\ \ri\}\ ,\cr
B^\pi_3(i,j,a,b)&=\int_{\Ic_\pi} dz_4\ \vev{c(z_1)c(z_2)c(z_3)}\ \Ec\  \fc{1}{z_{ab}\ z_{ij}}\ 
\lf\{\ \fc{(\xi_ak_i)(\xi_bk_j)}{z_{ai}z_{bj}}-\fc{(\xi_ak_j)(\xi_bk_i)}{z_{aj}z_{bi}}\ \ri\}\ ,}}
with the three positions $z_1,z_2$ and $z_3$ fixed due to $PSL(2,\IR)$ invariance
on the disk. In addition, we have introduced $\Ec=\prod\limits_{r<s} |z_{rs}|^{k_rk_s}$.
The region $\Ic_\pi$ of the $z_4$--integration \region\ is determined by the group structure
$\pi$ under consideration. {From} the sum \Gives, the meaning of the above five functions 
becomes clear:
While the functions $A^\pi_r$ appear in front of the $(\xi\xi)(\xi\xi)$ contractions
\eqn\Appear{
A^\pi_1(a,b,i,j)\ (\xi_a\xi_i)(\xi_b\xi_j)\ \ \ ,\ \ \ 
A^\pi_2(a,b,i,j)\ (\xi_a\xi_b)(\xi_i\xi_j)\ ,}
the functions $B^\pi_s$ capture the contractions $(\xi\xi)(\xi k)(\xi k)$:
\eqn\Appeari{
B^\pi_1(a,i,b,j)\ (\xi_a\xi_i)\ \ ,\ \ B^\pi_2(a,b,i,j)\ (\xi_a\xi_b)\ \ ,\ \ 
B^\pi_3(i,j,a,b)\ (\xi_i\xi_j)\ .}
The crucial difference between the two functions $A^\pi_2$ and $A^\pi_1$ is, whether 
the polarizations $\xi_a, \xi_b$ of the two vertices $a$ and $b$ in the $(-1)$--ghost 
picture are contracted among
themselves (\ie leading to the kinematics $(\xi_a\xi_b)$) or contracted with the other 
polarizations $\xi_i,\xi_j$, respectively.
Similarly, the two functions $B^\pi_2$ and $B^\pi_1$ account for the two different cases,
whether the polarization $\xi_a$ is contracted with $\xi_b$ or one of $\xi_i,\xi_j$, respectively.
Moreover, the function $B^\pi_3$ captures the case, when both $\xi_a$ and $\xi_b$
are contracted with momenta.

In each amplitude $A^\pi(a,b,i,j)$, for a given permutation $(a,b,i,j)\in \overline{(1,2,3,4)}$
all possible space--time kinematics show up once. {\it E.g.} the kinematics
$(\xi_1\xi_2)(\xi_3\xi_4),\ (\xi_1\xi_3)(\xi_2\xi_4)$ and 
$(\xi_1\xi_4)(\xi_2\xi_3)$ show up once. However, 
the five functions $A_r,B_s$ in front of those kinematics are  different depending 
on the permutation $(a,b,i,j)\in \overline{(1,2,3,4)}$ under consideration, \ie
which vertex pair $(a,b)$ we have put into the $(-1)$--ghost picture.
More concretely, for the given kinematics 
\eqn\Considerkin{
(\xi_A\xi_B)\ (\xi_C\xi_D)}
for one and the same group contraction $\pi$
we obtain the three seemingly different looking expressions
\eqn\threfunctions{
A^\pi_1(A,C,B,D)\ \ \ ,\ \ \ A^\pi_1(A,D,B,C)\ \ \ ,\ \ \ A^\pi_2(A,B,C,D)\ ,}
depending on which pair of  vertex operator has been put into the $(-1)$ ghost picture.
Recall, $A^\pi_1(A,C,B,D)$ is the function with the gluon vertices $A$ and $C$ in the 
$(-1)$--ghost picture, while
$A^\pi_1(A,D,B,C)$ with $A$ and $D$ and $A^\pi_2(A,B,C,D)$ with $A$ and $B$ singled out.
In other words, the three functions $A^\pi_r$ are associated to taking the 
pairs $(A,C),(A,D)$ or $(A,B)$ of vertex operators in
the $(-1)$--ghost picture, respectively. However, they all refer to the kinematical contraction 
$(\xi_A\xi_B)(\xi_C\xi_D)$ with the same group structure $\pi$. Hence, they  all have to be equal
as a result of just choosing different pairs of vertex operators in the $(-1)$--ghost picture.
In total, for the contractions $(\xi\xi)(\xi\xi)$ we obtain the following set of equations
(for a given group contraction $\pi$):
\eqn\seteq{\eqalign{
(\xi_1\xi_2)(\xi_3\xi_4):&\ \ \  A^\pi_2(1,2,3,4)=A^\pi_1(1,3,2,4)=A^\pi_1(1,4,2,3)\ ,\cr
(\xi_1\xi_3)(\xi_2\xi_4):&\ \ \  A^\pi_2(1,3,2,4)=A^\pi_1(1,2,3,4)=A^\pi_1(1,4,3,2)\ ,\cr
(\xi_1\xi_4)(\xi_2\xi_3):&\ \ \  A^\pi_2(1,4,2,3)=A^\pi_1(1,2,4,3)=A^\pi_1(1,3,4,2)\ .}}
Let us focus on the group theoretical factor 
$\Tr(\lambda^1\lambda^2\lambda^4\lambda^3)$, \ie $\pi=(1,2,4,3)$. We fix the first three vertex 
positions to $z_1=-z_\infty, z_2=0$ and $z_3=1$, \ie 
$\vev{c(z_1)c(z_2)c(z_3)}=-z_\infty^2$, with $z_\infty=\infty$.
According to \region, for $\Ic_\pi$ we integrate $z_4$ from $z_4=0$ until $z_4=1$.
Then, the integrals $A^\pi_r$ in \function\ can be written for $\pi=(1,2,4,3)$
\eqn\functions{\eqalign{
A^\pi_2(1,2,3,4)&=(s-1)\ F_4\ \ \ ,\ \ \ A^\pi_1(1,3,2,4)=-t\ F_3\ \ \ ,\ \ \ 
A^\pi_1(1,4,2,3)=u\ F_1\ ,\cr
A^\pi_2(1,3,2,4)&=(t-1)\ F_5\ \ \ ,\ \ \ A^\pi_1(1,2,3,4)=-s\ F_3\ \ \ ,\ \ \ 
A^\pi_1(1,4,3,2)=-u\ F_2\ ,\cr
A^\pi_2(1,4,2,3)&=(u-1)\ F_0\ \ \ ,\ \ \ A^\pi_1(1,2,4,3)=s\ F_1\ \ \ ,\ \ \ 
A^\pi_1(1,3,4,2)=-t\ F_2\ ,}}
with the integrals:
\eqn\veneziano{
F_j=\int_0^1 dx\ P_j\ x^t (1-x)^s\ ,}
and the polynomials:
\eqn\polynoms{\eqalign{
P_0&=1\ \ \ ,\ \ \ P_1=\fc{1}{x-1}\ \ \ ,\ \ \ P_2=\fc{1}{x}\ ,\cr
P_3&=\fc{1}{x(x-1)}\ \ \ ,\ \ \ P_4=\fc{1}{(x-1)^2}\ \ \ ,\ \ \ P_5=\fc{1}{x^2}\ .}}
Equating the functions \functions\ according to \seteq\ yields the following system
of six equations for the six functions $F_j$:
\eqn\system{\eqalign{
(s-1)\ F_4&=u\ F_1\ \ \ ,\ \ \ (s-1)\ F_4=-t\ F_3\cr
(t-1)\ F_5&=-s\ F_3\ \ \ ,\ \ \ (t-1)\ F_5=-u\ F_2\cr
(u-1)\ F_0&=s\ F_1\ \ \ ,\ \ \ (u-1)\ F_0=-t\ F_2\ .}}
Their solution may be written:
\eqn\singel{\eqalign{
F_1&=\fc{u-1}{s}\ F_0\ \ \ ,\ \ \ F_2=\fc{1-u}{t}\ F_0\ ,\cr
F_3&=\fc{u\ (1-u)}{st}\ F_0\ \ \ ,\ \ \ F_4=\fc{u\ (1-u)}{s\ (1-s)}\ F_0\ \ \ ,\ \ \ 
F_5=\fc{u\ (1-u)}{t\ (1-t)}\ F_0\ .}}
Hence, all functions $F_j$ may be expressed by one single function $F_0$
\eqn\BOXIV{\hskip-0.75cm{
\vbox{\offinterlineskip
\halign{\strut\vrule#
%%%%%%%%%%%%%%%%%%
&~$#$~\hfil
&\vrule# \cr
\noalign{\hrule}
%%%%%%%%%%%%%%%%%%
&\ \        \  \ &\cr
&F_j=\Lambda_{j}(s,t,u)\ \ F_0&\cr
&\ &\cr
%%%%%%%%%%%%%%%%%%
\noalign{\hrule}}} } }
with some momentum dependent factor $\Lambda_j$ like it is given in {\it Eqs.} \singel.
Of course, the integrals \veneziano\ are easy to determine with \eqq (4.1).
The relations \singel\ may be directly checked using the basic identity 
$\Gamma(x+1)=x\ \Gamma(x)$. More concretely,  we obtain:
\eqn\obtain{\eqalign{
F_0&=\fc{\Gamma(s+1)\ \Gamma(t+1)}{\Gamma(2+s+t)}\ \ \ ,\ \ \ 
F_1=-\fc{\Gamma(s)\ \Gamma(t+1)}{\Gamma(1+s+t)}\ \ \ ,\ \ \ 
F_2=\fc{\Gamma(s+1)\ \Gamma(t)}{\Gamma(1+s+t)}\ ,\cr
F_3&=-\fc{\Gamma(s)\ \Gamma(t)}{\Gamma(s+t)}\ \ \ ,\ \ \ 
F_4=\fc{\Gamma(s-1)\ \Gamma(t+1)}{\Gamma(s+t)}\ \ \ ,\ \ \ 
F_5=\fc{\Gamma(s+1)\ \Gamma(t-1)}{\Gamma(s+t)}\ .}}
However, what this exercise shows is, that with changing the ghost--picture of the
vertex operators we obtain different integrals \threfunctions\ for the kinematics 
$(\xi_A\xi_B)(\xi_C\xi_D)$.
These expressions have to be equal and we obtain a system of equations, which allows
us, to express all integrals $A^\pi_r$, given in \functions, in terms of a {\it single} 
integral.
In the concrete case \singel, this integral is $F_0$.
We have some freedom, which integral we want to choose to be left over.
The integral $F_3$ could also have been taken.
In fact, for higher gluon scattering, it proves to be very convenient to choose
an integral, which does not lead to poles in the kinematic invariants. For such an integral
its power series in the kinematic invariants may be easier derived.

Similar statements hold for the kinematics $(\xi\xi)(\xi k)(\xi k)$ referring to a fixed
group structure $\pi$.
More concretely, consider the kinematics 
\eqn\Considerkinnn{
(\xi_A\xi_B)\ (\xi_C k) (\xi_D \tilde k)\ ,}
with $k,\tilde k$ accounting for a set of momenta subject to momentum conservation.
In other words, the kinematics \Considerkinnn\ readily stands for four independent
space--time kinematics.
According to \Appeari\  the six functions
\eqn\SixCases{\eqalign{
&B^\pi_2(A,B,C,D)\ \ \ ,\ \ \ B^\pi_1(A,B,C,D)\ \ \ ,\ \ \ B^\pi_1(A,B,D,C)\ ,\cr
&B^\pi_1(B,A,C,D)\ \ \ ,\ \ \ B^\pi_1(B,A,D,C)\ \ \ ,\ \ \ B^\pi_3(A,B,C,D)}}
appear in \Gives\ and are relevant to the kinematics \Considerkinnn.
They refer to  the six cases, in which one of the pairs $(A,B),(A,C),(A,D),(B,C),(B,D)$ or $(C,D)$
of vertex operators is put into the $(-1)$--ghost picture, respectively.
Similar, as in \eqq \seteq\ all these six functions \SixCases\ have to be equated.
Since, after definition,  the latter still contain some $\xi k$ kinematical contractions,
what has to be equated, is the relevant piece of those functions referring to the same 
scalar product $(\xi k)(\xi k)$, after applying momentum conservation.
Hence, in what follows\foot{More concretely, \eg in the case $A=1,B=2,C=3$ and $D=4$ 
one reduces the functions
to the kinematics $(\xi_3k_1)\ (\xi_4k_2)$ and $(\xi_3k_2)\ (\xi_4k_1)$
[and $(\xi_3k_1)\ (\xi_4k_1)$ and $(\xi_3k_2)\ (\xi_4k_2)$] by eliminating the
combinations $\xi_3k_4$ and $\xi_4k_3$ as a result of momentum conservation.}
one obtains actually twenty equations, which give rise to relations
between different integrals of the type \veneziano.
After considering all possible kinematics, in addition to the polynomials \polynoms, the
following set
\eqn\addpyln{\eqalign{
P_6&=\fc{1}{x(x-1)^2}\ \ \ ,\ \ \ P_7=\fc{x}{(x-1)^2}\ \ \ ,\ \ \ P_8=\fc{1}{x^2(x-1)}\cr
P_9&=\fc{x-1}{x}\ \ \ ,\ \ \ P_{10}=\fc{x-1}{x^2}\ \ \ ,\ \ \ P_{11}=\fc{x}{x-1}}}
appears.
The equations following from equating \SixCases\ for all possible kinematics
provide the relations between the 
integrals \veneziano\ associated to \addpyln\ and $F_0$:
\eqn\providerel{\eqalign{
F_6&=F_4-F_3\ \ \ ,\ \ \ F_7=F_1+F_4\ \ \ ,\ \ \ F_8=F_3-F_5\ ,\cr
F_9&=F_0-F_2\ \ \ ,\ \ \ F_{10}=F_2-F_5\ \ \ ,\ \ \ F_{11}=F_0+F_2+F_3\ .}}
In addition, the identity $F_1-F_2=F_3$ appears.
Of course, the latter relation, which is satisfied by the solutions \obtain\  
is just a consequence of $P_1-P_2=P_3$, \ie $\fc{1}{x-1}-\fc{1}{x}=\fc{1}{x(x-1)}$.

Though, the identities \singel\ are easy to prove in the case at hand, 
the strategy, outlined above for the four--gluon scattering, yields non--trivial
identities in the case of gluon amplitudes involving more than four gluons.
{From} \eqq \singel\ we deduce, that those identities involve rational polynomials in the 
Mandelstam variables $s,t$ and $u$.
These polynomials account for the pole structure of the given kinematics, while
the function through which all other functions are expressed may be chosen to
be simple in the sense, that it does not contain poles.
{\it E.g.} in the case above, the function $F_0$ has the power series w.r.t. to
the Mandelstam variables $s,t,u$
\eqn\powerSS{
F_0=1-s-t+s^2 + 2\ s\ t - \zeta(2)\ s\ t + t^2+\ldots\ ,}
\ie it is finite w.r.t. to all invariants.
Hence all pole structure of the amplitude is encountered in the rational polynomials in front
of the function $F_0$, \ie they follow from solving the algebraic equations \system\
through expressing all functions $F_j, j\geq 1$ through $F_0$.

In the case of more than four gluons, those identities allow to express 
each integral accounting for a specific kinematics
through a linear combination of integrals without poles, such that the full
pole structure originates algebraically.
In fact, the case with six gluons shall be discussed in the next section, while
the five gluon case is revisted in section 6.

\newsec{Six gluon open superstring $S$--matrix}

In the following we study the string $S$--matrix describing the string
tree--level scattering of six gauge bosons on the disk ($N=6$ in \START):
\eqn\study{\eqalign{
&\Ac_6(k_1,\xi_1,a_1;k_2,\xi_2,a_2;k_3,\xi_3,a_3;k_4,\xi_4,a_4;k_5,\xi_5,a_5;k_6,\xi_6,a_6)\cr
&\hskip1cm=V_{\rm CKG}^{-1}\ \prod_{r=1}^6 \int d^2 z_r\
\vev{V_{A^{a_1}}^{(-1)}(z_1)\ V_{A^{a_2}}^{(-1)}(z_2)\  V_{A^{a_3}}^{(0)}(z_3)\
V_{A^{a_4}}^{(0)}(z_4)\ V_{A^{a_5}}^{(0)}(z_5)\  V_{A^{a_6}}^{(0)}(z_6)}\ .}}
We have total momentum conservation: 
\eqn\toalm{
\sum\limits_{i=1}^6\ k_i=0}
for the six external gluon momenta $k_i$.
Furthermore, we have chosen two gauge vertex operators
in the $(-1)$--ghost picture in order to guarantee a total ghost charge of $-2$ on the disk.
The basic ingredients to start with have been presented in subsection 2.1.
However, one has to keep in mind, that
the case $p=3$ is special, since in that case the six order Lorentz invariant $\tr F^6$ may
be expressed by the other two $(\tr F^2)^3$ and $\tr F^2\tr F^4$ invariants. 
In the scattering process of six gluons this difference manifests in a reduced number
of kinematical (space--time) invariants. Later we shall comment on this issue.

Recall, that the four--gluon string $S$--matrix $\Ac_4$, discussed
in the previous subsection,  gives rise to the following two kinematical structures
$(\xi\xi)(\xi\xi)$ and $(\xi\xi)(\xi k)(\xi k)$ entering \wehave. At order $\Oc(k^4)$ in the
external momenta, both kinematics comprise parts of the same higher order 
gauge interaction terms $F^4$ in the effective action as a result of gauge invariance. 
This means, that these interaction terms may be completely inferred from the first contractions 
$(\xi\xi)(\xi\xi)$ \TSE, while
the string $S$--matrix result for the second kinematics $(\xi\xi)(\xi k)(\xi k)$ may
be related to the first one.
The explicit evaluation of the correlator in \eqq \study\ leads to a sum of integrals 
with various different (space--time) kinematic contractions as coefficients.
More precisely, we encounter the three kinematical structures $(\xi\xi)(\xi\xi)(\xi\xi)$,
$(\xi\xi)(\xi\xi)(\xi k)(\xi k)$ and $(\xi\xi)(\xi k)(\xi k)(\xi k)(\xi k)$,
which in the following will be denoted by $A,B$ and $C$--kinematics, respectively.
Their contributions in the effective action are all related to each other through 
gauge--invariance.
In other words, the effective action at a given
order in the external momenta, may be reproduced from a particular class of those kinematics.

\subsec{General expression for the amplitude}

In \eqq \study\ we have chosen the first two gauge vertex operators in the $(-1)$--ghost picture.
We have already stressed in the previous subsection, that we are free to choose
which pair $(a,b)$ of the six gauge vertex operators we put into the $(-1)$--ghost picture.  
Hence, according to the definition \express, for a given group structure described
by the permutation $\pi$, instead of the particular choice in \study, 
we may generically consider the expression 
\eqn\Study{\eqalign{
\Tr(\lambda^{\pi(1)}\lambda^{\pi(2)}\lambda^{\pi(3)}\lambda^{\pi(4)}\lambda^{\pi(5)}
\lambda^{\pi(6)})&\ A^\pi(a,b,i,j,k,l)=V_{\rm CKG}^{-1}\  \int_{\Ic_\pi} \prod_{r=1}^6 d^2 z_r\cr
& \hskip-1cm\vev{V_{A^{a_a}}^{(-1)}(z_a)\ V_{A^{a_b}}^{(-1)}(z_b)\  V_{A^{a_i}}^{(0)}(z_i)\
V_{A^{a_j}}^{(0)}(z_j)\ V_{A^{a_k}}^{(0)}(z_k)\  V_{A^{a_l}}^{(0)}(z_l)}\ ,}}
with some permutation $(a,b,i,j,k,l)\in \overline{(1,2,3,4,5,6)}$ of vertex operators.
In \eqq \formally,  we have explained, how this term contributes to the full string $S$--matrix
\study. 

After performing the Wick contractions
the correlator in \Study\ decomposes into products of various two--point functions, given 
in \eqq \First:
\eqn\gives{\eqalign{
A^\pi(a,b,& i,j,k,l)=V_{\rm CKG}^{-1}\  \int_{\Ic_\pi} \prod_{r=1}^6 d^2 z_r
\sum_{(i_1,i_2,i_3,i_4)\atop\in\overline{(i,j,k,l)}}\ \lf\{\h\ 
\Ac_1(a,b,i_1,i_2,i_3,i_4)\ (\xi_a\xi_{i_1})\ (\xi_b\xi_{i_4})\ (\xi_{i_2}\xi_{i_3})\ri.\cr
&+\fc{1}{8}\ \Ac_2(a,b,i_1,i_2,i_3,i_4)\ (\xi_a\xi_b)\ (\xi_{i_1}\xi_{i_2})\ (\xi_{i_3}\xi_{i_4})
+\fc{1}{4}\ \Bc_2(a,b,i_1,i_2,i_3,i_4)\ (\xi_a\xi_b)\ (\xi_{i_1}\xi_{i_2})\cr
&+\h\ \Bc_1(a,b,i_1,i_2,i_3,i_4)\ (\xi_a\xi_{i_1})\ (\xi_b\xi_{i_2})
+\fc{1}{8}\ \Bc_4(i_1,i_2,i_3,i_4,a,b)\ (\xi_{i_1}\xi_{i_2})\ (\xi_{i_3}\xi_{i_4})\cr
&+\h\ \Bc_3(a,i_1,i_2,i_3,b,i_4)\ (\xi_a\xi_{i_1})\ (\xi_{i_2}\xi_{i_3})
+\h\ \Bc_3(b,i_1,i_2,i_3,a,i_4)\ (\xi_b\xi_{i_1})\ (\xi_{i_2}\xi_{i_3})\cr
&+\fc{1}{24}\ \Cc_1(a,b,i_1,i_2,i_3,i_4)\ (\xi_a\xi_b)+\fc{1}{4}\ 
\Cc_3(i_1,i_2,a,b,i_3,i_4)\ (\xi_{i_1}\xi_{i_2})\cr
&\lf.+\fc{1}{6}\ \Cc_2(a,i_1,b,i_2,i_3,i_4)\ (\xi_a\xi_{i_1})+
\fc{1}{6}\ \Cc_2(b,i_1,a,i_2,i_3,i_4)\ (\xi_b\xi_{i_1})\ri\}\ .}}
In \eqq \gives\ in front of the contractions 
$\xxi{a}{i}\ \xxi{j}{k}\ \xxi{l}{b}$ and $\xxi{a}{b}\ \xxi{i}{j}\ \xxi{k}{l}$ there appear 
the two basic functions $\Ac_1(a,b,i,j,k,l)$ and $\Ac_2(a,b,i,j,k,l)$, given by
\eqn\KIN{\eqalign{
\Ac_1(a,b,i,j,k,l)&=-\fc{\Ec}{z_{ab}\ z_{ai}}\ 
\lf\{\fc{(k_ik_j)\ (k_kk_l)}{z_{lb}\ z_{jk}\ z_{ij}\ z_{kl}}-
\fc{(k_ik_k)\ (k_jk_l)}{z_{lb}\ z_{jk}\ z_{ik}\ z_{jl}}
-\fc{(k_ik_l)\ (1-k_jk_k)}{z_{lb}\ z_{il}\ z_{jk}^2}\ri\}\ ,\cr
\Ac_2(a,b,i,j,k,l)&= 
-\fc{\Ec}{z_{ab}^2}\ \lf\{\fc{(1-k_ik_j)\ (1-k_kk_l)}{z_{ij}^2\ z_{kl}^2}-
\fc{(k_ik_k)\ (k_jk_l)}{z_{ij}\ z_{kl}\ z_{ik}\ z_{jl}}+
\fc{(k_ik_l)\ (k_jk_k)}{z_{ij}\ z_{kl}\ z_{il}\ z_{jk}}\ri\}\ ,}}
respectively, with:
\eqn\Expsix{ 
{\cal E}=\prod\limits_{r<s}^6 |z_{rs}|^{k_rk_s}\ .}
In the following, we in particular shall  need the integrated expressions
\eqn\KINALL{
A_i^\pi=V_{\rm CKG}^{-1}\ \int_{\Ic_\pi} \prod_{r=1}^6 d^6z_r\  \Ac_i\ \ ,\ \ 
B_i^\pi=V_{\rm CKG}^{-1}\ \int_{\Ic_\pi} \prod_{r=1}^6 d^6z_r\  \Bc_i\ \ ,\ \ 
C_i^\pi=V_{\rm CKG}^{-1}\ \int_{\Ic_\pi} \prod_{r=1}^6 d^6z_r\  \Cc_i\ ,}
which enter the expression \gives.  Furthermore  we have: 
\def\ss#1{{{#1}}}
$$\hskip-1.4cm{\eqalign{
\ss{\Bc_1(a,b,i,j,k,l)}&=\ss{ 
-\Ec\ \lf\{
\fc{z_{ai}(\xi_kk_i)(\xi_lk_i)}{z^2_{ab}z_{ij}z_{ik}z_{il}}\lf(\fc{z_{ai}}{z_{ak}z_{al}z_{ij}}-
\fc{z_{ai}k_ik_j}{z_{ak}z_{al}z_{ij}}+\fc{k_jk_k}{z_{al}z_{jk}}+
\fc{k_jk_l}{z_{ak}z_{jl}}\ri) \ri.}\cr
&\ss{-\fc{(\xi_kk_l)(\xi_lk_j)}{z^2_{ab}z_{ij}z_{jl}z_{kl}}\lf(\fc{z_{aj}(1-k_ik_j)}{z_{ak}z_{ij}}-
\fc{k_ik_k}{z_{ik}}-\fc{z_{al}k_ik_l}{z_{ak}z_{il}}\ri)+
\fc{(\xi_kk_i)(\xi_lk_b)}{z_{ab}z_{al}z_{bl}z_{ij}z_{ik}}\lf(\fc{z_{ai}(1-k_ik_j)}{z_{ak}z_{ij}}+
\fc{k_jk_k}{z_{jk}}\ri) }\cr
&\ss{+\fc{(\xi_kk_i)(\xi_lk_k)}{z^2_{ab}z_{ij}z_{ik}z_{kl}}\lf(\fc{z_{ai}(1-k_ik_j)}{z_{al}z_{ij}}+
\fc{z_{ak}k_jk_k}{z_{al}z_{jk}}+\fc{k_jk_l}{z_{jl}}\ri)+
\fc{(\xi_kk_b)(\xi_lk_j)}{z_{ab}z_{ak}z_{bk}z_{ij}z_{jl}}\lf(\fc{z_{aj}(1-k_ik_j)}{z_{al}z_{ij}}-
\fc{k_ik_l}{z_{il}}\ri)}\cr
&\ss{+\fc{(\xi_kk_j)(\xi_lk_i)}{z^2_{ab}z_{ij}z_{il}z_{jk}}
\lf(\fc{z_{ai}z_{aj}(1-k_ik_j)}{z_{ak}z_{al}z_{ij}}-
\fc{z_{ai}k_ik_k}{z_{al}z_{ik}}+\fc{z_{aj}k_jk_l}{z_{ak}z_{jk}}+\fc{k_kk_l}{z_{kl}}\ri)+
\fc{(\xi_kk_b)(\xi_lk_k)(1-k_ik_j)}{z_{ab}z_{al}z_{bk}z^2_{ij}z_{kl}}}\cr
&\ss{+\fc{(\xi_kk_j)(\xi_lk_b)}{z_{ab}z_{al}z_{bl}z_{ij}z_{jk}}
\lf(\fc{z_{aj}(1-k_ik_j)}{z_{ak}z_{ij}}-
\fc{k_ik_k}{z_{ik}}\ri)+
\fc{(\xi_kk_b)(\xi_lk_i)}{z_{ab}z_{ak}z_{bk}z_{ij}z_{il}}\lf(\fc{z_{ai}(1-k_ik_j)}{z_{al}z_{ij}}+
\fc{k_jk_l}{z_{jl}}\ri)}\cr
&\ss{+\fc{z_{aj}(\xi_kk_j)(\xi_lk_j)}{z^2_{ab}z_{ij}z_{jk}z_{jl}}
\lf(\fc{z_{aj}(1-k_ik_j)}{z_{ak}z_{al}z_{ij}} -
\fc{k_ik_k}{z_{al}z_{ik}}+\fc{k_ik_l}{z_{ak}z_{il}}\ri)+ 
\fc{(\xi_kk_b)(\xi_lk_b)(1-k_ik_j)}{z_{ak}z_{al}z_{bk}z_{bl}z^2_{ij}}}\cr
&\ss{+\fc{(\xi_kk_j)(\xi_lk_k)}{z^2_{ab}z_{ij}z_{jk}z_{kl}}\lf(\fc{z_{aj}(1-k_ik_j)}{z_{al}z_{ij}}-
\fc{z_{ak}k_ik_k}{z_{al}z_{ik}}-\fc{k_ik_l}{z_{il}}\ri)-
\fc{(\xi_kk_l)(\xi_lk_b)(1-k_ik_j)}{z_{ab}z_{ak}z_{bl}z^2_{ij}z_{kl}}}\cr
&\ss{+\fc{(\xi_kk_i)(\xi_lk_j)}{z^2_{ab}z_{ij}z_{ik}z_{jl}}\ 
\lf(\fc{z_{ai}z_{aj}(1-k_ik_j)}{z_{ak}z_{al}z_{ij}}-
\fc{z_{ai}k_ik_l}{z_{ak}z_{il}}+\fc{z_{aj}k_jk_k}{z_{al}z_{jk}}-\fc{k_kk_l}{z_{kl}}\ri)}\cr
&\ss{-\lf.\fc{(\xi_kk_l)(\xi_lk_i)}{z^2_{ab}z_{ij}z_{il}z_{kl}}\ 
\lf(\fc{z_{ai}(1-k_ik_j)}{z_{ak}z_{ij}}+
\fc{k_jk_k}{z_{jk}}+\fc{z_{al}k_jk_l}{z_{ak}z_{jl}}\ri)\ri\}}\ ,\cr
%%%%%%%%%%%%%%%%%%%%%%%%%%%%%%%%%%%%%%%%%%%%%%%%%%%%%%%%%%%%%%%%%%%%
\ss{\Bc_2(a,b,i,j,k,l)}&=\ss{
-\Ec\ \lf\{ 
 \fc{(\xi_kk_j)(\xi_lk_i)}{z_{ab}z_{bj}z_{il}z_{jk}}\lf(\fc{z_{aj}k_ik_j}{z_{ak}z_{al}z_{ij}}+
\fc{k_ik_k}{z_{al}z_{ik}}-\fc{z_{aj}k_jk_l}{z_{ai}z_{ak}z_{jl}}-
\fc{k_kk_l}{z_{ai}z_{kl}} \ri)\ri.}\cr
&\ss{+\fc{(\xi_kk_j)(\xi_lk_b)}{z_{ai}z_{al}z_{bj}z_{bl}z_{jk}}\lf(\fc{z_{aj}k_ik_j}{z_{ak}z_{ij}}+
\fc{k_ik_k}{z_{ik}} \ri)+
\fc{(\xi_kk_l)(\xi_lk_i)}{z_{ab}z_{bj}z_{il}z_{kl}}\lf(\fc{z_{al}k_jk_l}{z_{ai}z_{ak}z_{jl}}+
\fc{k_jk_k}{z_{ai}z_{jk}}-\fc{k_ik_j}{z_{ak}z_{ij}} \ri)}\cr
&\ss{+\fc{(\xi_kk_b)(\xi_lk_j)}{z_{ai}z_{ak}z_{bj}z_{bk}z_{jl}}\lf(\fc{z_{aj}k_ik_j}{z_{al}z_{ij}}-
\fc{k_ik_l}{z_{il}} \ri)
+\fc{(\xi_kk_i)(\xi_lk_i)}{z_{ab}z_{bj}z_{ik}z_{il}}\lf(\fc{z_{ai}k_ik_j}{z_{ak}z_{al}z_{ij}}-
\fc{k_jk_k}{z_{al}z_{jk}}-\fc{k_jk_l}{z_{ak}z_{jl}} \ri)}\cr
&\ss{+\fc{(\xi_kk_b)(\xi_lk_i)}{z_{ak}z_{bj}z_{bk}z_{il}}\lf(\fc{k_ik_j}{z_{al}z_{ij}}-
\fc{k_jk_l}{z_{ai}z_{jl}} \ri)+
\fc{(\xi_kk_j)(\xi_lk_j)}{z_{ab}z_{ai}z_{bj}z_{jk}z_{jl}}
\lf(\fc{z^2_{aj}k_ik_j}{z_{ak}z_{al}z_{ij}}+
\fc{z_{aj}k_ik_k}{z_{al}z_{ik}}+\fc{z_{aj}k_ik_l}{z_{ak}z_{il}} \ri)}}}$$
%%%%%%%%%%%%%%%%%%%%%%%%%%%%%%%%%%%%%%%%%%%%%%%%%%%%%%%%%%%%%%%%%%%%%%%%%%%%%%
\eqn\KINB{\hskip-0.7cm\eqalign{
&\ss{-\fc{(\xi_kk_l)(\xi_lk_j)}{z_{ab}z_{ai}z_{bj}z_{jl}z_{kl}}\lf(\fc{z_{aj}k_ik_j}{z_{ak}z_{ij}}+
\fc{k_ik_k}{z_{ik}}+\fc{z_{al}k_ik_l}{z_{ak}z_{il}} \ri)+
\fc{(\xi_kk_i)(\xi_lk_b)}{z_{al}z_{bj}z_{bl}z_{ik}}\lf(\fc{k_ik_j}{z_{ak}z_{ij}}-
\fc{k_jk_k}{z_{ai}z_{jk}} \ri)}\cr
&\ss{+\fc{(\xi_kk_i)(\xi_lk_j)}{z_{ab}z_{bj}z_{ik}z_{jl}}\lf(\fc{z_{aj}k_ik_j}{z_{ak}z_{al}z_{ij}}+
\fc{k_ik_l}{z_{ak}z_{il}}-\fc{z_{aj}k_jk_k}{z_{ai}z_{al}z_{jk}}+
\fc{k_kk_l}{z_{ai}z_{kl}} \ri)
\fc{(\xi_kk_b)(\xi_lk_k)}{z_{ai}z_{al}z_{bj}}\fc{k_ik_j}{z_{bk}z_{ij}z_{kl}}}\cr
&\ss{+\fc{(\xi_kk_i)(\xi_lk_k)}{z_{ab}z_{bj}z_{ik}z_{kl}}\lf(\fc{k_ik_j}{z_{al}z_{ij}}-
\fc{z_{ak}k_jk_k}{z_{ai}z_{al}z_{jk}}-\fc{k_jk_l}{z_{ai}z_{jl}} \ri)+
\fc{(\xi_kk_b)(\xi_lk_b)}{z_{ai}z_{ak}z_{al}}\fc{z_{ab}k_ik_j}{z_{bj}z_{bk}z_{bl}z_{ij}}}\cr
&\ss{+\lf.\fc{(\xi_kk_j)(\xi_lk_k)}{z_{ab}z_{ai}z_{bj}z_{jk}z_{kl}}\ 
\lf(\fc{z_{aj}k_ik_j}{z_{al}z_{ij}}+
\fc{z_{ak}k_ik_k}{z_{al}z_{ik}}+\fc{k_ik_l}{z_{il}} \ri)+
\fc{(\xi_kk_l)(\xi_lk_b)}{z_{ai}z_{ak}z_{bj}}\fc{k_ik_j}{z_{bl}z_{ij}z_{kl}}\ri\}}\ ,\cr
\ss{\Bc_3(a,i,j,k,b,l)}&=\ss{ 
-\Ec\ \lf\{\ 
\fc{(k_jk_k-1)}{z^2_{jk}}\ \lf(\fc{(\xi_bk_i)(\xi_lk_b)}{z_{ai}z_{al}z_{bi}z_{bl}}+
\fc{(\xi_bk_i)(\xi_lk_i)}{z_{ab}z_{al}z_{bi}z_{il}}+
\fc{(\xi_bk_l)(\xi_lk_i)}{z_{ab}z_{ai}z_{bl}z_{il}}\ri)\ri.}\cr
&\ss{+\fc{(\xi_bk_i)(\xi_lk_j)}{z_{ab}z_{ai}z_{jk}z_{jl}}\ 
\lf(\fc{(z_{aj})(k_jk_k-1)}{z_{al}z_{bi}z_{jk}}+\fc{k_kk_l}{z_{bi}z_{kl}}\ri)+
\fc{(\xi_bk_k)(\xi_lk_i)}{z_{ab}z_{bk}z_{il}z_{jk}}\ 
\lf(\fc{k_ik_j}{z_{al}z_{ij}}-\fc{k_jk_l}{z_{ai}z_{jl}} \ri)}\cr
&\ss{+\fc{(\xi_bk_j)(\xi_lk_i)}{z_{ab}z_{bj}z_{il}z_{jk}}\ \lf(\fc{k_kk_l}{z_{ai}z_{kl}}
-\fc{k_ik_k}{z_{al}z_{ik}} \ri) +\fc{(\xi_bk_i)(\xi_lk_k)}{z_{ab}z_{ai}z_{bi}z_{kl}}\ 
\lf(\fc{(z_{ak})(k_jk_k-1)}{z_{al}z_{jk}}+\fc{k_jk_l}{z_{jl}}\ri)}\cr
&\ss{+\fc{k_ik_j}{z_{ai}z_{ij}z_{jk}}\ \lf(\fc{(\xi_bk_k)(\xi_lk_b)}{z_{al}z_{bk}z_{bl}}+
\fc{z_{ak}(\xi_bk_k)(\xi_lk_k)}{z_{ab}z_{al}z_{bk}z_{kl}}+
\fc{(\xi_bk_l)(\xi_lk_k)}{z_{ab}z_{bl}z_{kl}} \ri)}\cr
&\ss{-\fc{k_ik_k}{z_{ai}z_{ik}z_{jk}}\ \lf(\fc{(\xi_bk_j)(\xi_lk_b)}{z_{al}z_{bj}z_{bl}}+
\fc{z_{aj}(\xi_bk_j)(\xi_lk_j)}{z_{ab}z_{al}z_{bj}z_{jl}}+
\fc{(\xi_bk_l)(\xi_lk_j)}{z_{ab}z_{bl}z_{jl}} \ri)}\cr
&\ss{+\lf.\fc{(\xi_bk_k)(\xi_lk_j)}{z_{ab}z_{ai}z_{bk}z_{jk}z_{jl}}\ 
\lf(\fc{z_{aj}k_ik_j}{z_{al}z_{ij}}+\fc{k_ik_l}{z_{il}} \ri)-
\fc{(\xi_bk_j)(\xi_lk_k)}{z_{ab}z_{ai}z_{bj}z_{jk}z_{kl}}\ \lf(\fc{z_{ak}k_ik_k}{z_{al}z_{ik}}
+\fc{k_ik_l}{z_{il}}\ri)\ \ri\}\ ,}\cr
\ss{\Bc_4(i,j,k,l,a,b)}&=\ss{-\fc{\Ec}{z_{ab}\ z_{ij}\ z_{kl}}  
\ \lf\{\ \fc{(1-k_kk_l)}{z_{kl}}\ \lf(\ \fc{(\xi_ak_i)\ (\xi_bk_j)}{z_{ai}\ z_{bj}}-
\fc{(\xi_ak_j)\ (\xi_bk_i)}{z_{aj}\ z_{bi}}\ \ri)\ri.}\cr 
&\ss{+\fc{k_ik_k}{z_{ik}}\ \lf( \fc{(\xi_ak_j)(\xi_bk_l)}{z_{aj}z_{bl}} -  
\fc{(\xi_ak_l)(\xi_bk_j)}{z_{al}z_{bj}}\ri)
+\fc{k_ik_l}{z_{il}}\ \lf(\fc{(\xi_ak_k)(\xi_bk_j)}{z_{ak}z_{bj}} -
\fc{(\xi_ak_j)(\xi_bk_k)}{z_{aj}z_{bk}}\ri)}\cr
&\ss{+\fc{k_jk_k}{z_{jk}}\ \lf( \fc{(\xi_ak_l)(\xi_bk_i)}{z_{al}z_{bi}}-
\fc{(\xi_ak_i)(\xi_bk_l)}{z_{ai}z_{bl}} \ri)  
 +\fc{k_jk_l}{z_{jl}}\ \lf(\fc{(\xi_ak_i)(\xi_bk_k)}{z_{ai}z_{bk}}-
\fc{(\xi_ak_k)(\xi_bk_i)}{z_{ak}z_{bi}} \ri)}\cr
&\ss{+\lf.\fc{(1-k_ik_j)}{z_{ij}}\ \lf( \fc{(\xi_ak_k)(\xi_bk_l)}{z_{ak}z_{bl}}- 
\fc{(\xi_ak_l)(\xi_bk_l)}{z_{al}z_{bk}}\ri)\ \ri\}\  .}}}
Besides, the three functions:  \def\ss#1{{\scriptstyle{#1}}}
$${\eqalign{\hskip-0.3cm
\ss{\Cc_1(a,b,i,j,k,}&\ss{l)}=\ss{-\Ec \lf.
\fc{(\xi_jk_b)(\xi_kk_b)(\xi_lk_b)z_{ab}}{z_{ai}z_{bj}z_{bk}z_{bl}}
\lf(\fc{(\xi_ik_b)z_{ab}}{z_{aj}z_{ak}z_{al}z_{bi}}-\fc{(\xi_ik_j)}{z_{ak}z_{al}z_{ij}}-
\fc{(\xi_ik_k)}{z_{aj}z_{al}z_{ik}}-\fc{(\xi_ik_l)}{z_{aj}z_{ak}z_{il}}\ri)  \ri. } \cr
+&\ss{\fc{(\xi_jk_i)(\xi_kk_b)(\xi_lk_b)}{z_{aj}z_{bk}z_{bl}z_{ij}}
\lf(\fc{(\xi_ik_b)z_{ab}}{z_{ak}z_{al}z_{bi}}-\fc{(\xi_ik_k)}{z_{al}z_{ik}}-
\fc{(\xi_ik_l)}{z_{ak}z_{il}}\ri)+
\fc{(\xi_jk_b)(\xi_kk_i)(\xi_lk_b)}{z_{ak}z_{bj}z_{bl}z_{ik}}
\lf(\fc{(\xi_ik_b)z_{ab}}{z_{aj}z_{al}z_{bi}}-\fc{(\xi_ik_j)}{z_{al}z_{ij}}-
\fc{(\xi_ik_l)}{z_{aj}z_{il}}\ri)}\cr
+&\ss{\fc{(\xi_jk_k)(\xi_kk_b)(\xi_lk_b)}{z_{ai}z_{bk}z_{bl}z_{jk}}
\lf(\fc{(\xi_ik_j)}{z_{al}z_{ij}}-\fc{(\xi_ik_b)z_{ab}}{z_{aj}z_{al}z_{bi}}+
\fc{(\xi_ik_k)z_{ak}}{z_{aj}z_{al}z_{ik}}+\fc{(\xi_ik_l)}{z_{aj}z_{il}}\ri)+
\fc{(\xi_jk_i)(\xi_kk_i)(\xi_lk_b)z_{ai}}{z_{aj}z_{ak}z_{bl}z_{ij}z_{ik}}
\lf(\fc{(\xi_ik_b)}{z_{al}z_{bi}}-\fc{(\xi_ik_l)}{z_{ab}z_{il}}\ri)  }\cr
+&\ss{\fc{(\xi_jk_l)(\xi_kk_b)(\xi_lk_b)}{z_{ai}z_{bk}z_{bl}z_{jl}}
\lf(\fc{(\xi_ik_j)}{z_{ak}z_{ij}}-\fc{(\xi_ik_b)z_{ab}}{z_{aj}z_{ak}z_{bi}}+
\fc{(\xi_ik_k)}{z_{aj}z_{ik}}+\fc{(\xi_ik_l)z_{al}}{z_{aj}z_{ak}}\ri)+
\fc{(\xi_jk_k)(\xi_kk_i)(\xi_lk_b)}{z_{aj}z_{bl}z_{ik}z_{jk}}
\lf(\fc{(\xi_ik_l)}{z_{ab}z_{il}}-\fc{(\xi_ik_b)}{z_{al}z_{bi}}\ri) }\cr
+&\ss{\fc{(\xi_jk_l)(\xi_kk_i)(\xi_lk_b)}{z_{ak}z_{bl}z_{ik}z_{jl}}
\lf(\fc{(\xi_ik_j)}{z_{ab}z_{ij}}-\fc{(\xi_ik_b)}{z_{aj}z_{bi}}+
\fc{(\xi_ik_l)z_{al}}{z_{ab}z_{aj}z_{il}}\ri)+
\fc{(\xi_jk_b)(\xi_kk_j)(\xi_lk_b)}{z_{ai}z_{bj}z_{bl}z_{jk}}
\lf(\fc{(\xi_ik_b)z_{ab}}{z_{ak}z_{al}z_{bi}}-\fc{(\xi_ik_j)z_{aj}}{z_{ak}z_{al}z_{ij}}-
\fc{(\xi_ik_k)}{z_{al}z_{ik}}-\fc{(\xi_ik_l)}{z_{ak}z_{il}}\ri)}}}$$
%%%%%%%%%%%%%%%%%%%%%%%%%%%%%%%%%%%%%%%%%%%%%%%%%%%%%%%%%%%
$${\hskip-0.7cm\eqalign{
+&\ss{\fc{(\xi_jk_l)(\xi_kk_j)(\xi_lk_b)}{z_{ai}z_{bl}z_{jk}z_{jl}}
\lf(\fc{(\xi_ik_j)z_{aj}}{z_{ab}z_{ak}z_{ij}}-\fc{(\xi_ik_b)}{z_{ak}z_{bi}}+
\fc{(\xi_ik_k)}{z_{ab}z_{ik}}+\fc{(\xi_ik_l)z_{al}}{z_{ab}z_{ak}z_{il}}\ri)+
\fc{(\xi_jk_i)(\xi_kk_b)(\xi_lk_i)z_{ai}}{z_{aj}z_{al}z_{bk}z_{ij}z_{il}}
\lf(\fc{(\xi_ik_b)}{z_{ak}z_{bi}}-\fc{(\xi_ik_k)}{z_{ab}z_{ik}}\ri)}\cr
+&\ss{\fc{(\xi_jk_b)(\xi_kk_l)(\xi_lk_b)}{z_{ai}z_{bj}z_{bl}z_{kl}}
\lf(\fc{(\xi_ik_j)}{z_{ak}z_{ij}}-\fc{(\xi_ik_b)z_{ab}}{z_{aj}z_{ak}z_{bi}}+
\fc{(\xi_ik_k)}{z_{aj}z_{ik}}+\fc{(\xi_ik_l)z_{al}}{z_{aj}z_{ak}z_{il}}\ri)+
\fc{(\xi_jk_i)(\xi_kk_l)(\xi_lk_b)}{z_{aj}z_{bl}z_{ij}z_{kl}}
\lf(\fc{(\xi_ik_k)}{z_{ab}z_{ik}}-\fc{(\xi_ik_b)}{z_{ak}z_{bi}}+
\fc{(\xi_ik_l)z_{al}}{z_{ab}z_{ak}z_{il}}\ri)}\cr
+&\ss{\fc{(\xi_jk_k)(\xi_kk_l)(\xi_lk_b)}{z_{ai}z_{bl}z_{jk}z_{kl}}
\lf(\fc{(\xi_ik_b)}{z_{aj}z_{bi}}-\fc{(\xi_ik_j)}{z_{ab}z_{ij}}-
\fc{(\xi_ik_k)z_{ak}}{z_{ab}z_{aj}z_{ik}}-\fc{(\xi_ik_l)z_{al}}{z_{ab}z_{aj}z_{il}}\ri)+
\fc{(\xi_jk_b)(\xi_kk_b)(\xi_lk_i)}{z_{al}z_{bj}z_{bk}z_{il}}
\lf(\fc{(\xi_ik_b)}{z_{aj}z_{ak}z_{bi}}-\fc{(\xi_ik_j)}{z_{ak}z_{ij}}-
\fc{(\xi_ik_k)}{z_{aj}z_{ik}}\ri)}\cr
+&\ss{\fc{(\xi_jk_l)(\xi_kk_l)(\xi_lk_b)z_{al}}{z_{ai}z_{bl}z_{jl}z_{kl}}
\lf(\fc{(\xi_ik_b)}{z_{aj}z_{ak}z_{bi}}-\fc{(\xi_ik_j)}{z_{ab}z_{ak}z_{ij}}-
\fc{(\xi_ik_k)}{z_{ab}z_{aj}z_{ik}}-\fc{(\xi_ik_l)z_{al}}{z_{ab}z_{aj}z_{ak}z_{il}}\ri)+
\fc{(\xi_jk_i)(\xi_kk_j)(\xi_lk_b)}{z_{ak}z_{bl}z_{ij}z_{jk}}
\lf(\fc{(\xi_ik_b)}{z_{al}z_{bi}}-\fc{(\xi_ik_l)}{z_{ab}z_{il}}\ri)}\cr
+&\ss{\fc{(\xi_jk_k)(\xi_kk_b)(\xi_lk_i)}{z_{al}z_{bk}z_{il}z_{jk}}
\lf(\fc{(\xi_ik_j)}{z_{ab}z_{ij}}-\fc{(\xi_ik_b)}{z_{aj}z_{bi}}+
\fc{(\xi_ik_k)z_{ak}}{z_{ab}z_{aj}z_{ik}}\ri)+
\fc{(\xi_jk_l)(\xi_kk_b)(\xi_lk_i)}{z_{aj}z_{bk}z_{il}z_{jl}}
\lf(\fc{(\xi_ik_k)}{z_{ab}z_{ik}}-\fc{(\xi_ik_b)}{z_{ak}z_{bi}}\ri)}\cr
+&\ss{\fc{(\xi_jk_b)(\xi_kk_i)(\xi_lk_i)z_{ai}}{z_{ak}z_{al}z_{bj}z_{ik}z_{il}}
\lf(\fc{(\xi_ik_b)}{z_{aj}z_{bi}}-\fc{(\xi_ik_j)}{z_{ab}z_{ij}}\ri)+
\fc{(\xi_ik_b)(\xi_jk_i)(\xi_kk_i)(\xi_lk_i)z^2_{ai}}
{z_{ab}z_{aj}z_{ak}z_{al}z_{bi}z_{ij}z_{ik}z_{il}}-
\fc{(\xi_kk_i)(\xi_lk_i)z_{ai}}{z_{ab}z_{aj}z_{bi}z_{ik}z_{il}}
\lf(\fc{(\xi_ik_b)(\xi_jk_k)}{z_{al}z_{jk}}+\fc{(\xi_ik_b)(\xi_jk_l)}{z_{ak}z_{jl}}\ri)}\cr
+&\ss{\fc{(\xi_jk_b)(\xi_kk_j)(\xi_lk_i)}{z_{al}z_{bj}z_{il}z_{jk}}
\lf(\fc{(\xi_ik_b)}{z_{al}z_{bi}}-\fc{(\xi_ik_j)z_{aj}}{z_{ab}z_{ak}z_{ij}}-
\fc{(\xi_ik_k)}{z_{ab}z_{ik}}\ri)+
\fc{(\xi_kk_j)(\xi_lk_i)}{z_{ab}z_{ak}z_{bi}z_{il}z_{jk}}
\lf(\fc{(\xi_ik_b)(\xi_jk_i)z_{ai}}{z_{al}z_{ij}}-\fc{(\xi_ik_b)(\xi_jk_l)}{z_{jl}}\ri)}\cr
+&\ss{\fc{(\xi_jk_b)(\xi_kk_l)(\xi_lk_i)}{z_{ak}z_{bj}z_{il}z_{kl}}
\lf(\fc{(\xi_ik_j)}{z_{ab}z_{ij}}-\fc{(\xi_ik_b)}{z_{aj}z_{bi}}\ri)+
\fc{(\xi_kk_l)(\xi_lk_i)}{z_{ab}z_{aj}z_{bi}z_{il}z_{kl}}
\lf(\fc{(\xi_ik_b)(\xi_jk_k)}{z_{jk}}-\fc{(\xi_ik_b)(\xi_jk_i)z_{ai}}{z_{ak}z_{ij}}+
\fc{(\xi_ik_b)(\xi_jk_l)z_{al}}{z_{ak}z_{jl}}\ri)}\cr
+&\ss{\fc{(\xi_jk_b)(\xi_kk_b)(\xi_lk_j)}{z_{aj}z_{bj}z_{bk}z_{jl}}
\lf(\fc{(\xi_ik_b)z_{ab}}{z_{ak}z_{al}z_{bi}}-\fc{(\xi_ik_j)z_{aj}}{z_{ak}z_{al}}-
\fc{(\xi_ik_k)}{z_{al}z_{ik}}-\fc{(\xi_ik_l)}{z_{ak}z_{il}}\ri)+
\fc{(\xi_jk_i)(\xi_kk_b)(\xi_lk_j)}{z_{al}z_{bk}z_{ij}z_{jl}}
\lf(\fc{(\xi_ik_b)}{z_{ak}z_{bi}}-\fc{(\xi_ik_k)}{z_{ab}z_{ik}}\ri)}\cr
+&\ss{\fc{(\xi_jk_k)(\xi_kk_b)(\xi_lk_j)}{z_{ai}z_{bk}z_{jk}z_{jl}}
\lf(\fc{(\xi_ik_j)z_{aj}}{z_{ab}z_{al}z_{ij}}-\fc{(\xi_ik_b)}{z_{al}z_{bi}}+
\fc{(\xi_ik_k)z_{ak}}{z_{ab}z_{al}z_{ik}}+\fc{(\xi_ik_l)}{z_{ab}z_{il}}\ri)+
\fc{(\xi_jk_b)(\xi_kk_i)(\xi_lk_j)}{z_{ak}z_{bj}z_{ik}z_{jl}}
\lf(\fc{(\xi_ik_b)}{z_{al}z_{bi}}-\fc{(\xi_ik_j)z_{aj}}{z_{ab}z_{al}z_{ij}}-
\fc{(\xi_ik_l)}{z_{ab}z_{il}}\ri)}\cr
+&\ss{\fc{(\xi_kk_i)(\xi_lk_j)}{z_{ab}z_{al}z_{bi}z_{ik}z_{jl}}
\lf(\fc{(\xi_ik_b)(\xi_jk_i)z_{ai}}{z_{ak}z_{ij}}-\fc{(\xi_ik_b)(\xi_jk_k)}{z_{jk}}\ri)+
\fc{(\xi_jk_b)(\xi_kk_l)(\xi_lk_j)}{z_{ai}z_{bj}z_{jl}z_{kl}}
\lf(\fc{(\xi_ik_j)z_{aj}}{z_{ab}z_{ak}z_{ij}}-\fc{(\xi_ik_b)}{z_{ak}z_{bi}}+
\fc{(\xi_ik_k)}{z_{ab}z_{ik}}+\fc{(\xi_ik_l)z_{al}}{z_{ab}z_{ak}z_{il}}\ri)}\cr
+&\ss{\fc{(\xi_jk_b)(\xi_kk_j)(\xi_lk_j)z_{aj}}{z_{ai}z_{bj}z_{jk}z_{jl}}
\lf(\fc{(\xi_ik_b)}{z_{ak}z_{al}z_{bi}}-\fc{(\xi_ik_j)z_{aj}}{z_{ab}z_{ak}z_{al}z_{ij}}-
\fc{(\xi_ik_k)}{z_{ab}z_{al}z_{ik}}-\fc{(\xi_ik_l)}{z_{ab}z_{ak}z_{il}}\ri)+
\fc{(\xi_ik_b)(\xi_jk_i)(\xi_kk_j)(\xi_lk_j)z_{aj}}{z_{ab}z_{ak}z_{al}z_{bi}z_{ij}z_{jk}z_{jl}}}\cr
-&\ss{\fc{(\xi_ik_b)(\xi_jk_i)(\xi_kk_l)(\xi_lk_j)}{z_{ab}z_{ak}z_{bi}z_{ij}z_{jl}z_{kl}}+
\fc{(\xi_jk_b)(\xi_kk_b)(\xi_lk_k)}{z_{ai}z_{bj}z_{bk}z_{kl}}
\lf(\fc{(\xi_ik_b)z_{ab}}{z_{aj}z_{al}z_{bi}}-\fc{(\xi_ik_j)}{z_{al}z_{ij}}-
\fc{(\xi_ik_k)z_{ak}}{z_{aj}z_{al}z_{ik}}-\fc{(\xi_ik_l)}{z_{aj}z_{il}}\ri)}\cr
+&\ss{\fc{(\xi_jk_i)(\xi_kk_b)(\xi_lk_k)}{z_{aj}z_{bk}z_{ij}z_{kl}}
\lf(\fc{(\xi_ik_b)}{z_{al}z_{bi}}-\fc{(\xi_ik_k)z_{ak}}{z_{ab}z_{al}z_{ik}}-
\fc{(\xi_ik_l)}{z_{ab}z_{il}}\ri)+
\fc{(\xi_jk_b)(\xi_kk_i)(\xi_lk_k)}{z_{al}z_{bj}z_{ik}z_{kl}}
\lf(\fc{(\xi_ik_b)}{z_{aj}z_{bi}}-\fc{(\xi_ik_j)}{z_{ab}z_{ij}}\ri)}\cr
+&\ss{\fc{(\xi_jk_k)(\xi_kk_b)(\xi_lk_k)z_{ak}}{z_{ai}z_{bk}z_{ik}z_{kl}}
\lf(\fc{(\xi_ik_j)}{z_{ab}z_{al}z_{ij}}-\fc{(\xi_ik_b)}{z_{aj}z_{al}z_{bi}}+
\fc{(\xi_ik_k)z_{ak}}{z_{ab}z_{aj}z_{al}z_{ik}}+\fc{(\xi_ik_l)}{z_{ab}z_{aj}z_{il}}\ri)-
\fc{(\xi_ik_b)(\xi_jk_l)(\xi_kk_i)(\xi_lk_j)}{z_{ab}z_{aj}z_{bi}z_{ik}z_{jl}z_{kl}}}\cr
+&\ss{\fc{(\xi_jk_l)(\xi_kk_b)(\xi_lk_k)}{z_{ai}z_{bk}z_{jl}z_{kl}}
\lf(\fc{(\xi_ik_j)}{z_{ab}z_{ij}}-\fc{(\xi_ik_b)}{z_{aj}z_{bi}}+
\fc{(\xi_ik_k)z_{ak}}{z_{ab}z_{aj}z_{ik}}+\fc{(\xi_ik_l)z_{al}}{z_{ab}z_{aj}z_{il}}\ri)+
\fc{(\xi_ik_b)(\xi_kk_i)(\xi_lk_k)}{z_{ab}z_{aj}z_{al}z_{bi}z_{ik}z_{kl}}
\lf(\fc{(\xi_jk_i)z_{ai}}{z_{ij}}-\fc{(\xi_jk_k)z_{ak}}{z_{jk}}\ri)}\cr
+&\ss{\fc{(\xi_jk_b)(\xi_kk_j)(\xi_lk_k)}{z_{ai}z_{bj}z_{jk}z_{kl}}
\lf(\fc{(\xi_ik_b)}{z_{al}z_{bi}}-\fc{(\xi_ik_j)z_{aj}}{z_{ab}z_{al}z_{ij}}-
\fc{(\xi_ik_k)z_{ak}}{z_{ab}z_{al}z_{ik}}-\fc{(\xi_ik_l)}{z_{ab}z_{il}}\ri)+
\fc{(\xi_ik_b)(\xi_jk_i)(\xi_kk_j)(\xi_lk_k)}{z_{ab}z_{al}z_{bi}z_{ij}z_{jk}z_{kl}}}\ ,\cr
%%%%%%%%%%%%%%%%%%%%%%%%%%%%%%%%%%%%%%%%%%%%%%%%%%%%%%%%%%%%%%%%%%
\ss{\Cc_2(a,i,b,j,k,l)}&=\ss{-\Ec\ \lf\{ 
\fc{(\xi_kk_b)(\xi_lk_b)z_{ab}}{z_{ai}z_{aj}z_{bi}z_{bk}z_{bl}}\ 
\lf(\fc{(\xi_bk_i)(\xi_jk_l)}{z_{ak}z_{jl}}
+\fc{(\xi_bk_i)(\xi_jk_k)}{z_{al}z_{jk}}-\fc{(\xi_bk_i)(\xi_jk_b)z_{ab}}{z_{ak}z_{al}z_{bj}}\ri)
\ri.}\cr
+&\ss{\fc{(\xi_jk_l)(\xi_kk_i)(\xi_lk_b)}{z_{aj}z_{bl}z_{ik}z_{jl}}\ 
\lf(\fc{(\xi_bk_i)}{z_{ak}z_{bi}}+\fc{(\xi_bk_k)}{z_{ai}z_{bk}}\ri)+
\fc{(\xi_bk_i)(\xi_kk_l)(\xi_lk_b)}{z_{ai}z_{aj}z_{bi}z_{bl}z_{kl}}
\ \lf(\fc{(\xi_jk_b)z_{ab}}{z_{ak}z_{bj}}
-\fc{(\xi_jk_k)}{z_{jk}}-\fc{(\xi_jk_l)}{z_{ak}z_{jl}}\ri)}\cr  
-&\ss{\fc{(\xi_kk_i)(\xi_lk_b)}{z_{al}z_{bl}z_{ik}}\ 
\lf(\fc{(\xi_bk_i)(\xi_jk_b)z_{ab}}{z_{aj}z_{ak}z_{bi}z_{bj}}
+\fc{(\xi_bk_k)(\xi_jk_b)z_{ab}}{z_{ai}z_{aj}z_{bj}z_{bk}}+
\fc{(\xi_bk_i)(\xi_jk_i)z_{ai}}{z_{aj}z_{ak}z_{bi}z_{ij}}+
\fc{(\xi_bk_j)(\xi_jk_i)}{z_{bj}z_{ij}}+\fc{(\xi_bk_k)(\xi_jk_i)}{z_{aj}z_{bk}z_{ij}}\ri)}\cr
+&\ss{\fc{(\xi_jk_k)(\xi_kk_i)(\xi_lk_b)}{z_{bl}z_{ik}z_{jk}}\ 
\lf(\fc{(\xi_bk_i)}{z_{aj}z_{al}z_{bi}}
+\fc{(\xi_bk_j)}{z_{ai}z_{al}z_{bj}}+\fc{(\xi_bk_k)z_{ak}}{z_{ai}z_{aj}z_{al}z_{bk}}\ri)+
\fc{(\xi_jk_i)(\xi_kk_l)(\xi_lk_b)}{z_{ak}z_{bl}z_{ij}z_{kl}}\ 
\lf(\fc{(\xi_bk_i)}{z_{aj}z_{bi}}+\fc{(\xi_bk_j)}{z_{ai}z_{bj}}\ri)}\cr
-&\ss{\fc{(\xi_jk_i)(\xi_kk_b)(\xi_lk_b)z_{ab}}{z_{ak}z_{al}z_{bk}z_{bl}z_{ij}}\ 
\lf(\fc{(\xi_bk_i)}{z_{aj}z_{bi}}+\fc{(\xi_bk_j)}{z_{ai}z_{bj}}\ri)-
\fc{(\xi_jk_i)(\xi_kk_j)(\xi_lk_b)}{z_{al}z_{bl}z_{ij}z_{jk}}\ 
\lf(\fc{(\xi_bk_i)}{z_{ak}z_{bi}}+\fc{(\xi_bk_j)z_{aj}}{z_{ai}z_{ak}z_{bj}}+
\fc{(\xi_bk_k)}{z_{ai}z_{bk}}\ri)}\cr
-&\ss{\fc{(\xi_jk_b)(\xi_kk_b)(\xi_lk_i)z_{ab}}{z_{aj}z_{ak}z_{bj}z_{bk}z_{il}}\ 
\lf(\fc{(\xi_bk_i)}{z_{al}z_{bi}}+\fc{(\xi_bk_l)}{z_{ai}z_{bl}}\ri)+
\fc{(\xi_jk_b)(\xi_kk_l)(\xi_lk_i)}{z_{aj}z_{bj}z_{il}z_{kl}}\ 
\lf(\fc{(\xi_bk_i)}{z_{ak}z_{bi}}+\fc{(\xi_bk_k)}{z_{ai}z_{bk}}+
\fc{(\xi_bk_l)z_{al}}{z_{ai}z_{ak}z_{bl}}\ri)}\cr
-&\ss{\fc{(\xi_jk_i)(\xi_kk_b)(\xi_lk_i)}{z_{ak}z_{bk}z_{ij}z_{il}}\ 
\lf(\fc{(\xi_bk_i)z_{ai}}{z_{aj}z_{al}z_{bi}}+\fc{(\xi_bk_j)}{z_{al}z_{bj}z_{il}}+
\fc{(\xi_bk_l)}{z_{aj}z_{bl}z_{il}}\ri)+
\fc{(\xi_jk_k)(\xi_kk_b)(\xi_lk_i)}{z_{aj}z_{bk}z_{il}z_{jk}}\ 
\lf(\fc{(\xi_bk_i)}{z_{al}z_{bi}}+\fc{(\xi_bk_l)}{z_{ai}z_{bl}}\ri)}\cr
+&\ss{\fc{(\xi_jk_l)(\xi_kk_b)(\xi_lk_i)}{z_{ak}z_{bk}z_{il}z_{jl}}\ 
\lf(\fc{(\xi_bk_i)}{z_{aj}z_{bi}}+\fc{(\xi_bk_j)}{z_{ai}z_{bj}}+
\fc{(\xi_bk_l)z_{al}}{z_{ai}z_{aj}z_{bl}}\ri)-
\fc{(\xi_jk_b)(\xi_kk_j)(\xi_lk_i)}{z_{ak}z_{bj}z_{il}z_{jk}}\ 
\lf(\fc{(\xi_bk_i)}{z_{al}z_{bi}}+\fc{(\xi_bk_l)}{z_{ai}z_{bl}}\ri)}}}$$
%%%%%%%%%%%%%%%%%%%%%%%%%%%%%%%%%%%%%%%%%%%%%%%%%%%%%%%%%%%%%%%%%%%%%%%%
$${\hskip-0.7cm\eqalign{
-&\ss{\fc{(\xi_jk_b)(\xi_kk_i)(\xi_lk_i)}{z_{aj}z_{bj}z_{ik}z_{il}}\ 
\lf(\fc{(\xi_bk_i)z_{ai}}{z_{ak}z_{al}z_{bi}}+
\fc{(\xi_bk_k)}{z_{al}z_{bk}}+\fc{(\xi_bk_l)}{z_{ak}z_{bl}}\ri)
+\fc{(\xi_bk_i)(\xi_kk_j)(\xi_lk_b)}{z_{ai}z_{ak}z_{bi}z_{bl}z_{jk}}\ \lf(\fc{(\xi_jk_l)}{z_{jl}}
-\fc{(\xi_jk_b)z_{ab}}{z_{al}z_{bj}}\ri)}\cr
-&\ss{\fc{(\xi_jk_i)(\xi_kk_i)(\xi_lk_i)z_{ai}}{z_{ab}z_{ij}z_{ik}z_{il}}\ 
\lf(\fc{(\xi_bk_i)z_{ai}}{z_{aj}z_{ak}z_{al}z_{bi}}+
\fc{(\xi_bk_j)}{z_{ak}z_{al}z_{bj}}+\fc{(\xi_bk_k)}{z_{aj}z_{al}z_{bk}}+ 
\fc{(\xi_bk_l)}{z_{aj}z_{ak}z_{bl}}\ri)}\cr
+&\ss{\fc{(\xi_jk_k)(\xi_kk_i)(\xi_lk_i)}{z_{ab}z_{ik}z_{il}z_{jk}}\ 
\lf(\fc{(\xi_bk_i)z_{ai}}{z_{aj}z_{al}z_{bi}}
+\fc{(\xi_bk_j)}{z_{al}z_{bj}}+\fc{(\xi_bk_k)z_{ak}}{z_{aj}z_{al}z_{bk}}+
\fc{(\xi_bk_l)}{z_{aj}z_{bl}}\ri)+
\fc{(\xi_bk_i)(\xi_jk_k)(\xi_kk_b)(\xi_lk_j)}{z_{ai}z_{al}z_{bi}z_{bk}z_{jk}z_{jl}}}\cr
+&\ss{\fc{(\xi_jk_l)(\xi_kk_i)(\xi_lk_i)}{z_{aj}z_{ik}z_{il}z_{jl}}\ 
\lf(\fc{(\xi_bk_i)z_{ai}}{z_{aj}z_{ak}z_{bi}}
+\fc{(\xi_bk_j)}{z_{ak}z_{bj}}+\fc{(\xi_bk_k)}{z_{aj}z_{bk}}+
\fc{(\xi_bk_l)z_{al}}{z_{aj}z_{ak}z_{bl}}\ri)-
\fc{(\xi_bk_i)(\xi_jk_b)(\xi_kk_b)(\xi_lk_j)z_{ab}}{z_{ai}z_{ak}z_{al}z_{bi}z_{bj}z_{bk}z_{jl}}}\cr
-&\ss{\fc{(\xi_jk_i)(\xi_kk_j)(\xi_lk_i)}{z_{ab}z_{ij}z_{il}z_{jk}}\ 
\lf(\fc{(\xi_bk_i)z_{ai}}{z_{ak}z_{al}z_{bi}}+
\fc{(\xi_bk_j)z_{aj}}{z_{ak}z_{al}z_{bj}}+\fc{(\xi_bk_k)}{z_{al}z_{bk}}+
\fc{(\xi_bk_l)}{z_{ak}z_{bl}}\ri)-
\fc{(\xi_bk_i)(\xi_jk_b)(\xi_kk_j)(\xi_lk_j)z_{aj}}{z_{ai}z_{ak}z_{al}z_{bi}z_{bj}z_{jk}z_{jl}}}\cr
+&\ss{\fc{(\xi_jk_l)(\xi_kk_j)(\xi_lk_i)}{z_{ab}z_{il}z_{jk}z_{jl}}\ 
\lf(\fc{(\xi_bk_i)}{z_{ak}z_{bi}}+
\fc{(\xi_bk_j)z_{aj}}{z_{ai}z_{ak}z_{bj}}+\fc{(\xi_bk_k)}{z_{ai}z_{bk}}+
\fc{(\xi_bk_l)z_{al}}{z_{ai}z_{ak}z_{bl}}\ri)+
\fc{(\xi_bk_i)(\xi_jk_b)(\xi_kk_l)(\xi_lk_j)}{z_{ai}z_{ak}z_{bi}z_{bj}z_{jl}z_{kl}}}\cr
+&\ss{\fc{(\xi_jk_i)(\xi_kk_l)(\xi_lk_i)}{z_{ab}z_{ij}z_{il}z_{kl}}\ 
\lf(\fc{(\xi_bk_i)z_{ai}}{z_{aj}z_{ak}z_{bi}}+\fc{(\xi_bk_j)}{z_{ak}z_{bj}}+
\fc{(\xi_bk_k)}{z_{aj}z_{bk}}+\fc{(\xi_bk_l)z_{al}}{z_{aj}z_{ak}z_{bl}}\ri)
-\fc{(\xi_bk_i)(\xi_jk_b)(\xi_kk_b)(\xi_lk_k)
z_{ab}}{z_{ai}z_{aj}z_{al}z_{bi}z_{bj}z_{bk}z_{kl}}}\cr
-&\ss{\fc{(\xi_jk_k)(\xi_kk_l)(\xi_lk_i)}{z_{ab}z_{il}z_{jk}z_{kl}}\ 
\lf(\fc{(\xi_bk_i)}{z_{aj}z_{bi}}+\fc{(\xi_bk_j)}{z_{ai}z_{bj}}+
\fc{(\xi_bk_k)z_{ak}}{z_{ai}z_{aj}z_{bk}}+
\fc{(\xi_bk_l)z_{al}}{z_{ai}z_{aj}z_{bl}}\ri)+
\fc{(\xi_bk_i)(\xi_jk_k)(\xi_kk_b)(\xi_lk_k)z_{ak}}{z_{ai}z_{aj}z_{al}z_{bi}z_{bk}z_{jk}z_{kl}}}\cr
-&\ss{\fc{(\xi_jk_l)(\xi_kk_l)(\xi_lk_i)z_{al}}{z_{ab}z_{il}z_{jl}z_{kl}}\ 
\lf(\fc{(\xi_bk_i)}{z_{aj}z_{ak}z_{bi}}+\fc{(\xi_bk_j)}{z_{ai}z_{ak}z_{bj}}+
\fc{(\xi_bk_k)}{z_{ai}z_{aj}z_{bk}}+\fc{(\xi_bk_l)z_{al}}{z_{ai}z_{aj}z_{ak}z_{bl}}\ri)}\cr
-&\ss{\fc{(\xi_jk_i)(\xi_kk_b)(\xi_lk_j)}{z_{ak}z_{bk}z_{ij}z_{jl}}\ 
\lf(\fc{(\xi_bk_i)}{z_{al}z_{bi}}+\fc{(\xi_bk_j)z_{aj}}{z_{ai}z_{al}z_{bj}}+
\fc{(\xi_bk_l)}{z_{ai}z_{bl}}\ri)-\fc{(\xi_jk_b)(\xi_kk_i)(\xi_lk_j)}{z_{al}z_{bj}z_{ik}z_{jl}}
\lf(\fc{(\xi_bk_i)}{z_{ak}z_{bi}}+\fc{(\xi_bk_k)}{z_{ai}z_{bk}}\ri)}\cr
-&\ss{\fc{(\xi_jk_i)(\xi_kk_i)(\xi_lk_j)}{z_{ab}z_{ij}z_{ik}z_{jl}}
\lf(\fc{(\xi_bk_i)z_{ai}}{z_{ak}z_{al}z_{bi}}
+\fc{(\xi_bk_j)z_{aj}}{z_{ak}z_{al}z_{bj}}+\fc{(\xi_bk_k)}{z_{al}z_{bk}}+
\fc{(\xi_bk_l)}{z_{ak}z_{bl}}\ri)+
\fc{(\xi_bk_i)(\xi_jk_l)(\xi_kk_b)(\xi_lk_k)}{z_{ai}z_{aj}z_{bi}z_{bk}z_{jl}z_{kl}}}\cr
+&\ss{\fc{(\xi_jk_k)(\xi_kk_i)(\xi_lk_j)}{z_{ab}z_{ik}z_{jk}z_{jl}}\ 
\lf(\fc{(\xi_bk_i)}{z_{al}z_{bi}}+\fc{(\xi_bk_j)z_{aj}}{z_{ai}z_{al}z_{bj}}+
\fc{(\xi_bk_k)z_{ak}}{z_{ai}z_{al}z_{bk}}+\fc{(\xi_bk_l)}{z_{ai}z_{bl}}\ri)-
\fc{(\xi_bk_i)(\xi_jk_i)(\xi_kk_j)(\xi_lk_k)}{z_{ab}z_{al}z_{bi}z_{ij}z_{jk}z_{kl}}}\cr
-&\ss{\fc{(\xi_jk_i)(\xi_kk_j)(\xi_lk_j)z_{aj}}{z_{ab}z_{ij}z_{jk}z_{jl}}\ 
\lf(\fc{(\xi_bk_i)}{z_{ak}z_{al}z_{bi}}+\fc{(\xi_bk_j)z_{aj}}{z_{ai}z_{ak}z_{al}z_{bj}}+
\fc{(\xi_bk_k)}{z_{ai}z_{al}z_{bk}}+\fc{(\xi_bk_l)}{z_{ai}z_{ak}z_{bl}}\ri)}\cr
+&\ss{\fc{(\xi_jk_i)(\xi_kk_l)(\xi_lk_j)}{z_{ab}z_{ij}z_{jl}z_{kl}}\ 
\lf(\fc{(\xi_bk_i)}{z_{ak}z_{bi}}+\fc{(\xi_bk_j)z_{aj}}{z_{ai}z_{ak}z_{bj}}+
\fc{(\xi_bk_k)}{z_{ai}z_{bk}}+\fc{(\xi_bk_l)}{z_{ai}z_{ak}z_{bl}}\ri)-
\fc{(\xi_bk_j)(\xi_jk_i)(\xi_kk_j)(\xi_lk_k)}{z_{ab}z_{ai}z_{al}z_{bj}z_{ij}z_{jk}z_{kl}}}\cr
-&\ss{\fc{(\xi_jk_i)(\xi_kk_b)(\xi_lk_k)}{z_{al}z_{bk}z_{ij}z_{kl}}\ 
\lf(\fc{(\xi_bk_i)}{z_{aj}z_{bi}}+\fc{(\xi_bk_j)}{z_{ai}z_{bj}}\ri)-
\fc{(\xi_bk_i)(\xi_jk_b)(\xi_kk_j)(\xi_lk_k)}{z_{ai}z_{al}z_{bi}z_{bj}z_{jk}z_{kl}}} \cr
-&\ss{\fc{(\xi_jk_b)(\xi_kk_i)(\xi_lk_k)}{z_{aj}z_{bj}z_{ik}z_{kl}}\ 
\lf(\fc{(\xi_bk_i)}{z_{al}z_{bi}}+\fc{(\xi_bk_k)z_{ak}}{z_{ai}z_{al}z_{bk}}+
\fc{(\xi_bk_l)}{z_{ai}z_{bl}} \ri)-
\fc{(\xi_bk_k)(\xi_jk_i)(\xi_kk_j)(\xi_lk_k)}{z_{ab}z_{ai}z_{al}z_{bk}z_{ij}z_{jk}z_{kl}}}\cr
-&\ss{\fc{(\xi_jk_i)(\xi_kk_i)(\xi_lk_k)}{z_{ab}z_{ij}z_{ik}z_{kl}}\ 
\lf(\fc{(\xi_bk_i)z_{ai}}{z_{aj}z_{al}z_{bi}}
+\fc{(\xi_bk_j)}{z_{al}z_{bj}}+\fc{(\xi_bk_k)z_{ak}}{z_{aj}z_{al}z_{bk}}+
\fc{(\xi_bk_l)}{z_{aj}z_{bl}}\ri)-
\fc{(\xi_bk_l)(\xi_jk_i)(\xi_kk_j)(\xi_lk_k)}{z_{ab}z_{ai}z_{bl}z_{ij}z_{jk}z_{kl}}}\cr
+&\ss{\fc{(\xi_jk_k)(\xi_kk_i)(\xi_lk_k)z_{ak}}{z_{ab}z_{ik}z_{jk}z_{kl}}\ 
\lf(\fc{(\xi_bk_i)z_{aj}}{z_{al}z_{bi}}
+\fc{(\xi_bk_j)}{z_{ai}z_{al}z_{bj}}+\fc{(\xi_bk_k)z_{ak}}{z_{ai}z_{aj}z_{al}z_{bk}}+ 
\fc{(\xi_bk_l)}{z_{ai}z_{aj}z_{bl}}\ri)}\cr
+&\ss{\fc{(\xi_jk_l)(\xi_kk_i)(\xi_lk_k)}{z_{ab}z_{ik}z_{jl}z_{kl}}\ 
\lf(\fc{(\xi_bk_i)}{z_{aj}z_{bi}}+
\fc{(\xi_bk_j)}{z_{ai}z_{bj}}+\fc{(\xi_bk_k)z_{ak}}{z_{ai}z_{aj}z_{bk}}+ 
\fc{(\xi_bk_l)z_{al}}{z_{ai}z_{aj}z_{bl}}\ri)\ ,}\cr
\ss{\Cc_3(i,j,a,b,k,l)}&=\ss{-\fc{\Ec}{z_{ij}}\ 
\lf\{\fc{(\xi_ak_i)(\xi_bk_j)(\xi_kk_b)(\xi_lk_b)z_{ab}}{z_{ai}z_{ak}z_{al}z_{bj}z_{bk}z_{bl}} + 
 \fc{(\xi_ak_j)(\xi_bk_i)(\xi_kk_b)(\xi_lk_b)z_{ab}}{z_{aj}z_{ak}z_{al}z_{bi}z_{bk}z_{bl}} \ri.}\cr
&\ss{-\fc{(\xi_ak_j)(\xi_bk_i)(\xi_kk_i)(\xi_lk_b)z_{ai}}{z_{aj}z_{ak}z_{al}z_{bi}z_{bl}z_{ik}}-
\fc{(\xi_ak_j)(\xi_bk_k)(\xi_kk_i)(\xi_lk_b)}{z_{aj}z_{al}z_{bk}z_{bl}z_{ik}}+ 
\fc{(\xi_ak_i+\xi_ak_k)(\xi_bk_j)(\xi_kk_i)(\xi_lk_b)}{z_{al}z_{bj}z_{bl}z_{ik}}}\cr
&\ss{-\fc{(\xi_ak_j+\xi_ak_k)(\xi_bk_i)(\xi_kk_j)(\xi_lk_b)}{z_{ak}z_{al}z_{bi}z_{bl}z_{jk}}+
\fc{(\xi_ak_i)(\xi_bk_j)(\xi_kk_j)(\xi_lk_b)z_{aj}}{z_{ai}z_{ak}z_{al}z_{bj}z_{bl}z_{jk}}+
\fc{(\xi_ak_i)(\xi_bk_k)(\xi_kk_j)(\xi_lk_b)}{z_{ai}z_{al}z_{bk}z_{bl}z_{jk}}}\cr
&\ss{+\fc{(\xi_ak_j)(\xi_bk_i)(\xi_kk_l)(\xi_lk_b)}{z_{aj}z_{ak}z_{bi}z_{bl}z_{kl}}
+\fc{(\xi_ak_i)(\xi_bk_j)(\xi_kk_l)(\xi_lk_b)}{z_{ai}z_{ak}z_{bj}z_{bl}z_{kl}}
+\fc{(\xi_ak_i+\xi_ak_k)(\xi_bk_j)(\xi_kk_b)(\xi_lk_i)}{z_{ak}z_{al}z_{bj}z_{bk}z_{il}} }\cr
&\ss{-\fc{(\xi_ak_j)(\xi_bk_l)(\xi_kk_b)(\xi_lk_i)}{z_{aj}z_{ak}z_{bk}z_{bl}z_{il}}+
\fc{(\xi_kk_i)(\xi_lk_i)z_{ai}}{z_{ab}z_{ak}z_{al}z_{ik}z_{il}}\lf(\fc{1}{z_{bj}}-
\fc{z_{ai}}{z_{aj}z_{bi}}  \ri)-
\fc{(\xi_ak_j)(\xi_bk_i)(\xi_kk_b)(\xi_lk_i)z_{ai}}{z_{aj}z_{ak}z_{al}z_{bi}z_{bk}z_{il}}}\cr
&\ss{+\fc{(\xi_ak_k+\xi_ak_l)(\xi_bk_j)(\xi_kk_i)(\xi_lk_i)
z_{ai}}{z_{ab}z_{ak}z_{al}z_{bj}z_{ik}z_{il}}-
\fc{(\xi_ak_j)(\xi_kk_i)(\xi_lk_i)z_{ai}}{z_{ab}z_{aj}z_{ik}z_{il}}\lf(\fc{\xi_bk_k}{z_{al}z_{bk}}+
\fc{\xi_bk_l}{z_{ak}z_{bl}}  \ri)}\cr
&\ss{-\fc{(\xi_ak_j+\xi_ak_k)(\xi_bk_i)(\xi_kk_j)(\xi_lk_i)
z_{ai}}{z_{ab}z_{ak}z_{al}z_{bi}z_{il}z_{jk}}+
\fc{(\xi_ak_i+\xi_ak_l)(\xi_bk_j)(\xi_kk_j)(\xi_lk_i)z_{aj}}{z_{ab}z_{ak}z_{al}z_{bj}
z_{il}z_{jk}}}}}$$
%%%%%%%%%%%%%%%%%%%%%%%%%%%%%%%%%%%%%%%%%%%%%%%%%%%
\eqn\KINC{\eqalign{
&\ss{+\fc{(\xi_ak_i+\xi_ak_l)(\xi_bk_k)(\xi_kk_j)(\xi_lk_i)}{z_{ab}z_{al}z_{bk}z_{il}z_{jk}}-
\fc{(\xi_ak_j+\xi_ak_k)(\xi_bk_l)(\xi_kk_j)(\xi_lk_i)}{z_{ab}z_{ak}z_{bl}z_{il}z_{jk}}-
\fc{(\xi_ak_j)(\xi_bk_i)(\xi_kk_i)(\xi_lk_k)z_{ai}}{z{ab}z_{aj}z_{al}z_{bi}z_{ik}z_{kl}}}\cr
&\ss{+\fc{(\xi_ak_j)(\xi_bk_i)(\xi_kk_l)(\xi_lk_i)z_{ai}}{z_{ab}z_{aj}z_{ak}z_{bi}z_{il}z_{kl}}-
\fc{(\xi_ak_i+\xi_ak_k+\xi_ak_l)(\xi_bk_j)(\xi_kk_l)(\xi_lk_i)}{z_{ab}z_{ak}z_{bj}z_{il}z_{kl}}}\cr
&\ss{+\fc{(\xi_ak_j)(\xi_kk_l)(\xi_lk_i)}{z_{ab}z_{aj}z_{il}z_{kl}}\lf(\fc{\xi_bk_k}{z_{bk}}+
\fc{\xi_bk_l}{z_{ak}z_{bl}} \ri) + 
\fc{(\xi_kk_b)(\xi_lk_j)}{z_{ai}z_{ak}z_{bk}z_{jl}}\lf(\fc{(\xi_ak_i)(\xi_bk_j)
z_{aj}}{z_{al}z_{bj}}+\fc{(\xi_ak_i)(\xi_bk_l)}{z_{bl}} \ri)}\cr
&\ss{-\fc{(\xi_ak_j+\xi_ak_l)(\xi_bk_i)(\xi_kk_b)(\xi_lk_j)}{z_{ak}z_{al}z_{bi}z_{bk}z_{jl}}+
\fc{(\xi_ak_i)(\xi_lk_j)z_{aj}}{z_{ab}z_{ai}z_{ak}z_{jl}}\lf(\fc{(\xi_bk_l)(\xi_kk_j)z_{bl}}{z_{jk}}-
\fc{(\xi_bk_j)(\xi_kk_l)}{z_{bj}z_{kl}} \ri)}\cr
&\ss{+\fc{(\xi_kk_i)(\xi_lk_j)}{z_{ab}z_{ak}z_{al}z_{ik}z_{jl}}\lf(\fc{(\xi_ak_i)(\xi_bk_j)z_{aj}}{z_{bj}}-
\fc{(\xi_ak_l)(\xi_bk_i)z_{ai}}{z_{bi}}-\fc{(\xi_ak_j)(\xi_bk_i)z_{ai}}{z_{bi}} \ri)}\ ,\cr
&\ss{+\fc{(\xi_kk_i)(\xi_lk_j)}{z_{ab}z_{ik}z_{jl}}\lf(\fc{(\xi_ak_k)(\xi_bk_j)z_{aj}}{z_{ak}z_{al}z_{bj}}-
\fc{(\xi_ak_j)(\xi_bk_k)}{z_{al}z_{bk}}-\fc{(\xi_ak_l)(\xi_bk_k)}{z_{al}z_{bk}}+
\fc{(\xi_ak_i)(\xi_bk_l)}{z_{ak}z_{bl}}+\fc{(\xi_ak_k)(\xi_bk_l)}{z_{ak}z_{bl}} \ri)}\cr
&\ss{+\fc{(\xi_bk_i)(\xi_kk_l)(\xi_lk_j)}{z_{ab}z_{ak}z_{bi}z_{jl}z_{kl}}\lf((\xi_ak_j)+(\xi_ak_k)+
(\xi_ak_l)\ri)- 
\fc{(\xi_ak_i)(\xi_kk_l)(\xi_lk_j)}{z_{ab}z_{ai}z_{jl}z_{kl}}\lf(\fc{\xi_bk_k}{z_{bk}}+
\fc{(\xi_bk_l)z_{al}}{z_{ak}z_{bl}} \ri)}\cr
&\ss{+\fc{(\xi_kk_j)(\xi_lk_j)z_{aj}}{z_{ab}z_{al}z_{jk}z_{jl}}\lf(\fc{(\xi_ak_i)(\xi_bk_k)}{z_{bk}}+
\fc{(\xi_ak_i)(\xi_bk_j)z_{aj}}{z_{ai}z_{ak}z_{bj}}-\fc{(\xi_ak_l)(\xi_bk_i)}{z_{ak}z_{bi}}-
\fc{(\xi_ak_k)(\xi_bk_i)}{z_{ak}z_{bi}}-\fc{(\xi_ak_j)(\xi_bk_i)}{z_{ak}z_{bi}} \ri)}\cr
&\ss{+\fc{(\xi_kk_b)(\xi_lk_k)}{z_{al}z_{bk}z_{kl}}\lf(\fc{(\xi_ak_i)(\xi_bk_j)}{z_{ai}z_{bj}}-
\fc{(\xi_ak_j)(\xi_bk_i)}{z_{aj}z_{bi}} \ri)+
\fc{(\xi_bk_j)(\xi_kk_i)(\xi_lk_k)}{z_{ab}z_{al}z_{bj}z_{ik}z_{kl}}\lf((\xi_ak_i)+(\xi_ak_k)+
(\xi_ak_l)\ri)}\cr
&\ss{-\fc{(\xi_ak_j)(\xi_kk_i)(\xi_lk_k)}{z_{ab}z_{aj}z_{ik}z_{kl}}\lf(\fc{(\xi_bk_k)z_{ak}}{z_{al}z_{bk}}+
\fc{(\xi_bk_l)}{z_{bl}} \ri)-
\fc{(\xi_bk_i)(\xi_kk_j)(\xi_lk_k)}{z_{ab}z_{al}z_{bi}z_{jk}z_{kl}}\lf((\xi_ak_j)+(\xi_ak_k)+
(\xi_ak_l)\ri)}\cr
&\ss{+\lf.
\fc{(\xi_ak_i)(\xi_kk_j)(\xi_lk_k)}{z_{ab}z_{ai}z_{jk}z_{kl}}\lf(\fc{(\xi_bk_j)z_{aj}}{z_{al}z_{bj}}+
\fc{(\xi_bk_k)z_{ak}}{z_{al}z_{bk}}+\fc{(\xi_bk_l)}{z_{bl}} \ri) \ri\}\ .}}}
From \eqqs \gives\ we now extract the various kinematics, classified according to their amount
of  self--contractions $\xi_i\xi_j$ of polarization vectors $\xi_k$.

\br
$\underline{Kinematics\ (\xi\xi)\ (\xi\xi)\ (\xi\xi)}$\br

The total number of the kinematics $(\xi\xi)(\xi\xi)(\xi\xi)$ 
is $15$. They are given as follows:
\eqn\XIS{\eqalign{
\Xi_1&:=   \xxi{1}{2}\ \xxi{3}{4}\ \xxi{5}{6}\ \ \ ,\ \ \
\Xi_2:=    \xxi{1}{2}\ \xxi{3}{5}\ \xxi{4}{6}\ \ \ ,\ \ \
\Xi_3:=    \xxi{1}{2}\ \xxi{3}{6}\ \xxi{4}{5},\cr
\Xi_4&:=   \xxi{1}{3}\ \xxi{2}{4}\ \xxi{5}{6}\ \ \ ,\ \ \
\Xi_5:=    \xxi{1}{3}\ \xxi{2}{5}\ \xxi{4}{6}\ \ \ ,\ \ \
\Xi_6:=    \xxi{1}{3}\ \xxi{2}{6}\ \xxi{4}{5},\cr
\Xi_7&:=   \xxi{1}{4}\ \xxi{2}{3}\ \xxi{5}{6}\ \ \ ,\ \ \
\Xi_8:=    \xxi{1}{4}\ \xxi{2}{5}\ \xxi{3}{6}\ \ \ ,\ \ \
\Xi_9:=    \xxi{1}{4}\ \xxi{2}{6}\ \xxi{3}{5},\cr
\Xi_{10}&:=\xxi{1}{5}\ \xxi{2}{3}\ \xxi{4}{6}\ \ \ ,\ \ \
\Xi_{11}:= \xxi{1}{5}\ \xxi{2}{4}\ \xxi{3}{6}\ \ \ ,\ \ \
\Xi_{12}:= \xxi{1}{5}\ \xxi{2}{6}\ \xxi{3}{4},\cr
\Xi_{13}&:=\xxi{1}{6}\ \xxi{2}{3}\ \xxi{4}{5}\ \ \ ,\ \ \
\Xi_{14}:= \xxi{1}{6}\ \xxi{2}{4}\ \xxi{3}{5}\ \ \ ,\ \ \
\Xi_{15}:= \xxi{1}{6}\ \xxi{2}{5}\ \xxi{3}{4}.}}
The two integrands \KIN, which appear in \gives\ in the combinations
$\Ac_1(a,b,i,j,k,l)$\  $(\xi_a\xi_{i})\ (\xi_b\xi_{l})\ (\xi_{j}\xi_{k})$
and $\Ac_2(a,b,i,j,k,l)\ (\xi_a\xi_b)\ (\xi_{i}\xi_{j})\ (\xi_{k}\xi_{l})$, 
account for the two cases, whether the two vertices $a,b$ 
in the $(-1)$--ghost picture are contracted among themselves or with other vertices.
The function $\Ac_1(a,b,i_1,i_2,i_3,i_4)$ is   
invariant under permutations of $i_2,i_3$, while the function $\Ac_2(a,b,i_1,i_2,i_3,i_4)$ 
shares permutation symmetry within the pairs of indices $i_1,i_2$ and  $i_3,i_4$ 
or mutually exchange of the latter pair. Moreover, the function 
$\Ac_2(a,b,i_1,i_2,i_3,i_4)$ is invariant under
permutation of $a$ and $b$. On the other hand, the function $\Ac_1(a,b,i_1,i_2,i_3,i_4)$ stays
invariant under the latter permutation, if in addition the indices $i_1$ and $i_4$ are exchanged.
It is easy to convince oneself, that under the above symmetries the kinematics
$(\xi_a\xi_{i_1})(\xi_b\xi_{i_4})(\xi_{i_2}\xi_{i_3})$ and 
$(\xi_a\xi_b)(\xi_{i_1}\xi_{i_2})(\xi_{i_3}\xi_{i_4})$, to which
the functions $\Ac_1,\Ac_2$ refer, respectively, do not change.

Due to  those symmetries it is obvious, that 
within the sum $\{\}$ of $A^\pi(a,b,i,j,k,l)$ 
any permutation of the four indices 
$i,j,k,l$, which label the four vertices in the zero--ghost picture, yields the same 
expression \gives\ as a result of Bose--symmetry of the $NS$--operators.
Moreover, the symmetry under the permutation of $a$ and $b$ is evident.
In  each amplitude $A^\pi(a,b,i,j,k,l)$, as given in \eqq \gives, all 
$15$ space--time kinematical contractions \XIS\ show up once.
Exchanging indices from the two sets $\{a,b\}$ and $\{i,j,k,l\}$, \ie
putting instead of the operators $a,b$ other operators in the $(-1)$--ghost picture,
leads to seemingly different expressions.  
In other words, putting different vertex operators in the $(-1)$--ghost picture,
provides a different function $A^\pi_r$ in front of a given kinematics $\Xi$.
The six--point amplitude allows for $15$ possibilities to choose which vertex pair $(a,b)$ 
to put into the $(-1)$--ghost picture.
Therefore, for the given kinematics \eqn\considerkin{
(\xi_A\xi_B)\ (\xi_C\xi_D)\ (\xi_E\xi_F)}
we obtain $15$ --a priori-- different expressions $A_r$, when we examine
\gives\ for the choices
\eqn\possib{ \eqalign{
(a,b)\in \{\ &(A,B),\ (A,C),\ (A,D),\ (A,E),\ (A,F),\ (B,C),\ (B,D),\ (B,E),\cr
&\ (B,F),\ (C,D),\ (C,E),\ (C,F),\ (D,E),\ (D,F),\ (E,F)\ \}\ .}}
The latter may be read off from \gives:
\eqn\gleichi{\eqalign{
&A^\pi_1(A,C,B,E,F,D)\ \ \ ,\ \ \ A^\pi_1(A,D,B,E,F,C)\ \ \ ,\ \ \ A^\pi_1(A,E,B,C,D,F)\ ,\cr
&A^\pi_1(A,F,B,C,D,E)\ \ \ ,\ \ \ A^\pi_1(B,C,A,E,F,D)\ \ \ ,\ \ \ A^\pi_1(B,D,A,E,F,C)\ ,\cr
&A^\pi_1(B,E,A,C,D,F)\ \ \ ,\ \ \ A^\pi_1(B,F,A,C,D,E)\ \ \ ,\ \ \ A^\pi_1(C,E,D,A,B,F)\ ,\cr
&A^\pi_1(C,F,D,A,B,E)\ \ \ ,\ \ \ A^\pi_1(D,E,C,A,B,F)\ \ \ ,\ \ \ A^\pi_1(D,F,C,A,B,E)\ ,\cr
&A^\pi_2(A,B,C,D,E,F)\ \ \ ,\ \ \ A^\pi_2(C,D,A,B,E,F)\ \ \ ,\ \ \ A^\pi_2(E,F,A,B,C,D)\ .}}
They all have to be equated, which gives rise to $14$ equations for each kinematics
\considerkin\ under consideration. {\it E.g.} the kinematics $\Xi_1$, \ie $A=1,B=2,C=3,D=4,E=5$
and $F=6$ yields the system of equations:
\eqn\gleich{\eqalign{
&\hskip1cm A_2^\pi(1,2,3,4,5,6)=A^\pi_2(3,4,1,2,5,6)=A^\pi_2(5,6,1,2,3,4)\cr
&=A_1^\pi(1,3,2,5,6,4)=A^\pi_1(1,4,2,5,6,3)=A^\pi_1(1,5,2,3,4,6)=A^\pi_1(1,6,2,3,4,5)\cr
&=A_1^\pi(2,3,1,5,6,4)=A^\pi_1(2,4,1,5,6,3)=A^\pi_1(2,5,1,3,4,6)=A^\pi_1(2,6,1,3,4,5)\cr
&=A_1^\pi(3,5,4,1,2,6)=A^\pi_1(3,6,4,1,2,5)=A^\pi_1(4,5,3,1,2,6)=A^\pi_1(4,6,3,1,2,5)\ .}}
This system of equations involves $15$ different functions $\tilde F$. Each of them appears
(linearly) three times. Not all of the equations are independent.

\br
$\underline{Kinematics\ (\xi\xi)\ (\xi \xi)\ (\xi k)\ (\xi k)}$\br

The kinematics $(\xi\xi)(\xi \xi)(\xi k)(\xi k)$, of which there exist 
the total number of $\h\lf(6\atop 4\ri)\lf(4 \atop 2\ri)\times 4^2=45 \times 16$ 
on--shell, appears in \gives\ with the four different functions $B_1,\ldots,B_4$, given 
in \eqqs\KINALL\ and \KINB.
The latter distinguish the way, how the polarizations of the two vertex operators
$a$ and $b$, which are put in the $(-1)$--ghost picture, are contracted with the other
vectors labeled by $i,j,k$ and $l$. More precisely, they account for the kinematics:
\eqn\fourcases{\eqalign{
B^\pi_2(a,b,i,j,k,l)\ (\xi_a\xi_b)\ (\xi_i\xi_j)   &\ra 
(\xi_a\xi_b)\ (\xi_i\xi_j)\ (\xi_k k_r)\ (\xi_l k_s)\ ,\cr
B^\pi_1(a,b,i,j,k,l)\ (\xi_a\xi_i)\ (\xi_b\xi_j)&\ra
(\xi_a\xi_i)\ (\xi_b\xi_j)\ (\xi_k k_r)\ (\xi_l k_s)\ ,\cr
B^\pi_3(a,i,j,k,b,l)\  (\xi_a\xi_i)\ (\xi_j\xi_k)&\ra
(\xi_a\xi_i)\ (\xi_j\xi_k)\ (\xi_b k_r)\ (\xi_l k_s)\ ,\cr
B^\pi_4(i,j,k,l,a,b)\ (\xi_i\xi_j)\ (\xi_k\xi_l)&\ra
(\xi_i\xi_j)\ (\xi_k\xi_l)\ (\xi_a k_r)\ (\xi_b k_s)\ .}}
As before, we may obtain $15$ different expressions for one given kinematics
\eqn\considerkini{
(\xi_A\xi_B)\ (\xi_C \xi_D)\ (\xi_E k)\ (\xi_F k)}
depending on which pair $(a,b)$ (running over the set \possib) 
of vertex operators we put into the $(-1)$ ghost picture.
Hence, the given kinematics \considerkini\ may be calculated from the 
following $15$ functions:
\eqn\gleichi{\eqalign{
&B^\pi_2(A,B,C,D,E,F)\ \ \ ,\ \ \ B^\pi_2(C,D,A,B,E,F)\ \ \ ,\ \ \ B^\pi_1(A,C,B,D,E,F)\ ,\cr
&B^\pi_1(A,D,B,C,E,F)\ \ \ ,\ \ \ B^\pi_1(B,C,A,D,E,F)\ \ \ ,\ \ \ B^\pi_1(B,D,A,C,E,F)\ ,\cr
&B^\pi_3(A,B,C,D,E,F)\ \ \ ,\ \ \ B^\pi_3(B,A,C,D,E,F)\ \ \ ,\ \ \ B^\pi_3(C,D,A,B,E,F)\ ,\cr
&B^\pi_3(D,C,A,B,E,F)\ \ \ ,\ \ \ B^\pi_3(A,B,C,D,F,E)\ \ \ ,\ \ \ B^\pi_3(B,A,C,D,F,E)\ ,\cr
&B^\pi_3(C,D,A,B,F,E)\ \ \ ,\ \ \ B^\pi_3(D,C,A,B,F,E)\ \ \ ,\ \ \ B^\pi_4(A,B,C,D,E,F)\ .}}
They all have to be equated, which gives rise to $14$ equations for each kinematics
\considerkin\ under consideration. More precisely, since the functions $B_i^\pi$
contain the contractions $\xi_E k$ and $\xi_F k$, of which we have $4\times 4 =16$ on--shell,
we obtain $14\times 16=224$ equations for the kinematics \considerkini.
Hence in total, after taking into account all $B$--kinematics we obtain 
$45 \times 224=10,080$ non--trivial relations, of which many are the same.
In fact, the length of the set boils down to $5,464$.

\br 
$\underline{Kinematics\ (\xi\xi)\ (\xi k)\ (\xi k)\ (\xi k)\ (\xi k)}$\br

The kinematics $(\xi\xi)(\xi k)(\xi k)(\xi k)(\xi k)$, of which there exist 
the total number $\lf(6 \atop 2\ri)\times 4^4=15 \times 256$ on--shell, appears 
in \gives\ with the three different functions $C^\pi_1,C^\pi_2,C^\pi_3$, given in \eqqs \KINC\ and \KINALL.
The latter distinguish the way, how the polarizations of the two vertex operators
$a$ and $b$, which are put in the $(-1)$--ghost picture, are contracted with the other
vectors labeled by $i,j,k$ and $l$:
\eqn\fourcases{\eqalign{
C_1^\pi(a,b,i,j,k,l)\ (\xi_a\xi_b)&\ra (\xi_a\xi_b)\ (\xi_ik)\ (\xi_jk)\ (\xi_k k)\ (\xi_lk)\ ,\cr
C_2^\pi(a,i,b,j,k,l)\ (\xi_a\xi_i)&\ra (\xi_a\xi_i)\ (\xi_bk)\ (\xi_jk)\ (\xi_k k)\ (\xi_lk)\ ,\cr
C_3^\pi(i,j,a,b,k,l)\ (\xi_i\xi_j)&\ra (\xi_i\xi_j)\ (\xi_ak)\ (\xi_bk)\ (\xi_k k)\ (\xi_lk)\ .}}
As before, we may obtain $15$ different expressions for one given kinematics
\eqn\considerkinii{
(\xi_A\xi_B)\ (\xi_C k_m)\ (\xi_D k_n)\ (\xi_E k_p)\ (\xi_F k_q)}
depending on which pair $(a,b)$ (running over the set \possib)  
of vertex operators is put into the $(-1)$ ghost picture.
Hence, the given kinematics \considerkini\ may be calculated from the 
following $15$ functions:
\eqn\gleichii{\eqalign{
&C^\pi_1(A,B,C,D,E,F)\ \ \ ,\ \ \ C^\pi_2(A,B,C,D,E,F)\ \ \ ,\ \ \ C^\pi_2(A,B,D,C,E,F)\ ,\cr
&C^\pi_2(A,B,E,C,D,F)\ \ \ ,\ \ \ C^\pi_2(A,B,F,C,D,E)\ \ \ ,\ \ \ C^\pi_2(B,A,C,D,E,F)\ ,\cr
&C^\pi_2(B,A,D,C,E,F)\ \ \ ,\ \ \ C^\pi_2(B,A,E,C,D,F)\ \ \ ,\ \ \ C^\pi_2(B,A,F,C,D,E)\ ,\cr
&C^\pi_3(A,B,C,D,E,F)\ \ \ ,\ \ \ C^\pi_3(A,B,C,E,,D,F)\ \ \ ,\ \ \ C^\pi_3(A,B,C,F,D,E)\ ,\cr
&C^\pi_3(A,B,D,E,C,F)\ \ \ ,\ \ \ C^\pi_3(A,B,D,F,C,E)\ \ \ ,\ \ \ C^\pi_3(A,B,E,F,C,D)\ .}}
They all have to be equated, which gives rise to $14$ equations for each kinematics
\considerkin\ under consideration. More precisely, since the functions $C_i^\pi$
contain the contractions $\xi_C k,\ \xi_D k,\ \xi_Ek$ and  $\xi_F k$, of which we have 
$4^4=256$ on--shell,
we obtain $14\times 256=3,584$ equations for the kinematics \considerkini.
Hence in total, after taking into account all $C$--kinematics we obtain 
$15 \times 3,584=53,760$ non--trivial relations, of which many are the same.
In fact, the length of the set boils down to $6,727$.
{From} the structure \KINC\ of the functions $C_i$, namely they do not involve 
self--contracted momenta, we deduce, that all those relations lead to  polynomial identities.

\subsec{Specifying the integration region $\Ic_\pi$}

Due to the $PSL(2,\IR)$--invariance on the disk, we may fix three positions of the vertex 
operators. A convenient choice in \study\ or in \KIN\ is
\eqn\Choice{
z_1=-z_\infty\ \ ,\ \ z_2=0\ \ ,\ \ z_3=1\ ,}
which implies the ghost factor $\vev{c(z_1)c(z_2)c(z_3)}=-z_\infty^2$.
A six--point amplitude involving massless external particles, \ie $k_i^2=0$, allows for
nine\foot{An $N$--point amplitude with $N$ external particles
has $\h N(N-3)$ independent kinematic invariants (quadratic in the momenta).
However, in $D=4$ space--time dimensions this number is reduced to $3N-10$ \nielsen.
This leads to qualitative different results for $D3$--branes.}
independent kinematic invariants $s_i$.
We may choose\foot{The remaining scalar products of momenta are expressed by these
nine invariants:
\eqn\eliminate{\eqalign{
k_1k_2&=s_4 + s_5 + s_6 + s_7 + s_8 + s_9\ ,\cr
k_1k_3&=s_1 + s_2 + s_3 +s_7+ s_8 + s_9\ ,\cr
k_1k_4&=-s_1 - s_4 - s_7 - s_8\ ,\cr
k_1k_5&=-s_2 - s_5 - s_7 - s_9\ ,\cr
k_1k_6&=-s_3 - s_6 - s_8 - s_9\ ,\cr
k_2k_3&=-s_1 - s_2 - s_3 - s_4 - s_5 - s_6 - s_7 - s_8 - s_9\ .}}}
the scalar products:
\eqn\Mandelstamm{\eqalign{
s_1&=k_2k_4\ \ ,\ \ s_2=k_2k_5\ \ ,\ \ s_3=k_2k_6\ ,\cr
s_4&=k_3k_4\ \ ,\ \ s_5=k_3k_5\ \ ,\ \ s_6=k_3k_6\ ,\cr
s_7&=k_4k_5\ \ ,\ \ s_8=k_4k_6\ ,\cr
s_9&=k_5k_6\ .}}
Note, that this particular choice of invariants corresponds to our choice of vertex
operator positions \Choice, \ie it results from fixing translation invariance on the disk.
In other words, our amplitude \study\ will only depend on these nine invariants $s_i$.

The  general structure of all the integrals \KINALL\ is then given by:
\eqn\genstr{\eqalign{
\tilde F\lf[{n_{24},n_{25},n_{26},n_{34},n_{35}\atop n_{36},n_{45},n_{46},n_{56}}\ri]:=
&\int_{\Ic_\pi} dz_4\ dz_5\  dz_6\ \ 
|z_4|^{\al_{24}}\ |z_5|^{\al_{25}}\ |z_6|^{\al_{26}}\cr
&\hskip-2.5cm\times |1-z_4|^{\al_{34}}\ |1-z_5|^{\al_{35}}\ |1-z_6|^{\al_{36}}\ 
|z_4-z_5|^{\al_{45}}\
|z_4-z_6|^{\al_{46}}\ |z_5-z_6|^{\al_{56}}\ .}}
Here, the powers $\al_{ij}$ are real numbers. More precisely they are given by
\eqn\poweri{\eqalign{
\al_{24}&=s_1+n_{24}\ \ ,\ \ \al_{25}=s_2+n_{25}\ \ ,\ \ \al_{26}=s_3+n_{26}\ ,\cr
\al_{34}&=s_4+n_{34}\ \ ,\ \ \al_{35}=s_5+n_{35}\ \ ,\ \ \al_{36}=s_6+n_{36}\ ,\cr
\al_{45}&=s_7+n_{45}\ \ ,\ \ \al_{46}=s_8+n_{46}\ ,\cr
\al_{56}&=s_9+n_{56}\ ,}}
with some negative or zero integers $n_{ij}$. The latter are elements  
$n_{ij}\in \{\pm 2,\pm 1,0\}$ 
depending on how many fermion or/and bosonic two--point functions are involved.
As we shall see in the next section, the integral \genstr\ represents a 
multiple hypergeometric function $\tilde F$.

The vertex positions $z_4,z_5$ and $z_6$ take arbitrary values on the real axis, \ie
$z_4,z_5,z_6\in\IR$. It will prove to be convenient to take the following choice
\eqn\convenient{
z_4=x\ \ ,\ \ z_5=xy\ \ ,\ \ z_6=xyz\ ,}
which has the Jacobian $\det\lf(\fc{\p (z_4,z_5,z_6)}{\p (x,y,z)}\ri)=x^2y$. 
Then the string $S$--matrix \study\ with  six external gluons becomes:
\eqn\studyi{\eqalign{
&\Ac_6(k_1,\xi_1,a_1;k_2,\xi_2,a_2;k_3,\xi_3,a_3;k_4,\xi_4,a_4;k_5,\xi_5,a_5;k_6,\xi_6,a_6)\cr
&=\int_{\IR} dx\ \int_{\IR} dy\ \int_{\IR} dz\
x^2y\ \vev{V_{A^{a_1}}^{(-1)}(-\infty)\ V_{A^{a_2}}^{(-1)}(0)\  V_{A^{a_3}}^{(0)}(1)\
V_{A^{a_4}}^{(0)}(x)\ V_{A^{a_5}}^{(0)}(xy)\  V_{A^{a_6}}^{(0)}(xyz)}\ .}}
According to \region, the arrangement of the positions $z_4,z_5,z_6$, \ie $x,y,z$, along 
the real axis implies an ordering $\pi$ of the group factors $\lambda^{a_i}$. 
This way the full integral in \studyi\ 
splits into $60$ different regions $\Ic_\pi$. Each of them gives rise to a different
ordering $\pi$ of the Chan--Paton factors $\lambda^{a_i}$ in the group trace $\Tr$
according to the general expression \formally.
In fact, there are $60$ different possibilities\foot{For a given hexagonal diagram, which
contracts the Chan--Paton factors $\lambda^{a_i}$ in the order $(i,j,k,l,m,n)$,
there exist 5 equivalent diagrams: $(j,k,l,m,n,i),\ldots,
(n,i,j,k,l,m)$, corresponding to the cyclic permutations.
Furthermore, changing their orientation results in twelve equivalent
diagrams. Thus, from the $6!$ possible permutations we only take into
account $720/12=60$ hexagonal diagrams.} how the Chan--Paton factors $\lambda^a$
may be contracted in a closed loop.
In the following, let us focus on the group contraction 
$\Tr(\lambda^1\lambda^2\lambda^3\lambda^4\lambda^5\lambda^6)$, \ie the permutation
$\pi=(1,2,3,4,5,6)$. According to the choice \Choice\ and \region, this ordering corresponds
to taking the region 
\eqn\Region{
\Ic_\pi=\{\ z_4,z_5,z_6\in\IR\ |\ 1<z_4<z_5<z_6 < \infty\}\ ,}
\ie $1<x,y,z<\infty$ along the real axis.
After subsection 2.1 and \eqq \Study\ we denote that particular part of the full 
string $S$--matrix  $\Ac_6$ by $A^\pi\equiv A^{(1,2,3,4,5,6)}$.
All other $59$ possible group theoretical contractions $\pi$
may be obtained from that particular expression $A^{(1,2,3,4,5,6)}$
by relabeling the quantum numbers of the vertex operators.
Therefore, in the following we concentrate on that particular group structure $\pi$
and omit the index $\pi$ in the following.
For that case, the amplitude \Study\ becomes:
\eqn\studyii{\eqalign{
&\Tr(\lambda^1\lambda^2\lambda^3\lambda^4\lambda^5\lambda^6)\ A^{(1,2,3,4,5,6)}\equiv
\Tr(\lambda^1\lambda^2\lambda^3\lambda^4\lambda^5\lambda^6)\ A^{(1,2,3,4,5,6)}(1,2,3,4,5,6)\cr
&=\int\limits_1^\infty dx\int\limits_1^\infty dy\int\limits_1^\infty dz\ x^2y\ 
\vev{V_{A^{a_1}}^{(-1)}(-\infty)\ V_{A^{a_2}}^{(-1)}(0)\  V_{A^{a_3}}^{(0)}(1)\
V_{A^{a_4}}^{(0)}(x)\ V_{A^{a_5}}^{(0)}(xy)\  V_{A^{a_6}}^{(0)}(xyz)}\ .}}
With the choices \convenient\ and \Region\ the integral \genstr\ simplifies drastically.
In fact, after performing the change of variables $x\ra 1/x,y\ra 1/y, z\ra 1/z$ 
the integral \genstr\ assumes the generic form
\eqn\genstri{
\tilde F\lf[{n_{24},n_{25},n_{26},n_{34},n_{35}\atop n_{36},n_{45},n_{46},n_{56}}\ri]
\longrightarrow
F\lf[{a,b,d,e,g\atop c,f,h,j}\ri]\ ,}
with
\eqn\todo{\eqalign{
F\lf[{a,b,d,e,g\atop c,f,h,j}\ri]&:=\int\limits_0^1 dx\ \int\limits_0^1 dy\ \int\limits_0^1 dz\cr
&\times x^a\ y^b\ z^c\ (1-x)^d\ (1-y)^e\ (1-z)^f\ (1-xy)^g\ (1-yz)^h\ (1-xyz)^j\ ,}}
and the new powers
\eqn\power{\eqalign{
a&=-4-\al_{24}-\al_{25}-\al_{26}-\al_{34}-\al_{35}-\al_{36}-\al_{45}-\al_{46}-\al_{56}\ ,\cr
b&=-3-\al_{25}-\al_{26}-\al_{35}-\al_{36}-\al_{45}-\al_{46}-\al_{56}\ ,\cr
c&=-2-\al_{26}-\al_{36}-\al_{46}-\al_{56}\ ,\cr
d&=\al_{34}\ \ \ ,\ \ \ e=\al_{45}\ \ \ ,\ \ \ f=\al_{56}\cr
g&=\al_{35}\ \ \ ,\ \ \ h=\al_{46}\ \ \ ,\ \ \ j=\al_{36}\ ,}}
given in terms of the integers $n_{ij}$ and invariants $s_i$ \poweri.
To this end, the integrands \KIN, \KINB\ and \KINC\ 
become certain products of powers of the nine polynomials \ninepolynomials. 
Of course, the number of these polynomials just reflects the number of kinematic invariants $s_i$.
{\it E.g.} for the kinematics $(\xi\xi)(\xi\xi)(\xi\xi)$, with \KIN\ the expressions \KINALL\ 
may be casted into the form
\eqn\KINN{\eqalign{
A^\pi_1(a,b,i,j,k,l)&=\si_{lb}\si_{ai}\si_{ab}\ \lf\{\si_{jk}\si_{ij}\si_{kl}\ (k_ik_j)\ (k_kk_l)\ 
\tilde 
F\lf[\ov n_{lb},\ov n_{jk}=-1,\ov n_{ai}=-1\atop \ov n_{ij},\ov n_{kl}=-1,\ov n_{ab}=-1\ri]\ri.\cr
&-\si_{jk}\si_{ik}\si_{jl}\ (k_ik_k)\ (k_jk_l)\ \tilde
F\lf[\ov n_{lb},\ov n_{jk}=-1,\ov n_{ai}=-1\atop \ov n_{ik},\ov n_{jl}=-1,\ov n_{ab}=-1\ri]\cr
&\lf.-\si_{il}\ (k_ik_l)\ (1-k_jk_k)\ 
\tilde F\lf[\ov n_{lb},\ov n_{il}=-1,\ov n_{ai}=-1\atop \ov n_{jk}=-2,\ov n_{ab}=-1\ri]\ri\},\cr
A^\pi_2(a,b,i,j,k,l)&=(1-k_ik_j)\ (1-k_kk_l)\ 
\tilde F\lf[{\ov n_{ij}=-2\atop \ov n_{kl}=-2},\ov n_{ab}=-2\ri]\cr
&-\si_{ij}\si_{kl}\si_{ik}\si_{jl}\ (k_ik_k)\ (k_jk_l)\ \tilde
F\lf[{\ov n_{ij},\ov n_{kl}=-1\atop \ov  n_{ik},\ov n_{jl}=-1},\ov n_{ab}=-2\ri]\cr
&+\si_{ij}\si_{kl}\si_{il}\si_{jk}\ (k_ik_l)\ (k_jk_k)\ \tilde
F\lf[{\ov n_{ij},\ov n_{kl}=-1\atop \ov n_{il},\ov n_{jk}=-1},\ov n_{ab}=-2\ri]\ ,}}
with the function 
$\ss{\tilde F\lf[{n_{24},n_{25},n_{26},n_{34},n_{35}\atop n_{36},n_{45},n_{46},n_{56}}\ri]}$
introduced  in \eqq \genstr\ and rewritten in \genstri. 
Furthermore, we have defined $\si_{ij}={\rm sign}(i-j)$, 
and $\ov n_{ij}:=\cases{n_{ij},& $i<j$ \cr 
                                                  n_{ji},& $i>j$}.$ 
The entries $\ov n _{ij}$ in the functions $\tilde F$ are to be understood such, 
that they are only of relevance, if they contribute in \genstri\ or \poweri. Otherwise
they are meaningless as a matter of the choice \Choice.
Similar expressions follow for the other kinematics $B$ and $C$, given in
\eqqs \KINB\ and \KINC. The same is true for the other group  contractions.

The full string $S$--matrix is invariant under cyclic permutation of the numbering of the six
gauge bosons. There are five possible cyclic symmetries, which do not affect the group factor
$\Tr(\lambda^1\lambda^2\lambda^3\lambda^4\lambda^5\lambda^6)$. The latter are
\eqn\cyclic{
(i):\ 1\ra 2,\ 2\ra 3,\ 3\ra 4,\ 4\ra 5,\ 5\ra 6,\ 6\ra 1\ \ \ ,\ \ \ \cases{s_1\ra s_5 \cr  
                                                              s_2\ra s_6\cr
                                                              s_3\ra s_1+s_2+s_3+s_7+s_8+s_9\cr
s_4\ra s_7\cr
s_5\ra s_8\cr
s_6\ra -s_1-s_4-s_7-s_8\cr
s_7\ra s_9\cr
s_8\ra -s_2-s_5-s_7-s_9\cr
s_9\ra -s_3-s_6-s_8-s_9\ ,}}
and
$$\eqalign{
(ii):&\ 1\ra 3,\ 2\ra 4,\ 3\ra 5,\ 4\ra 6,\ 5\ra 1,\ 6\ra 2\ ,\cr
(iii):&\ 1\ra 4,\ 2\ra 5,\ 3\ra 6,\ 4\ra 1,\ 5\ra 2,\ 6\ra 3\ ,\cr
(iv):&\ 1\ra 5,\ 2\ra 6,\ 3\ra 1,\ 4\ra 2,\ 5\ra 3,\ 6\ra 4\ ,\cr
(v):&\ 1\ra 6,\ 2\ra 1,\ 3\ra 2,\ 4\ra 3,\ 5\ra 4,\ 6\ra 5\ ,}$$
with some linear transformation on the nine invariants $s_i$.
The choice of positions \Choice\ puts the functions $\tilde F$ in a certain form \todo, 
which no longer  reflects the cyclic invariance of the amplitude \study.
In other words, the maps \cyclic\ have a non--trivial impact on the appearance of the string $S$--matrix
\studyii. However, cyclic symmetry is still present. 
It only requires some non--trivial change of the integration variables $x,y$ and $z$ in \todo. 
Generically, the integrand of \todo\ looks like 
$$\eqalign{
&x^{k_2k_3+\ldots}\ y^{k_2k_3+k_2k_4+k_3k_4+\ldots}\ z^{k_1k_6+\ldots}\ (1-x)^{k_3k_4+\ldots}\ 
(1-y)^{k_4k_5+\ldots}\ (1-z)^{k_5k_6+\ldots}\cr
&(1-xy)^{k_3k_5+\ldots}\ (1-yz)^{k_4k_6+\ldots}\ (1-xyz)^{k_3k_6+\ldots}\ ,}$$ 
with the dots accounting for combinations of integers $n_{ij}$ as given in \power.
The transformations \cyclic\ change the form of the integrand. 
However, for each case $(i)-(v)$ there exists a change of variables $x,y,z$ such, that the 
integrand of \todo, may be brought back into its original form \todo 
$$\eqalign{
&\tilde x^{k_2k_3+\ldots}\ \tilde y^{k_2k_3+k_2k_4+k_3k_4+\ldots}\ \tilde z^{k_1k_6+\ldots}\ 
(1-\tilde x)^{k_3k_4+\ldots}\ (1-\tilde y)^{k_4k_5+\ldots}\ (1-\tilde z)^{k_5k_6+\ldots}\cr 
&\times (1-\tilde x\tilde y)^{k_3k_5+\ldots}\ (1-\tilde y\tilde z)^{k_4k_6+\ldots}\ 
(1-\tilde x\tilde y\tilde z)^{k_3k_6+\ldots}\ ,}$$
with some new variables $\tilde x,\tilde y,\tilde z$ and new combinations of integers $m_{ij}$.
In fact, these coordinate transformations may be found:
\def\ss#1{{\scriptstyle{#1}}}
\eqn\hessian{\eqalign{
(i)\ \ &\ \tilde x=1-xyz\ \ \ ,\ \ \ \tilde y=\fc{1-yz}{1-xyz}\ \ \ ,\ \ \ 
\tilde z=\fc{1-z}{1-yz}\ ,\cr
(ii)\ \ &\ \tilde x=z\ \ \ ,\ \ \ \tilde y=\fc{1-xy}{1-xyz}\ \ \ ,\ \ \ 
\tilde z=\fc{1-y}{1-yz}\ \fc{1-xyz}{1-xy}\ ,\cr
(iii)\ \ &\ \tilde x=\fc{1-z}{1-yz}\ \ \ ,\ \ \ \tilde y=y\ \ \ ,\ \ \ 
\tilde z=\fc{1-x}{1-xy}\ ,\cr
(iv)\ \ &\ \tilde x=\fc{1-y}{1-xy}\ \fc{1-xyz}{1-yz}\ \ \ ,\ \ \ \tilde y=\fc{1-yz}{1-xyz}
\ \ \ ,\ \ \ \tilde z=x\ ,\cr
(v)\ \ &\ \tilde x=\fc{1-x}{1-xy}\ \ \ ,\ \ \ \tilde y=\fc{1-xy}{1-xyz}\ \ \ ,\ \ \ 
\tilde z=1-xyz\ .}}
After taking into account their Jacobians for a given function 
$\ss{\tilde F\lf[{n_{24},n_{25},n_{26},n_{34},n_{35}\atop n_{36},n_{45},n_{46},n_{56}}\ri]}$, with its 
integral representation given in \eqq \todo\ and its parameter detailed in \power, we find the 
following  five maps under the actions \cyclic:
\eqn\findmap{
\tilde F\lf[{n_{24},n_{25},n_{26},n_{34},n_{35}\atop n_{36},n_{45},n_{46},n_{56}}\ri]\longrightarrow
\tilde F\lf[{m_{24},m_{25},m_{26},m_{34},m_{35}\atop m_{36},m_{45},m_{46},m_{56}}\ri]\ ,}
with:
\eqn\changecyclic{\eqalign{
(i)\ \ \ &m_{24} = 2 + n_{24} + n_{25} + n_{26} + n_{45} + n_{46} + n_{56},\  
      m_{25} = -2 - n_{24} - n_{34} - n_{45} - n_{46}\ ,\cr 
     &m_{26} = -2 - n_{25} - n_{35} - n_{45} - n_{56}\ ,\cr 
     &m_{34} = -4 - n_{24} - n_{25} - n_{26} - n_{34} - n_{35} - n_{36} - n_{45} - n_{46} - n_{56}\ ,\cr 
     &m_{35} = n_{24},\ m_{36} = n_{25}, \ m_{45} = n_{34}, \ m_{46} = n_{35}, \ m_{56} = n_{45}\ ,\cr
(ii)\ \ \ &m_{24} = n_{26}, \ m_{25} = n_{36},\ m_{26} = n_{46}, \ 
       m_{34} = 2 + n_{34} + n_{35} + n_{36} + n_{45} + n_{46} + n_{56}\ ,\cr 
      &m_{35} = 2 + n_{24} + n_{25} + n_{26} + n_{45} + n_{46} + n_{56}, \ 
       m_{36} = -2 - n_{24} - n_{34} - n_{45} - n_{46}\ ,\cr 
      &m_{45} = -4 - n_{24} - n_{25} - n_{26} - n_{34} - n_{35} - n_{36} - n_{45} - n_{46} - n_{56}\ ,\cr 
      &m_{46} = n_{24}\ ,m_{56} = n_{34}\ ,\cr
(iii)\ \ \ &m_{24} = -2 - n_{25} - n_{35} - n_{45} - n_{56},\ m_{25} = n_{25},\ m_{26} = n_{35}\ ,\cr
       &m_{34} = -2 - n_{26} - n_{36} - n_{46} - n_{56},\ m_{35} = n_{26},\ m_{36} = n_{36}\ ,\cr 
       &m_{45} = 2 + n_{34} + n_{35} + n_{36} + n_{45} + n_{46} + n_{56}\ ,\cr 
       &m_{46} = 2 + n_{24} + n_{25} + n_{26} + n_{45} + n_{46} + n_{56}\ ,\cr 
       &m_{56} = -4 - n_{24} - n_{25} - n_{26} - n_{34} - n_{35} - n_{36} - n_{45} - n_{46} - n_{56}\ ,\cr
(iv)\ \ \ &m_{24} = n_{46},\ m_{25} = -2 - n_{24} - n_{34} - n_{45} - n_{46},\ m_{26} = n_{24},\ 
       m_{34} = n_{56}\ ,\cr 
      &m_{35} = -2 - n_{25} - n_{35} - n_{45} - n_{56},\ m_{36} = n_{25},\ 
       m_{45} =-2-n_{26}-n_{36} - n_{46} - n_{56}\ ,\cr
      &m_{46} = n_{26},\ m_{56} = 2 + n_{34} + n_{35} + n_{36} + n_{45} + n_{46} + n_{56}\ ,\cr
(v)\  \ \ &m_{24} = n_{35},\ m_{25} = n_{36},\ 
       m_{26} = 2 + n_{24} + n_{25} + n_{26} + n_{45} + n_{46} + n_{56}\ ,\cr
      &m_{34} = n_{45},\ m_{35} = n_{46},\ m_{36} = -2 - n_{24} - n_{34} - n_{45} - n_{46},\ 
       m_{45} = n_{56}\ ,\cr 
     &m_{46} = -2 - n_{25}  - n_{35} - n_{45} - n_{56},\ m_{56} = -2 - n_{26} - n_{36}-n_{46} -n_{56}\ .}}
{From} these equations we find the two parameter class of cyclically invariant functions:
\eqn\findtwoparameter{
\tilde F\lf[{n_{46}, \ -2 - 2\ n_{46} - 2\ n_{56},\ n_{46},\ n_{56},\ n_{46}\atop
  -2 - 2\ n_{46} - 2\ n_{56},\ n_{56},\ n_{46},\ n_{56}}\ri]\ \ \ ,\ \ \ 
n_{46},\ n_{56}\in \IZ\ .}
After having fixed three vertex positions as \eg in \Choice, the meaning of the cyclic symmetries 
\cyclic\ on the amplitude \studyii\ is, that we essentially may fix three other vertex positions $z_i$ 
while keeping the ordering $\pi$. This is why the choice 
\Choice\ does not represent any restriction, since we may easily move to an other choice, 
\eg $z_2=-\infty,\ z_3=0,\ z_4=1$ through applying the cyclic symmetry $(i)$.
Of course, in total  
the string $S$--matrix result for a given group structure does not change as long as
we keep the ordering $\pi$ of vertex positions, which corresponds to a certain group structure
or integration region $\Ic_\pi$.

{From} the amplitude $A^\pi(1,2,4,3,6,5)$, given in \gives\ or from \eqq \KINN, the contribution
to the five kinematics $\Xi_1,\Xi_2,\Xi_5,\Xi_7$ and $\Xi_8$ may be read off:
\eqn\dan{\hskip-0.5cm\eqalign{
&-\lf\{(1-s_4)(1-s_9)\tilde F\lf[{0, 0, 0, -2, 0\atop 0, 0, 0, -2}\ri]
  +s_6s_7\ \tilde F\lf[{0, 0, 0, -1, 0\atop -1, -1, 0, -1}\ri]
  -s_5s_8\ \tilde F\lf[{0, 0, 0, -1, -1\atop 0, 0, -1, -1}\ri]\ri\}\Xi_{1}\cr
&-\lf\{(1-s_5)(1-s_8)\tilde F\lf[{0, 0, 0, 0, -2\atop 0, 0, -2, 0}\ri]
-s_6s_7\ \tilde F\lf[{0, 0, 0, 0, -1\atop -1, -1, -1, 0}\ri]
-s_4s_9\ \tilde F\lf[{0, 0, 0, -1, -1\atop 0, 0, -1, -1}\ri]\ri\}\Xi_2\cr
&-\lf\{s_5(1-s_8)\tilde F\lf[{0, -1, 0, 0, -1\atop 0, 0, -2, 0}\ri]
 +s_6s_7\ \tilde F\lf[{0, -1, 0, 0, 0\atop -1, -1, -1, 0}\ri]
 +s_4s_9\ \tilde F\lf[{0, -1, 0, -1, 0\atop 0, 0, -1, -1}\ri]\ri\}\Xi_5\cr
&-\lf\{-s_4(1-s_9)\ \tilde F\lf[{0, 0, 0, -1, 0\atop 0, 0, 0, -2}\ri]
 +s_6s_7\ \tilde F\lf[{0, 0, 0, 0, 0\atop -1, -1, 0, -1}\ri]
  -s_5s_8\ \tilde F\lf[{0, 0, 0, 0, -1\atop 0, 0, -1, -1}\ri]\ri\}\Xi_7\cr
&+\lf\{-s_7\ (1-s_6)\ \tilde F\lf[{0, -1, 0, 0, 0\atop -2, -1, 0, 0}\ri]
-s_5s_8\ \tilde F\lf[{0, -1, 0, 0, -1\atop -1, 0, -1, 0}\ri]
+s_4s_9\ \tilde F\lf[{0, -1, 0, -1, 0\atop -1, 0, 0, -1}\ri]\ri\}\ \Xi_8\ .}}
The tensor $\Xi_8$ stays invariant under all five cyclic transformations \cyclic.
Hence, this symmetry has also to be respected by the last line of \dan\ and
the kinematics $\Xi_8$ represents a kind of one--dimensional irreducible representation
of the cyclic symmetries \cyclic. 
This may be verified using \cyclic\ and \changecyclic.
On the other hand, through applying the cyclic symmetries \cyclic\ 
on the four kinematics $\Xi_1,\Xi_2,\Xi_5,\Xi_7$ we may generate all remaining 
$10$ kinematics \XIS\ from those.  Hence, \dan\ represents the full content for the $A$--kinematics \XIS.
Similar symmetries hold  for the other two classes of kinematics $B$ and $C$.

According to \gives\ the expression \dan\ refers to the case, where the first and the second
gauge boson vertices are chosen to be in the $(-1)$--ghost picture, \ie $a=1$ and $b=2$.
Any other choice would lead to a seemingly different expression than \dan. 
Only after a detailed knowledge about the functions $\tilde F$ and their
various relations (\cf section 4) we would be able to see, that the expressions for the other choices 
\eqn\SET{\eqalign{
(a,b)\in \{(1,2),(1,3),(1,4),(1,5),(1,6),(2,3),(2,4),(2,5),(2,6),&\cr
(3,4),(3,5),(3,6),(4,5),\br(4,6),(5,6)\}&}}
are indeed the same. There are $\lf(6\atop 2\ri)=15$ such choices.
However, it is this lack of knowledge about the other choices \SET\ we precisely want to make use of.
We have already explained in subsection 3.1, that for the kinematics \dan\ 
we shall now derive fourteen other expressions with different vertex pairs $(a,b)$ 
in the $(-1)$--ghost picture.
We shall equate all these expressions  with \dan\ and obtain
striking identities between all the functions $\tilde F$, similar as described in subsection 
2.2 for the four gluon scattering case (\cf \eqq \system).
This information will give us sufficient knowledge about all functions $\tilde F$, which are
involved in \eqq \dan.
Without that information we would have to really calculate the integrals \todo\ for all 
the  functions $\tilde F$, which appear in \dan. These would be $14$ functions.
In addition, there appear a lot more from the kinematics $B$ and $C$.
This program shall be outlined in the next subsection.

\subsec{System of equations and its six--element solution}

{From} equating all expressions \gleichii\ for the $C$--kinematics we obtain 
$1,041$ linear relations between different hypergeometric functions \genstri.
One example for such a relation is:
\eqn\TRIPLE{
\tilde F\lf[{0,-1,0,0, -1\atop -1,-1, -1,1}\ri] =
\tilde F\lf[{0,0,-1,0, -1\atop -1,-1, -1,1}\ri] - 
\tilde F\lf[{0,-1, -1,0, -1\atop -1, -1,-1, 2}\ri]\ .}
After looking closer to the structure \todo\ of the functions involved,
we realize, that this identity is just a manifestation of the simple relation 
\eqn\triple{\eqalign{
&\fc{y\ (1-z)}{(1-y)\ (1-xy)\ z\ (1-yz)\ (1-xyz)}\cr
&=\fc{y\ (1 - z)}{(1 - y)\ (1 - x y)\ (1 - y z)\ (1 - x y z)}+
\fc{y\ (1 - z)^2}{(1 - y)\ (1 - x y)\ z\ (1 - y z)\ (1 - x y z)}\ .}}
between the three polynomials appearing in the integrands of those three functions.
An other more involved example is:
\eqn\TRIPLEI{\eqalign{
     \tilde F\lf[{-1, -1,-1,0,0\atop -1,0,0,0}      \ri]
&=   \tilde F\lf[{-1,0, -1,0,-1\atop -1, -1,-1,0}   \ri]- 
     \tilde F\lf[{-1,0, -1, 0,-1\atop -1,-1,0,1}    \ri] - 
     \tilde F\lf[{-1,0,0,2, -1\atop -1, -1,-1,0}    \ri]\cr 
&-   \tilde F\lf[{1, -1, -1, -1, 0\atop 0,-1,-1,0}  \ri] + 
     \tilde F\lf[{2, -1, -1, -1, 0\atop 0,-1,-1,0}  \ri] - 
2\   \tilde F\lf[{0,-1, -1, -1, 1 \atop -1,-1,0,0}  \ri]\cr
&+2\ \tilde F\lf[{0,-1, -1, -1,0\atop -1, -1, 1,0}  \ri] - 
2\   \tilde F\lf[{0,-1, -1, -1, 0\atop -1, -1, 0,1} \ri] - 
     \tilde F\lf[{0,-1, -1,0, -1\atop -1, 0,-1, 1}  \ri]\cr 
&+   \tilde F\lf[{0,-1, -1,0,0\atop -1, -1, -1, 1}  \ri]\ .}}
Again, this relation originates from a simple polynomial relation:
\eqn\triplei{\eqalign{
\ss{\fc{1}{1-xyz}}&=\ss{-\fc{y (1-x)^2}{x (1-y) (1-x y) (1-y z) (1-x y z)}+
\fc{1-y z}{(1-y) (1-x y) z (1-x y z)}-\fc{1-z}{(1-y) (1-x y) z (1-x y z)}}\cr
&\ss{+\fc{y (1-z)}{(1-y) (1-y z) (1-xy z)}-
\fc{y (1-z)}{(1-x y) (1-y z) (1-x y z)}+
\fc{y}{x (1-y) (1-y z) (1-x)}-\frac{y}{(1-y) (1-y z) (1-x)}}\cr
&\ss{-2\ \fc{1-x y}{(1-y) (1-x y z) (1-x)}+
2\ \fc{1-y z}{(1-y) z (1-x y z) (1-x)}-
2\ \fc{1-z}{(1-y) z (1-x y z) (1-x)}\ .}}}
All relations, which we find from solving the system of $6,727$ equations 
for the $C$--kinematics, have their origin in polynomial relations.
These identities are to be compared with a similar type of identities \providerel, 
which appear in the case of four--gluon scattering.

Let us now move on to the equations following from the $A$ and $B$--kinematics, \ie
following from equating \gleich\ and \gleichii\ for each kinematics.
In \eqq \gleich\ we have listed all expressions, which calculate the kinematics $\Xi_1$.
{\it E.g.} we have:
\eqn\weHAVE{\eqalign{
A_2(1,2,3,4,5,6)&=-(1-s_4)\ (1-s_9)\ \tilde F\lf[{0, 0, 0, -2, 0\atop 0, 0, 0, -2}\ri]
-s_6\ s_7\ \tilde F\lf[{0, 0, 0, -1, 0\atop -1, -1, 0, -1}\ri]\cr
&+s_5\ s_8\ \tilde F\lf[{0, 0, 0, -1, -1\atop 0, 0, -1, -1}\ri]\ ,\cr
A_1(1,3,2,5,6,4)&=-s_1\ (1-s_9)\ \tilde F\lf[{-1, 0, 0, -1, 0\atop 0, 0, 0, -2}\ri]   
-s_2\ s_8\ \tilde F\lf[{0, -1, 0, -1, 0\atop 0, 0, -1, -1}\ri]  \cr
&+ s_3\ s_7\ \tilde F\lf[{0, 0, -1, -1, 0\atop 0, -1, 0, -1}\ri]\ ,\cr 
A_1(3,5,4,1,2,6)&=-s_8\ (1-s_4-s_5-s_6-s_7-s_8-s_9)\ 
\tilde F\lf[{0, 0, 0, -1, -1\atop 0, 0, -1, -1}\ri] \cr  
&-s_1\ (s_3+s_6+s_8+s_9)\ \tilde F\lf[{-1, 0, 0, -1, -1\atop 0, 0, 0, -1}\ri]    \cr
&+s_3\ (s_1+s_4+s_7+s_8) \ \tilde F\lf[{0, 0, -1, -1, -1\atop 0, 0, 0, -1}\ri]\ ,\cr
&\hskip8cm \ldots\ .}}
All these equations have to be identical, because each of them determines the kinematics $\Xi_1$.
Hence, after equating them we obtain a set of linear equations with coefficients depending
on the kinematic invariants $s_i$:
\eqn\IE{
A_2(1,2,3,4,5,6)=A_1(1,3,2,5,6,4)=A_1(3,5,4,1,2,6)=\ldots\ .}
Above we have only depicted three expressions for 
the kinematics $\Xi_1$. Writing down all of them for all $A$-- and $B$--kinematics
we obtain a huge set of equations, which qualitatively looks like \IE.
This system gives rise to various non--trivial relations among different hypergeometric
functions. {\it E.g.} we obtain:
\eqn\PARTIAL{\eqalign{
&(1 - s_6)\ \lf(\ \tilde F\lf[{-1,0,0,0,0\atop -2, -1,0,0}\ri]-
      \tilde F\lf[{0,-1,0,0,0\atop -2,-1,0,0}\ri]\ \ri)-s_8\ 
      \tilde F\lf[{-1, -1,0,0,0\atop -1, 0,-1,0}\ri] \cr 
&=s_9 \tilde F\lf[{-1, -1,0,0,0\atop -1,0,0,-1}\ri] +s_3\ \lf(\   
      \tilde F\lf[{-1, 0,-1,0,0\atop -1,-1,0,0}\ri] -  
      \tilde F\lf[{0,-1, -1,0,0\atop -1, -1,0,0}\ri]\ \ri)\ ,}}
$${\eqalign{(1 + s_4)\  \tilde F\lf[{0,-1, -1, -2,0\atop 0,0,0,0}\ri]&= + s_1\  
\lf(\  \tilde F\lf[{-1,-1,0, -1,0\atop 0,0,0, -1}\ri] - 
       \tilde F\lf[{-1, 0,-1, -1,0\atop 0,0,0, -1}\ri]\ \ri)\cr 
&-s_7\ \tilde F\lf[{0,-1, -1, -1,0\atop 0, -1,0,0}\ri] - s_8
       \tilde F\lf[{0,-1, -1,-1,0\atop 0,0,-1,0}\ri]\ ,}}$$
$${\hskip-0.6cm\eqalign{&s_2\  
      \tilde F\lf[{0,-1,0,-1,0,\atop 0,0, -1,0}\ri]=-(s_4 + s_5 + s_6 + s_7 + s_8 + s_9)\  
\lf(\ \tilde F\lf[{0,0,0,-1,0\atop 0,-1,0, -1}\ri] - 
      \tilde F\lf[{0,0,0,-1,0\atop 0,0, -1,-1}\ri]\ \ri)\cr 
&-s_3\ \tilde F\lf[{0,0,-1, -1,0\atop 0,-1,0,0}\ri] -
(1 + s_1 + s_2 + s_3 + s_4 + s_5 + s_6 + s_7 + s_8 + s_9)\ 
      \tilde F\lf[{0,0,0,0,0\atop 0, -1, -1,0}\ri] \ ,}}$$
$$\eqalign{s_2\ \tilde F\lf[{0,0,-1,0,-1\atop 0, -1, -1, 1}\ri]&=  
  s_3\ \tilde F\lf[{0,-1, -1, -1,0 \atop -1, -1, 0,1}\ri] + (s_1 + s_2 - s_7)\ 
\tilde F\lf[{0,-1, -1, 0,0\atop -1, -1,-1, 1}\ri]\cr 
&+ (s_2 - s_3 - s_4 + s_9)\  \tilde F\lf[{0,-1,-1,0,-1\atop -1, -1, -1, 2}\ri]\ .}$$
After inserting the definitions \genstr\ and \todo\ into the above equations
we see, that these relations may be proven through performing partial integration.
We have found a huge amount of identities of that type.

Moreover, after inserting the $1,041$ solutions from the $C$--kinematics into the whole system, which 
we obtain from the $A$ and $B$--kinematics \gleichi\ and \gleichii, \ie after eliminating
$1,041$ functions $\tilde F$  from the $A$ and $B$--system, we are left over with $366+210$ equations.
The latter only contain $1270-1041=229$  functions $\tilde F$. 
This system of equations is undetermined.
We realize that all equations may be solved in terms of a basis of six functions.
Hence, any of our $1,270$ hypergeometric function 
$\ss{\tilde F\lf[{n_{24},n_{25},n_{26},n_{34},n_{35}\atop n_{36},n_{45},n_{46},n_{56}}\ri]}$
may be represented as a linear combination of six (multiple) hypergeometric functions.
This is in complete analogy with the four--gluon string $S$--matrix (\cf \eqq \BOXIV), where
the basis was just one--dimensional and with the five--gluon string $S$--matrix with
a two--dimensional basis (\cf appendix \appA).
Recall, that in the four--gluon case (and five--gluon case), we were free, 
which functions $F_j$ we choose for our basis. In that case we had chosen $F_0$ from 
the set \veneziano. The same was true in the five--gluon case. There 
we had singled out $\Phi_1$ and $\Phi_2$. In both cases, $F_0$ and $\Phi_1,\Phi_2$
share some special properties in view of their momentum expansions: \cf \eqqs \powerSS\ and (A.27).
Here again, for our six--dimensional basis we may choose a convenient basis of six functions.
Following the discussion after \eqq \providerel\ one simple choice for these six functions
are integrals, which do not involve poles in the Mandelstamm variables $s_i$. As basis we may
choose the following six functions
$\ss{\tilde F\lf[{-1,-1,-2,0,0 \atop 0,0,0,0}\ri]}$,
$\ss{\tilde F\lf[{-1,-1,0,0,0 \atop -2,0,0,0}\ri]}$, 
$\ss{\tilde F\lf[{-1,0,-2,0,-1 \atop 0,0,0,0}\ri]}$,
$\ss{\tilde F\lf[{-1,0,-2,0,0 \atop -1,0,0,0}\ri]}$,
$\ss{\tilde F\lf[{0,-2,-1,0,0 \atop -1,0,0,0}\ri]}$
and 
$\ss{\tilde F\lf[{-2,-1,-1,0,0 \atop -1,0,0,0}\ri]}$ and express all remaining $1,264$
multiple hypergeometric functions
$\ss{\tilde F\lf[{n_{24},n_{25},n_{26},n_{34},n_{35}\atop n_{36},n_{45},n_{46},n_{56}}\ri]}$,
which show up in the string $S$--matrix, as linear combination of those six:
\eqn\RESULT{\hskip0cm{
\vbox{\offinterlineskip
\halign{\strut\vrule#
%%%%%%%%%%%%%%%%%%
&~$#$~\hfil
&\vrule# \cr
\noalign{\hrule}
%%%%%%%%%%%%%%%%%%
&\ \ \        \  \ \ &\cr
&\ \tilde F\lf[{n_{24},n_{25},n_{26},n_{34},n_{35}\atop n_{36},n_{45},n_{46},n_{56}}\ri]=
\Lambda^1_{\{n_{ij}\}}(s_i)\ \tilde F\lf[{-1,-1,-2,0,0 \atop 0,0,0,0}\ri]+ 
\Lambda^2_{\{n_{ij}\}}(s_i)\ \tilde F\lf[{-1,-1,0,0,0 \atop -2,0,0,0}\ri]\ &\cr
&\ \   \ &\cr
&\ \   \ &\cr
&\hskip3.5cm +\Lambda^3_{\{n_{ij}\}}(s_i)\ \tilde F\lf[{-1,0,-2,0,-1 \atop 0,0,0,0}\ri]+
\Lambda^4_{\{n_{ij}\}}(s_i)\ \tilde F\lf[{-1,0,-2,0,0 \atop -1,0,0,0}\ri]\ &\cr
&\ \  \ &\cr
&\ \  \ &\cr
&\ \hskip3.5cm +\Lambda^5_{\{n_{ij}\}}(s_i)\ \tilde F\lf[{0,-2,-1,0,0 \atop -1,0,0,0}\ri]+
\Lambda^6_{\{n_{ij}\}}(s_i)\ \tilde F\lf[{-2,-1,-1,0,0 \atop -1,0,0,0}\ri]\ .\ &\cr
&\ \ \ \ &\cr
%%%%%%%%%%%%%%%%%%
\noalign{\hrule}}} } }
In other words, our system of equations may be solved through this six--dimensional
ansatz \RESULT. Each function 
$\ss{\tilde F\lf[{n_{24},n_{25},n_{26},n_{34},n_{35}\atop n_{36},n_{45},n_{46},n_{56}}\ri]}$
has its six coefficients $\Lambda^l_{\{n_{ij}\}}(s_i)\ \ ,\ \ l=1,\ldots,6$ depending
non--trivially on the nine kinematic invariants $s_i$. 
Note, that the result \RESULT\ states quite non--trivial mathematical relations between the 
functions 
$\ss{\tilde F\lf[{n_{24},n_{25},n_{26},n_{34},n_{35}\atop n_{36},n_{45},n_{46},n_{56}}\ri]}$, 
introduced in  \genstr\ and \todo.
The identities like \TRIPLE, \TRIPLEI\ and \PARTIAL, which are unified in the striking
solution \RESULT, are interesting by there own, as they give non--trivial relations
between different multiple hypergeometric functions 
$\ss{\tilde F\lf[{n_{24},n_{25},n_{26},n_{34},n_{35}\atop n_{36},n_{45},n_{46},n_{56}}\ri]}$.
These relations are to be compared with the identities 
for ordinary hypergeometric functions $\F{p}{q}$, as \eg in {\it Ref.} \bailey.

Let us now turn to our preferred\foot{An alternative choice would replace one of the
six functions by:
\eqn\onealter{
\tilde F\lf[{-1,-1,-1,0,0\atop -1,0,0,0}\ri]=\int\limits_0^1 dx\ \int\limits_0^1 dy\ 
\int\limits_0^1 dz\  \fc{\Pc(x,y,z)}{1-xyz}\ .}}
basis of six functions. According to \genstr\ and \todo\ the
six functions, which serve as our basis in \RESULT, are given by the following integrals:
\eqn\Usebasis{\eqalign{
\tilde F\lf[{-1,-1,-2,0,0 \atop 0,0,0,0}\ri]&=\int\limits_0^1 dx\ \int\limits_0^1 dy\ 
\int\limits_0^1 dz\  \Pc(x,y,z)\ ,\cr
\tilde F\lf[{-1,-1,0,0,0 \atop -2,0,0,0}\ri]&=\int\limits_0^1 dx\ \int\limits_0^1 dy\ 
\int\limits_0^1 dz\  \fc{\Pc(x,y,z)}{(1-xyz)^2}\ ,\cr
\tilde F\lf[{-1,0,-2,0,-1 \atop 0,0,0,0}\ri]&=\int\limits_0^1 dx\ \int\limits_0^1 dy\ 
\int\limits_0^1 dz\  \fc{\Pc(x,y,z)}{1-xy}\ ,\cr
\tilde F\lf[{-1,0,-2,0,0 \atop -1,0,0,0}\ri]&=\int\limits_0^1 dx\ \int\limits_0^1 dy\ 
\int\limits_0^1 dz\  \fc{z\ \Pc(x,y,z)}{1-xyz}\ ,\cr
\tilde F\lf[{0,-2,-1,0,0 \atop -1,0,0,0}\ri]&=\int\limits_0^1 dx\ \int\limits_0^1 dy\ 
\int\limits_0^1 dz\  \fc{y\ \Pc(x,y,z)}{1-xyz}\ ,\cr
\tilde F\lf[{-2,-1,-1,0,0 \atop -1,0,0,0}\ri]&=\int\limits_0^1 dx\ \int\limits_0^1 dy\ 
\int\limits_0^1 dz\  \fc{x\ \Pc(x,y,z)}{1-xyz}\ .}}
with
\eqn\expdef{\eqalign{
\Pc(x,y,z)&=
x^{-s_1-s_2-s_3-s_4-s_5-s_6-s_7-s_8-s_9}\ y^{-s_2-s_3-s_5-s_6-s_7-s_8-s_9}\  
z^{-s_3-s_6-s_8-s_9}\cr 
&\times (1-x)^{s_4}\ (1-y)^{s_7}\ (1-z)^{s_9}\ (1-xy)^{s_5}\ (1-yz)^{s_8}\ (1-xyz)^{s_6}\cr
&=x^{k_2k_3}\ y^{k_2k_3+k_2k_4+k_3k_4}\  
z^{k_1k_6}\ (1-x)^{k_3k_4}\ (1-y)^{k_4k_5}\ (1-z)^{k_5k_6}\cr
&\times (1-xy)^{k_3k_5}\ (1-yz)^{k_4k_6}\ (1-xyz)^{k_3k_6} ,}}
stemming from the contractions $\Ec$ of the exponentials \Expsix.

The result \RESULT\ is the key to express the whole string six--gluon $S$--matrix 
in terms of the six functions \Usebasis. 
By knowing the momentum ($\ap$--) expansion of those six functions, we immediately obtain the 
momentum expansions of the remaining
functions through the identity \RESULT. In practice, after having obtained the relations \RESULT\
for all of the $1,264$ functions,  for each kinematics
one inserts those relations into the corresponding expressions
$A^\pi$ (\cf \eqqs \KINN\ or \dan) and similarly for $B^\pi$ and $C^\pi$.
Hence, by calculating the momentum expansion
of \Usebasis\ we are able to write down the $\ap$--expansion for the full string $S$--matrix
(\cf next subsection).

Let us point out, that we are free in choosing the 
six--dimensional basis, and with \eqq \RESULT\ we may easily switch to different basis 
representations.
The advantage of our basis \Usebasis\ is, that we have only to deal with six functions,
whose momentum ($\ap$--) expansions do not have poles, just like $F_0$ in \powerSS\ for
the four--gluon amplitude. This drastically simplifies
the way, how we obtain those expansions (\cf the next section).
Eventually, one may wish to choose a basis, which encodes the structure of the irreducible
gluon interactions to all orders in $\ap$. This may be achieved with finding a cyclically invariant basis
\findtwoparameter. For further details see {\it Ref.}  \MHV.

In addition, we remark, that the relations \RESULT\ are general, \ie independent on the gauge
structure $\pi$ and the integral region $\Ic_\pi$.
Though we have worked for a given group structure, resulting in the representation 
\todo\ for the functions, an other group structure $\pi$ would only result in a different 
integral representation \todo, specified by the integral region $\Ic_\pi$ and its parameterization.
However, the relations \RESULT\ stay the same as a matter
of changing coordinates of the integrands. Hence, the relations \RESULT\ are completely 
general and hold also for other integral representations. This is just like in the 
case of relations for ordinary hypergeometric functions, which also hold for different integral
representations.

We have already pointed out in the introduction, that the equations we impose follow from 
world--sheet supersymmetry and not (directly) from the cyclic invariance of a gluon amplitude.
In fact, the equations, which derive from the cyclic symmetry are automatically 
contained in the system of equations we have studied. Nevertheless this means, that
qualitative different results are to be expected for writing 
the final expression for the string $S$--matrix of the bosonic string compared to the superstring.
By that we mean, that one should expect a smaller basis of functions than \RESULT\ for the 
much fewer subset of functions, which appear in a bosonic amplitude.
On the other hand, the various relations for the triple hypergeometric functions \RESULT\ 
following from the superstring amplitude can be used to also simplify the corresponding bosonic 
string $S$--matrix. The latter clearly involves much less triple hypergeometric functions
from the beginning.

\subsec{Momentum expansion of the string $S$--matrix}

In the following, to display our result we shall adapt to a different basis \Mandelstamm\
of kinematical invariants. We choose\foot{We may easily
switch from the basis \Mandelstamm\ to this new basis through: $s_1\ra s_8-s_2-s_3,\ 
s_2\ra s_3+s_6-s_8-s_9,\ s_3\ra -s_1-s_6+s_9,\ s_4\ra s_3,\ s_5\ra s_9-s_3-s_4,\ s_6\ra s_1+s_4-s_7-
s_9,\ 
s_7\ra s_4,\ s_8\ra -s_4-s_5+s_7\ ,s_9\ra s_5$.} the more physical basis \Mangano:
\eqn\mang{\eqalign{
s_1=k_1k_2\ \ \ ,\ \ \ s_2&=k_2k_3\ \ \ ,\ \ \ s_3=k_3k_4\ ,\cr
s_4=k_4k_5\ \ \ ,\ \ \ s_5&=k_5k_6\ \ \ ,\ \ \ s_6=k_6k_1\ ,\cr
s_7=\h\ (k_1+k_2+k_3)^2\ \ ,\ \ s_8&=\h\ (k_2+k_3+k_4)^2\ \ ,\ \ s_9=\h\ (k_3+k_4+k_5)^2\ .}}
In this basis \eg the cyclic symmetries \cyclic\ act as simple permutation symmetry 
within the two sets $\{s_1,s_2,s_3,s_4,s_5,s_6\}$ and $\{s_7,s_8,s_9\}$.
The momentum (or $\ap$)--expansion of the six functions \Usebasis, 
which will be the subject of subsection 4.5, is given\foot{For the function
\onealter\ we obtain:
\eqn\alterexp{
\tilde F\lf[{-1,-1,-1,0,0\atop -1,0,0,0}\ri]=\zeta(3)-\fc{1}{4}\ (s_1+4\ s_2+3\ s_3+2\ s_4+
3\ s_5+4\ s_6+s_7+4\ s_8+\ s_9)\ \zeta(4)+\ldots\ .}}
in terms of these new invariants:
\eqn\MOMENTUMEXP{\eqalign{
\Phi_1=\tilde F\lf[{-1,-1,-2,0,0 \atop 0,0,0,0}\ri]&=
1-3\ s_1 - s_2 + s_3 + s_5 - s_6 + s_7 - s_8 + s_9\cr
&+(s_1 - s_3 - s_4 - s_5)\ \zeta(2)+(s_1 + s_4 - s_7 - s_9)\ \zeta(3)+\ldots\ ,\cr
\Phi_2=\tilde F\lf[{-1,-1,0,0,0 \atop -2,0,0,0}\ri]&=(1+s_1+s_4-s_7-s_9)\ \zeta(2)\cr
&-(2\ s_1+s_2+s_3+2\ s_4+s_5+s_6-s_7+s_8-s_9)\ \zeta(3)+\ldots\ ,\cr
\Phi_3=\tilde F\lf[{-1,0,-2,0,-1 \atop 0,0,0,0}\ri]&=
(1 - s_1 + s_2 + s_3 + 3\ s_4 + s_5 + s_7 - s_8 - s_9)\ \zeta(2)\cr
&+(s_1 - s_2 - 2\ s_3 - 4\ s_4 - s_5 - s_7 + 2\ s_8 + 2\ s_9)\ \zeta(3)+\ldots\ ,\cr
\Phi_4=\tilde F\lf[{-1,0,-2,0,0 \atop -1,0,0,0}\ri]&=
-1+\zeta(2)+s_1 - s_2 - s_4 + s_5 + 3\ s_6 + s_7 - s_8 - s_9\cr
&+(s_2 - s_3 - s_4 - s_5 - 2\ s_6 + s_8 + 2\ s_9)\ \zeta(2)\cr
&-(s_1 + s_2 - s_3 - 2\ s_4 + s_7 + s_8 + 2\ s_9)\ \zeta(3)\ldots\ ,\cr
\Phi_5=\tilde F\lf[{0,-2,-1,0,0 \atop -1,0,0,0}\ri]&=-1+\zeta(2)+
s_1 - s_2 - 2\ s_3 - 2\ s_5 - s_6 + s_7 + 3\ s_8 + s_9\cr
&+(s_2 + s_3 - s_4 + s_5 + s_6 - 2\ s_8)\ \zeta(2)\cr
&-(s_1 + s_2 - s_4 + s_6 + s_7 + s_9)\ \zeta(3)+\ldots\ ,}}
$${\eqalign{
\Phi_6=\tilde F\lf[{-2,-1,-1,0,0 \atop -1,0,0,0}\ri]&=-1+\zeta(2)+s_1 + 3\ s_2 + s_3 - s_4 - s_6 
-s_7 - s_8 + s_9\cr
&-(2\ s_2 + s_3 + s_4 + s_5 - s_6 - 2\ s_7 - s_8)\ \zeta(2)\cr
&-(s_1 - 2\ s_4 - s_5 + s_6 + 2\ s_7 + s_8 + s_9)\ \zeta(3)+\ldots\ .}}$$
The higher orders, which are denoted by the dots, may be depicted from subsection 4.5.
With these six functions the six--gluon superstring $S$--matrix \study\
may be (formally) written for the group structure 
$\Tr(\lambda^1\lambda^2\lambda^3\lambda^4\lambda^5\lambda^6)$:
\eqn\FINAL{\hskip-0.75cm{
\vbox{\offinterlineskip
\halign{\strut\vrule#
%%%%%%%%%%%%%%%%%%
&~$#$~\hfil
&\vrule# \cr
\noalign{\hrule}
%%%%%%%%%%%%%%%%%%
&\ \        \  \ &\cr
&\Ac^{(1,2,3,4,5,6)}=\sum\limits_{j=1}^6   \Pc^j(s_1,\ldots,s_9)\ \ \Phi_j\ .&\cr
&\ &\cr
%%%%%%%%%%%%%%%%%%
\noalign{\hrule}}} } }
Here $\Pc^j(s_1,\ldots,s_9)$ are  meromorphic polynomials in the invariants $s_i$, which 
capture the set of kinematics $(\xi\xi)(\xi\xi)(\xi\xi)$, 
$(\xi\xi)(\xi \xi)(\xi k)(\xi k)$ and $(\xi\xi)(\xi k)(\xi k)(\xi k)(\xi k)$.
The polynomials $\Pc^j$ are quite huge expressions, to be compared with the corresponding
five--gluon case in \eqqs (A.28) and (A.29).
It is this form \FINAL\ of the six--gluon string $S$--matrix \study, which
allows  to arrange in the effective action the higher order $\ap$ gauge 
interaction terms with six external legs (like $F^6, D^4F^4, D^2 F^5, D^6F^4, D^2F^6,\ldots$)
thanks to the properties of the functions
$\Phi_j$ and coefficients $\Pc^j$. 
While the latter $\Pc^j$ keep track about the pole structure of the reducible diagrams,
the functions $\Phi_j$ organize the various terms according to their zeta--values
\doubref\MHV\progress.

Let us now present our final result for the six gluon string $S$--matrix \study, listed
according to the kinematics $A,B$ and $C$.
As explained before, we have focused on the group structure 
$\Tr(\lambda^1\lambda^2\lambda^3\lambda^4\lambda^5\lambda^6)$, \ie on the piece \studyii, 
since all other $59$ group orderings may be obtained by relabeling the gauge vertices, \ie 
the polarization vectors $\xi_i$ and momenta $k_i$.

\br
$\underline{Kinematics\ (\xi\xi)\ (\xi\xi)\ (\xi\xi)}$

We start with the kinematics $A$, \ie self--contractions of all six polarization vectors. 
Their expressions are given in \eqq \dan. There we have also explained,
that from the set \XIS\ we have to display only the five kinematics $\Xi_1,\Xi_2,\Xi_5,\Xi_7$ 
and $\Xi_8$ as a result of cyclic invariance. We obtain the following $\ap$--expansion
for that part of the six--gluon string $S$--matrix \study:
\eqn\sixpointA{\eqalign{
&\Xi_1\ \lf\{-\frac{s_9 s_2}{s_1 s_3
   s_5}-\frac{s_2}{s_1
   s_3}+\frac{s_2}{s_1
   s_5}+\frac{s_2}{s_3
   s_5}-\frac{s_2}{s_1
   s_7}-\frac{s_4 s_2}{s_1 s_5
   s_7}-\frac{s_6 s_2}{s_3 s_5
   s_8}-\frac{s_2}{s_3
   s_8}+\frac{s_4}{s_1
   s_3}+\frac{s_6}{s_1
   s_3}\ri. \cr
&- \frac{s_6}{s_1
   s_5}+\frac{s_6}{s_3
   s_5}-\frac{s_6 s_7}{s_1 s_3
   s_5}-\frac{s_7}{s_1
   s_5}+\frac{s_7 s_8}{s_1 s_3
   s_5}-\frac{s_4 s_8}{s_1 s_3
   s_5}-\frac{s_8}{s_3
   s_5}+\frac{s_7 s_9}{s_1 s_3
   s_5}+\frac{s_8 s_9}{s_1 s_3
   s_5}-\frac{s_9}{s_1
   s_3}\cr
&+\frac{1}{s_1}+\frac{1}{s_3}+\frac{
   s_4}{s_1 s_5}-\frac{s_4}{s_3
   s_5}+\frac{1}{s_5}-\frac{s_4}{s_5
   s_7}-\frac{1}{s_7}-\frac{s_6}{s_5
   s_8}-\frac{1}{s_8}-\frac{s_4}{s_3
   s_9}-\frac{s_4 s_6}{s_1 s_3
   s_9}-\frac{s_6}{s_1 
   s_9}-\frac{1}{s_9}\cr
&+\lf(\ \frac{
   s_2^2}{s_7}-\frac{ s_2^2}{s_5}+\frac{s_4  s_2^2}{s_5 s_7}+\frac{s_6  s_2^2}{s_5 s_8}+\frac{
   s_2^2}{s_8}+\frac{s_9^2  s_2}{s_1 s_3}+\frac{s_3 
   s_2}{s_1}-\frac{s_4  s_2}{s_1}-\frac{s_5 
   s_2}{s_1}-\frac{s_5  s_2}{s_3}-\frac{s_6 
   s_2}{s_3}+\frac{s_1 s_6  s_2}{s_3 s_5}\ri.\cr
&-\frac{s_6 
   s_2}{s_5}+\frac{s_7  s_2}{s_1}-\frac{s_3 s_7 
   s_2}{s_1 s_5}+\frac{s_8  s_2}{s_3}-\frac{s_1 s_8 
   s_2}{s_3 s_5}+\frac{s_5 s_9  s_2}{s_1 s_3}+\frac{s_7
   s_9  s_2}{s_1 s_5}+\frac{s_8 s_9  s_2}{s_3
   s_5}-\frac{s_9  s_2}{s_1}-\frac{s_9 
   s_2}{s_3}\cr
&-\frac{s_9  s_2}{s_5}+\frac{s_1 
   s_2}{s_3}+\frac{s_3 s_4  s_2}{s_1 s_5}-\frac{s_4 
   s_2}{s_5}+\frac{s_4^2  s_2}{s_1 s_7}+\frac{s_1 
   s_2}{s_7}+\frac{s_4 s_5  s_2}{s_1 s_7}+\frac{s_1 s_4
    s_2}{s_5 s_7}+\frac{s_6^2  s_2}{s_3
   s_8}+\frac{s_3  s_2}{s_8}\cr
&+\frac{s_5 s_6  s_2}{s_3
   s_8}+\frac{s_3 s_6  s_2}{s_5 s_8}-2 
   s_2-\frac{s_4^2 }{s_1}-\frac{s_6^2 }{s_3}+\frac{s_6
   s_7^2 }{s_1 s_5}+\frac{s_3 s_7^2 }{s_1
   s_5}-\frac{s_7 s_8^2 }{s_3 s_5}+\frac{s_4 s_8^2
   }{s_3 s_5}+\frac{s_1 s_8^2 }{s_3 s_5}+\frac{s_5
   s_9^2 }{s_1 s_3}\cr
&-\frac{s_7 s_9^2 }{s_1
   s_3}-\frac{s_8 s_9^2 }{s_1 s_3}-\frac{s_1 s_4
   }{s_3}-2 s_4 +\frac{s_4 s_5 }{s_3}-\frac{s_3
   s_6 }{s_1}-\frac{s_4 s_6 }{s_1}-\frac{s_4 s_6
   }{s_3}+\frac{s_4 s_5 s_6 }{s_1 s_3}+\frac{s_5 s_6
   }{s_1}+\frac{s_1 s_6 }{s_5}\cr
&-\frac{s_3 s_6
   }{s_5}-2 s_6 -\frac{s_3 s_7 }{s_1}-\frac{s_6
   s_7 }{s_1}-\frac{s_6 s_7 }{s_3}+\frac{s_3 s_6 s_7
   }{s_1 s_5}-\frac{s_6 s_7 }{s_5}-\frac{s_3 s_7
   }{s_5}-\frac{s_3 s_4 s_7 }{s_1 s_5}+\frac{s_4 s_7
   }{s_5}\cr
&-\frac{s_7^2 s_8 }{s_1 s_5}-\frac{s_4 s_8
   }{s_1}-\frac{s_4 s_8 }{s_3}-\frac{s_1 s_6 s_8
   }{s_3 s_5}+\frac{s_6 s_8 }{s_5}+\frac{s_6 s_7 s_8
   }{s_3 s_5}+\frac{s_7 s_8 }{s_1}+\frac{s_7 s_8
   }{s_3}+\frac{s_4 s_7 s_8 }{s_1 s_5}+\frac{2 s_7
   s_8 }{s_5}\cr
&-\frac{s_1 s_8 }{s_3}-\frac{s_1 s_8
   }{s_5}+\frac{s_1 s_4 s_8 }{s_3 s_5}-\frac{s_4 s_8
   }{s_5}-\frac{s_7^2 s_9 }{s_1 s_5}-\frac{s_8^2 s_9
   }{s_3 s_5}+\frac{s_4 s_9 }{s_3}-\frac{s_4 s_5 s_9
   }{s_1 s_3}-\frac{s_5 s_9 }{s_1}-\frac{s_5 s_9
   }{s_3}\cr
&-\frac{s_5 s_6 s_9 }{s_1 s_3}+\frac{s_6 s_9
   }{s_1}+\frac{s_6 s_7 s_9 }{s_1 s_3}+\frac{2 s_7
   s_9 }{s_1}+\frac{s_7 s_9 }{s_3}+\frac{s_7 s_9
   }{s_5}+\frac{s_4 s_8 s_9 }{s_1 s_3}-\frac{s_7 s_8
   s_9 }{s_1 s_3}-\frac{s_7 s_8 s_9 }{s_1
   s_5}\cr
&-\frac{s_7 s_8 s_9 }{s_3 s_5}+\frac{s_8 s_9
   }{s_1}+\frac{2 s_8 s_9 }{s_3}+\frac{s_8 s_9
   }{s_5}-\frac{s_1 s_4 }{s_5}+\frac{s_3 s_4
   }{s_5}+\frac{s_4^2 }{s_7}+\frac{s_4 s_5
   }{s_7}+\frac{s_6^2 }{s_8}+\frac{s_5 s_6
   }{s_8}+\frac{s_4^2 }{s_9}\cr
&+\lf.\lf.\frac{s_4 s_6^2 }{s_3
   s_9}+\frac{s_6^2 }{s_9}+\frac{s_3 s_4 }{s_9}+\frac{s_4^2
   s_6 }{s_1 s_9}+\frac{s_1 s_6 }{s_9}+\frac{s_3 s_4
   s_6 }{s_1 s_9}+\frac{s_1 s_4 s_6}{s_3 s_9}\ \ri)\ \zeta(2) \ri\}\cr
&+\Xi_2\ \lf\{\frac{s_2}{s_1 s_7}-\frac{1}{s_1}+\frac{1}{s_7}+
\frac{s_6}{s_1 s_9}+\frac{1}{s_9}\ri.\cr
&+\lf(\ -\frac{s_2^2}{s_7}-\frac{s_3 s_2}{s_1}+
\frac{s_5 s_2}{s_1}-\frac{s_1 s_2}{s_7}-
\frac{s_4 s_5 s_2}{s_1
   s_7}+s_2+s_1+s_4+\frac{s_3 s_6}{s_1}-
\frac{s_5 s_6}{s_1}+s_6+\frac{s_3 s_7}{s_1}\ri.\cr
&\lf.\lf.-s_7+\frac{s_4s_8}{s_1}-\frac{s_7 s_8}{s_1}+
\frac{s_5 s_9}{s_1}-\frac{s_8 s_9}{s_1}-s_9-
\frac{s_4
   s_5}{s_7}-\frac{s_6^2}{s_9}-\frac{s_3 s_4}{s_9}-\frac{s_1 s_6}{s_9}-\frac{s_3 s_4 s_6}{s_1 s_9}
\ \ri)\ \zeta(2)\ri\}\cr
&+\Xi_5\ \lf\{-\frac{1}{s_7}+\lf(\ \frac{s_2 s_1}{s_7}-s_1-s_2-s_4-s_5+s_7+s_8+s_9+\frac{s_4 s_5}{s_7}\ 
\ri)\ \zeta(2)\ri\}}}
%%%%%%%%%%%%%%%%%%%%%%%%%%%%%%%%%%%%%%%%%%%%%%%%%%%%%%%%%%%%%%%%%%%%%%%%%%%%%%%%%%%%%%%
$${\eqalign{
&+\Xi_7\ \lf\{
-\frac{s_1}{s_2 s_7}-\frac{s_4 s_1}{s_2 s_5 s_7}+\frac{s_9}{s_2
   s_5}+\frac{1}{s_2}+\frac{1}{s_5}-\frac{s_4}{s_5 s_7}-\frac{1}{s_7}-\frac{s_3}{s_2
   s_8}-\frac{s_3 s_6}{s_2 s_5 s_8}-\frac{s_6}{s_5 s_8}-\frac{1}{s_8}\ri.\cr
&+\lf[\frac{s_4 s_1^2}{s_5 s_7}+\frac{s_1^2}{s_7}-\frac{s_4 s_1}{s_2}+\frac{s_6
   s_1}{s_2}+\frac{s_6 s_1}{s_5}+\frac{s_4 s_8 s_1}{s_2 s_5}-\frac{s_8
   s_1}{s_5}-\frac{s_9 s_1}{s_5}+\frac{s_3 s_1}{s_5}-\frac{s_4
   s_1}{s_5}+\frac{s_4^2 s_1}{s_2 s_7}\ri.\cr
&+\frac{s_2 s_1}{s_7}+\frac{s_4 s_5
   s_1}{s_2 s_7}+\frac{s_2 s_4 s_1}{s_5 s_7}-s_1-s_2-s_3+\frac{s_3
   s_4}{s_2}-s_4-s_5-\frac{s_3 s_6}{s_2}+\frac{s_4 s_6}{s_2}-\frac{s_3
   s_6}{s_5}-s_6\cr
&-\frac{s_6 s_7}{s_2}+\frac{s_3 s_6 s_7}{s_2 s_5}-
\frac{s_3
   s_7}{s_5}+s_7-\frac{s_4 s_8}{s_2}+s_8-\frac{s_4 s_9}{s_2}-\frac{s_5
   s_9}{s_2}-\frac{s_6 s_9}{s_2}+\frac{s_7 s_9}{s_2}+\frac{s_7
   s_9}{s_5}-\frac{s_7 s_8 s_9}{s_2 s_5}\cr
&+\frac{s_8 s_9}{s_2}+\frac{s_8
   s_9}{s_5}-\frac{s_2 s_9}{s_5}-\frac{s_3 s_9}{s_5}+\frac{s_3
   s_4}{s_5}+\frac{s_4^2}{s_7}+\frac{s_4 s_5}{s_7}+\frac{s_3^2}{s_8}+\frac{s_3
   s_6^2}{s_2 s_8}+\frac{s_6^2}{s_8}+\frac{s_2 s_3}{s_8}+\frac{s_3 s_5
   s_6}{s_2 s_8}\cr
&\lf.\lf.+\frac{s_5 s_6}{s_8}+\frac{s_3^2 s_6}{s_5 s_8}+
\frac{s_2 s_3   s_6}{s_5 s_8}\ri]\ \zeta(2)\ri\}\cr
&+\Xi_8\ (s_1 + s_2 + s_3 + s_4 + s_5 + s_6 - s_7 - s_8 - s_9)\ \zeta(2) +\Oc(k^4)\ .}}$$
To keep the expressions short we do not write the orders beyond $k^4$.
Clearly, the momentum expansions of the kinematics $\Xi_8$ shows cyclic invariance.
There are no $k^0$--contributions, a result, which has impact on the possible reducible diagrams
contributing to these kinematics (\cf section 5).

\br
$\underline{Kinematics\ (\xi\xi)\ (\xi \xi)\ (\xi k)\ (\xi k)}$

Let us now present the $B$--kinematics. As outlined in subsection 3.1. there are many  
independent contractions of the form $(\xi\xi)\ (\xi \xi)\ (\xi k)\ (\xi k)$ even 
after making use of the cyclic symmetry \cyclic. It is not possible to write down
all of them. Instead we pick up a representative example.
{\it E.g.} in the string $S$--matrix \study\ after eliminating $\xi_5k_6$ and $\xi_6k_5$ on--shell
the kinematics $(\xi_1\xi_2)(\xi_3\xi_4)(\xi_5k_2)(\xi_6k_1)$ 
comes along with the momentum dependent expression:
\def\text#1{#1}
\eqn\sixpointB{\eqalign{
(\xi_1 \xi_2)&(\xi_3\xi_4)(\xi_5k_2)(\xi_6k_1)\cr
\times&\lf\{- \frac{\text{s_2}}{\text{s_3} \text{s_5}
   \text{s_8}}-\frac{\text{s_2}}{\text{s_3} \text{s_6}
   \text{s_8}}-\frac{\text{s_7}}{\text{s_1} \text{s_3}
   \text{s_5}}+\frac{1}{\text{s_1} \text{s_3}}-\frac{1}{\text{s_1}
   \text{s_5}}+\frac{1}{\text{s_3} \text{s_5}}+\frac{1}{\text{s_3}
   \text{s_6}}-\frac{1}{\text{s_5} \text{s_8}}-\frac{1}{\text{s_6}
   \text{s_8}}\ri. \cr
&- \frac{\text{s_4}}{\text{s_1} \text{s_3}
   \text{s_9}}-\frac{1}{\text{s_1}
   \text{s_9}}-\frac{\text{s_4}}{\text{s_3} \text{s_6}
   \text{s_9}}-\frac{1}{\text{s_6} \text{s_9}} \cr
&+\lf(\frac{\text{s_2}^2}{\text{s_5}
   \text{s_8}}+\frac{\text{s_2}^2}{\text{s_6}
   \text{s_8}}+\frac{\text{s_9} \text{s_2}}{\text{s_3}
   \text{s_6}}-\frac{\text{s_2}}{\text{s_3}}+\frac{\text{s_1}
   \text{s_2}}{\text{s_3}
   \text{s_5}}-\frac{\text{s_2}}{\text{s_5}}-\frac{\text{s_2}}{\text{s_6}}+
\frac{\text{s_5} \text{s_2}}{\text{s_3}
   \text{s_8}}+\frac{\text{s_6} \text{s_2}}{\text{s_3}
   \text{s_8}}+\frac{\text{s_3} \text{s_2}}{\text{s_5}
   \text{s_8}}+\frac{\text{s_3} \text{s_2}}{\text{s_6}
   \text{s_8}}\ri.\cr
&+\frac{\text{s_7}^2}{\text{s_1}
   \text{s_5}}-\frac{\text{s_3}}{\text{s_1}}-\frac{\text{s_4}}{\text{s_1}}-
\frac{\text{s_4}}{\text{s_3}}+\frac{\text{s_4}
   \text{s_5}}{\text{s_1}
   \text{s_3}}+\frac{\text{s_5}}{\text{s_1}}-\frac{\text{s_6}}{\text{s_3}}-
\frac{\text{s_7}}{\text{s_1}}-\frac{\text{s_7}}{\text{
   s_3}}+\frac{\text{s_3} \text{s_7}}{\text{s_1}
   \text{s_5}}-\frac{\text{s_7}}{\text{s_5}}+\frac{\text{s_7}
   \text{s_8}}{\text{s_3} \text{s_5}}-\frac{\text{s_1}
   \text{s_8}}{\text{s_3}
   \text{s_5}}\cr
&+\frac{\text{s_8}}{\text{s_5}}+\frac{\text{s_4}
   \text{s_8}}{\text{s_3}
   \text{s_6}}+\frac{\text{s_8}}{\text{s_6}}-\frac{\text{s_5}
   \text{s_9}}{\text{s_1} \text{s_3}}+\frac{\text{s_7}
   \text{s_9}}{\text{s_1} \text{s_3}}-\frac{\text{s_8}
   \text{s_9}}{\text{s_3}
   \text{s_6}}+\frac{\text{s_9}}{\text{s_1}}+\frac{\text{s_9}}{\text{s_6}}+
\frac{\text{s_1}}{\text{s_5}}-\frac{\text{s_3}}{\text{
   s_5}}-\frac{\text{s_3}}{\text{s_6}}-\frac{\text{s_4}}{\text{s_6}
   }+\frac{\text{s_5}}{\text{s_8}}\cr
&+\lf.\frac{\text{s_6}}{\text{s_8}}+
\frac{\text{s_4}^2}{\text{s_1}
   \text{s_9}}+\frac{\text{s_1}}{\text{s_9}}+\frac{\text{s_3}
   \text{s_4}}{\text{s_1} \text{s_9}}+\frac{\text{s_1}
   \text{s_4}}{\text{s_3} \text{s_9}}+\frac{\text{s_4}
   \text{s_6}}{\text{s_3}
   \text{s_9}}+\frac{\text{s_6}}{\text{s_9}}+
\frac{\text{s_4}^2}{\text{s_6} \text{s_9}}+
\frac{\text{s_3} \text{s_4}}{\text{s_6}
   \text{s_9}}-2\ri)\zeta(2)\cr
&+\lf(-\frac{\text{s_2}^3}{\text{s_5}
   \text{s_8}}-\frac{\text{s_2}^3}{\text{s_6} \text{s_8}}-\frac{2
   \text{s_3} \text{s_2}^2}{\text{s_5} \text{s_8}}-\frac{2
   \text{s_3} \text{s_2}^2}{\text{s_6}
   \text{s_8}}-\frac{\text{s_9}^2 \text{s_2}}{\text{s_3}
   \text{s_6}}+\frac{3 \text{s_7}
   \text{s_2}}{\text{s_5}}+\frac{\text{s_8}
   \text{s_2}}{\text{s_3}}-\frac{\text{s_1} \text{s_8}
   \text{s_2}}{\text{s_3} \text{s_5}}+\frac{\text{s_8}
   \text{s_2}}{\text{s_5}}\ri.\cr
&+\frac{\text{s_8}
   \text{s_2}}{\text{s_6}}-\frac{\text{s_8} \text{s_9}
   \text{s_2}}{\text{s_3} \text{s_6}}+\frac{\text{s_9}
   \text{s_2}}{\text{s_3}}-\frac{\text{s_9}
   \text{s_2}}{\text{s_6}}+\frac{\text{s_1}
   \text{s_2}}{\text{s_3}}-\frac{\text{s_1}
   \text{s_2}}{\text{s_5}}+\frac{2 \text{s_3}
   \text{s_2}}{\text{s_5}}-\frac{\text{s_1}^2
   \text{s_2}}{\text{s_3} \text{s_5}}+\frac{2 \text{s_3}
   \text{s_2}}{\text{s_6}}\cr
&+\frac{3 \text{s_4}
   \text{s_2}}{\text{s_6}}-\frac{\text{s_5}^2
   \text{s_2}}{\text{s_3} \text{s_8}}-\frac{\text{s_6}^2
   \text{s_2}}{\text{s_3} \text{s_8}}-\frac{2 \text{s_5} \text{s_6}
   \text{s_2}}{\text{s_3} \text{s_8}}-\frac{\text{s_3}^2
   \text{s_2}}{\text{s_5} \text{s_8}}-\frac{\text{s_3}^2
   \text{s_2}}{\text{s_6} \text{s_8}}-2
   \text{s_2}-\frac{\text{s_7}^3}{\text{s_1}
   \text{s_5}}+\frac{\text{s_3}^2}{\text{s_1}}\cr
&-\frac{\text{s_4}
   \text{s_5}^2}{\text{s_1}
   \text{s_3}}-\frac{\text{s_5}^2}{\text{s_1}}+\frac{\text{s_6}^2}
   {\text{s_3}}-\frac{2 \text{s_3} \text{s_7}^2}{\text{s_1}
   \text{s_5}}-\frac{\text{s_7} \text{s_8}^2}{\text{s_3}
   \text{s_5}}+\frac{\text{s_1} \text{s_8}^2}{\text{s_3}
   \text{s_5}}-\frac{\text{s_8}^2}{\text{s_5}}-\frac{\text{s_4}
   \text{s_8}^2}{\text{s_3}
   \text{s_6}}-\frac{\text{s_8}^2}{\text{s_6}}+\frac{\text{s_5}
   \text{s_9}^2}{\text{s_1} \text{s_3}}\cr
&-\frac{\text{s_7}
   \text{s_9}^2}{\text{s_1} \text{s_3}}+\frac{\text{s_8}
   \text{s_9}^2}{\text{s_3}
   \text{s_6}}-\frac{\text{s_9}^2}{\text{s_1}}-\frac{\text{s_9}^2}
   {\text{s_6}}+\text{s_1}-2 \text{s_3}+\frac{2 \text{s_3}
   \text{s_4}}{\text{s_1}}-2 \text{s_4}-\frac{\text{s_4}
   \text{s_5}}{\text{s_1}}+\frac{\text{s_4}
   \text{s_5}}{\text{s_3}}-\frac{\text{s_1}
   \text{s_5}}{\text{s_3}}\cr
&+\text{s_5}+2 \text{s_6}+\frac{2
   \text{s_3} \text{s_7}}{\text{s_1}}+\frac{3 \text{s_4}
   \text{s_7}}{\text{s_1}}+\frac{\text{s_5}
   \text{s_7}}{\text{s_1}}+\frac{\text{s_5}
   \text{s_7}}{\text{s_3}}+\frac{2 \text{s_6}
   \text{s_7}}{\text{s_3}}+\frac{\text{s_1}
   \text{s_7}}{\text{s_3}}-\frac{\text{s_3}^2
   \text{s_7}}{\text{s_1} \text{s_5}}+\frac{\text{s_1}
   \text{s_7}}{\text{s_5}}\cr
&+\frac{2 \text{s_3}
   \text{s_7}}{\text{s_5}}+\text{s_7}+\frac{\text{s_4}
   \text{s_8}}{\text{s_3}}-\frac{\text{s_1} \text{s_7}
   \text{s_8}}{\text{s_3} \text{s_5}}-\frac{\text{s_7}
   \text{s_8}}{\text{s_5}}-\frac{\text{s_1}
   \text{s_8}}{\text{s_5}}+\frac{\text{s_1}^2
   \text{s_8}}{\text{s_3} \text{s_5}}-\frac{\text{s_4}
   \text{s_8}}{\text{s_6}}+2 \text{s_8}+\frac{\text{s_5}^2
   \text{s_9}}{\text{s_1} \text{s_3}}\cr
&+\frac{\text{s_8}^2
   \text{s_9}}{\text{s_3} \text{s_6}}+\frac{\text{s_4}
   \text{s_9}}{\text{s_1}}+\frac{\text{s_4}
   \text{s_9}}{\text{s_3}}-\frac{\text{s_4} \text{s_5}
   \text{s_9}}{\text{s_1} \text{s_3}}-\frac{\text{s_5}
   \text{s_9}}{\text{s_1}}-\frac{\text{s_5} \text{s_7}
   \text{s_9}}{\text{s_1} \text{s_3}}-\frac{\text{s_7}
   \text{s_9}}{\text{s_1}}-\frac{\text{s_8}
   \text{s_9}}{\text{s_3}}-\frac{\text{s_4} \text{s_8}
   \text{s_9}}{\text{s_3} \text{s_6}}\cr
&-\frac{\text{s_8}
   \text{s_9}}{\text{s_6}}+\frac{\text{s_4}
   \text{s_9}}{\text{s_6}}+2
   \text{s_9}-\frac{\text{s_1}^2}{\text{s_5}}+\frac{\text{s_3}^2}{
   \text{s_5}}+\frac{\text{s_3}^2}{\text{s_6}}+\frac{2 \text{s_3}
   \text{s_4}}{\text{s_6}}-\frac{\text{s_5}^2}{\text{s_8}}-\frac{
   \text{s_6}^2}{\text{s_8}}-\frac{2 \text{s_5}
   \text{s_6}}{\text{s_8}}\cr
&-\frac{\text{s_4}^3}{\text{s_1}
   \text{s_9}}-\frac{\text{s_1}^2}{\text{s_9}}-\frac{2 \text{s_3}
   \text{s_4}^2}{\text{s_1} \text{s_9}}-\frac{\text{s_4}
   \text{s_6}^2}{\text{s_3}
   \text{s_9}}-\frac{\text{s_6}^2}{\text{s_9}}-\frac{\text{s_3}^2
   \text{s_4}}{\text{s_1} \text{s_9}}-\frac{\text{s_1}^2
   \text{s_4}}{\text{s_3} \text{s_9}}-\frac{2 \text{s_1}
   \text{s_6}}{\text{s_9}}-\frac{2 \text{s_1} \text{s_4}
   \text{s_6}}{\text{s_3}
   \text{s_9}}\cr
&-\lf.\lf.\frac{\text{s_4}^3}{\text{s_6} \text{s_9}}-\frac{2
   \text{s_3} \text{s_4}^2}{\text{s_6}
   \text{s_9}}-\frac{\text{s_3}^2 \text{s_4}}{\text{s_6}
   \text{s_9}}\ri)\zeta(3)\ \ri\}+\Oc(k^6)\ .}}
Again, there are no $k^0$--order contributions for this type of kinematics.

\br 
$\underline{Kinematics\ (\xi\xi)\ (\xi k)\ (\xi k)\ (\xi k)\ (\xi k)}$

Let us now present the $C$--kinematics. Here again, as outlined in subsection 3.1, 
there are many  
independent contractions of the form $(\xi\xi)\ (\xi k)\ (\xi k)\ (\xi k)\ (\xi k)$
 even after applying the cyclic symmetry \cyclic. Therefore, it is impossible to 
write down
all of them. Instead, we pick up two representative examples: One
with triple poles in the kinematic invariants and an other without any poles.
{\it E.g.} in the string $S$--matrix \study\ 
after eliminating $\xi_1k_2,\ \xi_4k_2,\ \xi_5k_2$ and $\xi_6k_2$ on--shell
the kinematics $(\xi_2\xi_3)(\xi_1 k_3)(\xi_4k_1)(\xi_5k_1)(\xi_6k_1)$ 
comes along with the momentum dependent expression:
\eqn\sixpointCC{\eqalign{
(\xi_2\xi_3)(\xi_1 k_3)(\xi_4k_1)(\xi_5k_1)(\xi_6k_1)\ &
\lf\{\fc{1}{s_2\ s_6\ s_8}-\lf(\ \frac{s_4}{s_2\,s_6}+ \frac{s_5}{s_2\,s_8} + 
  \frac{s_3}{s_6\,s_8}\ \ri)\ \zeta(2)\ri.\cr
&+\lf(\ \frac{s_4}{s_2} + 
  \frac{s_5}{s_2} + 
  \frac{s_3}{s_6} + 
  \frac{s_4}{s_6} + 
  \frac{{s_4}^2}
   {s_2\,s_6} - 
  \frac{s_7}{s_2} + 
  \frac{{s_5}^2}
   {s_2\,s_8} + 
  \frac{s_2\,s_3}
   {s_6\,s_8}  \ri.\cr
&\lf.\lf.+ 
  \frac{{s_3}^2}
   {s_6\,s_8}+ 
  \frac{s_5\,s_6}
   {s_2\,s_8} + 
  \frac{s_4\,s_8}
   {s_2\,s_6} - 
  \frac{s_9}{s_6}\ \ri)\ \zeta(3)\ \ri\}+\Oc(k^6)\ .}}
Similarly, the kinematics $(\xi_1\xi_2)(\xi_3 k_6)(\xi_4k_2)(\xi_5k_2)(\xi_6k_2)$ 
appears with the momentum dependent expansion
\eqn\sixpointC{\eqalign{
(\xi_1\xi_2)(\xi_3 k_6)(\xi_4k_2)(\xi_5k_2)(\xi_6k_2)\ &
\lf\{\zeta(3)-\lf(\ \fc{1}{4}\ s_1+s_2+\fc{3}{4}\ s_3+\h\ s_4+\fc{3}{4}\ s_5+s_6\ri.\ri.\cr
&\lf.\lf.+\fc{1}{4}\ s_7
+s_8+\fc{1}{4}\ s_9\ \ri)\ \zeta(4)\ \ri\}+\Oc(k^8)\ ,}}
after eliminating $\xi_3k_1,\ \xi_4k_1,\ \xi_5k_1$ and $\xi_6k_1$ on--shell.

We have evaluated all kinematical configurations $A,B$ and $C$ up to  including
the $k^6$--order, \ie $\ap^4$ in the string tension. The expressions become rather long.
The interested reader may obtain more results from the authors upon request.

\break

\newsec{Multiple hypergeometric functions, generalized Kamp\'e de F\'eriet function and
Euler--Zagier sums}

The integrals, which show up in any scattering process of (only) open strings are
quite generic. Just as any four--point open string amplitude is given by the Euler--Beta
function $B(a,b)$, higher point amplitudes are described by hypergeometric functions or
generalizations thereof. The latter are also known as {\it 
multiple Gaussian hypergeometric functions}.
A multiple Gaussian hypergeometric series is a hypergeometric series in two or more variables,
which boils down  to the famous Gaussian hypergeometric series $\F{2}{1}$ if only one 
variable is non--zero.

\subsec{Generalized hypergeometric functions $_pF_q$}

Before moving to the six--point amplitude we review the structure of the
integrals appearing for four-- and five--point open string amplitudes.

The Euler--Beta function $B(a,b)$, relevant for four open string processes, is
given by
\eqn\euler{\eqalign{
B(a,b)&=\int_0^1 dx\ x^a\ (1-x)^b=\fc{1}{1+a}\ \F{2}{1}\lf[1+a,-b,2+a;1\ri]\cr
&=\fc{\Gamma(1+a)\ \Gamma(1+b)}{\Gamma(2+a+b)}\ \ ,\ \ \re\ a>-1\ ,\ \re\ b>-1\ .}}
Moreover, a scattering of five open strings boils down to the integral\foot{
In that case, we have five kinematic invariants, encoded in the powers $a,b,c,d$ and $e$
(\cf \eqq (A.15) for more details}:
\eqn\five{
C(a,b,c,d,e):=\int_0^1 dx\int_0^1 dy\ x^a\ y^b\ (1-x)^c\ (1-y)^d\ (1-xy)^e\ .}
It is straightforward to evaluate this integral by making use of the integral
representation of the generalized hypergeometric function $\F{p}{q}$  \Grad:
\eqn\HYPERGEO{\eqalign{
\F{2}{1}[-c,1+a,2+a+b;\ y\ ]\ \fc{\Gamma(1+a)\ \Gamma(1+b)}{\Gamma(2+a+b)}
&=\int^1_0\ dx\ x^a\ (1-x)^b\ (1-xy)^c\ ,\cr
\hskip-0.5cm\F{p+1}{q+1}\lf[{1+a,a_1,\ldots,a_p\atop 2+a+b, b_1, \ldots, b_q} ; \lambda
\ri]\ \fc{\Gamma(1+a)\ \Gamma(1+b)}{\Gamma(2+a+b)}&=\int^1_0 dx\ x^a\
(1-x)^b\ \F{p}{q}\lf[{a_1,\ldots,a_p\atop b_1,\ldots,b_q}; \lambda x\ri],}}
with $\re\ a>-1,\ \re\ b>-1$, and $p\leq q+1$. In addition $|\lambda|<1$ for the case $p=q+1$.
To this end one obtains
\eqn\Five{
C(a,b,c,d,e)=\fc{\Ga(1+a)\ \Gamma(1+b)\ \Ga(1+c)\ \Ga(1+d)}{\Ga(2+a+c)\
\Ga(2+b+d)}\ \F{3}{2}\lf[{1+a,\ 1+b,\ -e\atop 2+a+c,\ 2+b+d}\ ;\ 1\ri]\ ,}
with $\re(a),\re(b),\re(c),\re(d)>-1$.
Later, we will need the power series representation of the hypergeometric
functions $\F{p}{q}$ \slater:
\eqn\powerhyper{
\F{p}{q}\lf[{a_1,\ldots,a_p \atop b_1,\ldots,b_q}\ ;\ x\ri]=\sum_{s=0}^\infty
\fc{1}{s!}\ \fc{(a_1,s)\cdot\ldots\cdot(a_p,s)}{(b_1,s)\cdot\ldots\cdot(b_q,s)}\ x^s\ .}
Here we have introduced the Pochhammer symbol
$(a,m)=\fc{\Gamma(a+m)}{\Gamma(a)}$. The parameters $a_i,b_j$ and the variable $x$ 
assumed complex values, except $b_j\neq 0,-1,-2,\ldots$. 
We shall be only concerned with the case $p=q+1$. For that case, the series
is convergent for $|x|<1.$ With
\eqn\converg{
\omega:=\re\lf(\sum\limits_{i=1}^q b_i-\sum\limits_{i=1}^p a_i\ri)\ ,}
at $|x|=1$ the series is absolutely convergent for $\omega>0$
and divergent for $\omega\leq -1$.

\subsec{Generalized Kamp\'e de Feriet  function $F^{A:B}_{C:D}$ and triple 
hypergeometric function $F^{(3)}$}

After this warm up we now turn our attention to the  integral \todo:
\eqn\serious{\eqalign{
F\lf[{a,b,d,e,g\atop c,f,h,j}\ri]&:=\int_0^1 dx\int_0^1 dy\int_0^1 dz\cr
&\times\  x^a\ y^b\ z^c\ (1-x)^d\ (1-y)^e\ (1-z)^f\ (1-xy)^g\ (1-yz)^h\ (1-xyz)^j\ .}}
Any tree--level scattering process with six open strings involves such type of integral.
Hence, the latter is as elementary as the expressions \euler\ or \five\
are generic to open string four or five--point amplitudes, respectively.
The function, which eventually will turn to represent some generalized Kamp\'e
de F\'eriet function shows among others the symmetry:
\eqn\symmetry{
F\lf[{a,b,d,e,g\atop c,f,h,j}\ri]=F\lf[{c,b,f,e,h\atop a,d,g,j}\ri]\ .}
To proceed, we use the series expansion:
\eqn\series{
(1-xy)^g\ (1-yz)^h=\sum_{m,n=0}^\infty \fc{(-g,m)\ (-h,n)}{(1,m)\ (1,n)}\ x^m\
z^n\ y^{m+n}\ \ \ ,\ \ \ |x|<1\ \ ;\ \ |y|<1\ \ ;\ \ |z|<1\ .}
After replacing in \serious\ the polynomials $(1-xy)^g\ (1-yz)^h$ by the above
double sum we are able to perform the integrals with the help of the formulas
\HYPERGEO, provided $\re(d),\re(e),\re(f)>-1$ and $m+\re(a)>-1,\ n+\re(c)>-1,\
\re(b)+m+n>-1$:
\eqn\arrive{\eqalign{
F\lf[{a,b,d,e,g\atop c,f,h,j}\ri]&=\Ga(1+d)\ \Ga(1+e)\ \Ga(1+f)\
\sum_{m,n=0}^\infty \fc{(-g,m)\ (-h,n)}{(1,m)\ (1,n)}\cr
&\times \fc{\Ga(1+m+n+b)\ \Ga(1+m+a)\ \Ga(1+n+c)}{\Ga(2+m+n+b+e)\ \Ga(2+m+a+d)\ \Ga(2+n+c+f)}\cr
&\times \F{4}{3}\lf[{1+m+n+b,\ 1+m+a,\ 1+n+c,\ -j\atop 2+m+n+b+e,\ 2+m+a+d,\ 2+n+c+f}\
;\ 1\ri]\ . }}
Evidently, the result \arrive\ shows the symmetry \symmetry.

Before we undertake to write \arrive\ as a single analytic function
we shall discuss the three special cases $(i)$\  $h=0=g$, $(ii)$\ $h=0$ and $(iii)$\ $g=0$.
For the special case $(i)$, \ie $g=0=h$, the sum becomes trivial. In that case, only the
terms with $m=0=n$ give a non--vanishing contribution.
This can be seen from the identity $\fc{(-g,m)}{(1,m)}\ra \delta_m$ for $g\ra 0$
and similarly for $\fc{(-h,n)}{(1,n)}$.
Hence we find:
\eqn\simplify{\eqalign{
F\lf[{a,b,d,e,0\atop c,f,0,j}\ri]&=
\fc{\Ga(1+a)\ \Ga(1+b)\ \Ga(1+c)\ \Ga(1+d)\ \Ga(1+e)\ \Ga(1+f)}
{\Ga(2+b+e)\ \Ga(2+a+d)\ \Ga(2+c+f)}\cr
&\times \F{4}{3}\lf[{1+b,\ 1+a,\ 1+c,\ -j\atop 2+b+e,\ 2+a+d,\ 2+c+f}\
;\ 1\ri]\ .}}
Furthermore, for the case $(ii)$, with $h=0$:
\eqn\simplifyi{\eqalign{
F\lf[{a,b,d,e,g\atop c,f,0,j}\ri]&=\Ga(1+d)\ \Ga(1+e)\ \Ga(1+f)\
\sum_{m=0}^\infty \fc{(-g,m)}{(1,m)}\cr
&\times \fc{\Ga(1+m+b)\ \Ga(1+m+a)\ \Ga(1+c)}{\Ga(2+m+b+e)\ \Ga(2+m+a+d)\ \Ga(2+c+f)}\cr
&\times \F{4}{3}\lf[{1+m+b,\ 1+m+a,\ 1+c,\ -j\atop 2+m+b+e,\ 2+m+a+d,\ 2+c+f}\ ;\ 1\ri]\ .}}
With the power series representation for the hypergeometric function \powerhyper\
we may express \simplifyi\ through a generalized Kamp\'e de F\'eriet function $F^{A:B}_{C:D}$
\PARIS. These are functions in $N$ variables $x_1,\ldots,x_N$ and have the power series
representation:
\eqn\kampe{\eqalign{
&F^{A:B}_{C:D}\lf[{a_1,\ldots,a_A:b_{1,1},\ldots,b_{1,B};\ b_{2,1},\ldots,b_{2,B};\ \ldots\ ;\
b_{N,1},\ldots,b_{N,B}\atop c_1,\ldots,c_C:d_{1,1},\ldots,d_{1,D};\ d_{2,1},\ldots,d_{2,D};\
\ldots\ ;\ d_{N,1},\ldots,d_{N,D}}\ ;\ x_1,\ldots,x_N\ri]\cr
&=\sum_{m_1,\ldots,m_N=0}^\infty\ \fc{\prod\limits_{j=1}^A(a_j,m_1+\ldots m_N)\
\prod\limits_{j=1}^B (b_{1,j},m_1)\cdot\ldots(b_{N,j},m_N)}
{\prod\limits_{j=1}^C(c_j,m_1+\ldots m_N)\
\prod\limits_{j=1}^D (d_{1,j},m_1)\cdot\ldots(d_{N,j},m_N)}\
\fc{x_1^{m_1}\cdot\ldots\cdot x^{m_N}_N}{m_1!\cdot\ldots\cdot m_N!}\ .}}
Kamp\'e de F\'eriet functions are generalizations of the four Lauricella hypergeometric
functions\foot{The latter correspond to the following four special cases of $F^{A:B}_{C:D}$\ :
\eqn\APPELL{\eqalign{
F_A&:\ A=1,\ B=1,\ C=0,\ D=1\ ,\cr
F_B&:\ A=0,\ B=2,\ C=1,\ D=0\ ,\cr
F_C&:\ A=2,\ B=0,\ C=0,\ D=1\ ,\cr
F_D&:\ A=1,\ B=1,\ C=1,\ D=0\ .}}} $F_A,F_B,F_C$ and $F_D$.
The latter are extensions of Appell's hypergeometric functions $F_2,F_3,F_4$ and $F_1$
from two variables to $N$ variables \appell, respectively.
To this end we may rewrite \simplifyi:
\eqn\simplifyii{\eqalign{
F\lf[{a,b,d,e,g\atop c,f,0,j}\ri]&=\fc{\Ga(1+a)\ \Ga(1+b)\ \Ga(1+c)\ \Ga(1+d)\ \Ga(1+e)\ \Ga(1+f)}
{\Ga(2+a+d)\ \Ga(2+b+e)\ \Ga(2+c+f)}\cr
&\times
F^{2:2}_{2:1}\lf[{\ \ 1+a,1+b\ :\ 1+c,\ -j\ ;\ -g,\ 1
\atop 2+a+d,2+b+e\ :\ 2+c+f\ ;\ 1}\ ;\ 1,1\ri]\ .}}
We may derive a similar expression for the third case $(iii)$.

After having discussed the three special cases we now want to turn to the general case \arrive
\eqn\general{\eqalign{
F\lf[{a,b,d,e,g\atop c,f,h,j}\ri]&=\fc{\Ga(1+d)\ \Ga(1+e)\ \Ga(1+f)}{\Ga(-g)\ \Ga(-h)\ \Ga(-j)}
\sum_{m_i=0}^\infty \fc{\Ga(-g+m_1)\ \Ga(-h+m_2)\ \Ga(-j+m_3)}{m_1!\ m_2!\ m_3!}\cr
&\hskip-1.2cm\times \fc{\Ga(1+m_1+m_2+m_3+b)}{\Ga(2+m_1+m_2+m_3+b+e)}\
\fc{\Ga(1+m_1+m_3+a)}{\Ga(2+m_1+m_3+a+d)}\ \fc{\Ga(1+m_2+m_3+c)}{\Ga(2+m_2+m_3+c+f)}\ .}}
A convenient expression for the triple sum \general\ may be also 
\eqn\General{\eqalign{
F\lf[{a,b,d,e,g\atop c,f,h,j}\ri]&=\fc{\Ga(1+d)\ \Ga(1+e)\ \Ga(1+f)}{\Ga(-g)\ \Ga(-h)\ \Ga(-j)}\cr
&\times\sum_{n_1=0}^\infty \sum_{n_2=0}^{n_1}\sum_{n_3=n_1-n_2}^{n_1}
\fc{\Ga(-g+n_1-n_3)}{\Ga(1+n_1-n_3)}\fc{\Ga(-h+n_1-n_2)}{\Ga(1+n_1-n_2)}
\fc{\Ga(-j-n_1+n_2+n_3)}{\Ga(1-n_1+n_2+n_3)}\cr
&\times \fc{\Ga(1+n_1+b)}{\Ga(2+n_1+b+e)}\
\fc{\Ga(1+n_2+a)}{\Ga(2+n_2+a+d)}\ \fc{\Ga(1+n_3+c)}{\Ga(2+n_3+c+f)}}}
with two terminating sums.
We want to express the triple sum \general\ through a generalized Kamp\'e de F\'eriet function. 
In fact, Srivastava has introduced a general
{\it triple hypergeometric function} $F^{(3)}[x,y,z]$ depending on three variables, \cf 
last {\it Ref.} of \PARIS
\eqn\SRIVASTAVA{\hskip-0.3cm
F^{(3)}[x,y,z]\equiv F^{(3)}\lf[{(a)::(b);(b');(b''):(c);(c');(c'')\atop 
(e)::(g);(g');(g''):(h);(h');(h'')}\ ;\ x,y,z\ri]=\sum_{m,n,p=0}^\infty
\Lambda(m,n,p)\ \fc{x^m}{m!}\ \fc{y^n}{n!}\ \fc{z^p}{p!}\ ,}
with 
\eqn\SRIVAST{\eqalign{
\Lambda(m,n,p)&=\fc{\prod\limits_{j=1}^A (a_j,m+n+p)\ \prod\limits_{j=1}^B (b_j,m+n)
\ \prod\limits_{j=1}^{B'} (b'_j,n+p)\ \prod\limits_{j=1}^{B''} (b''_j,m+p)}
{\prod\limits_{j=1}^E (e_j,m+n+p)\ \prod\limits_{j=1}^G (g_j,m+n)
\ \prod\limits_{j=1}^{G'} (g'_j,n+p)\ \prod\limits_{j=1}^{G''} (g''_j,m+p)}\cr
&\times \fc{\prod\limits_{j=1}^C (c_j,m)\ \prod\limits_{j=1}^{C'} (c'_j,n)
\ \prod\limits_{j=1}^{C''} (c''_j,p)}{\prod\limits_{j=1}^H (h_j,m)\ \prod\limits_{j=1}^{H'} (h'_j,n)
\ \prod\limits_{j=1}^{H''} (h''_j,p)}\ ,}}
with $(a)$ denoting a set of $A$ parameter $a_1,\ldots,a_A$ and similarly for $(b),(b'),\ldots$.
With this nice expression we may write our integral \serious, which is expressed through \general,
in the following closed form:
\eqn\KAMPE{\eqalign{
F\lf[{a,b,d,e,g\atop c,f,h,j}\ri]&=\fc{\Ga(1+a)\ \Ga(1+b)\ \Ga(1+c)\ \Ga(1+d)\ \Ga(1+e)\ \Ga(1+f)}
{\Ga(2+a+d)\ \Ga(2+b+e)\ \Ga(2+c+f)}\cr
&\times F^{(3)}\lf[{1+b::1;1+c;1+a:-g,1;-h,1;-j,1\atop 
2+b+e::1;2+c+f;2+a+d:1;1;1}\ ;\ 1,1,1\ri]\ .}}

\subsec{Harmonic Number and Euler sums}

To analyze the string $S$--matrix results, we need their momentum expansion.
Hence, we need to expand the hypergeometric series \powerhyper\ w.r.t.
small entries $a_i,b_i$.
A formal expression for the generalized hypergeometric function \powerhyper\ may be given
\eqn\Exphar{\eqalign{
\F{p}{q}\lf[a_1+\ep\ \al_1,\ldots,a_p+\ep\ \al_p\atop b_1+\ep\ \beta_1,\ldots,b_q+
\ep\ \beta_q\ri]&=
\sum_{s=0}^\infty \fc{1}{s!}\ 
\fc{\prod\limits_{i=1}^p(a_i+\ep\ \al_i,s)}{\prod\limits_{j=1}^q (b_j+\ep\ \beta_j,s)}\cr
\fc{(m+\alpha\ \epsilon,s)}{(m,s)}&=e^{-\sum\limits_{k=1}^\infty 
\fc{(-\alpha\ \epsilon)^k}{k}\ \lf(H_{m+s-1,k}-H_{m-1,k}\ri)}\ .}}
Strictly speaking, in the last formula, $m$ has to be an integer positive number, with $m>1$.
This condition limits the range of parameters $a_i,b_j$. However, by analytic continuation,
the above expression keeps its form.
At any rate, from this formula we see, that quite generically a class of infinite sums appears.
These sums are known as {\it Euler sums} and involve the {\it harmonic number} $H_n$ 
or the {\it generalized harmonic number} $H_{n,a}$.
The harmonic number $H_n$ and the generalized harmonic number $H_{n,a}$ are defined:
\eqn\harmonic{
H_n=\sum_{k=1}^n \fc{1}{k}\ \ \ ,\ \ \ H_{n,a}=\sum_{k=1}^n \fc{1}{k^a}\ .}
Clearly, we have $H_{n,1}=H_n$ and $H_n=\gamma_E+\psi(n+1)$.
The harmonic number is related to the polygamma function $\psi$:
\eqn\rewriting{\eqalign{
\psi(n)&=-\gamma_E+H_{n-1}\ ,\cr
\psi^{(1)}(n)&=\zeta(2)-H_{n-1,2}\ .}}
The sums, which appear after expanding the exponential \Exphar\ w.r.t. small $\epsilon$, 
generically lead to Euler sums.
The following two classes of Euler sums are relevant to us:
\eqn\EULER{\eqalign{
s_h(m,n)&=\sum_{k=1}^\infty\fc{H_k^m}{(k+1)^n}\ ,\cr
\sigma_h(m,n)&=\sum_{k=1}^\infty\fc{H_{k,m}}{(k+1)^n}\ .}}
for $m\geq 1,\ n\geq 2$.
We have the following relations:
\eqn\have{\eqalign{
\sum_{k=1}^\infty\fc{H_{k,m}}{k^n}&=\sigma_h(m,n)+\zeta(m+n)\ ,\cr
\sum_{k=1}^\infty\fc{H_k^m}{k^n}&=s_h(m,n)-\sum_{j=0}^{m-1}\lf(m\atop j\ri)\sum_{k=1}^\infty
(-1)^{m-j}\ \fc{H_k^j}{k^{m+n-j}}\ .}}
{From} the last equation we obtain the identity:
\eqn\identity{
\sum_{k=1}^\infty\fc{H_k^2}{k^n}=s_h(2,n)+2\ s_h(1,n+1)+\zeta(2+n)\ .}
It has been already pointed out by Euler, that in some cases the sums may be reduced 
to a rational  linear combination of product of single sums. 
There is recent work on the explicit evaluation of some Euler sums in {\it Refs.}
\threeref\borwein\BaileyBorwein\BBG. In the following let us report some of their results:
\eqn\report{\eqalign{
\sigma_h(1,2)&=\zeta(3)\ ,\cr
\sigma_h(1,3)&=\fc{3}{2}\ \zeta(4)-\h\ \zeta(2)^2\ ,\cr
\sigma_h(2,2)&=\h\ \zeta(2)^2-\h\ \zeta(4)\ ,\cr
\sigma_h(1,4)&=2\ \zeta(5)-\zeta(2)\ \zeta(3)\ ,\cr
\sigma_h(2,3)&=-\fc{11}{2}\ \zeta(5)+3\ \zeta(2)\ \zeta(3)\ ,\cr
\sigma_h(2,4)&=-6\ \zeta(6)+\fc{8}{3}\ \zeta(2)\ \zeta(4)+\zeta(3)^2\ ,\cr
\sigma_h(1,5)&=\fc{5}{2}\ \zeta(6)-\zeta(2)\ \zeta(4)-\h\ \zeta(3)^2\ ,\cr
\sigma_h(4,2)&=5\ \zeta(6)-\fc{5}{3}\ \zeta(2)\ \zeta(4)-\zeta(3)^2\ ,}}
and:
\eqn\reportiii{\eqalign{
s_h(2,2)&=\fc{3}{2}\ \zeta(4)+\h\ \zeta(2)^2=\fc{11}{4}\ \zeta(4)\ ,\cr
s_h(3,2)&=\fc{15}{2}\ \zeta(5)+\zeta(2)\ \zeta(3)\ ,\cr
s_h(2,4)&=\fc{2}{3}\ \zeta(6)-\fc{1}{3}\ \zeta(2)\ \zeta(4)+\fc{1}{3}\ \zeta(2)^3-\zeta(3)^2\ .}}
In addition, we have \BaileyBorwein:
\eqn\reporti{
s_h(1,n)=\sigma_h(1,n)=\h\ n\ \zeta(n+1)-\h\ \sum_{k=1}^{n-2}\zeta(n-k)\ \zeta(k+1)\ .}
An useful identity is the {\it reflection formula} \BBG
\eqn\duplication{
\sigma_h(s,t)+\sigma_h(t,s)=\zeta(s)\ \zeta(t)-\zeta(s+t)\ \ \ ,\ \ \ s,t\geq 2\ ,}
which leads to
\eqn\Generalharm{
\sigma_h(a,a)=\h\ \zeta(a)^2-\h\ \zeta(2a)\ \ \ ,\ \ \ a\geq 2\ ,}
or: 
\eqn\generalharm{
\sum_{n=1}^\infty \fc{H_{n,a}}{n^a}=\h\ \zeta(a)^2+\h\ \zeta(2a)\ \ \ ,\ \ \ a\geq 2\ .}
So far, we have reviewed very useful identities about Euler sums from the literature.
Essentially we have quoted results from \threeref\borwein\BaileyBorwein\BBG.
Let us now apply these results to infinite sums involving the harmonic number series \harmonic\
as they appear in the expansion of hypergeometric series:
First we shall need the class of sums:
\eqn\Borwein{\eqalign{
&\sum_{n=1}^\infty\fc{H_n}{n^2}=2\ \zeta(3)\ \ \ ,\ \ \ 
\sum_{n=1}^\infty\fc{H_n^2}{n^2}=\fc{11}{2}\ \zeta(4)-\h\ \zeta(2)^2=\fc{17}{4}\ \zeta(4)
\ \ \ ,\ \ \ 
\sum_{n=1}^\infty\fc{H_n}{n^3}=\fc{5}{4}\ \zeta(4)\ ,\cr
&\sum_{n=1}^\infty\fc{H_n}{n^4}=3\ \zeta(5)-\zeta(2)\ \zeta(3)\ \ \ ,\ \ \ 
\sum_{n=1}^\infty\fc{H_n}{n^5}=\zeta(2)\ \zeta(4)-\h\ \zeta(3)^2\ .}}
Later, we shall need the following series involving the generalized harmonic number. They
may be easily derived from the above identities:
\eqn\Simone{\eqalign{
\sum_{n=1}^\infty \fc{H_{n,2}}{n^4}&=\sigma_h(2,4)+\zeta(6)=
-5\ \zeta(6)+\fc{8}{3}\ \zeta(2)\ \zeta(4)+\zeta(3)^2\ ,\cr
\sum_{n=1}^\infty \fc{H_{n,2}}{n^3}&=\sigma_h(2,3)+\zeta(5)=
-\fc{9}{2}\ \zeta(5)+3\ \zeta(2)\ \zeta(3)\ ,\cr
\sum_{n=1}^\infty \fc{H_{n,4}}{n^2}&=\sigma_h(4,2)+\zeta(6)=
6\ \zeta(6)-\fc{5}{3}\ \zeta(2)\ \zeta(4)-\zeta(3)^2\ ,}}

In the following we shall be concerned to extend the above list to our needs in the subsections
4.4. and 4.5.
First, using \generalharm\ let us evaluate the sum:
\eqn\first{
\sum_{k=1}^\infty\fc{1}{k^{a}}\ \psi^{(b)}(k)=(-1)^{b+1}\ b!\ 
\lf[\ \zeta(a)\ \zeta(1+b)-\sigma_h(1+b,a)\ \ri]\ .}
Here we have \Grad:
\eqn\PSI{
\psi^{(b)}(x)=(-1)^{b+1}\ b!\ \sum_{k=0}^\infty\fc{1}{(x+k)^{b+1}}=(-1)^{b+1}\ b!\ \zeta(b+1,x)\ ,}
which for $x\in {\bf N}$ may be also conveniently written: 
\eqn\NICEHARM{
\psi^{(b-1)}(n)=(-1)^b\ (b-1)!\ [\ \zeta(b)-H_{n-1,b}\ ]\ \ \ ,\ \ \ b>1\ .} 
With \Generalharm\ \eqq \first\ boils down to
$$\sum_{k=1}^\infty\fc{1}{k^{1+b}}\ \psi^{(b)}(k)=
\h\ (-1)^{b+1}\ b!\ \lf[\ \zeta(2b+2)+\zeta(b+1)^2\ \ri]\ ,$$
for the case $a=1+b$.

Before we proceed, we shall present the series:
\eqn\proceed{
\sum_{n=1}^\infty\fc{1}{n(n+1)^\alpha}=\alpha-\sum_{i=2}^\alpha\zeta(i)\ .}
This identity may be proven by induction, with noting:
$\sum\limits_{n=1}^\infty\fc{1}{n(n+1)}=\psi(1)+\gamma_E+1=1$; $\psi(1)=-\gamma_E$.
Next, we present the identity:
\eqn\nextid{\eqalign{
\sum_{n=1}^\infty\fc{1}{n(n+1)^\al} H_{n-1}&=
\h\al(\al+1)-\h\sum_{i=2}^\al i\ \zeta(i+1)-\sum_{i=2}^\al(\al-i+1)\ \zeta(i)\cr
&+\h\sum_{k=0}^{\al-3}
\sum_{i\geq 0}^k\zeta(2+i)\ \zeta(2+k-i)\ .}}
Again, this identity may be proven by induction, with noting 
$\sum\limits_{n=1}^\infty\fc{1}{n(n+1)} H_{n-1}=1$.
Some special cases are:
\eqn\Katze{\eqalign{
\sum_{n=1}^\infty\fc{1}{(n+1)^4} H_{n-1}&=-4+\zeta(2)+\zeta(3)+\zeta(4)-\zeta(2)\ \zeta(3)+
2\ \zeta(5)\ ,\cr
\sum_{n=1}^\infty\fc{1}{n(n+1)^4} H_{n-1}&=10-3\ \zeta(2)-3\ \zeta(3)-\fc{5}{4}\ \zeta(4)
+\zeta(2)\ \zeta(3)-2\ \zeta(5)\ ,\cr
\sum_{n=1}^\infty\fc{1}{n^2(n+1)^4} H_{n-1}&=-20+6\ \zeta(2)+7\ \zeta(3)+\fc{3}{2}\ \zeta(4)
-\zeta(2)\ \zeta(3)+2\ \zeta(5)\ .}}

For the power series expansion of the hypergeometric functions $_4F_3$ we shall 
need the two series:
\eqn\putting{\eqalign{
(i)&\ \ \ \sum_{n=1}^\infty\fc{\psi^{(1)}(n)}{n\ (n+1)^3}=10-3\ \zeta(2)
-\h\ \zeta(2)^2-\zeta(3)-\h\ \zeta(4)+2\ \zeta(2)\ \zeta(3)-\fc{11}{2}\ \zeta(5)\ ,\cr
(ii)&\ \ \ \sum_{n=1}^\infty\fc{H_{n-1}^2}{n\ (n+1)^3}
=10-\h\ \zeta(2)^2-5\ \zeta(3)-2\ \zeta(4)-\zeta(2)\ \zeta(3)+\fc{3}{2}\ \zeta(5)\ .}}
Both series may be determined by making use of the decomposition:
\eqn\decomp{
\fc{1}{n(n+1)^3}=\fc{1}{n(n+1)}-\fc{1}{(n+1)^2}-\fc{1}{(n+1)^3}\ .}

Let us first prove the series:
\eqn\firstprove{
\sum_{n=1}^\infty\fc{H_{n,2}}{n(n+1)^3}=\zeta(3)-\h\ \zeta(2)^2+\h\ \zeta(4)+\fc{11}{2}\ \zeta(5)
-3\ \zeta(2)\ \zeta(3)\ .}
According to the decomposition \decomp\ we need the series:
\eqn\turn{\eqalign{
\sum_{n=1}^\infty \fc{H_{n,2}}{n(n+1)}&=\sum_{n=1}^\infty\lf(\fc{1}{n}-\fc{1}{n+1}\ri)\ H_{n,2}
=\sum_{n=1}^\infty \lf(\fc{H_{n,2}}{n}-\fc{H_{n-1,2}}{n}\ri)=\sum_{n=1}^\infty \fc{1}{n^3}=
\zeta(3)\  \ ,\cr
\sum_{n=1}^\infty \fc{H_{n,2}}{(n+1)^2}&= \sum_{n=1}^\infty \fc{H_{n-1,2}}{n^2}=
\sum_{n=1}^\infty \fc{H_{n,2}}{n^2}-\sum_{n=1}^\infty\fc{1}{n^4}=\h\ \zeta(2)^2-\h\ \zeta(4)\ ,\cr
\sum_{n=1}^\infty \fc{H_{n,2}}{(n+1)^3}&=\sigma_h(2,3)=
-\fc{11}{2}\ \zeta(5)+3\ \zeta(2)\ \zeta(3)  \ .\cr}}
After putting together all three series along \decomp\ we obtain \firstprove.

To prove the series $(i)$ of \eqq \putting\ we write:
\eqn\writee{\eqalign{
\sum_{n=1}^\infty\fc{\psi^{(1)}(n)}{n(n+1)^3}&=\sum_{n=1}^\infty\fc{1}{n(n+1)^3}\ 
\sum_{k=0}^\infty\fc{1}{(n+k)^2}=\sum_{n=1}^\infty\fc{1}{n(n+1)^3}
\lf(\sum_{k=1}^\infty\fc{1}{k^2}-\sum_{k=1}^{n-1}\fc{1}{k^2}\ri)\cr
&=\zeta(2)\ \sum_{n=1}^\infty\fc{1}{n(n+1)^3}-
\sum_{n=1}^\infty\fc{H_{n-1,2}}{n(n+1)^3}\cr
&=\zeta(2)\ [3-\zeta(2)-\zeta(3)]-\sum_{n=1}^\infty\fc{H_{n,2}}{n(n+1)^3}+
\sum_{n=1}^\infty\fc{1}{n^3(n+1)^3}\ .}}
The first term is obtained from \proceed. The second term corresponds to \firstprove\ 
and the last sum is straightforward to calculate: 
$$\sum\limits_{n=1}^\infty \fc{1}{n^3(n+1)^3}=10-\pi^2\ .$$ 
Hence, after putting all pieces together we obtain the series $(i)$ of \eqq \putting.

Before we undertake to evaluate series $(ii)$, we shall prove the series:
\eqn\secondprove{
\sum_{n=1}^\infty\fc{H_n^2}{n(n+1)^3}=3\ \zeta(3)-\fc{3}{2}\ \zeta(4)-\h\ \zeta(2)^2+\fc{3}{2}\ 
\zeta(5)-\zeta(2)\ \zeta(3)\ .}
Similar as before, we shall calculate the series:
\eqn\focus{\eqalign{
\sum_{n=1}^\infty \fc{H_n^2}{n(n+1)}&=\sum_{n=1}^\infty\lf(\fc{1}{n}-\fc{1}{n+1}\ri)\ H_n^2
=\sum_{n=1}^\infty \lf(\fc{H_n^2}{n}-\fc{H_{n-1}^2}{n}\ri)\cr
&=2\sum_{n=1}^\infty \fc{H_n}{n^2}-\sum_{n=1}^\infty\fc{1}{n^3}=3\ \zeta(3)\  \ ,\cr
\sum_{n=1}^\infty \fc{H_n^2}{(n+1)^2}&=s_h(2,2)=\fc{3}{2}\ \zeta(4)+\h\ \zeta(2)^2\ ,\cr
\sum_{n=1}^\infty \fc{H_n^2}{(n+1)^3}&=s_h(2,3)\ .}}
The Euler sum $s_h(2,3)$ may not be found in the list \reportiii. Instead we may determine it 
by using some identities from \Girgensohn. In this work, the relation
\eqn\triplesum{
\sum_{n=1}^\infty \fc{1}{n^a}\ H_{n-1,b}\ H_{n-1,c}=\zeta(a,b,c)+\zeta(a,c,b)+\sigma_h(b+c,a)\ ,}
involving the so--called {\it triple Euler sum}
\eqn\ZETA{
\zeta(a,b,c)=\sum_{n=1}^\infty\sum_{m=1}^{n-1}\sum_{k=1}^{m-1}\fc{1}{n^am^bk^c}\ ,}
has been proven. Furthermore, in {\it Ref.} \Girgensohn\ a list of values for $\zeta(a,b,c)$ 
has been given, in particular $\zeta(3,1,1)=2\ \zeta(5)-\zeta(2)\ \zeta(3)$.
Hence we have: 
\eqn\Eulerzweidrei{\eqalign{
s_h(2,3)&=\sum_{n=1}^\infty \fc{H_n^2}{(n+1)^3}=\sum_{n=1}^\infty \fc{H^2_{n-1}}{n^3}
=2\ \zeta(3,1,1)+\sigma_h(2,3)=-\fc{3}{2}\ \zeta(5)+\zeta(2)\ \zeta(3)  \ .}}
{From} \focus, together with the decomposition \decomp, we obtain \secondprove.

Finally, we undertake to prove $(ii)$ of \eqq \putting. We first write:
\eqn\writei{
\sum_{n=1}^\infty\fc{H_{n-1}^2}{n(n+1)^3}=\sum_{n=1}^\infty\fc{H_n^2}{n(n+1)^3}+
\sum_{n=1}^\infty\fc{1}{n^3(n+1)^3}-2\sum_{n=1}^\infty\fc{H_n}{n^2(n+1)^3}\ .}
The first two sums have been presented above, while for the last sum we have:
$$\sum_{n=1}^\infty \fc{H_n}{n^2(n+1)^3}=-3\ \zeta(2)+4\ \zeta(3)+\fc{1}{4}\ \zeta(4)\ .$$
After inserting all sums into \writei\ we arrive at the final result $(ii)$ of \eqq
\putting.

The relation  \triplesum\ is very useful. For $b=c$ it boils down to:
\eqn\IMport{
\sum_{n=1}^\infty\fc{H_{n-1,b}^2}{n^a}=2\ \zeta(a,b,b)+\sigma_h(2b,a)\ .}
{From} {\it Ref.} \Girgensohn\ we have
\eqn\newhave{
\zeta(a,a,a)=\fc{1}{6}\ \zeta(a)^3-\h\ \zeta(a)\ \zeta(2a)+\fc{1}{3}\ \zeta(3a)\ ,}
which allows to derive the following identity from \IMport:
\eqn\IMporti{
\sum_{n=1}^\infty\fc{H_{n-1,2}^2}{n^2}=2\ \zeta(2,2,2)+\sigma_h(4,2)
=\fc{1}{3}\ \zeta(2)^3-\zeta(3)^2-\fc{8}{3}\ \zeta(2)\ \zeta(4)+\fc{17}{3}\ \zeta(6)\ .}

\subsec{Momentum expansion of hypergeometric functions $\F{3}{2}$ and $\F{4}{3}$}

In the previous subsection we have compiled material, which one needs
to expand Gaussian hypergeometric functions \powerhyper\ w.r.t. to small parameter
$a_i$ and $b_j$. To obtain the $\ap$--expansion of a five--point string 
$S$--matrix one needs the power series expansion of \five\ w.r.t. the small parameter $a,\ldots,e$.
Let us present some important examples, which prove to be useful in any five--point 
scattering process on the disk (\cf appendix \appA).
With the previous results and appendix \appB\ we may derive the following two expansions
% \foot{
% Note, that with an identity from \bailey, we may also write:
% \eqn\Bailey{\eqalign{
% \F{3}{2}\left[{2+\alpha_1,1+\alpha_2,1+\alpha_3  
% \atop 3+\beta_1,2+\beta_2} \right]&=\fc{\Gamma(3+\beta_1)\Gamma(2+\beta_2)
% \Gamma(1-\alpha_1-\alpha_2-\alpha_3+\beta_1+\beta_2)}{\Gamma(2+\alpha_1)
% \Gamma(2-\alpha_1-\alpha_3+\beta_1+\beta_2)
% \Gamma(2-\alpha_1-\alpha_2+\beta_1+\beta_2)}\cr
% &\times \F{3}{2}\left[ {1-\alpha_1+\beta_1,\beta_2-\alpha_1,
% 1-\alpha_1-\alpha_2-\alpha_3+\beta_1+\beta_2  \atop 
% 2-\alpha_1-\alpha_3+\beta_1+\beta_2,
% 2-\alpha_1-\alpha_2+\beta_1+\beta_2} \right]\ .}}} 
w.r.t. small parameter $a,\ldots e$
\eqn\MOMEXP{\eqalign{
&C(a,b,c,d,e)=\fc{\Gamma(1+a)\ \Gamma(1+b)\ \Gamma(1+c)\ \Gamma(1+d)}{\Gamma(2+a+c)\ \Gamma(2+b+d)}\ 
\ \F{3}{2}\lf[{1+a,\ 1+b,\ -e\atop 2+a+c,\ 2+b+d}\ri]\cr
&=1 - a - b - c - d -2\ e+e\ \zeta(2)
+a^2 + a\,b + b^2 + b\,c + c^2 + a\,d + c\,d + d^2\cr
&+ 3\,a\,e + 3\,b\,e + 3\,c\,e + 3\,d\,e + 3\,e^2+2\ ac+2\ bd\cr
&-(a\ c + b\ d + a\ e + b\ e + e^2)\ \zeta(2)-(a\ e + b\ e + 2\ c\ e + 2\ d\ e + e^2)\ \zeta(3)\cr
&-a^3 - a^2\,b - a\,b^2 - b^3 - 3\,a^2\,c - 2\,a\,b\,c - b^2\,c - 3\,a\,c^2 - b\,c^2 - c^3 - a^2\,d - 
2\,a\,b\,d- 3\,b^2\,d\cr
&- 2\,a\,c\,d - 2\,b\,c\,d - c^2\,d - a\,d^2 - 3\,b\,d^2 - c\,d^2 - d^3 - 4\,a^2\,e - 4\,a\,b\,e - 
4\,b^2\,e - 8\,a\,c\,e\cr 
&- 4\,b\,c\,e - 4\,c^2\,e - 4\,a\,d\,e - 8\,b\,d\,e - 4\,c\,d\,e - 4\,d^2\,e - 
  6\,a\,e^2 - 6\,b\,e^2 - 6\,c\,e^2 - 6\,d\,e^2 - 4\,e^3\cr
&+(a^2\,c + a\,b\,c + a\,c^2 + a\,b\,d + b^2\,d + a\,c\,d + b\,c\,d + b\,d^2 + a^2\,e + a\,b\,e + b^2\,e 
+ 3\,a\,c\,e + 3\,b\,d\,e \cr
&- c\,d\,e + 2\,a\,e^2 + 2\,b\,e^2 + e^3)\ \zeta(2)
+(a^2\,c + a\,c^2 + b^2\,d + b\,d^2 + a^2\,e + a\,b\,e + b^2\,e + 2\,a\,c\,e \cr
&+ 2\,b\,c\,e + 2\,a\,d\,e + 2\,b\,d\,e + 2\,a\,e^2 + 2\,b\,e^2 + 2\,c\,e^2 + 2\,d\,e^2 + 
  e^3)\ \zeta(3)\cr
&+\lf(a^2\,e + a\,b\,e + b^2\,e + 3\,a\,c\,e + \frac{5\,b\,c\,e}{4} + 3\,c^2\,e + \frac{5\,a\,d\,e}{4} 
+ 3\,b\,d\,e + \frac{17\,c\,d\,e}{4} + 3\,d^2\,e + 
  \frac{a\,e^2}{4}\ri. \cr
&+\lf. \frac{b\,e^2}{4} + 3\,c\,e^2 + 3\,d\,e^2 + e^3\ri)\ \zeta(4)+\ldots\ ,}}
and
\eqn\MOMEXPi{\eqalign{
&C(a,b,c,d,e-1)=\fc{\Gamma(1+a)\ \Gamma(1+b)\ \Gamma(1+c)\ \Gamma(1+d)}{\Gamma(2+a+c)\ \Gamma(2+b+d)}\ 
\ \F{3}{2}\lf[{1+a,\ 1+b,\ 1-e\atop 2+a+c,\ 2+b+d}\ri]\cr
&=\zeta(2)-(a+b+2\ c+2\ d+e)\ \zeta(3)\cr
&+\lf(a^2 + a\,b + b^2 + \frac{a\,c}{2} + \frac{5\,b\,c}{4} + 3\,c^2 + \frac{5\,a\,d}{4} + 
  \frac{b\,d}{2} +  3\,d^2 + \frac{a\,e}{4} + \frac{b\,e}{4} + 
  3\,c\,e\ri.\cr
&\lf. \frac{17\,c\,d}{4} + 3\,d\,e + e^2\ri)\ \zeta(4)+\ldots\ ,}}
for the two integrals $C(a,b,c,d,e)$ and $C(a,b,c,d,e-1)$, defined in \eqq \Five, respectively.
We now want to propose an interesting identity 
\eqn\JAPAN{\eqalign{
&\F{3}{2}\lf[\al_1,\al_2,\al_3\atop\be_1,\be_2\ri]=
\fc{1}{\al_1\ \al_2\ (\al_1-\be_1)\ (\al_2-\be_2)\ (\al_1+\al_2+\al_3-\be_1-\be_2)}\cr
&=\lf\{\ \fc{}{}(1-\al_3+\be_1)\ (1-\al_3+\be_2)\ \lf[\ \be_1\ \be_2\ 
(\al_1+\al_2+\al_3-\be_1-\be_2)-\al_1\ \al_2\ \al_3\ \ri]\ri.\cr 
&\times\ C(\al_1,\ \al_2,\ \be_1-\al_1,\ \be_2-\al_2,\ -\al_3)\cr
&-\al_3\ \lf[\ \al_1^2\ (\al_2-\be_1)\ (\al_2-\be_2)-\be_1\ \be_2\ (\al_2+\al_3-\be_1-\be_2)\ 
(1-\al_2-\al_3+\be_1+\be_2)\ri.+\cr
&+\al_1\ \al_2\ \al_3-\al_1\ \al_2\ \al_3^2-\al_1\ \al_2^2\ \be_1+\al_1\ \al_2\ \be_1^2-
\al_1\ \al_2^2\ \be_2-\al_1\ \be_1\ \be_2\cr
&\lf.+3\ \al_1\ \al_2\ \be_1\ \be_2+2\ \al_1\ \al_3\ \be_1\ \be_2-2\ \al_1\ \be_1^2\ \be_2
+\al_1\ \al_2\ \be_2^2-2\ \al_1\ \be_1\ \be_2^2\ \ri]\cr
&\lf.\times\ C(\al_1,\ \al_2,\ \be_1-\al_1,\ \be_2-\al_2,\ -\al_3-1)\ \fc{}{}\ri\}\ 
\fc{\Gamma(\beta_1)\ \Gamma(\beta_2)}{\Gamma(\alpha_1)\ \Gamma(\alpha_2)\ \Gamma(\beta_1-\alpha_1)\ 
\Gamma(\beta_2-\alpha_2)}\cr
&=1-\fc{\al_1\al_2\al_3}{\be_1\ \be_2\ (\al_1+\al_2+\al_3-\be_1-\be_2)}\cr
&\times\lf\{1+\lf[\ 
(\al_2-\be_1)\ (\al_3-\be_1)+\al_1\ (\al_2+\al_3-\be_1-\be_2)-(\al_2+\al_3-\be_1)\ \be_2
+\be_2^2\ \ri]\ \zeta(2)\ri.\cr
&+\lf[\ \al_1^2\,\al_2+ \al_1\,\al_2^2 + \al_1^2\,\al_3 + 4\,\al_1\,\al_2\,\al_3 + 
  \al_2^2\,\al_3 + \al_1\,\al_3^2 + \al_2\,\al_3^2 - \al_1^2\,\be_1 - 
  4\,\al_1\,\al_2\,\be_1 \ri.\cr
&\hskip0.4cm - \al_2^2\,\be_1 - 
  4\,\al_1\,\al_3\,\be_1 - 
  4\,\al_2\,\al_3\,\be_1 - 
  \al_3^2\,\be_1 + 
  3\,\al_1\,\be_1^2 + 
  3\,\al_2\,\be_1^2 + 
  3\,\al_3\,\be_1^2 - 
  2\,\be_1^3-\al_1^2\,\be_2 \cr
&\hskip0.4cm   - 4\,\al_1\,\al_2\,\be_2 - 
  \al_2^2\,\be_2 - 
  4\,\al_1\,\al_3\,\be_2 - 
  4\,\al_2\,\al_3\,\be_2 - 
  \al_3^2\,\be_2 + 
  4\,\al_1\,\be_1\,\be_2 + 
  4\,\al_2\,\be_1\,\be_2 \cr
&\hskip0.4cm  \lf.
  +4\,\al_3\,\be_1\,\be_2-3\,\be_1^2\,\be_2 + 
  3\,\al_1\,\be_2^2 + 
  3\,\al_2\,\be_2^2 + 
  3\,\al_3\,\be_2^2 - 
  3\,\be_1\,\be_2^2 - 2\,\be_2^3\ \ri]\ \zeta(3)+\ldots\ ,}}
which will be proven in section 5.
In fact, it pops up automatically as a byproduct of our five--gloun scattering amplitude, 
once we perform the calculation in the way described in section 2.
The above equation relates the hypergeometric function 
$\ss{\F{3}{2}\lf[\al_1,\al_2,\al_3\atop\be_1,\be_2\ri]}$ to the two expressions 
\MOMEXP\ and \MOMEXPi\ from before.
The power series of the hypergeometric function
\eqn\divFIVE{\eqalign{
\F{3}{2}\lf[{\alpha_1,\ \alpha_2,\ \alpha_3\atop\beta_1,\ \beta_2}\ri]&=
\fc{\Gamma(\beta_1)\ \Gamma(\beta_2)}{\Gamma(\alpha_1)\ \Gamma(\alpha_2)\ \Gamma(\beta_1-\alpha_1)\ 
\Gamma(\beta_2-\alpha_2)}\cr
&\times C(\alpha_1-1,\alpha_2-1,\beta_1-\alpha_1-1,\beta_2-\alpha_2-1,-\alpha_3)\ .}}
has meromorphic poles in $\be_1,\be_2$ and $\al_1+\al_2+\al_3-\be_1-\be_2$,
while the functions \MOMEXP\ and \MOMEXPi\ are finite.
Hence the above identity yields a non--trivial relation between a singular hypergeometric
function and two non--singular ones. From the previous results \MOMEXP\ and \MOMEXPi\ we may easily deduce
the power series expansion of the singular hypergeometric function 
$\ss{\F{3}{2}\lf[\al_1,\al_2,\al_3\atop\be_1,\be_2\ri]}$.
In fact, from \Exphar\ one deduces, that in that case non--convergent sums would appear, if
one performed a naive expansion. The techniques, how to deal with such cases, have been
pioneered in {\it Ref.} \noerlund. However, it is quite intriguing (\cf section 5), 
that from string--theory we find powerful identities to avoid this problem.

Finally, the momentum expansion of 
$\F{4}{3}\lf[{1+b,\ 1+a,\ 1+c,\ -j\atop 2+b+e,\ 2+a+d,\ 2+c+f}\ri]$, relevant 
to six open string amplitudes, will be given in appendix \appC.

\subsec{Multiple zeta values, Euler--Zagier sums and momentum expansion of 
$F^{(3)}$}

In this subsection we present the tools relevant for the momentum expansion of
the triple hypergeometric function $F^{(3)}$, given in \eqq\general.
Expanding the function $F^{(3)}$ w.r.t. the parameter $a,\ldots,j$ 
generically leads to triple infinite sums over positive integers.
These sums represent generalizations of the multiple zeta values of length $k$:
\eqn\genzeta{
\zeta(s_1,\ldots,s_k)=\sum_{n_1>\ldots>n_k>0}\ \prod_{j=1}^k\ \fc{1}{n_j^{s_j}}=
\sum_{n_1,\ldots,n_k=1}^\infty\ \prod_{j=1}^k\ \lf(\sum\limits_{i=j}^k n_i\ri)^{-s_j}\ ,}
with $s_1\geq 2\ ,\ s_2,\ldots,s_k\geq 1$.
The multiple zeta value are related to the multiple polylogarithm
\eqn\polylog{
\Li_{s_1,\ldots,s_k}(x_1,\ldots,x_k)=\sum_{n_1>\ldots>n_k>0}\ \prod_{j=1}^k\ 
\fc{x_j^{n_j}}{n_j^{s_j}}\ ,}
for $x_j=1$, \ie $\zeta(s_1,\ldots,s_k)=\Li_{s_1,\ldots,s_k}(1,\ldots,1)$. 
The latter reduces to the Riemann--Zeta function for $k=1$, \ie $\Li_a(1)=\zeta(a)$.
The case $k=3$ has already occurred in \ZETA\ and representative examples may be found
in {\it Ref.} \Girgensohn.
Furthermore, the multiple zeta values \genzeta\ have the integral representation \BBBL
\eqn\polyintegral{
\zeta(s_1,\ldots,s_k)=\int_1^\infty\ \fc{dx_1}{x_1}\ \ldots\int_1^\infty\ \fc{dx_k}{x_k}\ 
\prod_{j=1}^k\ \fc{1}{\Gamma(s_j)}\ \fc{(\ln x_j)^{s_j-1}}{\prod\limits_{i=1}^jx_i-1}\ .}
Our case of interest is $k=3$:
\eqn\Polyintegral{
\zeta(s_1,s_2,s_3)=-\fc{(-1)^{s_1+s_2+s_3}}{\Gamma(s_1)\ \Gamma(s_2)\ \Gamma(s_3)}
\ \int_0^1dx\int^1_0dy\int^1_0dz\  x^2\ y\ \fc{(\ln x)^{s_1-1}\ (\ln y)^{s_2-1}\ (\ln z)^{s_3-1}}
{(1-x)\ (1-xy)\ (1-xyz)}\ .}
It is this relation \Polyintegral, which gives (formally) the link between the two 
expressions
\todo\ and \general\ after expanding them  w.r.t. to small parameter $a,\ldots, j$.

On the other hand, after looking closer at the nine polynomials showing up in \todo\
we realize, that \Polyintegral\ is not of considerable help.
In fact, In the following we shall deal with some generalizations of the sum \genzeta.
These sums are called  
{\it Euler--Zagier double series} or 
 {\it Witten zeta function\foot{These zeta functions represent sums over all 
finite--dimensional representations $\rho$ of a semi--simple Lie algebra $\rm g$: 
$\zeta_{\rm g}(s)=\sum\limits_\rho\ [{\rm dim}(\rho)]^{-s}$.
{\it E.g.:} $\zeta_{sl(2)}(s)=\zeta(s)$ and $\zeta_{sl(3)}(s)=2^s\ W(s,s,s)$.
Special values of these zeta functions calculate the volume
of certain moduli spaces of vector bundles of curves in analogy as $\zeta(2g)$ is proportional
to the Euler characteristic of the moduli space of Riemann surfaces of genus $g$.} } 
\doubref\ZAGIER\Crandall.
One example is the following double sum:
\eqn\zagier{
W(r,s,t)=\sum_{m,n=1}^\infty \fc{1}{n^r\ m^s\ (m+n)^t}\ .}
It enjoys the {\it Pascal triangle recurrence} relation 
\eqn\triangle{
W(r,s,t)=W(r-1,s,t+1)+W(r,s-1,t+1)\ ,}
and the useful relations \Tornheim:
\eqn\useftorn{\eqalign{
2\ W(a-2,1,1)-W(1,1,a-2)&=2\ \zeta(a)\ ,\cr
W(1,1,a-2)&=(a-1)\ \zeta(a)-\sum_{i=2}^{a-2}\zeta(i)\ \zeta(a-i)\ ,\cr
W(a-2,1,1)&=\h\ W(1,1,a-2)+\zeta(a)\ ,\cr
W(1,0,a-1)&=\h\ W(1,1,a-2)\ .}}
{From} the latter relations we may deduce \eg
\eqn\remark{
W(1,1,1)=2\ \zeta(3)\ \ \ ,\ \ \ W(1,1,2)=\h\ \zeta(4)}
among other remarkable identities.
An immediate consequence of \useftorn\ is the identity ($\alpha\geq 1$):
\eqn\remarkim{
\sum_{m,n=1}^\infty\fc{1}{m\ (1+n)\ (1+m+n)^\alpha}=-\alpha+(\alpha+1)\ \zeta(\alpha+2)+
\sum_{i=2}^\alpha\zeta(i)-\sum_{i=2}^\alpha\zeta(i)\ \zeta(\alpha+2-i)\ ,}
which may be proven by noting $\sum\limits_{m,n=1}^\infty\fc{1}{m\ (1+n)\ (1+m+n)^\alpha}=
W(1,1,\alpha)-\sum\limits_{m=1}^\infty\fc{1}{m\ (m+1)^\alpha}$ and using \proceed.
We may derive an other important series, similar to  \remarkim, namely
\eqn\similarremark{\eqalign{
\sum_{m,n=1}^\infty \fc{1}{m\ (1+m+n)^\alpha}&=\sum_{m,n=1\atop n>m}^\infty\fc{1}{m\ (n+1)^\alpha}
=\sum_{n=1}^\infty \fc{H_{n-1}}{(n+1)^\alpha}=\sum_{n=1}^\infty \fc{H_n}{(n+1)^\alpha}-
\sum_{n=1}^\infty \fc{1}{n\ (n+1)^\alpha}\cr
&=-\alpha+\h\ \alpha\ \zeta(\alpha+1)+ 
\sum_{i=2}^\alpha\zeta(i)-\h\ \sum_{i=1}^{\alpha-2}\zeta(\alpha-i)\ \zeta(i+1)\ ,}}
where we have used \eqqs \proceed\ and \reporti.
Similarly, we may deduce:
\eqn\sidas{
\sum_{m,n=1}^\infty\fc{1}{(1+m)\ (1+m+n)^\alpha}=1+\h\ \alpha\ \zeta(\alpha+1)-
\zeta(\alpha)-\h\ \sum_{i=1}^{\alpha-2}\zeta(\alpha-i)\ \zeta(i+1)\ .}
After subtracting the two equations \similarremark\  and \sidas, we obtain:
\eqn\Sidas{
\sum_{m,n=1}^\infty\fc{1}{m\ (1+m)\ (1+m+n)^\alpha}=-\alpha-1+2\ \zeta(\alpha)+
\sum_{i=2}^{\alpha-1}\zeta(i)\ .}

The next type of sum we shall be concerned with is:
\eqn\good{
\sum_{m,n=1}^\infty\fc{1}{n\ (1+m)\ (1+m+n)\ (2+m+n)}=
2\ \zeta(3)-\fc{9}{4}\ .}
This equality may be proven after performing the partial fraction decomposition\br
$\fc{1}{(1+m+n)\ (2+m+n)}=
\fc{1}{(1+m+n)}-\fc{1}{(2+m+n)}$ and determining the two series
\eqn\twoseries{\eqalign{
\sum_{m,n=1}^\infty\fc{1}{n\ (1+m)\ (1+m+n)}&=-1+2\ \zeta(3)\ ,\cr
\sum_{m,n=1}^\infty\fc{1}{n\ (1+m)\ (2+m+n)}&=\sum_{m,n=1}^\infty\fc{1}{n\ m\ (1+m+n)}-
\sum_{n=1}^\infty\fc{1}{n(n+2)}=2-\fc{3}{4}=\fc{5}{4}\ ,}}
which follow from \remarkim\ and:
\eqn\addwitten{
\sum_{m,n=1}^\infty \fc{1}{n\ m\ (1+m+n)}=\sum_{m,n=1}^\infty
\lf(\fc{1}{n}-\fc{1}{1+m+n}\ri)\ \fc{1}{m(m+1)}=\sum_{m=1}^\infty \fc{H_{m+1}}{m(m+1)}=2\ ,}
respectively. Furthermore, we present the identity
\eqn\Present{
\sum_{m,n=1}^\infty\fc{1}{m\ n\ (1+m)\ (1+n)\ (1+m+n)}=5-4\ \zeta(3)\ ,}
which may be proven after partial fraction decomposition
$$\fc{1}{m\ n\ (1+m)\ (1+n)\ (1+m+n)}=\lf(\fc{1}{m}-\fc{1}{1+m}\ri)\ \lf(\fc{1}{n}-\fc{1}{1+n}\ri)\ 
\fc{1}{1+m+n}$$
and determining the three sums:
\eqn\foursums{\eqalign{
\sum_{m,n=1}^\infty\fc{1}{m\ n\ (1+m+n)}&=2\ ,\cr
\sum_{m,n=1}^\infty\fc{1}{m\ (1+n)\ (1+m+n)}&=W(1,1,1)-\sum_{m=1}^\infty\fc{1}{m\ (m+1)}=
2\ \zeta(3)-1\ ,\cr
\sum_{m,n=1}^\infty
\fc{1}{(1+m)\ (1+n)\ (1+m+n)}&=\sum_{m,n=1}^\infty\fc{1}{n\ (n+1)}\ \lf(\fc{1}{m+1}-
\fc{1}{m+n+1}\ri)\cr
&=\sum_{n=1}^\infty \fc{1}{n\ (n+1)}\ (H_{n+1}-1)=1\ .}}
The first sum follows from \addwitten\ and  the second from \remark\ or \remarkim.
{From} \eqqs \foursums\  we may also derive
\eqn\alsofind{
\sum_{m,n=1}^\infty\fc{1}{(1+m)\ n\ (1+n)\ (1+m+n)}=2\ \zeta(3)-2\ ,}
after performing the partial fraction decomposition:
$$\fc{1}{(1+m)\ n\ (1+n)\ (1+m+n)}=\lf(\fc{1}{n}-\fc{1}{1+n}\ri)\ 
\fc{1}{(1+m)\ (1+m+n)}\ .$$
Moreover, we prove the two identities:
\eqn\moreoverr{\eqalign{
\sum_{m,n=1}^\infty \fc{1}{n\ m\ (1+m)\ (1+m+n)^2}&=
8-3\ \zeta(2)-2\ \zeta(3)-\h\ \zeta(4)\ ,\cr
\sum_{m,n=1}^\infty \fc{1}{m\ (2+n)\ (1+m+n)^2}&=
-\fc{3}{2}\ \zeta(2)+3\ \zeta(3)-1\ .}}
After relabeling $m$  and performing partial fraction decomposition the first sum 
may be casted into:
$$\hskip-0.3cm\sum_{n=1\atop m\geq 2}^\infty \fc{1}{n\ m\ (m-1)\ (m+n)^2}=
\sum_{m,n=1}^\infty \fc{1}{m\ n\ (1+m+n)^2}-\sum_{m,n=1}^\infty\fc{1}{m\ n\ (m+n)^2}+
\sum_{n=1}^\infty\fc{1}{n\ (n+1)^2}\ .$$
According to \remark\ and \proceed, the last two sums give $W(1,1,2)=\h\zeta(4)$ and 
$2-\zeta(2)$, respectively, while the first sum may be written:
$$\eqalign{
&\sum_{m,n=1}^\infty \fc{1}{m\ n\ (1+m+n)^2}=\sum_{m,n=1}^\infty
\lf(\fc{1}{m}+\fc{1}{n}\ri)\ \fc{1}{(m+n)\ (1+m+n)^2}\cr
&=2\ \sum_{m,n=1}^\infty
\fc{1}{m\ (m+n)\ (1+m+n)^2}=2\ \sum_{n=1}^\infty\fc{H_{n-1}}{n(n+1)^2}=
2\ \lf[\ 3-\zeta(3)-\zeta(2)\ \ri]\ .}$$
On the other hand, the second sum of \moreoverr\  may be written:
$$\eqalign{
&\sum_{m,n=1}^\infty \fc{1}{m\ (2+n)\ (1+m+n)^2}=2\ \sum_{m,n=1}^\infty
\fc{1}{m\ (m+n)\ (m+n-1)^2}-\sum_{m=1}^\infty\fc{1}{2\ m\ (1+m)^2}\cr
&-\sum_{m=1}^\infty\fc{1}{m^3}=-1+\h\ \zeta(2)-
\zeta(3)+2\ \sum_{n=1}^\infty\fc{H_n}{n^2(1+n)}=-\fc{3}{2}\ \zeta(2)+3\ \zeta(3)-1\ .}$$
Besides, let us check the following sum:
\eqn\whatsum{
\sum_{m,n=1}^\infty\fc{1}{(1+n)\ m^2\ (m+n)^2}=-\h\ \zeta(4)+\h\ \zeta(2)^2+5\ \zeta(3)
-4\ \zeta(2)\ .}
After writing the numerator $1=1+n+m-m-n$ we may divide the sum 
into a combination of the following three pieces:
\eqn\pieces{\eqalign{
\sum_{m,n=1}^\infty \fc{1}{m^2\ (m+n)^2}&=\sum_{m<n}
\fc{1}{m^2\ n^2}=\zeta(2,2)=\h\ \zeta(2)^2-\h\ \zeta(4)\ ,\cr
\sum_{m,n=1}^\infty \fc{1}{m\ (1+n)\ (m+n)^2}&=\sum_{m,n=1}^\infty
\fc{1}{m\ (2+n)\ (1+m+n)^2}+\h\ \sum_{m=1}^\infty\fc{1}{m\ (1+m)^2}\cr
&=-2\ \zeta(2)+3\ \zeta(3)\ ,\cr
\sum_{m,n=1}^\infty\fc{1}{m^2\ (1+n)\ (m+n)}&=\sum_{m,n=1}^\infty
\fc{1}{m^2}\ \fc{1}{n\ (1+n)}-\sum_{n=1}^\infty \fc{1}{n\ (1+n)^2}\cr
&-\sum_{m,n=1}^\infty\fc{1}{n\ (1+n)\ (1+m)\ (1+m+n)}=2\ \zeta(2)-2\ \zeta(3)\ .\cr}}
While the first sum of \pieces\ is standard (\cf \eqq \ZETA), for the second
sum the result \moreoverr\ and for the third sum the identity \alsofind\ 
have been used in addition to \eqq \proceed. After putting these sums
together we prove \whatsum.

So far we have listed all sums we shall need for expanding the six functions
\Usebasis\ up to first order in the momenta $s_i$.
To go beyond this order we have also calculated the following three triple sums:
\eqn\triplesums{\eqalign{
\sum\limits_{m_i=1}^{\infty} \frac{m_3}{m_1\ m_2\ (m_1+m_3)\ (m_2+m_3)\ (m_1+m_2+m_3)}&=
 \frac{7}{4}\ \zeta(4) \ ,\cr
\sum\limits_{m_i=1}^{\infty} \frac{1}{m_1\ m_2\ (1+m_1+m_3)\ (m_2+m_3)\ (m_1+m_2+m_3)}&=
\frac{19}{4}\ \zeta(4) -4\ \zeta(3)\ , \cr
\sum\limits_{m_i=1}^{\infty} \frac{1}{m_1\ m_2\ (m_1+m_3)\ (m_2+m_3)\ (1+m_1+m_2+m_3)}&=
 \frac{17}{4}\ \zeta(4)+2\ \zeta(3)-\zeta(2)-5\ .}}
These two sums may be proven, by first computing the following somewhat simpler triple sums:
\eqn\somewhatsimpler{\hskip-0.4cm\eqalign{
\sum\limits_{m_i=1}^{\infty} \frac{1}{(1+m_1+m_3)\ (m_2+m_3)\ (m_1+m_2+m_3)^2}&=
 -\frac{13}{4}\ \zeta(4) -3\ \zeta(3) +2\ \zeta(2) +\frac{3}{2}\ \zeta(2)^2 \ , \cr
\sum\limits_{m_i=1}^{\infty} \frac{1}{m_1\ (1+m_1+m_3)\ (m_2+m_3)\ (m_1+m_2+m_3)}&=
 -2\ \zeta(4) -2\ \zeta(3) +\zeta(2) +\frac{3}{2}\ \zeta(2)^2 \ , \cr
\sum\limits_{m_i=1}^{\infty} \frac{1}{m_1\ m_2\ (1+m_1+m_3)\ (m_2+m_3)}&=
 2\ \zeta(3) + \zeta(2)\ ,\cr
\sum\limits_{m_i=1}^{\infty} \frac{1}{m_1\ m_2\ (m_2+m_3)\ (m_1+m_2+m_3)}&=
 8\ \zeta(4) -2\ \zeta(2)^2 \ , \cr
\sum\limits_{m_i=1}^{\infty} \frac{1}{m_1\ m_2\ (m_2+m_3)\ (m_1+m_3)}&=
 \frac{11}{2}\ \zeta(4) - \zeta(2)^2 \ , \cr
\sum\limits_{m_i=1}^{\infty} \frac{1}{m_2\ (m_1+m_3)\ (m_2+m_3)\ (m_1+m_2+m_3)}&=
 -\frac{5}{2}\ \zeta(4) +\frac{3}{2}\ \zeta(2)^2 \ ,\cr
\sum\limits_{m_i=1}^{\infty} \frac{1}{(m_1+\ m_3)\ (m_2+m_3)\ (m_1+m_2+m_3)^2}&=
\fc{17}{4}\ \zeta(4)-2\ \zeta(2)\ ,\cr
\sum_{m_i=1}^\infty\fc{1}{m_1\ m_2\ (m_1 + m_3)\ (m_2 + m_3)\ (1 + m_1 + m_2 + m_3)}&=
\fc{17}{4}\ \zeta(4)+2\ \zeta(3)-\zeta(2)-5\ ,\cr
\sum_{m_i=1}^\infty\fc{1}{m_1\ m_2\ (1 + m_2)\ (m_2 + m_3)\ (1 + m_1 + m_2 + m_3)}&=
5-2\ \zeta(3)-\zeta(2)\ .}}
In addition, we have also derived the following sums involving the harmonic number
\eqn\harmtriple{\eqalign{
\sum\limits_{m_i=1}^{\infty} \frac{H_{m_1+m_3}}{m_1\ (1+m_1+m_3)^2}&= 
-\zeta(2)+\frac{1}{2}\ \zeta(2)^2 +\zeta(3) +\frac{3}{2}\ \zeta(4)\ , \cr
\sum\limits_{m_i=1}^{\infty} \frac{H_{m_3}}{m_1\ (1+m_1+m_3)^2}&= 
\frac{9}{2}\ \zeta(4) -  \zeta(2)^2   \ , \cr
\sum\limits_{m_i=1}^{\infty} \frac{H_{m_1-1}}{m_1\ (1+m_1+m_3)^2}&=-3+\zeta(2)+
\zeta(3)+\zeta(4)\ ,\cr
\sum\limits_{m_i=1}^{\infty} \frac{H_{m_3}}{m_1\ (1+m_3)\ (1+m_1+m_3)\ (2+m_1+m_3)}&= 
1+ \zeta(2)-2\ \zeta(3)  \ ,}}
$${\eqalign{
\sum\limits_{m_i=1}^{\infty} \frac{H_{m_1-1}}{m_1\ (1+m_3)\ (1+m_1+m_3)\ (2+m_1+m_3)}&= 
\frac{15}{4}-\zeta(2) -2\ \zeta(3) \ , \cr
\sum\limits_{m_i=1}^{\infty} \frac{H_{m_1+m_3}}{m_1\ (1+m_3)\ (1+m_1+m_3)\ (2+m_1+m_3)}&= 
-\frac{3}{2}- \frac{5}{2}\ \zeta(2) + \frac{11}{2}\ \zeta(4)  \ ,  \cr
\sum\limits_{m_i=1}^{\infty} \frac{H_{m_1+m_2-1}}{m_1\ (m_1+m_2)^2}&=
 \frac{3}{2}\ \zeta(4)+\frac{1}{2}\ \zeta(2)^2 \ , \cr
\sum\limits_{m_i=1}^{\infty} \frac{H_{m_2}}{m_1\ (m_1+m_2)^2}&=
 5\ \zeta(4)-\zeta(2)^2 \ ,\cr
\sum\limits_{m_i=1}^{\infty} \frac{H_{m_2}}{m_1\ m_2\ (m_1+m_2)}&=
 \frac{11}{2}\ \zeta(4) -\frac{1}{2}\ \zeta(2)^2  \ , \cr
\sum\limits_{m_i=1}^{\infty} \frac{H_{m_2}}{m_2\ (m_1+m_2)^2}&= 
\frac{1}{2}\ \zeta(4)+\frac{1}{2}\ \zeta(2)^2 \ ,\cr
\sum\limits_{m_i=1}^{\infty} \frac{H_{m_1+m_2-1}}{(1+m_1)\ (m_1+m_2)^2}&=
 \frac{7}{4}\ \zeta(4)+\frac{1}{2}\ \zeta(2)^2-\zeta(3) \ , \cr
\sum\limits_{m_i=1}^{\infty} \frac{H_{m_2}}{(1+m_1)\ (m_1+m_2)^2}&=
 5\ \zeta(4) - \zeta(2)^2 +2\ \zeta(3) -2\ \zeta(2) \ ,\cr
\sum\limits_{m_i=1}^{\infty} \frac{H_{m_3}}{m_2\ (2+m_3)\ (1+m_2+m_3)^2}&=
-2\ \zeta(2)-\zeta(2)^2 + \frac{23}{4}\ \zeta(4) \ , \cr
\sum\limits_{m_i=1}^{\infty} \frac{H_{m_2-1}}{m_2\ (2+m_3)\ (1+m_2+m_3)^2}&=
 -\frac{3}{2} + \frac{1}{2}\ \zeta(2)-\zeta(2)^2-\frac{5}{2}\ \zeta(3) +6\ \zeta(4) \ , \cr
\sum\limits_{m_i=1}^{\infty} \frac{H_{m_2+m_3}}{m_2\ (2+m_3)\ (1+m_2+m_3)^2}&= 
\fc{23}{4}\ \zeta(4)-\fc{5}{2}\ \zeta(2)-\fc{3}{2}\ \zeta(3)\ ,}}$$
and finally the following triple sums:
\eqn\alsotriple{\eqalign{
\sum\limits_{m_i=1}^{\infty} \frac{1}{m_1\ (1+m_1+m_3)^3}&=
 -3 + \frac{3}{2}\ \zeta(4) - \frac{1}{2}\ \zeta(2)^2 + \zeta(2) + \zeta(3) \ , \cr
\sum\limits_{m_i=1}^{\infty} \frac{1}{m_1\ (1+m_3)\ (1+m_1+m_3)^2}&=
 -2 + 3\ \zeta(4) -  \zeta(2)^2 + \zeta(2)  \ , \cr
\sum\limits_{m_i=1}^{\infty} \frac{1}{m_1\ (1+m_3)\ (1+m_1+m_3)\ (2+m_1+m_3)}&=
 -\frac{9}{4} + 2\ \zeta(3) \ , \cr
\sum\limits_{m_i=1}^{\infty} \frac{1}{m_1\ (1+m_3)^2\ (1+m_1+m_3)\ (2+m_1+m_3)}&=
 \frac{11}{4} -\frac{\pi^2}{6}+\frac{\pi^4}{72}-2\ \zeta(3)  \ ,}}
$${\eqalign{
\sum\limits_{m_i=1}^{\infty} \frac{1}{m_1\ (1+m_3)\ (1+m_1+m_3)\ (2+m_1+m_3)^2}&= 
-\frac{29}{4}+\frac{\pi^2}{4}+4\ \zeta(3) \ , \cr
\sum\limits_{m_i=1}^{\infty} \frac{1}{m_1\ (1+m_3)\ (1+m_1+m_3)^2\ (2+m_1+m_3)}&= 
\frac{1}{4}+ 3\ \zeta(4)+\zeta(2) -  \zeta(2)^2 -2\ \zeta(3)  \ ,  \cr
\sum\limits_{m_i=1}^{\infty} \frac{1}{m_2\ (2+m_3)^2\ (1+m_2+m_3)^3}&=
-\frac{1}{2}\ -\frac{15}{4}\ \zeta(2) + \frac{1}{2}\ \zeta(2)^2 +4\ \zeta(3) 
+ \frac{1}{2}\ \zeta(4)\ , \cr
\sum\limits_{m_i=1}^{\infty} \frac{1}{m_2\ (2+m_3)\ (1+m_1+m_2+m_3)^3}&=
 -\frac{3}{2} +\frac{5}{2}\ \zeta(2) -\frac{1}{2}\ \zeta(2)^2 
-\frac{7}{2}\ \zeta(3) + \frac{11}{4}\ \zeta(4) \ , \cr
\sum\limits_{m_i=1}^{\infty} \frac{1}{(1+\ m_3)\ m_2\ (2+m_3)\ (1+m_2+m_3)^2}&=
 -1+3\ \zeta(4) + \frac{5}{2}\ \zeta(2) -3\ \zeta(3) -\zeta(2)^2\ .}}$$
These results may be proven through multiple applying partial fractioning and using
relations derived above and in subsection 4.3.

Let us move on to the power series expansion w.r.t. small parameter $a,\ldots,j$
of the function 
$\ss{F\lf[{a+m_{24},b+m_{25},d+m_{26},e+m_{34},g+m_{35}\atop c+m_{36},f+m_{45},h+m_{46},j+m_{56}}\ri]}$, 
given in \eqq \serious. However, the parameter $m_{ij}$ are some integers 
in the range $m_{ij}\in \{0,\pm 1,\pm 2\}$.
According to the definition \genstr, this corresponds to the integral \todo, with
the parameter $a,\ldots,b$ being some linear combination of kinematic invariants $s_i$ and $n_{ij}$, 
given by \eqq \power.
We shall expand the triple sum \general\ w.r.t. small parameter $a,\ldots,j$. Generically this 
leads to a power series in the parameter $a,\ldots,j$ with its coefficients given by some triple,
double or single Euler sum of the type we have discussed above.
Let demonstrate, how it works for the function $\ss{F\lf[{a,b,d,e,g\atop c,f,h,j-2}\ri]}$.
To keep the expressions short, we shall only  write up, how the $g$ and $h$--dependent terms
appear up to the second order.
With this example the procedure  should become clear, how the power series expansion of
\serious\ is obtained and how at each order generically some multiple Euler--Zagier sum appears.  
Of course working out all the dependence in all parameter $a,\ldots,j$ is quite cumbersome, 
since at each order some triple sums have to be evaluated.
For  the case at hand we obtain from \eqq \general:
\eqn\demonstrate{\hskip-0.5cm\eqalign{
F\lf[{a,b,d,e,g\atop c,f,h,j-2}\ri]&
=\fc{\Ga(1+d)\ \Ga(1+e)\ \Ga(1+f)}{\Ga(-g)\ \Ga(-h)\ \Ga(2-j)}
\sum_{m_i=0}^\infty \fc{\Ga(-g+m_1)\ \Ga(-h+m_2)\ \Ga(2-j+m_3)}{m_1!\ m_2!\ m_3!}\cr
&\hskip-1.5cm\times \fc{\Ga(1+m_1+m_2+m_3+b)}{\Ga(2+m_1+m_2+m_3+b+e)}\
\fc{\Ga(1+m_1+m_3+a)}{\Ga(2+m_1+m_3+a+d)}\ \fc{\Ga(1+m_2+m_3+c)}{\Ga(2+m_2+m_3+c+f)}\cr
&\hskip-1.5cm=1+\underbrace{\sum_{m_3=1}^\infty\fc{1}{(1+m_3)^2}}_{=\zeta(2)-1}
-(g+h)\ [\underbrace{\sum_{m_1,m_3=1}^\infty \fc{1}{m_1\ (1+m_1+m_3)^2}}_{=-2+\zeta(2)+\zeta(3)}
+\underbrace{\sum_{m_1=1}^\infty \fc{1}{m_1\ (1+m_1)^2}}_{=2-\zeta(2)}\ ]\cr
&\hskip-1.5cm
+gh\ \underbrace{\sum_{m_i=1}^\infty \fc{m_3}{m_1\ m_2\ (m_1+m_3)\ (m_2+m_3)\ (m_1+m_2+m_3)}}_{
=\fc{7}{4}\ \zeta(4)}\cr
&\hskip-1.5cm+(g^2+h^2)\ 
[\underbrace{\sum_{m_1,m_3=1}^\infty \fc{H_{m_1-1}}{m_1\ (m_1+m_3)^2}}_{=\zeta(2,1,1)=\zeta(4)}\ ]
+(a,b,c,d,e,f,j)-{\rm dependent\ terms}+\ldots\cr
&\hskip-1.5cm=\zeta(2)-(g+h)\ \zeta(3)+\lf(g^2+h^2+\fc{7}{4}\ gh\ri)\ \zeta(4)
+(a,b,c,d,e,f,j)-{\rm dep.\ terms}+\ldots\ .}}
We now present the full expansions for the functions \MOMENTUMEXP, which we need to obtain
the $\ap$--expansion of the six--gluon amplitude.
Above we have compiled all the sums necessary to perform this work. 
We obtain the following results:
\eqn\phizero{\eqalign{\hskip-0.8cm
F\lf[{a,b,d,e,g \atop c,f,h,j}\ri]&=1 - a - b - c - d - e - f - 2\ g-2\ h-3\ j+
(g+h+j)\ \zeta(2)+j\ \zeta(3)\cr
&+a^2 + a\,b + b^2 + a\,c + b\,c + c^2 + 2\,a\,d + b\,d + c\,d + d^2 + a\,e + 2\,b\,e + c\,e\cr 
&+d\,e + e^2 + a\,f + b\,f + 2\,c\,f + d\,f + e\,f + f^2 + 3\,a\,g + 3\,b\,g + 2\,c\,g +3\,d\,g \cr
&+ 3\,e\,g + 2\,f\,g + 3\,g^2 + 2\,a\,h + 3\,b\,h + 3\,c\,h + 2\,d\,h + 3\,e\,h +3\,f\,h+ 5\,g\,h \cr
&+ 3\,h^2 + 4\,a\,j + 4\,b\,j + 4\,c\,j + 4\,d\,j + 4\,e\,j + 4\,f\,j +8\,g\,j + 8\,h\,j + 6\,j^2\cr
&-\lf(a\,d + b\,e +c\,f +a\,g +b\,g + c\,g + f\,g + g^2 +a\,h + b\,h + c\,h +d\,h + h^2\ri.\cr
&\lf. + a\,j + b\,j + c\,j + 3\,g\,j + 3\,h\,j + 2\,j^2\ri)\ \zeta(2)\cr
&-\lf[\ g^2+2\ d g+2\ e g+4\ h g+h^2+2\ j^2+2\ e h+2\ f h\ri.\cr
&\lf.+2\ (d+e+f+g+h)\ j+a\ (g+j)+c\ (h+j)+b\ (g+h+j)\ \ri]\ \zeta(3)\cr
&-[\ j\ (a+b+c)+\fc{j}{4}\ (5\ d+5\ e+5\ f+2\ g+2\ h+j)\ ]\ \zeta(4)+\ldots\ .\cr}}
The other sums encountered above appear in the expansion of the following functions:
\eqn\tripleexpand{\eqalign{\hskip-0.8cm
F\lf[{a,b,d,e,g \atop c,f,h,j-2}\ri]&=(1+j)\ \zeta(2)
-(a+b+c+2\ d+2\ e+2\ f+g+h+2\ j)\ \zeta(3)\cr
&+j^2\ \zeta(2)-j\ (a + b + c + 2\ d + 2\ e + 2\ f + g + h + 2\ j)\ \zeta(3)\cr
&+\lf\{\fc{}{}a^2+b a+c a+b^2+c^2+3\ d^2+3 e^2+3\ f^2+g^2+h^2+3\ d g\ri.\cr
&+3\ e g+3\ e h+3\ f h+b c+\frac{d a}{2}+\frac{5 e a}{4}+\frac{5 f a}{4}+\frac{g a}{4}+
\frac{h a}{2}+\frac{5 j a}{4}\cr
&+\frac{5 j^2}{4}+\frac{5 b d}{4}+\frac{5 c d}{4}+\frac{b e}{2}+\frac{5 c e}{4}
+\frac{17 d e}{4}+\frac{5 b f}{4}+\frac{c f}{2}+\frac{17 d
   f}{4}\cr
&+\frac{17 e f}{4}+\frac{b g}{4}+\frac{c g}{2}
+\frac{5 f g}{2}+\frac{bh}{4}+\frac{c h}{4}+\frac{5 d h}{2}+\frac{7 g h}{4}+\frac{5 b j}{4}\cr
&\lf.+\frac{5 c j}{4}+\frac{17 d j}{4}+\frac{17 e j}{4}+
\frac{17 f j}{4}+\frac{5 g j}{2}+\frac{5 h j}{2}+\fc{5 j^2}{4}\ \ri\}\ \zeta(4)+\ldots\ ,\cr
F\lf[{a+1,b,d,e,g \atop c,f,h,j-1}\ri]&=-1+\zeta(2)+3\ a-b-c+3\ d+2\ g+j\cr
&-(2\ a-b-c+d-e-f-2\ h)\ \zeta(2)\cr
&-(b+c+2\ d+2\ e+2\ f+2\ g+3\ h+j)\ \zeta(3)\cr
&-6\ a^2+4\ (b+c-3\ d-2\ g-j)\ a-b^2-c^2-6\ d^2+\frac{31 g^2}{8}-j^2\cr
&+4\ c d+3\ c g-8\ d g+4\ f g+c j-4\ d j+gj+b\ (-c+4\ d+2\ g+j)\cr
&+\lf\{\ 3\ a^2+[-3\ b-3\ c+5\ d-2\ (e+f-g+2\ h)+j]\ a+b^2+c^2\ri.\cr
&+d^2-g^2-2\ c d+c e-2\ d e+2\ c f-2\ d f-c g-2\ e g+2\ ch\cr
&\lf.-4\ d h+b\ (c-2\ d+2\ e+f+2\ h)-(e+f-g+2\ h)\ j\ \ri\}\ \zeta(2)\cr
&+\lf\{a^2+[b+c+4\ d+3 e+3\ f+4\ (g+h)+2\ j]\ a-b^2-c^2+2\ d^2\ri.\cr
&-2\ e^2-2\ f^2-2\ g^2-3\ h^2-b c-b d-c d-3\ b e-2\ ce-2\ b f-3\ c f\cr
&-3\ e f-2\ b g-2\ c g+2\ d g-2\ e g-4\ f g-4\ b h-4\ c h+d h-6\ e h\cr
&\lf.-6\ f h-4\ g h-b j-c j+d j-e j-f j-2\ g j\ri\}\ \zeta(3)\cr
&+\lf\{\ b^2+c b+\frac{5 d b}{4}+\frac{e b}{2}+\frac{5 f b}{4}+\frac{g b}{2}+\frac{3 h b}{2}+\frac{j
   b}{4}+c^2+3\ d^2+3\ e^2\ri.\cr
&+3\ f^2+\frac{7 h^2}{2}+j^2-\frac{3 a d}{4}+\frac{5 c d}{4}+\frac{5 c
   e}{4}+\frac{17 d e}{4}+\frac{c f}{2}+\frac{17 d f}{4}+\frac{17 e f}{4}\cr
&+\frac{5 c g}{4}+6\ d g+6e g+\frac{5 f g}{4}+\frac{7 a h}{4}+\frac{3 c h}{2}+
\frac{11 d h}{2}+\frac{29 e h}{4}+\frac{29 f h}{4}\cr
&\lf.+\frac{19 g h}{4}+\frac{c j}{4}+3\ d j+3\ e j+3\ f j+\frac{13 h j}{4}\ \ri\}\ \zeta(4)
+\ldots\ \ .}}

\eqn\Tripleexpand{\eqalign{\hskip-0.5cm
F\lf[{a,b,d,e,g-1 \atop c,f,h,j}\ri]&=\zeta(2)-(c+f-h+j)\ \zeta(2)\cr
&-(a+b+2\ d+2\ e+g+2\ h-j)\ \zeta(3)+\ldots\ ,\cr
F\lf[{a,b,d,e,g \atop c+1,f,h,j-1}\ri]&=-1+\zeta(2)-a-b+3\ c+3\ f+2\ h+j\cr
&+(a+b-2\ c+d+e-f+2\ g)\ \zeta(2)\cr
&-(a+b+2\ d+2\ e+2\ f+3\ g+2\ h+j)\ \zeta(3)+\ldots\ ,\cr
F\lf[{a,b+1,d,e,g \atop c,f,h,j-1}\ri]&=-1+\zeta(2)-a+3\ b-c+3\ e+2\ g+2\ h+j\cr
&+(a-2\ b+c+d-e+f)\ \zeta(2)\cr
&-(a+c+2\ d+2\ e+2\ f+2\ g+2\ h+j)\ \zeta(3)+\ldots\ ,\cr
F\lf[{a,b,d,e,g\atop c,f,h,j-1}\ri]=\zeta(3) &
-\fc{1}{4}\ \lf(4\ a+4\ b+4\ c+5\ d+5\ e+5\ f+2\ g+2\ h+j\ri)\ \zeta(4)+\ldots\ .}}
The function $\ss{F\lf[{a,b,d,e,g \atop c+1,f,h,j-1}\ri]}$ may be obtained from 
$\ss{F\lf[{a+1,b,d,e,g \atop c,f,h,j-1}\ri]}$ thanks to the symmetry \symmetry, \ie
through the permutations  
$a\leftrightarrow c,\ d\leftrightarrow f,\ g\leftrightarrow h$ on the latter.

\subsec{Momentum expansion of singular triple hypergeometric functions $F^{(3)}$}

In this subsection we shall address the case when the integrand of \serious\ has poles.
One example is the integral
\eqn\onexample{ \eqalign{
F\lf[{a-1,b-1,d,e,g\atop c,f,h,j}\ri]&=\int_0^1 dx\int_0^1 dy\int_0^1 dz\ 
x^{a-1}\ y^{b-1}\ z^c\cr
&\times  (1-x)^d\ (1-y)^e\ (1-z)^f\ (1-xy)^g\ (1-yz)^h\ (1-xyz)^j\ ,}}
whose integrand diverges for $x,y\ra 0$. Hence we should expect a double pole $\fc{1}{ab}$
in the expansion of $\ss{F\lf[{a-1,b-1,d,e,g\atop c,f,h,j}\ri]}$ 
w.r.t. to small parameter $a,\ldots,j$.
For the case at hand the sums in the general expression \general\ are not convergent.
However, this problem is only a manifestation of the chosen integral representation \general\ 
and other representation would be well--behaved after proper analytic continuation.
For the case at hand we may not expand \general\ due to the lack of absolute convergence.
We do not undertake to obtain the appropriate convergent 
expression of \general\ for that case rather we follow the methods  anticipated in \noerlund.
We add and subtract a similar singular piece to $\ss{F\lf[{a-1,b-1,d,e,g\atop c,f,h,j}\ri]}$, 
such that we obtain 
two expressions, which may be handled easily. The integral 
$\ss{F\lf[{a-1,b-1,d,e,g\atop c,f,h,j}\ri]}$ is rearranged   
as a sum of two pieces: $\ss{F\lf[{a-1,b-1,d,e,g\atop c,f,h,j}\ri]}=\Ic^a_I+\Ic^b_I$, with:
\eqn\rearr{\eqalign{
\Ic^a_{I}&=\int_0^1 dx\int_0^1 dy\int_0^1 dz\ 
x^{a-1}\ y^{b-1}\ z^c\ (1-x)^d\ (1-y)^e\ (1-z)^f\ (1-yz)^h\cr
&\times\lf\{\ (1-xy)^g\ (1-xyz)^j-1\ \ri\}\cr
\Ic^b_I&=\underbrace{\lf(\int_0^1 dx\ x^{a-1}\ (1-x)^d\ri)}_{=\ B(a-1,d)}\ 
\underbrace{\lf(\int_0^1 dy\int_0^1 dz\ 
y^{b-1}\ z^c\ (1-y)^e\ (1-z)^f\ (1-yz)^h\ri)}_{=\ C(b-1,c,e,f,h)}\ .}}
The integrand of the first integral $\Ic^a_I$ stays finite for $x,y\ra 0$. 
Hence we may easily expand this integral w.r.t. to the parameter $a,\ldots,j$.
On the other hand, the two integrals of the second expression $\Ic_I^b$ have poles. 
However, the latter may be treated easily by using material of subsections 4.1.
and 4.4.
The double integral $C(b-1,c,e,f,h)$ may be obtained from the results of appendix \appA.
More precisely, the latter may be expressed as a linear combination of $\Phi_1$ and 
$\Phi_2$, given in \MOMEXP\ and \MOMEXPi, respectively:
$$C(a-1,b,c,d,e)=\fc{(1-a+b+d+e)\ (1+a+c+e)\ \Phi_1-e\ (1+c+d+e)\ \Phi_2}{a\ (1-a+b+d)}\ .$$
To this end, we find
\eqn\find{\eqalign{
\Ic_I^a&=j\ (\zeta(2)-1) -(g+j)\ \zeta(3)+\ldots\ ,  \cr
\Ic_I^b&=\lf(\ \fc{1}{a}-d\ \zeta(2)+\ldots\ri)\ \lf(\ \fc{1}{b}-\fc{c+f}{b}+h+
\fc{(c+f)^2}{b}-\zeta(2)\ (e+\fc{cf}{b}+h)+\ldots\ \ri)\ ,}}
and altogether:
\eqn\altog{
F\lf[{a-1,b-1,d,e,g\atop c,f,h,j}\ri]
=\fc{1}{a\ b}-\fc{c+f}{a\ b}+\fc{(c+f)^2+b\ h}{a\ b}-\lf(\fc{d}{b}+\fc{c\ f}{a\ b}+
\fc{e+h}{a}\ri)\ \zeta(2)+\Oc(\epsilon)\ .}
Our next example is the integral:
\eqn\onexamples{\eqalign{ 
F\lf[{a-1,b,d,e,g\atop c,f,h,j}\ri]&=\int_0^1 dx\int_0^1 dy\int_0^1 dz\ 
x^{a-1}\ y^{b}\ z^c\ \cr
&\times (1-x)^d\ (1-y)^e\ (1-z)^f\ (1-xy)^g\ (1-yz)^h\ (1-xyz)^j\ ,}}
which again may be split into two contributions:
\eqn\rearrs{\eqalign{
\Ic_{II}^a&=\int_0^1 dx\int_0^1 dy\int_0^1 dz\ 
x^{a-1}\ y^b\ z^c\ (1-x)^d\ (1-y)^e\ (1-z)^f\ (1-yz)^h\cr
&\times\lf\{\ (1-xy)^g\ (1-xyz)^j-1\ \ri\}\cr
\Ic_{II}^b&=\underbrace{\lf(\int_0^1 dx\ x^{a-1}\ (1-x)^d\ri)}_{=\ B(a-1,d)}\ 
\underbrace{\lf(\int_0^1 dy\int_0^1 dz\ 
y^b\ z^c\ (1-y)^e\ (1-z)^f\ (1-yz)^h\ri)}_{=\ C(b,c,e,f,h)}\ .}}
The first integrand of the first integral $\Ic_{II}^a$ 
stays finite. Moreover, the second integral of $\Ic_{II}^b$ 
has not any poles either and may be evaluated by
the results of section 4.4. It is equivalent to $\Phi_1$, given in \MOMEXP. With this information,
we find:
\eqn\finds{\eqalign{
\Ic^a_{II}&=g+3\ j\ -(g+2\ j)\ \zeta(2)+\ldots\ ,  \cr
\Ic^b_{II}&=\lf(\ \fc{1}{a}-d\ \zeta(2)+\ldots\ri)\ \lf(\ 1-b-c-e-f-2\ h+\zeta(2)\ h
+\ldots\ \ri)\ .}}
Finally, we obtain
\eqn\obtfin{
F\lf[{a-1,b,d,e,g\atop c,f,h,j}\ri]
=\fc{1}{a}-\fc{b+c+e+f+2h}{a}+\fc{h}{a}\ \zeta(2)+\Oc(\epsilon)\ .}
As a third example we want to discuss the following integral:
\eqn\onexampless{\eqalign{
F\lf[{a-1,b-1,d-1,e-1,g\atop c-1,f-1,h,j}\ri]&=\int_0^1 dx\int_0^1 dy\int_0^1 dz\ 
x^{a-1}\ y^{b-1}\ z^{c-1}\ (1-x)^{d-1}\cr 
&\times (1-y)^{e-1}\ (1-z)^{f-1}\ (1-xy)^g\ (1-yz)^h\ (1-xyz)^j\ .}}
The integrand has  meromorphic poles at $x,y,z=0,1$. Therefore, this
integral implies various triple poles after expanding w.r.t. to small $a,\ldots,j$.
There is not a simple way to disentangle the singular part of this integral in the way
applied before. On the other hand, from our system of equations for $F^{(3)}$
we may easily deduce the momentum expansion:
\eqn\deduce{\hskip-0.6cm\eqalign{
F\lf[{a-1,b-1,d-1,e-1,g\atop c-1,f-1,h,j}\ri]&=
\fc{(a+d)\ (c+f)}{a\ b\ c\ d\ f}+\fc{(a+d)\ (d+e)+d\ g}{a\ c\ d\ e\ (d+e+g)}
+\fc{e+f}{a\ e\ f\ (e+f+h)}\cr
&+\fc{(d + e)\ (e + f)\ (d + e + f + g) + [(d + e)\ (e + f) + e\ g]\ h}
{d\ e\ f\ (d + e + g)\ (e + f + h)\ (d + e + f + g + h + j)}+\Oc(\epsilon^{-1})\ .}}

\newsec{Reducible diagrams and contact interactions }

We have found, that thanks to the relations \eqq \RESULT\ 
the six--gluon string $S$--matrix \study\ at the disk--level
may be expressed as a linear combination of six triple hypergeometric functions.
The latter encode the $\ap$--dependence of the full string amplitude. In that linear combination the 
coefficients in front of those six functions are meromorphic polynomials in the 
kinematical invariants $s_i$.
In \eqqs \sixpointA, \sixpointB\ and \sixpointC\ the string $S$--matrix is given 
as a power series in $\ap$ up to the order $\ap^4$. 
In order to correctly interpret this result, we look on the latter from the field theory side. 
More precisely, the string $S$--matrix comprises 
many field--theoretical Feynman diagrams with six external legs. Out of this set of
diagrams we are only interested in the irreducible diagrams as they only represent new 
interaction terms in the low--energy effective action. The reducible diagrams may be built by the 
Feynman rules applied to the known terms in the effective action. The string $S$--matrix
result is organized as a power series in the momenta of the external states
(\cf \eqqs \sixpointA, \sixpointB\ and \sixpointC).
Each power in the momenta comprises a certain class of reducible and irreducible Feynman diagrams.
All terms in the expansions are weighted with special values of the Riemann zeta--function.
Formally, the six--gluon string $S$--matrix assumes the momentum ($\ap$)--expansion:
\eqn\serieszeta{
\Ac_6(k)\sim k^{-2}+0\ k^0+\zeta(2)\ k^2+\zeta(3)\ k^4+\zeta(4)\ k^6+\Oc(k^8)\ .}
Up to $k^6$ ($\ap^4$), there occurs the set of zeta--values: 
$\zeta(2),\ \zeta(3),\ \zeta(4)$ and $\zeta(2)^2$, together with $1$ accounting
for pure Yang--Mills diagrams (field--theory).
But only $\zeta(2)^2$ and $\zeta(4)$ appear at the order $\ap^4$ (\cf \eqq \sixpointC).
This has important consequences on the number of reducible diagrams, which the string $S$--matrix 
may comprise at this order in $\ap$. 
Each $\ap$--order brings a new set of zeta functions and combinations thereof. 
{\it E.g.} at the $\ap^5$--order $\zeta(5)$ and $\zeta(2)\ \zeta(3)$ show up, while up to
the $\ap^6$--order $\zeta(6),\zeta(2)\ \zeta(4), \zeta(2)^3$ and $\zeta(3)^2$ 
show up in addition to the previous zeta--values.
As we shall see, each zeta--value in the $\ap$--expansion gives
information about a class of underlying reducible and irreducible diagrams. 

Since we want to extract the six--point interactions\foot{Note, that
the $D^4F^4$ terms are in principle already known from the four-- and five--gluon string $S$--matrix.
Hence, its four--vertex $\Vc_4^8$ with four external gluons and its five--vertex $\Vc_5^7$ 
with five external gluons will also appear in the reducible diagrams. 
The same is true for $D^2F^5$: Its five--vertex $\Vc_5^7$
is already determined from a five gluon scattering amplitude.
However, the six--vertex $\Vc_6$ with six external gluons, build 
from $D^4F^4$ and $D^2F^5$, 
is determined by our six--gluon amplitude at $\ap^4$.} of the terms 
$F^6, D^2 F^5$ and $D^4F^4$ from the string $S$--matrix
\study, we shall be only interested in the irreducible diagrams at momentum order $\Oc(k^6)$,
\ie the contact interactions of six gluons,
as only those represent new interaction terms in the effective action at order $\ap^4$.
Therefore, we shall study the  $\ap^4$--order of \eqqs \sixpointA, \sixpointB\ and \sixpointC\
in more detail. Of course, the lower orders of the latter are also interesting
as they should correctly reproduce Yang--Mills interactions of Table 1 up to the order $\ap^3$
as a consequence of unitarity.
The lower momentum orders $k^{-2},\ k^2,\ k^4$ of the string $S$--matrix \study, 
represent reducible diagrams with six external legs. Because of unitarity they must be completely
determined by the Feynman rules applied to the effective action, valid up to $\Oc(\ap^3)$.
On the other hand, momentum powers of \study\ beyond $k^6$ represent gluon interactions
such as $D^2F^6, D^4F^5, D^6 F^4$,  which only become relevant at $\ap^5$ or higher.
To conclude, we shall only be interested 
in the reducible and irreducible diagrams of momentum order $k^6$.

Let $P$ denote the number of internal propagators inside a reducible Feynman graph and 
$V_n^k$ the number of vertices $\Vc_n^k$ with $n$ legs and energy dimension $k$. We
are then able to compute the total number of external lines $N$ of a given Feynman
graph 
\eqn\legs{\eqalign{
N=\sum_{n\geq 3\atop k\geq 0}nV_n^k-2P\ ,}} 
with $N$ the total number of external legs. Notice, that the sum starts at
$n=3$, in accordance that a vertex  cannot have less than three legs.
If two vertices are glued together, two legs are converted into an propagator. This way
one has to subtract the double number of propagators $P$. 
Let us now take into account, that we only consider tree--level
diagrams. This restricts the number of propagators for a given diagram to
\eqn\propagators{\eqalign{
P=\sum_{n\geq 3\atop k\geq 0}V_n^k-1\ ,}}
where the sum runs over the whole number of vertices.
After eliminating the number of propagators from \legs\ we arrive at:
\eqn\total{\eqalign{
N=\sum_{n\geq 3\atop k\geq 0}(n-2)\ V_n^k+2\ .}}
When studying a six--point string $S$--matrix, the vertices of their reducible (field--theory)
diagrams have at most five legs. In addition, the 
string $S$--matrix may contain contact interactions with six legs.
Hence, we shall only be concerned about vertices with at most six external legs and  
we may explicitly evaluate the sum over $n$ in \total:
\eqn\sixlegs{\eqalign{
6=2+\sum_{k\geq 0} \lf(V_3^k + 2V_4^k + 3V_5^k + 4V_6^k\ri)\ .}}
There are five possibilities to fulfill the constraint \sixlegs. The cases 
range from diagrams comprising  only four vertices $\Vc^k_3$ with three legs up to diagrams of only one
vertex $\Vc^6_6$ with six legs. Only the latter diagram, representing a contact interaction,
gives rise to an irreducible diagram. Hence, those are the interactions to be added
to the effective action at $\ap^4$, 
while the other diagrams involving vertices with fewer than six legs
are reducible and may be already obtained by the Feynman rules of the effective action, valid
up to $\ap^3$. More concretely, Table 2 displays the possibilities of how to met 
the condition \sixlegs.
\vskip0.5cm
{\vbox{\ninepoint{
$$
\vbox{\offinterlineskip\tabskip=0pt
\halign{\strut\vrule#
%%%%%%%%%%%%%%%%%%
&~$#$~\hfil
&\vrule#
&~$#$~\hfil
&~$#$~\hfil
&~$#$~\hfil
&~$#$~\hfil
&~$#$~\hfil
&~$#$~\hfil
&~$#$~\hfil
&~$#$~\hfil
&~$#$~\hfil
&~$#$~\hfil
&\vrule#&\vrule#
&~$#$~\hfil
&\vrule#  
\cr
%%%%%%%%%%%%%%%%%%
\noalign{\hrule}
&
\ 
&&&
V_3^k
&&
V_4^k
&&
V_5^k
&&
V_6^k
&&
P
&
\cr
%%%%%%%%%%%%%%%%%%
\noalign{\hrule}
\noalign{\hrule}
&
a
&&&
4
&&
0
&&
0
&&
0
&&
3
&
\cr
%%%%%%%%%%%%%%%%%%
% \noalign{\hrule}
&
b
&&&
2
&&
1
&&
0
&&
0
&&
2
&
\cr
%%%%%%%%%%%%%%%%%%
% \noalign{\hrule}
&
c
&&&
1
&&
0
&&
1
&&
0
&&
1
&
\cr
%%%%%%%%%%%%%%%%%%
%%%%%%%%%%%%%%%%%%
%\noalign{\hrule}
&
d
&&&
0
&&
2
&&
0
&&
0
&&
1
&
\cr
%%%%%%%%%%%%%%%%%%
% \noalign{\hrule}
&
e
&&&
0
&&
0
&&
0
&&
1
&&
0
&
\cr
%%%%%%%%%%%%%%%%%%
\noalign{\hrule}}}$$
\vskip-10pt
\centerline{\noindent{\bf Table 2:}
{\sl Number of vertices $V_n^k$ to }}
\centerline{\sl meet the condition \sixlegs.}
\vskip10pt}}}
\vskip-1cm \ \br
For the the five cases $a)$--$e)$, shown in Table 2, their corresponding 
Feynman diagrams are schematically drawn in Figure 1 and 2.
\fig{Feynman diagrams for the three cases $a), b)$ and $c)$.}{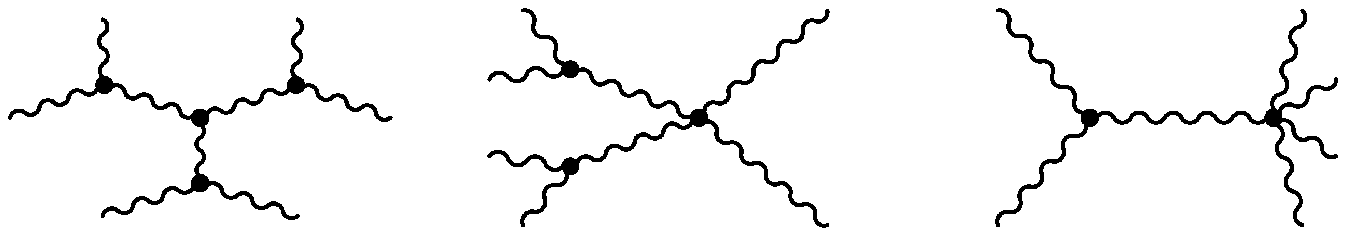}{15truecm}
\fig{Feynman diagrams for the two cases $d)$ and $e)$.}{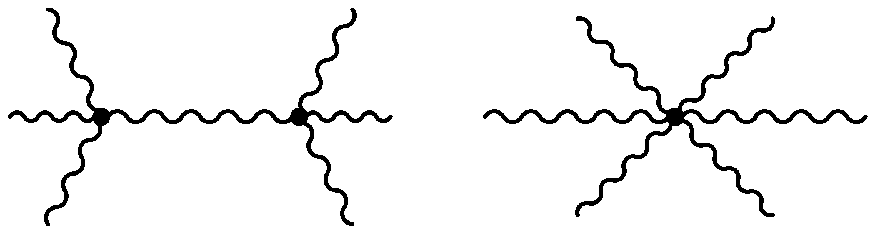}{10truecm}

The vertices $\Vc_n^k$ with at most six external legs following
from the $F^n$-- and $D^m F^n$--terms in the effective action (\cf Table 1)
are displayed in Table 3.
\vskip0.5cm
{\vbox{\ninepoint{
$$
\vbox{\offinterlineskip\tabskip=0pt
\halign{\strut\vrule#
%%%%%%%%%%%%%%%%%%
&~$#$~\hfil
&\vrule#
&~$#$~\hfil
&~$#$~\hfil
&~$#$~\hfil
&~$#$~\hfil
&~$#$~\hfil
&~$#$~\hfil
&\vrule#&\vrule#
\cr
%%%%%%%%%%%%%%%%%%
\noalign{\hrule}
&
1\ \ \ \ \ \ap^0\ \ F^2
&&&
\Vc_3^1
&&
\Vc_4^0
&&
\ 
&
\cr
%%%%%%%%%%%%%%%%%%
% \noalign{\hrule}
&
\zeta(2)\ \ap^2\ \  F^4
&&&
\Vc_4^4
&&
\Vc_5^3
&&
\Vc_6^2
&
\cr
%%%%%%%%%%%%%%%%%%
% \noalign{\hrule}
&
\zeta(3)\ \ap^3\ \ F^5
&&&
\Vc_5^5
&&
\Vc_6^4
&&
\ 
&
\cr
%%%%%%%%%%%%%%%%%%
% \noalign{\hrule}
&
\zeta(3)\ \ap^3\ \  D^2 F^4
&&&
\Vc_4^6
&&
\Vc_5^5
&&
\Vc_6^4
&
\cr
%%%%%%%%%%%%%%%%%%
%\noalign{\hrule}
&
\zeta(4)\ \ap^4\ \  F^6
&&&
\Vc_6^6
&&
\ 
&&
\ 
&
\cr
%%%%%%%%%%%%%%%%%%
% \noalign{\hrule}
&
\zeta(4)\ \ap^4\ \  D^4 F^4
&&&
\Vc_4^8
&&
\Vc_5^7
&&
\Vc_6^6
&
\cr
%%%%%%%%%%%%%%%%%%
% \noalign{\hrule}
&
\zeta(4)\ \ap^4\ \  D^2 F^5
&&&
\Vc_5^7
&&
\Vc_6^6
&&
\ 
&
\cr
%%%%%%%%%%%%%%%%%%
%%%%%%%%%%%%%%%%%%
% \noalign{\hrule}
&
\zeta(5)\ \ap^5\ \  D^6 F^4
&&&
\Vc_4^{10}
&&
\Vc_5^{9}
&&
\Vc_6^{8}
&
\cr
%%%%%%%%%%%%%%%%%%
% \noalign{\hrule}
&
\zeta(5)\ \ap^5\ \  D^2 F^6
&&&
\Vc_6^{8}
&&
\ 
&&
\ 
&
\cr
\noalign{\hrule}}}$$
\vskip-10pt
\centerline{\noindent{\bf Table 3:}{\sl \ \ Possible vertices $\Vc_n^k$}}
\centerline{\sl with at most six external legs }
\vskip10pt}}}
\vskip-0.5cm \ \br
Of course, $\Vc_3^1$ and $\Vc_4^0$ are just the standard non--Abelian Yang--Mills vertices.
Note, that we do not show any vertices (namely $\Vc_3^3, \Vc_4^2$ and $\Vc_5^1$) 
stemming from $\Tr F^3$ as this term is absent in
the superstring. This fact leads to somewhat fewer diagrams relevant to us than in the 
bosonic string.

So far, we have only discussed the number and kind of vertices, which appear in a 
tree--level Feynman diagram with six external legs.
We are interested in diagrams, which have momentum power $k^6$. The latter correspond to 
order $\ap^4$ and mix in the $\ap$--expansion of \study\ with contact interactions
of the same order.
Similarly as for the number of legs (\cf \eqq \legs) we may find an expression for the 
energy--dimension of a given graph. With the total energy--dimension $K$ given by 
$K=\sum\limits_{n\geq 3,\ k\geq 0} k V_n^k-2P$, we find: 
\eqn\energy{\eqalign{
K=\sum_{n\geq 3\atop k\geq 0}(k-2)\ V_n^k+2\ .}}
Demanding $K=6$ and using the fact, that there is only one three--vertex $\Vc^1_3$, which follows
from the standard Yang--Mills interaction $\Tr(\tr F^2)$, we obtain the condition:
\eqn\writenow{
4=4\ V_6^6-V_3^1+\sum_{k\geq 0} (k-2)\ V_4^k+\sum_{k\geq 0} (k-2)\ V_5^k\ .}
Now, we have to verify, which of the five cases, shown in Table 2, fulfill the constraint
\writenow. There is a fast answer to the two cases $a)$ and $e)$: Any Feynman 
tree--level diagram with only four three-vertices $\Vc_3^1$ leads to a 
total momentum power $k^{-2}$ and never meets the constraint \writenow. 
Among others this diagram contributes to the $k^{-2}$ powers 
in \eqqs \sixpointA, \sixpointB\ and \sixpointC. On the other hand,
any single contact vertex $\Vc_6^6$ of momentum order $k^6$ automatically fulfills the constraint
These vertices give rise to a new interaction terms in the effective action at $\ap^4$.
Next, the case $b)$ may only fulfill the requirement \writenow\ with the four--vertex 
$\Vc_4^8$ involved. This is shown in Figure 3.
\fig{Diagram $\sim \zeta(4)\ k^6$ 
with one four--vertex $\Vc_4^8$ and two three--vertices $\Vc_3^1$}{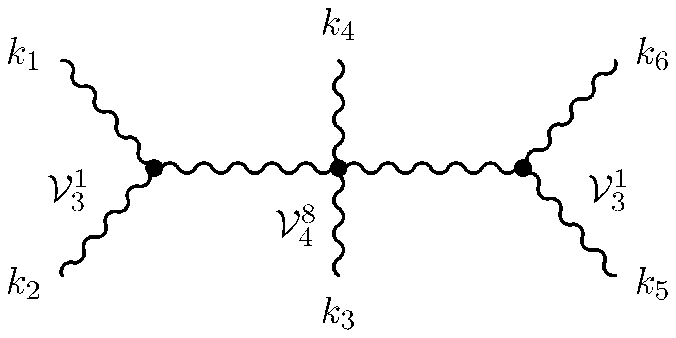}{6.5truecm}
\br
Furthermore, the case $c)$ needs the five--vertex $\Vc_5^7$, shown in Figure 4.
\fig{Diagram $\sim \zeta(4)\ k^6$
with one five--vertex $\Vc_5^7$ and one three--vertex $\Vc_3^1$}{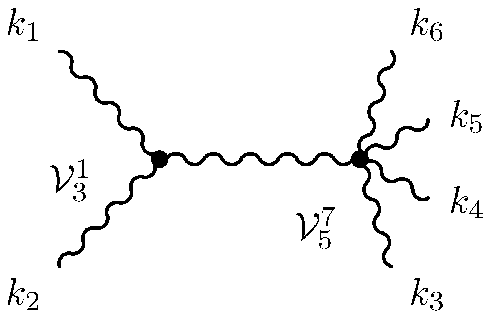}{5truecm}
\br
Finally, the case $d)$ allows for two four--vertices, with their energies $k,\tilde k$
fulfilling the constraint: $k+\tilde k=8$. This leads to two classes of possibilities:
one possibility involving two times the same vertex $\Vc_4^4$ or an other  with two different
four--vertices $\Vc_4^k$ and $\Vc_4^{\tilde k}$, with $k+\tilde k=8$.
According to Table 3 we have four different four--vertices $\Vc_4^k$, with $k=0,4,6,8$.
Hence, the case $d)$ allows for the following two pairs of four--vertices $\Vc_4^k$ and 
$\Vc_4^{\tilde k}$, with $(k,\tilde k)\in\{(4,4),(0,8) \}$, shown in Figure 5 and 6.
\fig{Diagram $\sim \zeta(2)^2\ k^6$ 
with two four--vertices $\Vc_4^4$ and one internal line}{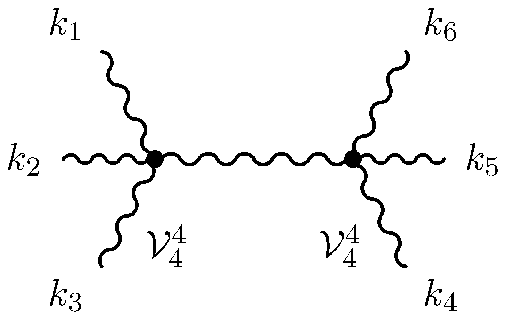}{5truecm}
\br
\fig{Diagram $\sim \zeta(4)\ k^6$
with the two four--vertices $\Vc_4^0$ and $\Vc_4^8$}{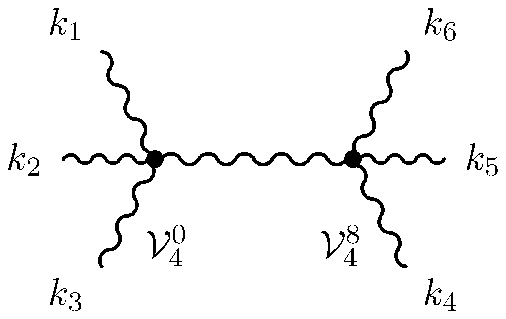}{5truecm}

To conclude, Figures Nr. 3--6 show the four reducible Feynman diagrams, which lead to total
energy dimension six, \ie order $k^6$ in the external momenta $k_i$.
These diagrams correspond to the cases $b)$--$d)$ of Table 2. 
Furthermore, the two vertices $\Vc^6_6$ from $F^6$ and $D^4 F^4$
give rise to two contact interactions of momentum order $k^6$, accounting for the case $e)$ of
Table 2.
All these six diagrams are contained in the $k^6$--part of our string $S$--matrix \study.
The latter is given in \eqqs \sixpointA, \sixpointB\ and \sixpointC.
Hence, to extract from the string $S$--matrix the information about the new 
interaction terms $F^6$ and $D^4 F^4$, to be written at $\ap^4$ in the effective action, 
one has to take from our string $S$--matrix results the $k^6$--part
and  subtract the  contribution of the above four diagrams, displayed in Figures Nr. 3--6. 
Hence the work to be still done is the following \progress: 
With the vertices from Table 3 we have to calculate the above described four reducible diagrams 
in field--theory. Then we subtract the result from the $k^6$ order of the string $S$--matrix \study.
The result, which has no poles, then accounts for the contact six--point interactions
following from $F^6, D^2 F^5$ and $D^4 F^4$ after inserting the linearized field strength
$F_{\mu\nu}=\xi_\mu k_\nu-\xi_\nu k_\mu$.
To obtain their precise form with the right coefficients
one makes an ansatz for the terms $F^6, D^2F^5$ and $D^4 F^4$ 
and matches this ansatz with the relevant piece
of the string $S$--matrix. This gives a system of linear equations for the coefficients.
Solving these equations  allows to write down the complete $\ap^4$--order of the effective 
gluon action \super, \cf \progress.
The calculation of the Feynman graphs, shown in Figure 3-6, consists of two parts: First extracting the 
Feynman rules for the vertices from either string--theory or field--theory. Second we
have to calculate the exchange diagram under consideration with respecting
permutation symmetries.
At this point we should remind about the ambiguity of relating  string $S$--matrix
results with corresponding terms in the effective action. It is notorious (see \eg 
{\it Ref.} \Sloan), that
the string $S$--matrix reproduces given terms in the effective action only up to field
redefinitions. Hence it is a lot more efficient, if we extract the Feynman rules for
the vertices from known string amplitudes with fewer than six external legs rather than extracting
them from field--theory terms, which are ambiguous.

Finally, let us point out, that we do not encounter reducible Feynman diagrams at order 
$k^6$ involving vertices from the $\ap^3$--terms.  This fact is related to the absence
of vertices from $F^3$--terms due to supersymmetry. Otherwise \eg the diagram $c$ of Figure 1
with the two vertices $\Vc_5^5$ and $\Vc_3^3$ connected by an internal line, 
would give rise to $k^6$. 
In other words, since in the string amplitude at the $\ap^4$--order we do not encounter any 
odd zeta--functions or combinations thereof, like $\zeta(3),\ \zeta(3)\ \zeta(2)$, no reducible diagrams 
involving the vertices from $F^5$ or $D^2F^4$ have to be considered at this order in $\ap$.
This may be of deeper origin related to the fact, that interaction terms
of odd powers in $\ap$ appear with the transcendental $\zeta$--values: $\zeta(3),\ \zeta(5),\ldots$.
As it can be seen from Table 3 any diagram involving a vertex from the $\ap^3$--terms would 
be proportional to $\zeta(3)$. However, our expansions \eqqs \sixpointA, \sixpointB\ and \sixpointC\
do not have $\zeta(3)$ at $\ap^4$. Hence, these vertices do not show up in the string
$S$--matrix \study\ at the $\ap^4$  order, neither in reducible nor in irreducible diagrams. 
However, the vertex $\Vc_6^4$ from $F^5$ and $D^2 F^4$ contributes at the order $\ap^3$ as  
terms proportional to $\zeta(3)\ k^4$ in the expansions 
\sixpointA, \sixpointB\ and \sixpointC.
On the other hand, since the $\ap^4$--order of the six--gluon string $S$--matrix
contains only $\zeta(4)$ and $\zeta(2)^2$, any reducible diagram contributing
at this order contains either two vertices from $F^4$ or one vertex from $D^4 F^4$
together with some vertices from the Yang--Mills interaction $F^2$.
This is precisely, what we have found above (\cf Figure 3-6).
Hence, in the $\ap$--expansion of a given gluon string $S$--matrix there seems to be an 
intriguing disentangling of terms  between $\zeta(2s)$ and $\zeta(2s+1)$, respectively
at each order.  
This means, that for a given $\ap$--order only a certain class of reducible and
irreducible diagrams contributes.
In addition there is a crucial difference between even and odd $F^n$--terms 
in the effective action,
which relies in the difference of even and odd zeta--values.
In the $\ap$--expansion of a string $S$--matrix, e.g. a term $\zeta(2)\ \zeta(3)$ may be 
immediately traced back to a contribution from a reducible diagram involving one vertex from the 
$F^5, D^2 F^4$ terms and an other from the $F^4$--terms. Similarly for $\zeta(3)^2$ or 
$\zeta(3)\ \zeta(5)$. Hence the effective action terms involving odd zeta--values may be much
easier extracted from a given string expansion than the terms proportional to
even zeta--values or combinations thereof. In fact due to the identity
\eqn\ZETAREL{
\zeta(2)^2=\fc{5}{2}\ \zeta(4)}
the contribution to the $\ap^4$--order of the string $S$--matrix of \eg Figure 5 may not be 
disentangled from the contribution from Figure 6.
Similar statements hold at the $\ap^6$--order due to the relations:
\eqn\ZETASECHS{\eqalign{
\zeta(2)^3&=\fc{35}{8}\    \zeta(6)\ ,\cr
\zeta(2)\ \zeta(4)&=\fc{7}{4}\ \zeta(6)\ .}}

\newsec{Concluding remarks}

In this work we have calculated the open superstring $S$--matrix with four, five and six 
external gluons on the disk by introducing a powerful method.
We have found a systematic and efficient way to calculate tree--level string amplitudes
by equating seemingly different expressions for one and the same string $S$--matrix.
An intriguing way of using world--sheet supersymmetry generates a system of non--trivial 
equations for string tree--level amplitudes.
These equations represent algebraic identities between different multiple  
hypergeometric functions or generalizations thereof.
Their solutions give the ingredients of the string $S$--matrix.
Generically, one-- or higher--loop string amplitudes give much more constraints on
the form of the effective action as a result of world--sheet supersymmetry manifested 
through Riemann identities in the string $S$--matrix calculation.
The latter impose quite strong conditions on the allowed heavy string states 
running the loops. On the other hand, what we find here is, that it is the superdiffeomorphism
invariance on the string world--sheet, which imposes constraints on the effective superstring 
tree--level action: by writing the tree--level string $S$--matrix in different, but 
equivalent forms, we obtain a system of equations, whose solutions capture 
the pieces of the full string $S$--matrix. Hence, in all cases, both string tree--level and 
string loop, it is world--sheet supersymmetry, which may be used to 
generate various non--trivial equations between different amplitudes.
Especially consult {\it Ref.} \STi\ for highlighting this method at two--loop. 
It would be very interesting, whether in the pure spinor formalism similar relations
may be obtained. See {\it Ref.} \berko\ for further information on this interesting question.

The full string $S$--matrix can be expressed by six (generalized) hypergeometric
functions as in \eqq \FINAL, which in the effective action play an important r\^ole in arranging 
the higher order $\ap$ gauge interaction terms (such as $F^6, D^2F^6, D^4F^4, D^6F^4,\ldots$) 
with six external legs. The final expression \FINAL\ 
is given by six triple hypergeometric functions $\Phi_j$,
which encode the full $\ap$--dependence and some polynomials
$\Pc^j$, which keep track about the pole structure of the reducible diagrams.
The functions $\Phi_j$ organize the various terms according to their zeta--values
\doubref\MHV\progress.

The general complication in constructing the effective action from string $S$--matrices
is, that one has to expand the result for the string amplitude w.r.t. to the
kinematical invariants (or $\ap$), in order to subtract the terms accounting for 
the reducible diagrams. Since generically any tree--level string amplitude is given
by some multiple hypergeometric functions (or generalizations thereof, \cf section 4), 
with the latter comprising
the $\ap$--dependence of the full amplitude, this problem 
turns to the mathematical task 
of expanding hypergeometric functions (or their generalizations) w.r.t. small parameters.
In the case, when these functions have poles, this is a very challenging work
(\cf subsection 4.6).
However, with the formalism developed in this article, 
it is generically possible to express those hypergeometric
functions, which have poles, as a linear combination of other hypergeometric functions, which
are simpler to expand (\cf section 4 for explicit results).
{\it Cf.} \eqq \RESULT\ for the six--gluon amplitude: There, the triple hypergeometric 
functions $\Phi_j$ comprise a basis of finite functions and the poles simply appear 
as coefficients $\Lambda^j$  in front of those simpler hypergeometric functions.
In fact, as we have demonstrated in subsection 3.3, these poles  follow from solving 
algebraic equations.

So far, the heterotic--type $I$ duality \witten\ has been checked up to certain $\ap^2$--terms
in the effective action \threeref\AA\paris\LSWF. 
The terms tested represent $1/2$ BPS saturated couplings, 
\ie through supersymmetry they are related to eight fermion terms in the effective
action. This duality maps certain (anomaly related) $\ap^2$--terms 
to each other. More precisely, the heterotic one--loop  corrections  to the terms 
$F^4,\ F^2R^2,\ (R^2)^2,\ R^4$ reproduce the analogous terms of the type $I$ $SO(32)$ at tree--level
\doubref\AA\paris. Hence, there is a map between the integral over the complex structure modulus of the 
heterotic one--loop world--sheet torus to the single position integration
of the disk--calculation of those $\ap^2$--terms.
It is certainly important, to go beyond that order to verify this duality, in
particular extending the checks to non--BPS saturated amplitudes.
Though heterotic--type $I$ duality is a strong--weak coupling duality, there have been given 
arguments due to non--renormalization theorems and underlying BPS--structure of certain amplitudes
\STii\ (see also {\it Refs.} \DP), that this duality should also work perturbatively, \ie
perturbative heterotic calculations map to analogous perturbative type $I$ results.
In this line we expect, that a perturbative heterotic two--loop calculation of the six gluon 
string $S$--matrix in $D=10$, captures the tree--level six gluon string $S$--matrix 
in the  type $I$ superstring \doubref\STi\STii. Hence, the irreducible part of the 
$\ap^4$--order of \eqqs \sixpointA, \sixpointB\ and \sixpointC\ should
agree with the two--loop results derived in \threeref\STi\STii\STiii. 
This check will be performed in {\it Ref.} \STiii.
At any rate, since a certain (BPS protected) kinematics of the heterotic two--loop 
amplitude (\cf the discussions in \doubref\STi\STii) is given by an integral over
the fundamental domain of the modular group of the $g=2$ torus \siegel
\eqn\modulispace{
\int_{\Fc_2}\fc{d^3\Omega\ d^3\ov\Omega}{\det \im(\Omega)} =\fc{1}{3\pi}\ \zeta(4)\ ,}
the duality seems to work, since the relevant terms from \sixpointC, 
indeed start with $\zeta(4)$ \STiii.
Hence, as in the heterotic one--loop case, there seems to be a link
between the integral over the moduli space of the heterotic two--loop torus \modulispace\ and 
the integral over positions of three vertex operators on the disk.
Qualitatively for a kinematics $(\xi\xi)\ (\xi\xi)\ (\xi k)(\xi k)$
the latter integral may be mimicked by: 
\eqn\May{
\lf.\int_0^1 dx\int_0^1 dy\int_0^1 dz\ \fc{(1-xy)^{\ap}}{(1-xyz)^2}\ \ \ri|_{\ap^2}=\zeta(4)\ .}
The duality then relates the two integrals \modulispace\ and \May.
Understanding this relation also for the non--BPS protected terms would be
certainly interesting for a deeper understanding of the heterotic--type $I$ duality.
In that case, the integrand is of the generic form \todo, 
while for the heterotic two--loop result the integrand over the $g=2$
fundamental region represents some gauged heterotic two--loop partition function, 
given by some combination of $g=2$ modular functions. 
Therefore, the duality turns into the non--trivial mathematical statement, that 
there is a link between the integral \todo\ and a two--loop torus integral
over a certain combination of $g=2$ modular functions. 
Perhaps the various identities between hypergeometric functions, that we 
have found (\cf subsection 3.3), 
may have a direct translation into two--loop Riemann identities on the heterotic side.
Certainly this would lead to quite non--trivial relations between generalized hypergeometric functions
and theta--functions, initiated by the duality between open and closed strings.

In perturbative gauge theories the maximally helicity violating (MHV) 
$N$--point amplitudes, with $N-2$ gluons of one helicity and two gluons of the other helicity,
show a remarkable simplicity, which is not apparent from the Feynman rules \PT.
This result has a nice explanation after Fourier transforming the momentum space 
amplitudes to twistor space \Witten.
Based on this observation recently there has been made progress 
in perturbative gauge theory by studying the underlying
twistor structure of scattering amplitudes. At tree--level this investigation has provided
recursion relations for on--shell amplitudes \twist, which give new and simple forms for many 
amplitudes in field--theory. The question one may ask is, whether after 
including $\ap$--corrections to the MHV--amplitudes 
would motivate a possible generalization of the duality
between perturbative gauge theory and twistor string theory \MHV.

% We believe, that the relation \RESULT\ and the fact, that any gluon string $S$--matrix may
% be written as a linear combination of some basic functions, which encode the full $\ap$--dependence
% may help to understand a possible generalization of the MHV results to include also $\ap$--corrections
% \MHV.

\ \
{\bf Acknowledgements:}
We wish to thank Tom Taylor and Christoph Sieg for many valuable and useful discussions.
Furthermore, we are very grateful to Wolfgang Lerche and Dieter L\"ust for useful discussions.
D.O. thanks the Ludwig--Maximilians--Universit\"at M\"unchen for hospitality.

%\break

\appendix\appA{Five gluon open superstring $S$--matrix}

In section 2 we have presented a powerful technique to calculate tree--level string
amplitudes. As an example we outlined the procedure  at the four gluon string $S$--matrix.
In section 3 this technique considerably helped to derive the full six--gluon string $S$--matrix.
For the reader, who still might not be convinced about the tremendous simplifications occurring
once this method  is used, we now apply this recipe to the five gluon string $S$--matrix.
Though this amplitude has been already calculated in {\it Ref.} \Brazil, we find it
quite enlightening to redo this calculation within the frame established in this
article. Moreover, this calculation allows to derive striking identities for
the hypergeometric function $\ss{\F{3}{2}\lf[{a,b,c\atop c,d}\ri]}$. 
In particular, we find interesting relations,
which allows to express one $\F{3}{2}$, which has a complicated pole structure, 
as a linear combination of two others, which do not have poles. Hence, in this linear 
combination the poles are completely
captured by the coefficients in front of those hypergeometric functions.
This is in lines of \eqq \RESULT, where one needs a basis of six functions, to express
a triple hypergeometric function 
$\ss{\tilde F\lf[{n_{24},n_{25},n_{26},n_{34},n_{35}\atop n_{36},n_{45},n_{46},n_{56}}\ri]}$
in terms of others.

We study the string $S$--matrix describing the string tree--level scattering of five 
gauge bosons on the disk:
\eqn\studyv{\eqalign{
&\Ac_5(k_1,\xi_1,a_1;k_2,\xi_2,a_2;k_3,\xi_3,a_3;k_4,\xi_4,a_4;k_5,\xi_5,a_5)\cr
&\hskip1cm=V_{\rm CKG}^{-1}\ \prod_{r=1}^5 \int d^2 z_r\
\vev{V_{A^{a_1}}^{(-1)}(z_1)\ V_{A^{a_2}}^{(-1)}(z_2)\  V_{A^{a_3}}^{(0)}(z_3)\
V_{A^{a_4}}^{(0)}(z_4)\ V_{A^{a_5}}^{(0)}(z_5)}\ .}}
We have total momentum conservation $\sum\limits_{i=1}^5 k_i=0$.
Furthermore, we have chosen two gauge vertex operators
in the $(-1)$--ghost picture in order to guarantee a total ghost charge of $-2$ on the disk.
The explicit evaluation of the correlator in \eqq \studyv\ leads to a sum of integrals 
with the two classes of (space--time) kinematic contractions: $(\xi\xi)(\xi\xi)(\xi k)$
and $(\xi\xi)(\xi k)(\xi k)(\xi k)$.  The first kinematics comprises 
the following $15\times 3$ contractions
\eqn\XXIS{\eqalign{
\Xi_1&:=   \xxi{1}{2}\ \xxi{3}{4}\ (\xi_5k_s)\ \ \ ,\ \ \
\Xi_2:=    \xxi{1}{3}\ \xxi{2}{4}\ (\xi_5k_s)\ \ \ ,\ \ \
\Xi_3:=    \xxi{1}{4}\ \xxi{2}{3}\ (\xi_5k_s),\cr
\Xi_4&:=   \xxi{1}{2}\ \xxi{3}{5}\ (\xi_4k_s)\ \ \ ,\ \ \
\Xi_5:=    \xxi{1}{3}\ \xxi{2}{5}\ (\xi_4k_s)\ \ \ ,\ \ \
\Xi_6:=    \xxi{1}{5}\ \xxi{2}{3}\ (\xi_4k_s),\cr
\Xi_7&:=   \xxi{1}{2}\ \xxi{4}{5}\ (\xi_3k_s)\ \ \ ,\ \ \
\Xi_8:=    \xxi{1}{4}\ \xxi{2}{5}\ (\xi_3k_s)\ \ \ ,\ \ \
\Xi_9:=    \xxi{1}{5}\ \xxi{2}{4}\ (\xi_3k_s),\cr
\Xi_{10}&:=\xxi{1}{3}\ \xxi{4}{5}\ (\xi_2k_s)\ \ \ ,\ \ \
\Xi_{11}:= \xxi{1}{4}\ \xxi{3}{5}\ (\xi_2k_s)\ \ \ ,\ \ \
\Xi_{12}:= \xxi{1}{5}\ \xxi{3}{4}\ (\xi_2k_s),\cr
\Xi_{13}&:=\xxi{2}{3}\ \xxi{4}{5}\ (\xi_1k_s)\ \ \ ,\ \ \
\Xi_{14}:= \xxi{2}{4}\ \xxi{3}{5}\ (\xi_1k_s)\ \ \ ,\ \ \
\Xi_{15}:= \xxi{2}{5}\ \xxi{3}{4}\ (\xi_1k_s),}}
with $k_j$ accounting for three space--time momenta as a result of applying the
on--shell constraint $\xi_i k_i=0$ and momentum conservation.

In \eqq \studyv\ we have chosen the first two gauge vertex operators in the $(-1)$--ghost picture.
As a matter of $PSL(2,\IR)$--invariance on the disk, we are free to chose
which pair $(a,b)$ of the six gauge vertex operators we put into the $(-1)$--ghost
picture.  
Hence, instead of the particular choice in \studyv, for a given group structure $\pi$, 
we may generically consider the 
expression 
\eqn\Studyv{\eqalign{
&\Tr(\lambda^{\pi(1)}\lambda^{\pi(2)}\lambda^{\pi(3)}\lambda^{\pi(4)}\lambda^{\pi(5)})\   
A^\pi(a,b,i,j,k)\cr
&\hskip2cm=V_{\rm CKG}^{-1}\  \int_{\Ic_\pi} \prod_{r=1}^5d^2 z_r\
\vev{V_{A^{a_a}}^{(-1)}(z_a)\ V_{A^{a_b}}^{(-1)}(z_b)\  V_{A^{a_i}}^{(0)}(z_i)\
V_{A^{a_j}}^{(0)}(z_j)\ V_{A^{a_k}}^{(0)}(z_k)}\ ,}}
with some permutation $(a,b,i,j,k)\in \overline{(1,2,3,4,5)}$ of vertex operators.
According to \formally\ from this piece we may obtain the whole string $S$--matrix \studyv.
After performing the Wick contractions and with the correlators \First\ 
we may evaluate \Studyv\ and obtain:
\eqn\givesv{\eqalign{
A^\pi(a,b,i,j,k)&=V_{\rm CKG}^{-1}\  \int_{\Ic_\pi} \prod_{r=1}^5d^2 z_r\ 
\sum_{(i_1,i_2,i_3)\atop\in\overline{(i,j,k)}} \lf\{
\h\ \Ac_1(a,b,i_1,i_2,i_3)\ (\xi_a\xi_b)\ (\xi_{i_1}\xi_{i_2})\ri.\cr
&+\Ac_2(a,i_1,b,i_2,i_3)\ (\xi_a\xi_{i_1})\ (\xi_b\xi_{i_2})\cr
&+\h\ \Ac_3(a,i_1,i_2,i_3,b)\ (\xi_a\xi_{i_1})\ (\xi_{i_2}\xi_{i_3})
+\h\ \Ac_3(b,i_1,i_2,i_3,a)\ (\xi_b\xi_{i_1})\ (\xi_{i_2}\xi_{i_3})\cr
&+\fc{1}{6}\ \Cc_1(a,b,i_1,i_2,i_3)\ (\xi_a\xi_b)+
\h\ \Cc_3(i_1,i_2,a,b,i_3)\ (\xi_{i_1}\xi_{i_2})\cr
&+\lf.\h\ \Cc_2(a,i_1,b,i_2,i_3)\ (\xi_a\xi_{i_1})
+\h\ \Cc_2(b,i_1,a,i_2,i_3)\ (\xi_b\xi_{i_1})\ri\}\ .}}
In the following, we shall  need the integrated expressions
\eqn\KINALL{
A_i^\pi=V_{\rm CKG}^{-1}\ \int_{\Ic_\pi} \prod_{r=1}^6 d^6z_r\  \Ac_i\ \ ,\ \ 
C_i^\pi=V_{\rm CKG}^{-1}\ \int_{\Ic_\pi} \prod_{r=1}^6 d^6z_r\  \Cc_i\ ,}
which enter the expression \givesv.
In front of the contractions 
$\xxi{a}{b}\ \xxi{i}{j}$,\ $\xxi{a}{i}\ \xxi{b}{j}$ and $\xxi{a}{i}\ \xxi{j}{k}$
there appear the three basic functions $\Ac_1(a,b,i,j,k)$,\ $\Ac_2(a,i,b,j,k)$ and
$\Ac_3(a,i,j,k,b)$,  respectively
\eqn\KINv{\eqalign{
\Ac_1(a,b,i,j,k)&=i\ \Ec\ \lf\{\fc{1-k_ik_j}{z_{ab}^2\ z_{ij}^2}\ 
\lf[(\xi_kk_a)\ \fc{z_{ja}}{z_{ak}z_{jk}}+(\xi_kk_b)\ \fc{z_{jb}}{z_{bk}z_{jk}}+(\xi_kk_i)\ 
\fc{z_{ji}}{z_{ik}z_{jk}}\ri]\ri.\cr
&\lf.+\fc{1}{z_{ab}^2z_{ik}z_{jk}z_{ij}}\ \lf[(k_jk_k)\ (\xi_kk_i)-(k_ik_k)\ (\xi_kk_j)\ri]\ri\}\ ,
\cr
\Ac_2(a,i,b,j,k)&=i\ \Ec\ \lf\{\fc{k_ik_j}{z_{ab}z_{ai}z_{bj}z_{ij}}\ 
\lf[(\xi_kk_a)\ \fc{z_{ja}}{z_{ak}z_{jk}}+(\xi_kk_b)\ \fc{z_{jb}}{z_{bk}z_{jk}}+(\xi_kk_i)\ 
\fc{z_{ji}}{z_{ik}z_{jk}}\ri]\ri.\cr
&\lf.+\fc{1}{z_{ab}z_{ai}z_{bj}z_{ik}z_{jk}}\ 
\lf[(k_ik_k)\ (\xi_kk_j)-(k_jk_k)\ (\xi_kk_i)\ri]\ri\}\ ,\cr
\Ac_3(a,i,j,k,b)&= \fc{i\ \Ec}{z_{ab}z_{ai}z_{jk}}\ \lf\{\fc{(k_jk_k)\ (\xi_bk_i)}{z_{bi}z_{jk}}-
\fc{(k_ik_k)\ (\xi_bk_j)}{z_{bj}z_{ik}}+\fc{(k_ik_j)\ (\xi_bk_k)}{z_{bk}z_{ij}}-
\fc{(\xi_bk_i)}{z_{bi}z_{jk}}\ri\}\ ,}}
with: 
$${\cal E}=\prod\limits_{r<s}^5 |z_{rs}|^{k_rk_s}\ .$$
Obviously, the first two integrals $A^\pi_1$ and $A^\pi_2$  account for the two cases 
whether the two vertices $a,b$ 
in the $(-1)$--ghost picture are contracted among themselves or with other vertices, respectively.
The integral $A_3^\pi$ describes the case if only the polarization $\xi_a$ of the vertex $a$ is
contracted with an other polarization $\xi_i$, while the polarization $\xi_b$ is contracted 
with momenta. 
The function $\Ac_1(a,b,i_1,i_2,i_3)$ is   
invariant under permutations of the indices $i_1,i_2$ and $a,b$, 
while the function $\Ac_2$ enjoys the property $\Ac_2(a,i_1,b,i_2,i_3)=\Ac_2(b,i_2,a,i_1,i_3)$.
Furthermore, the function $\Ac_3(a,i_1,i_2,i_3,b)$ is symmetric under permutations of
$i_2$ and $i_3$.
Due to  those symmetries it is obvious, that in the amplitude $A^\pi(a,b,i,j,k)$ 
any permutation of the three indices 
$i,j,k$, which label the three vertices in the zero--ghost picture, yields the same 
expression \gives\ as a result of Bose--symmetry of the $NS$--operators. The same is true
for the permutation of $a$ and $b$.
In  each amplitude $A^\pi(a,b,i,j,k)$, as given in \eqq \givesv, all 
$15$ space--time kinematical contractions \XXIS\ show up.
For the kinematics $(\xi\xi) (\xi k)(\xi k)(\xi k)$, which in the string $S$--matrix \givesv\
may appear in the combinations $C_1(a,b,i,j,k)\ (\xi_a\xi_b)$\ ,
$C_2(a,i,b,j,k)\ (\xi_a\xi_i)$ and $C_3(i,j,a,b,k)\ (\xi_i\xi_j)$, we have the
corresponding integrands
\eqn\KINv{\hskip-0.5cm\eqalign{
\Cc_1(a,b,i,j&,k)=i\ \Ec\ \lf\{  \frac{(\xi_jk_b)(\xi_kk_b)}{z_{ai}z_{bj}z_{bk}}
\lf(\frac{(\xi_ik_b)z_{ab}}{z_{aj}z_{ak}z_{bi}} -\frac{(\xi_ik_j)}{z_{ak}z_{ij}}-
\frac{(\xi_ik_k)}{z_{aj}z_{ik}}  \ri)   \ri. \cr
+&  \frac{(\xi_jk_i)(\xi_kk_b)}{z_{aj}z_{bk}z_{ij}}
\lf(\frac{(\xi_ik_b)}{z_{ak}z_{bi}} -\frac{(\xi_ik_k)}{z_{ab}z_{ik}}  \ri)+
\frac{(\xi_jk_k)(\xi_kk_b)}{z_{ai}z_{bk}z_{jk}}
\lf(\frac{(\xi_ik_j)}{z_{ab}z_{ij}} -\frac{(\xi_ik_b)}{z_{aj}z_{bi}}+
\frac{(\xi_ik_k)z_{ak}}{z_{ab}z_{aj}z_{ik}}  \ri)\cr
+&\frac{(\xi_ik_b)(\xi_kk_i)}{z_{aj}z_{bi}z_{ik}}
\lf(\frac{(\xi_jk_b)}{z_{ak}z_{bj}} -\frac{(\xi_jk_i)z_{ai}}{z_{ab}z_{ak}z_{ij}}-
\frac{(\xi_jk_k)}{z_{ab}z_{jk}} -
\frac{(\xi_ik_j)(\xi_jk_b)(\xi_kk_j)}{z_{ab}z_{ak}z_{bj}z_{ij}z_{ik}} \ri) \cr
+&\lf.\frac{(\xi_jk_b)(\xi_kk_j)}{z_{ai}z_{bj}z_{jk}}
\lf(\frac{(\xi_ik_b)}{z_{aj}z_{ak}z_{bi}} -\frac{(\xi_ik_j)z_{aj}}{z_{ab}z_{ak}z_{ij}}-
\frac{(\xi_ik_k)}{z_{ab}z_{ik}}  \ri)-
\frac{(\xi_ik_b)(\xi_jk_i)(\xi_kk_j)}{z_{ab}z_{ak}z_{bi}z_{ij}z_{jk}} \ri\}\ ,\cr
\Cc_2(a,i,b,j&,k)=i\ \Ec\ \lf\{  \frac{(\xi_bk_i)(\xi_kk_b)}{z_{aj}z_{bi}z_{bk}}
\lf(\frac{(\xi_jk_k)}{z_{ai}z_{jk}} -\frac{(\xi_jk_b)z_{ab}}{z_{ai}z_{ak}z_{bj}}-
\frac{(\xi_jk_i)}{z_{ak}z_{ij}}  \ri)   \ri. \cr
-&\frac{(\xi_jk_i)(\xi_kk_i)}{z_{ab}z_{ij}z_{ik}}
\lf(\frac{(\xi_bk_i)z_{ai}}{z_{aj}z_{ak}z_{bi}} +\frac{(\xi_bk_j)}{z_{ak}z_{bj}}+
\frac{(\xi_bk_k)}{z_{aj}z_{bk}}  \ri)-
\frac{(\xi_jk_b)(\xi_kk_\i)}{z_{aj}z_{bj}z_{ik}}
\lf(\frac{(\xi_bk_i)}{z_{ak}z_{bi}} +\frac{(\xi_bk_k)}{z_{ai}z_{bk}}  \ri)\cr
+&\frac{(\xi_jk_k)(\xi_kk_i)}{z_{ab}z_{ik}z_{jk}}
\lf(\frac{(\xi_bk_i)}{z_{aj}z_{bi}} +\frac{(\xi_bk_j)}{z_{ai}z_{bj}}+
\frac{(\xi_bk_k)z_{ak}}{z_{ai}z_{aj}z_{bk}}  \ri)-
\frac{(\xi_bk_j)(\xi_jk_i)(\xi_kk_b)}{z_{ai}z_{ak}z_{bj}z_{bk}z_{ij}}\cr
-&\lf. \frac{(\xi_jk_i)(\xi_kk_j)}{z_{ab}z_{ij}z_{jk}}
\lf(\frac{(\xi_bk_i)}{z_{ak}z_{bi}} +\frac{(\xi_bk_j)z_{aj}}{z_{ai}z_{ak}z_{bj}}+
\frac{(\xi_bk_k)}{z_{ai}z_{bk}}  \ri)-
\frac{(\xi_bk_i)(\xi_jk_b)(\xi_kk_j)}{z_{ai}z_{ak}z_{bi}z_{bj}z_{jk}} \ri\}\ ,\cr
\Cc_3(i,j,a,b&,k)=i\ \Ec\ \lf\{  
\frac{(\xi_kk_b)}{z_{ak}z_{bk}z_{ij}}\lf(\frac{(\xi_ak_i)(\xi_bk_j)}{z_{ai}z_{bj}} 
-\frac{(\xi_ak_j)(\xi_bk_i)}{z_{aj}z_{bi}}\ri)   \ri. \cr
-&\frac{(\xi_kk_i)}{z_{ab}z_{ak}z_{ij}z_{ik}}\lf(\frac{(\xi_ak_i)(\xi_bk_j)}{z_{bj}} -
\frac{(\xi_ak_j)(\xi_bk_i)z_{ai}}{z_{aj}z_{bi}}\ri)+
\frac{(\xi_ak_i)(\xi_kk_j)}{z_{ab}z_{ai}z_{ij}z_{jk}}\lf(\frac{(\xi_bk_j)z_{aj}}{z_{ak}z_{bj}} +
\frac{(\xi_bk_k)}{z_{bk}}\ri)\cr
-&\lf. \frac{(\xi_bk_i)(\xi_kk_j)}{z_{ab}z_{bi}z_{ij}z_{jk}}\lf(\frac{(\xi_ak_j)}{z_{ak}} +
\frac{(\xi_ak_k)}{z_{ak}}\ri)+
\frac{(\xi_kk_i)}{z_{ak}z_{ij}z_{ik}}\lf(\frac{(\xi_ak_k)(\xi_bk_j)}{z_{ak}z_{bj}} -
\frac{(\xi_ak_j)(\xi_bk_k)}{z_{aj}z_{bk}}\ri)  \ri\}\ ,}}
respectively.

Exchanging in the amplitude $A^\pi(a,b,i,j,k)$ indices from the two sets $\{a,b\}$ and $\{i,j,k\}$, \ie
putting instead of the operators $a,b$ other operators in the $(-1)$--ghost picture,
leads to seemingly different expressions.  
In other words, putting different vertex operators in the $(-1)$--ghost picture,
provides  different function $A^\pi_r,C_s^\pi$ in front of a given kinematics.
The five--point amplitude \Studyv\ allows for $\lf(5\atop 2\ri)=10$ possibilities to choose 
which vertex pair $(a,b)$ to put into the $(-1)$--ghost picture.
Therefore, for a given group structure $\pi$ and the given kinematics
\eqn\considerkinv{
(\xi_A\xi_B)(\xi_C\xi_D)(\xi_E k_s)}
we obtain ten  --a priori-- different expressions $A^\pi_r$, when we examine
\givesv\ for the choices
\eqn\possib{
(a,b)\in \{\ (A,B),\ (A,C),\ (A,D),\ (A,E),\ (B,C),\ (B,D),\ (B,E),\ (C,D),\ (C,E),\ (D,E)\ \}\ .}
The latter may be read off from \givesv:
\eqn\gleichiv{\eqalign{
&A^\pi_1(A,B,C,D,E)\ \ \ ,\ \ \ A^\pi_1(C,D,A,B,E)\ ,\cr
&A^\pi_2(A,B,C,D,E)\ \ \ ,\ \ \ A^\pi_2(B,A,C,D,E)\ \ \ ,\ \ \ 
A^\pi_2(B,A,D,C,E)\ \ \ ,\ \ \ A^\pi_2(A,B,D,C,E)\ ,\cr
&A^\pi_3(A,B,C,D,E)\ \ \ ,\ \ \ A^\pi_3(B,A,C,D,E)\ \ \ ,\ \ \ A^\pi_3(C,D,A,B,E)\ \ \ ,\ \ \ 
A^\pi_3(D,C,A,B,E)\ .}}
They all have to be equated, which gives rise to $9\times 3$ equations for each kinematics
\considerkinv\ under consideration. The additional factor of three arises, since $s=A,B,C$ due
to the on--shell constraint $\xi_E k_E=0$ and momentum conservation.
{\it E.g.} the kinematics $\Xi_1$, \ie $A=1,B=2,C=3,D=4$ and $E=5$
yields the system of equations:
\eqn\gleichv{\eqalign{
&\hskip0.5cm A^\pi_1(1,2,3,4,5)=A^\pi_1(3,4,1,2,5)=A^\pi_2(1,2,3,4,5)=A^\pi_2(2,1,3,4,5)=
A^\pi_2(2,1,4,3,5)\cr
&=A^\pi_2(1,2,4,3,5)=A^\pi_3(1,2,3,4,5)=A^\pi_3(2,1,3,4,5)=A^\pi_3(3,4,1,2,5)=
A^\pi_3(4,3,1,2,5)\ .}}
Taking into account all $15\times 3$ kinematics \XXIS, we obtain $27\times 15=405$ equations, which
give rise to non--trivial relations among the integrals of the type \KINv\ in the same way, as
we have seen before in the case of the six-- or four--gluon string $S$--matrix. However, again
many equations are equal. Actually, we are left with only $291$ equations.

Similarly, let us discuss the ten different expressions for a kinematics of the
structure:
\eqn\Considerkin{
(\xi_A\xi_B)\ (\xi_C k_r)\ (\xi_D k_s)\ (\xi_E k_t)\ .} 
From \givesv\ we read off the ten possibilities
\eqn\Gleichiv{\eqalign{
&C^\pi_1(A,B,C,D,E)\ \ \ ,\ \ \ C^\pi_2(A,B,C,D,E)\ \ \ ,\ \ \ C^\pi_2(B,A,C,D,E)\ \ \ ,\ \ \ 
C^\pi_2(A,B,D,C,E)\ ,\cr
&C^\pi_2(B,A,D,C,E)\ \ \ ,\ \ \ C^\pi_2(A,B,E,C,D)\ \ \ ,\ \ \ C^\pi_2(B,A,E,C,D)\ ,\cr 
&C^\pi_3(A,B,C,D,E)\ \ \ ,\ \ \ C^\pi_3(A,B,C,E,D)\ \ \ ,\ \ \ C^\pi_3(A,B,D,E,C)}}
to extract the kinematics \Considerkin. They all have to lead to the same result. After
equating them we obtain nine equations for each kinematics \Considerkin.
On--shell, there are $10\times 3^3=270$ kinematics
\Considerkin. Hence we obtain $2,430$ equations. However, many of them are equal and
and in fact this system reduces to only $151$ equations.

So far, we have not yet fixed any of the vertex operators. Due to the $PSL(2,\IR)$ invariance
on the disk, we may fix three positions of the vertex operators.
A convenient choice in \studyv\ or in \KINv\ is \Choice, 
which implies the ghost factor $\vev{c(z_1)c(z_2)c(z_3)}=-z_\infty^2$.
A five--point amplitude involving massless external particles, \ie $k_i^2=0$, allows for
five independent kinematic invariants $s_i$.
We may choose\foot{The remaining scalar products of momenta are expressed by these
nine invariants:
\eqn\eliminate{\eqalign{
k_1k_2&=s_3+s_4+s_5\ ,\cr
k_1k_3&=s_1+s_2+s_5\ ,\cr
k_1k_4&=-s_1-s_3-s_5\ ,\cr
k_1k_5&=-s_2-s_4-s_5\ ,\cr
k_2k_3&=-s_1 - s_2 - s_3 - s_4 - s_5\ .}}}
the scalar products:
\eqn\Mandelstam{\eqalign{
s_1&=k_2k_4\ \ ,\ \ s_2=k_2k_5\ \ ,\ \ s_3=k_3k_4\ ,\cr
s_4&=k_3k_5\ \ ,\ \ s_5=k_4k_5\ .}}
The  general structure of the integrals \KINv\ is then given by:
\eqn\genstriv{
\tilde F[n_{24},n_{25},n_{34},n_{35},n_{45}]:=
\int_{\Ic_\pi} dz_4\ dz_5\ \ |z_4|^{\al_{24}}\ |z_5|^{\al_{25}}\ 
|1-z_4|^{\al_{34}}\ |1-z_5|^{\al_{35}}\ |z_4-z_5|^{\al_{45}}\ .}
Here, the powers $\al_{ij}$ are real numbers, which are given by
\eqn\poweriv{\eqalign{
\al_{24}&=s_1+n_{24}\ \ ,\ \ \al_{25}=s_2+n_{25}\ \ ,\ \ \al_{34}=s_3+n_{34}\ ,\cr
\al_{35}&=s_4+n_{35}\ \ ,\ \ \al_{45}=s_5+n_{45}\ ,}}
with some integers $n_{ij}\in \{-2,-1,0\}$.
The integrals \genstriv\ may be transformed into the form \five.
Of course, this depends on the range of integration $\Ic_\pi$  chosen in \genstriv.
For instance, if we concentrate on the group contractions 
$\Tr(\lambda^1\lambda^2\lambda^3\lambda^4\lambda^5)$, which we shall do in the following, 
in \eqq \genstriv\ we have to perform the integral over the region:
\eqn\regionv{
\Ic_\pi=\{\ z_4,z_5\in \IR\ |\ 1<z_4<z_5<\infty\ \}\ .}
For this case we obtain:
\eqn\performgenstriv{\eqalign{
\tilde F[n_{24},n_{25},n_{34},n_{35},n_{45}]&=
\int_1^\infty dz_4\ \int_{z_4}^\infty dz_5\ |z_4|^{\al_{24}}\ |z_5|^{\al_{25}}\ 
|1-z_4|^{\al_{34}}\ |1-z_5|^{\al_{35}}\ |z_4-z_5|^{\al_{45}}\cr
&\hskip-3.5cm=\int_0^1dx\ \int_0^1 dy\ x^{-3-\al_{24}-\al_{25}-\al_{34}-\al_{35}-\al_{45}}
\ y^{-2-\al_{25}-\al_{35}-\al_{45}}
\ (1-x)^{\al_{34}}\ (1-y)^{\al_{45}}\ (1-xy)^{\al_{35}}\cr
&\hskip-3.5cm=\fc{\Ga(-2-\al_{24}-\al_{25}-\al_{34}-\al_{35}-\al_{45})\ 
\Gamma(-1-\al_{25}-\al_{35}-\al_{45})\ \Ga(1+\al_{34})\ \Ga(1+\al_{45})}{
\Ga(-1-\al_{24}-\al_{25}-\al_{35}-\al_{45})\ \Ga(-\al_{25}-\al_{35} )}\cr
&\hskip-3.5cm\times \F{3}{2}\lf[-2-\al_{24}-\al_{25}-\al_{34}-\al_{35}-\al_{45},\ 
-1-\al_{25}-\al_{35}-\al_{45},\ -\al_{35} \atop 
-1-\al_{24}-\al_{25}-\al_{35}-\al_{45},\ -\al_{25}-\al_{35}\ri]\ .}}
For the group structure $\pi$ under consideration we may cast the integrals \KINv\ into the form
\eqn\KINNv{\eqalign{
A^\pi_1(a,b,i,j,k)&=\si_{ij}\si_{jk}\ \lf\{(1-k_ik_j)\ \lf(
\si_{ij}\si_{ja}\si_{ak}\ 
(\xi_kk_a)\ \tilde F\lf[\ov n_{ab},\ov n_{ij}=-2\atop \ov n_{jk},\ov n_{ak}=-1,\ov n_{ja}=1\ri] 
\ri.\ri.\cr
&\lf.+
\si_{ij}\si_{jb}\si_{bk}\ (\xi_kk_b)\   
\tilde F\lf[\ov n_{ab},\ov n_{ij}=-2\atop \ov n_{jk},\ov n_{bk}=-1,\ov n_{jb}=1\ri] \ri)\cr
&\lf.+\si_{ik}\lf[(\xi_kk_i)(k_jk_k)-(\xi_kk_j)(k_ik_k)-(1-k_ik_j)(\xi_kk_i)\ri] 
\tilde F\lf[\ov n_{ab}=-2\atop \ov n_{ij},\ov n_{ik},\ov n_{jk}=-1\ri] \ri\},\cr
A^\pi_2(a,i,b,j,k)&=\si_{ab}\si_{ai}\si_{jk}\ \lf\{\si_{ij}\si_{bj}\si_{ja}\si_{ak}\ 
(\xi_kk_a)(k_ik_j)\   
\tilde F
\lf[\ov n_{ab},\ov n_{ai},\ov n_{ak}=-1\atop \ov n_{bj},\ov n_{ij},\ov n_{jk}=-1,\ov n_{ja}=1\ri] 
\ri.\cr
&-\si_{bk}\si_{ij}\ (\xi_kk_b)(k_ik_j)\   
\tilde F\lf[\ov n_{ab},\ov n_{ai},\ov n_{bk}=-1\atop \ov n_{ij},\ov n_{jk}=-1\ri]  \cr
&+\lf.\si_{bj}\si_{ik}\ \lf[
(\xi_kk_j)(k_ik_k)-(\xi_kk_i)(k_jk_k)-(\xi_kk_i)(k_ik_j)\ri]\   
\tilde F\lf[\ov n_{ab},\ov n_{ai},\ov n_{bj}=-1\atop \ov n_{ik},\ov n_{jk}=-1\ri]
\ri\},\cr
A^\pi_3(a,i,j,k,b)&=\si_{ab}\si_{ai}\si_{jk}\ \lf\{
-\si_{bi}\si_{jk}\ (\xi_bk_i)\ (1-k_jk_k)\   
\tilde F\lf[\ov n_{ab},\ov n_{ai},\ov n_{bi}=-1\atop \ov n_{jk}=-2,\ri] \ri.\cr
&-\si_{bj}\si_{ik}\ (\xi_bk_j)\ (k_ik_k)\  
\tilde F\lf[\ov n_{ab},\ov n_{ai},\ov n_{bj}=-1\atop \ov n_{ik},\ov n_{jk}=-1\ri] \cr
&\lf.+\si_{bk}\si_{ij}\ (\xi_bk_k)\ (k_ik_j)\  
\tilde F\lf[\ov n_{ab},\ov n_{ai},\ov n_{bk}=-1\atop \ov n_{ij},\ov n_{jk}=-1\ri]\ri\}\ .}}
with $\si_{ij}={\rm sign}(i-j)$, and $\ov n_{ij}:=\cases{n_{ij},& $i<j$ \cr 
                                                  n_{ji},& $i>j$}.$ 
The entries $\ov n _{ij}$ in the functions $\tilde F$ are to be understood such, 
that they are only of relevance, if they contribute in \genstriv\ or \poweriv. Otherwise
they are meaningless as a matter of the choice \Choice. Altogether, 
there appear $77$ different functions $\tilde F$ in the expressions \gleichiv\ and \Gleichiv.

Let us come back to the system of equations \gleichv, written down for all $3\times 15$
kinematics $(\xi\xi)(\xi\xi)(\xi k)$. With \KINNv, it gives rise to
$291$ relations among the functions $\tilde F$, very much in the sense we have encountered
in \system\ for the case of four--gluon $S$--matrix or in subsection 3.3 for the six--gluon case:
These relations are to be compared with \eqqs \PARTIAL, \ie they may be proven through partial
integration.
In addition, the $151$ equations from the kinematics $(\xi\xi)(\xi k)(\xi k)(\xi k)$
give rise to simpler identities comparable to \eqqs \TRIPLE\ and \TRIPLEI, \ie they 
follow from simple polynomial relations. Let us present a selection of these relations:
\eqn\somewhat{\eqalign{
\tilde F[-2,0,0,-2,1]&=\tilde F[-2,0,0,-1,0]-\tilde F[-2,0,1,-2,0]\ ,\cr
\tilde F[-2,-1,0,-1,0]&=\tilde F[-1,-1,0,0,-1]-\tilde F[-2,0,1,-1,-1]\ ,\cr
\tilde F[-1,-1,-1,-1,0]&=\tilde F[-1,0,-1,-1,0]-\tilde F[-1,-,1,-1,0,0]\ ,\cr
\tilde F[-1,0,0,-1,0]&=\tilde F[-1,0,0,0,-1]-\tilde F[-1,0,1,-1,-1]\ ,\cr
\tilde F[0,-2,-2,0,1]&=\tilde F[0,-2,-2,1,0]-\tilde F[0,-2,-1,0,0]\ .}}
After inserting for the $\tilde F$'s their definition \performgenstriv\ these equations appear 
like the theorems derived in {\it Ref.} \bailey.

As in the case of the six--gluon amplitude we may first solve for all equations following from 
the kinematics $(\xi\xi)(\xi k)(\xi k)(\xi k)$ and insert its solution into the equations
derived from the kinematics $(\xi\xi)(\xi\xi)(\xi k)$. In doing so, solving for all $151$ equations
provides eight solutions, \ie eight functions $\tilde F$ out of the $77$ may be expressed linearly
in terms of others as in \eqq \somewhat. Eliminating those eight functions in the second system
of $291$ equations leaves $275$ equations to be solved.
Of course, now these equations are more involved than in the four--gluon case
and they describe various relations between hypergeometric functions $\F{3}{2}$.
This way we obtain a whole set of identities relating various functions $\tilde F$.
These relations are similar to the identities for the
hypergeometric function $\F{3}{2}$, presented in \bailey. 
Recall, that in the four--gluon scattering case \singel\ we have twelve equations and 
eleven functions $F_j$. However not all of these twelve equations are linear independent from each 
other. In effect, the system could be solved by introducing one single function $F_0$ and
expressing all other eleven functions through $F_0$, \cf \eqq \BOXIV.
For the case at hand  we encounter similar properties:
Though the system of  $275$ equations is overdetermined, 
many of them are linear dependent on each other and it is possible to express $75$ functions 
$\tilde F$ in terms of a basis of two functions. In the following, the latter shall be 
denoted by $\Phi_1$ and $\Phi_2$.

In the four--gluon  
it proved to be convenient, to express all eleven functions 
$F_j$ ($j=1,\ldots 11$) through one single function $F_0$ with special properties.
The latter has no  poles in the Mandelstam variables $s,t$ or $u$, \cf the expansion
\powerSS. Similarly, in the six--gluon case we express $1,264$ functions in terms of a basis
of six finite functions $\Phi_j$ \MOMENTUMEXP.
All pole structure of the amplitudes has appeared through the algebraic equations \singel\ or
\RESULT.
Similarly here, out of the total $77$ functions $\F{3}{2}$ only a few of them share the 
same properties, namely they do not have any poles in the kinematic invariants $s_i$.
{\it E.g.} we may single out the following two 
integrals
\eqn\singleOUT{\eqalign{
\Phi_1:=\tilde F[-1,-2,0,0,0]&=\int_0^1dx\ \int_0^1 dy\ x^{-s_1-s_2-s_3-s_4-s_5}
\ y^{-s_2-s_4-s_5}\cr 
&\times (1-x)^{s_3}\ (1-y)^{s_5}\ (1-xy)^{s_4}\ ,\cr
\Phi_2:=\tilde F[-1,-1,0,-1,0]&=\int_0^1dx\ \int_0^1 dy\ x^{-s_1-s_2-s_3-s_4-s_5}
\ y^{-s_2-s_4-s_5}\cr
&\times \ (1-x)^{s_3}\ (1-y)^{s_5}\ (1-xy)^{s_4-1}}}
from the $77$ functions $\tilde F$. The functions $\Phi_1,\Phi_2$
do not have poles in the invariants $s_i$. In fact, their expansion w.r.t. small $s_i$
may be easily obtained from the expressions \MOMEXP\ and \MOMEXPi, \cf \eqq (A.27).
According to \performgenstriv\ 
they may be also written as hypergeometric functions:
\eqn\fsevten{\eqalign{
\Phi_1&=\fc{\Gamma(1-s_1-s_2-s_3-s_4-s_5)\ \Gamma(1-s_2-s_4-s_5)\ \Gamma(1+s_3)\ \Gamma(1+s_5)}{
\Gamma(2-s_1-s_2-s_4-s_5)\ \Gamma(2-s_2-s_4)}\cr 
&\times\F{3}{2}\lf[1-s_1-s_2-s_3-s_4-s_5,\ 1-s_2-s_4-s_5,\ -s_4 \atop 
2-s_1-s_2-s_4-s_5,\ 2-s_2-s_4\ri]\ ,\cr
\Phi_2&=\fc{\Gamma(1-s_1-s_2-s_3-s_4-s_5)\ \Gamma(1-s_2-s_4-s_5)\ \Gamma(1+s_3)\ \Gamma(1+s_5)}{
\Gamma(2-s_1-s_2-s_4-s_5)\ \Gamma(2-s_2-s_4)}\cr 
&\times\F{3}{2}\lf[1-s_1-s_2-s_3-s_4-s_5,\ 1-s_2-s_4-s_5,\ 1-s_4 \atop 
2-s_1-s_2-s_4-s_5,\ 2-s_2-s_4\ri]\ .}}
In complete analogy to \eqqs \BOXIV\ and \RESULT\ we may solve all remaining $275$ 
equations through introducing the two--dimensional basis of functions $\Phi_1$ and $\Phi_2$.
All other $75$ functions are given in terms of the latter
\eqn\BOXV{{
\vbox{\offinterlineskip
\halign{\strut\vrule#
%%%%%%%%%%%%%%%%%%
&~$#$~\hfil
&\vrule# \cr
\noalign{\hrule}
%%%%%%%%%%%%%%%%%%
&\ \        \  \ &\cr
&\ \ \tilde F[n_{24},n_{25},n_{34},n_{35},n_{45}]=\Lambda^{1}_{\{n_{ij}\}}(s_i)\ \ \tilde F[-1,-2,0,0,0]
+\Lambda^{2}_{\{n_{ij}\}}(s_i)\ \ \tilde F[-1,-1,0,-1,0]\ \ &\cr
&\ &\cr
%%%%%%%%%%%%%%%%%%
\noalign{\hrule}}} }}
with some polynomials $\Lambda^{1}_{\{n_{ij}\}}(s_i),\ \Lambda^{2}_{\{n_{ij}\}}(s_i)$, 
which depend non--trivially on the five kinematic invariants $s_i$.
Note, that due to the finiteness of $\Phi_1,\Phi_2$, the pole structure of each of the $75$
hypergeometric functions $\F{3}{2}$ is  determined by those prefactors 
$\Lambda^{j}_{\{n_{ij}\}}(s_i)$. The latter follow
algebraically from solving the system of $275$ equations.
To demonstrate the power of our method, let us present a few of those relations \BOXV, we have found: 
\eqn\listfew{\eqalign{
\tilde F[-1,-1,0,0,-1]&=\frac{1  - s_2 }{s_5}\ \Phi_1- \frac{s_4}{s_5}\ \Phi_2\ ,\cr
\tilde F[-1,-1,-1,0,0]&=-\frac{\left( 1 - s_2 \right) \,\left( 1 - s_1 - s_2 - s_5 \right) }
     {s_3\,\left( s_2 + s_4 + s_5 \right) } \ \Phi_1  + 
  \frac{\left( 1 - s_2 + s_3 \right) \,s_4}
   {s_3\,\left( s_2 + s_4 + s_5 \right) }\ \Phi_2\ ,\cr
\tilde F[-2,0,0,0,-1]&=\frac{\left( 1 - s_2 \right) \,\left[ \left( 1 - s_1 \right) \,s_2 + 
         s_4\,\left( 1 - s_1 - s_5 \right)  \right]}
{\left( 1 - s_1 \right) \,s_5\ \left( s_2 + s_4 + s_5 \right) } \ \Phi_1  \cr
&- \frac{s_4\,\left[ \left( 1 - s_1 \right) \,s_2 + s_4\,
\left( 1 - s_1 - s_5 \right)  - 
       s_5\,\left( s_1 + s_3 + s_5 \right)  \right] \,
}{\left( 1 - s_1 \right) \,s_5\,
     \left( s_2 + s_4 + s_5 \right)}\  \Phi_2\ ,\cr
\tilde F[0,-2,-1,0,0]&
=\frac{\left( 1 - s_1 - s_2 - s_5 \right) }{s_3} \ \Phi_1 - \frac{s_4}{s_3}\ \Phi_2\ ,\cr
\tilde F[0,-2,0,-1,0]&=\frac{1 - s_1 - s_2 - s_5 }{1 + s_1 + s_3 + s_5}\ \Phi_1  + 
  \frac{s_1 - s_4}{1 + s_1 + s_3 + s_5}\ \Phi_2\ ,\cr
\tilde F[-1,-1,-1,-1,1]&=-\frac{\left( 1 - s_2 \right) \,\left( 1 - s_1   - s_2 - s_5 \right)}
{s_3\ ( s_2 + s_4 + s_5) }\ \Phi_1  + 
  \frac{s_4 - s_2\,\left( s_3 + s_4 \right)  - 
s_3\,s_5 }{s_3\,\left( s_2 + s_4 + s_5 \right)}\ \Phi_2\ .}}
In addition, let us present the somewhat more involved relation:
\eqn\moreinvolved{\eqalign{
&\tilde F[0,0,-1,0,-1]\cr
&=\frac{\left( 1 - s_2 \right) \,\left( 1 - s_1 - s_2 - s_5 \right) \,
    \left[\ s_1\,s_3\ \left( s_2 + s_4 \right)  + s_1\,s_2\,s_5 + 
      s_2\,\left( s_3 + s_5 \right) \,\left( s_2 + s_4 + s_5 \right)\  \right] }{
    s_3\ s_5\,\left( s_2 + s_4 + s_5 \right) \,\left( s_3 + s_4 + s_5 \right) \,
\left( s_1 + s_2 + s_3 + s_4 + s_5 \right) }\ \Phi_1\cr
&-\lf\{\ \frac{ \left( 1 - s_1 - s_2 \right) \,\left( s_1 + s_2 \right) \ 
       \left( s_2 + s_4 \right)}{s_5\,\left( s_2 + s_4 + s_5 \right) \,
\left( s_3 + s_4 + s_5 \right) \,
    \left( s_1 + s_2  + s_3  + s_4 + s_5 \right) }\ri.\cr
&\hskip0.5cm\lf.+
\fc{\left( 1 - s_2 \right) \,s_2 - s_1\,s_3 }{s_3\ 
    \left( s_2 + s_4 + s_5 \right) \,\left( s_3 + s_4 + s_5 \right)}\ri\}\ s_4\ \Phi_2\ .}}
The importance of this identity becomes clear, once we insert \performgenstriv\ into
this equation. According to \eqq \performgenstriv\ the function $\tilde F[0,0,-1,0,-1]$ 
corresponds to the integral
$${\eqalign{\tilde F[0,0,-1,0,-1]&=\int_0^1dx\ \int_0^1 dy\ x^{-1-s_1-s_2-s_3-s_4-s_5}\ 
y^{-1-s_2-s_4-s_5}\cr
&\times  (1-x)^{-1+s_3}\ (1-y)^{-1+s_5}\ (1-xy)^{s_4}\ ,}}$$
which has several double poles in the invariants $s_i$. However, since the functions $\Phi_1,\Phi_2$
in the expansion \moreinvolved\ do not have any poles, all the poles of $\tilde F[0,0,-1,0,-1]$ 
are encoded in the fractions in front of those functions and we may easily obtain the full power series
of $\tilde F[0,0,-1,0,-1]$ by only knowing the series expansions of $\Phi_1,\Phi_2$.
Furthermore, this equation becomes \JAPAN. Hence, we have proven the non--trivial
equation \JAPAN\ just from relating equivalent parts of the five--gloun string $S$--matrix.
According to \performgenstriv\ the above identities \somewhat, \listfew\ and \moreinvolved\ 
give rise to non--trivial relations among three hypergeometric functions $\F{3}{2}$, which
only partly may be found in the literature \bailey.
We should stress, that the relations listed above are just twelve out of $75$ identities
we obtain from our method.

After having solved all equations, the full $\ap$--dependence of the five gluon
amplitude is captured by the two functions $\Phi_1,\Phi_2$, given in \eqqs \singleOUT\ or \fsevten.
Hence, to obtain the $\ap$--expansion of \studyv, we only have to know their
power series expansions w.r.t. to the kinematical invariants $s_i$.
With the material, derived in section 4, in particular setting
$a=-s_1-s_2-s_3-s_4-s_5,\ b=-s_2-s_4-s_5,\ c=s_3,\ d=s_5$ and $e=s_4$ in \MOMEXP\ and \MOMEXPi, we obtain:
\eqn\Fsevten{\eqalign{
\Phi_1&=1+s_1+2\ s_2+s_5+s_4\ \zeta(2)+
s_1^2 + 3\ s_1\ s_2 + 3\ s_2^2 + 2\ s_1\ s_5 + 3\ s_2\ s_5 + s_5^2\cr
&+(s_1\ s_3 + s_2\ s_3 + s_3^2 + s_1\ s_4 + 
2\ s_2\ s_4 + 2\ s_3\ s_4 + s_4^2 + s_2\ s_5 + s_3\ s_5 + 3\ s_4\ s_5 + s_5^2)\ \zeta(2)\cr
&+(s_1\ s_4+2\ s_2\ s_4-s_3\ s_4+s_4^2)\ \zeta(3)+\ldots\ ,\cr
\Phi_2&=\zeta(2)+(\ s_1+2\ s_2-s_3+s_4\ )\ \zeta(3)\cr
&+\lf(s_1^2+3\ s_1\ s_2+3\ s_2^2+\fc{3}{2}\ s_1\ s_3+\fc{5}{4}\ s_2\ s_3+\fc{7}{2}\ s_3^2
+\fc{11}{4}\ s_1\ s_4+\fc{11}{2}\ s_2\ s_4+4\ s_3\ s_4+\ri.\cr
&+\lf.\fc{7}{2}\ s_4^2+\fc{7}{4}\ s_1\ s_5
+\fc{17}{4}\ s_2\ s_5+\fc{17}{4}\ s_3\ s_5+\fc{27}{4}\ s_4\ s_5+\fc{17}{4}\ s_5^2\ri)+\ldots\ .}}
These two functions $\Phi_1,\Phi_2$  have no poles in the kinematic invariants $s_i$  
and their momentum expansion may be obtained by the methods presented in section 4.
This property is just in lines of \powerSS\ in the four--gluon case and \MOMENTUMEXP\ 
in the six--gluon case.

{\it E.g.} in the string $S$--matrix \studyv\ after eliminating $\xi_5k_4$ on--shell
the kinematics $(\xi_1\xi_2)(\xi_3\xi_4)(\xi_5k_1)$ comes along with the momentum dependent expression:
\eqn\comes{\eqalign{
&(\xi_1\xi_2)(\xi_3\xi_4)(\xi_5k_1)\ \ \ \fc{1}{s_3\,
    s_5\,\left( s_2 + s_4 + s_5 \right) \,\left( s_3 + s_4 + s_5 \right) }\cr
&\times\lf\{\fc{}{}\ \left( 1 - s_2 \right) \,\left( 1 - s_1 - s_2 - s_5 \right) \,
    \left[\ s_1\,s_3\,\left( s_2 + s_4 \right)  + s_1\,s_2\,s_5 + 
      s_2\,\left( s_3 + s_5 \right) \,\left( s_2 + s_4 + s_5 \right)  \ \right]\ \Phi_1\ri.\cr
&+s_4\,\left\{ \left( -1 + s_1 + s_2 \right) \,\left( s_1 + s_2 \right) \,s_3\,
       \left( s_2 + s_4 \right)  - \left[ \left( 1 - s_2 \right) \,s_2 - s_1\,s_3 \right] \,
       \left( s_1 + s_2 + s_3 + s_4 \right) \,s_5 \ri.\cr
&\lf.\lf.- \left[ \left( 1 - s_2 \right) \,s_2 - s_1\,s_3 \right] \,{s_5}^2 \right\} \ \Phi_2\ 
\fc{}{}\ri\}\ .}}
According to \eqq \considerkinv\ the above expression 
may be derived from $A_1(1,2,3,4,5)$.
On the other hand, in the string $S$--matrix \studyv\ after eliminating $\xi_5k_4, \xi_4k_5$ and $\xi_3k_5$
on--shell the kinematics $(\xi_1\xi_2)(\xi_3k_1)(\xi_4k_2)(\xi_5k_1)$ 
comes along with the momentum dependent expression:
\eqn\comesi{\eqalign{
&(\xi_1\xi_2)(\xi_3k_1)(\xi_4k_2)(\xi_5k_1)\ \ \ \ \fc{1}{s_5\ (s_3+s_4+s_5)}\cr
&\times \ \lf\{\ 
(1-s_2)\ (1-s_1-s_2-s_5)\ \Phi_1-\lf[\ s_4\ (1-s_1-s_2-s_5)\ -s_5\ (s_1+s_3+s_5)\ \ri]\ \Phi_2\ \ri\}\ .}}
According to \eqq \Considerkin\ the above expression 
may be derived from $C_1(1,2,3,4,5)$.
All other kinematics may be expressed similarly. The result \comes\ is to be compared with
the final result \FINAL\ of the six--gluon case.
To conclude, we may express the full five gluon kinematics \gleichiv\ through
the two functions $\Phi_1,\Phi_2$, which are rather easy 
to handle due to their pole structure.
In fact, after expanding \comes\ and \comesi\ up to 
second order in the Mandelstam variables, we find agreement with the results of \Brazil. 
In addition, we shall point out, that in the beautiful works \BRAZIL, it has also been found, that
the five--gluon string $S$--matrix may be expressed by two single hypergeometric functions $\F{3}{2}$.

\appendix{\appB}{Series involving the Harmonic Number}

For the power series expansion of the hypergeometric function $_3F_2$ we shall 
need the nine series:
\eqn\nine{\eqalign{
(i)&\ \ \ \sum_{n=0}^\infty\fc{1}{(n+1)(n+2)}\ \psi^{(1)}(n+1)=1\ ,\cr
(ii)&\ \ \ \sum_{n=0}^\infty\fc{1}{(n+1)(n+2)}\ \psi^{(1)}(n+2)=\zeta(2)-\zeta(3)\ ,\cr
(iii)&\ \ \ \sum_{n=0}^\infty\fc{1}{(n+1)(n+2)}\ \psi^{(1)}(n+3)=-3+2\ \zeta(2)\ ,\cr
(iv)&\ \ \ \sum_{n=0}^\infty\fc{1}{(n+1)(n+2)}\ H_{n+1}\ \psi(n+1)
=(1-\gamma_E)\ \zeta(2)+\zeta(3)\ ,\cr
(v)&\ \ \ \sum_{n=0}^\infty\fc{1}{(n+1)(n+2)}\ H_{n+2}\ \psi(n+1)
=3-2\ \gamma_E\ ,}}
$${\eqalign{
(vi)&\ \ \ \sum_{n=0}^\infty\fc{1}{(n+1)(n+2)}\ H_{n+2}\ \psi(n+2)
=-2\ \gamma_E+\zeta(2)+2\ \zeta(3)\ ,\cr
(vii)&\ \ \ \sum_{n=0}^\infty\fc{1}{(n+1)(n+2)}\ \psi(n+1)^2=(1-\gamma_E)^2+\zeta(2)\ ,\cr
(viii)&\ \ \ \sum_{n=0}^\infty\fc{1}{(n+1)(n+2)}\ \psi(n+2)^2=
\gamma_E^2-2\ \gamma_E\ \zeta(2)+3\ \zeta(3)\ ,\cr
(ix)&\ \ \ \sum_{n=0}^\infty\fc{1}{(n+1)(n+2)}\ \psi(n+3)^2=
3-4\ \gamma_E+\gamma_E^2+\zeta(2)\ .}}$$
These identities may be proven by using \rewriting\ and applying formulae shown 
in subsection 4.3.
Furthermore, we prove the three series:
\eqn\katze{\eqalign{
(i)&\ \ \ \sum_{n=0}^\infty\fc{1}{(n+1)^2}\ \psi(n+1)^2=
\fc{11}{4}\ \zeta(4)-2\ \gamma_E\ \zeta(3)+\gamma_E^2\ \zeta(2)\ ,\cr
(ii)&\ \ \ \sum_{n=0}^\infty\fc{1}{(n+1)^2}\ \psi(n+2)^2=
\fc{17}{4}\ \zeta(4)-4\ \gamma_E\ \zeta(3)+\gamma_E^2\ \zeta(2)\ ,\cr
(iii)&\ \ \ \sum_{n=0}^\infty\fc{1}{(n+1)^2}\ \psi^{(1)}(n+1)^2=
\fc{1}{3}\ \zeta(2)^3-\zeta(3)^2-\fc{5}{3}\ \zeta(2)\ \zeta(4)+\fc{17}{3}\ \zeta(6)\ .}}
After rewriting the sums with \rewriting, these identities may be proven by using various
relations presented in subsection 4.3, in particular \eqq \IMporti.

\appendix{\appC}{Momentum expansion of 
$\F{4}{3}\lf[{1+b,\ 1+a,\ 1+c,\ -j\atop 2+b+e,\ 2+a+d,\ 2+c+f}\ri]$}

\eqn\MOMEXPii{\eqalign{
&\fc{\Ga(1+d)\ \Ga(1+e)\ \Ga(1+f)\ \Ga(1+b)\ \Ga(1+a)\ \Ga(1+c)}
{\Ga(2+b+e)\ \Ga(2+a+d)\ \Ga(2+c+f)}\ 
\F{4}{3}\lf[{1+b,\ 1+a,\ 1+c,\ -j\atop 2+b+e,\ 2+a+d,\ 2+c+f}\ri]\cr
&=1-a-b-c-d-e-f-3 j+a^2+b a+c a+2 d a+e a+f a+4 j a+b^2+c^2+d^2\cr
&+e^2+f^2+6 j^2+b c+b d+c d+2 b e+c e+d e+bf+2cf+df+ef+4bj+4cj+4dj\cr
&+4ej+4fj-a^3-ba^2-ca^2-3da^2-ea^2-fa^2-5 j a^2-b^2 a-c^2 a-3 d^2 a-e^2 a\cr
&-f^2 a-10 j^2 a-b c a-2 b d a-2 c d a-2 b ea-c e a-2 d e a-b f a-2 c f a-2 d f a-efa\cr
&-5b j a-5 cj a-10 d j a-5 e j a-5 f j a-b^3-c^3-d^3-e^3-f^3-10j^3-b c^2-b d^2-c d^2\cr
&-3 b e^2-c e^2-d e^2-b f^2-3 cf^2-d f^2-e f^2-10 b j^2-10 c j^2-10 d j^2-10 e j^2-
10f j^2\cr
&-b^2 c-b^2 d-c^2 d-b c d-3 b^2 e-c^2 e-d^2 e-2 b ce-2 b d e-c d e-b^2 f-3 c^2 f-
d^2 f\cr
&-e^2 f-2 b c f-b df-2 c d f-2 b e f-2 c e f-d e f-5 b^2 j-5 c^2 j-5 d^2j-5 e^2 j
-5 f^2 j\cr
&-5 b c j-5 b d j-5 c d j-10 b e j-5 ce j-5 d e j-5 b f j-10 c f j-5 d f j-5 e f j\cr
&+\zeta(2)\ \lf\{j-2j^2-aj-bj-cj-ad-be-cf+3j^3+3aj^2+3bj^2+3cj^2+a^2j+b^2j\ri.\cr
&+c^2j+a b j+a c j+b c j+4 a d j+4 b ej-d e j+4 c f j-d f j-e f j+a d^2+b e^2+c f^2
+a^2 d\cr
&\lf.+a b d+a c d+b^2 e+a b e+bc e+a d e+b d e+c^2 f+acf+bcf+adf+c d f+b e f+cef\ri\}\cr
&+\fc{1}{4}\ \zeta(4)\ \lf\{-j^2-4 a j-4 b j-4 c j-5 d j-5 e j-5 f j+
5 j^3+5 a j^2+5 b j^2+5 c j^2+17 d j^2\ri.\cr
&+17 e j^2+17 f j^2+4 a^2 j+4 b^2 j+4 c^2j+12 d^2 j+12 e^2 j+
12 f^2 j+4 a b j+4 a c j+4 b c j\cr
&\lf.+2 a d j+5 b d j+5 c d j+5
a e j+2 b e j+5 c e j+17 d e j+5 a f j+5 b f j+2 c f j+17 d f j+17 e f j
\ri\}\cr
&+\zeta(3)\ 
\lf\{j-2j^2-a j-bj-c j-2 d j-2 e j-2 f j+3 j^3+3 a j^2+3 b j^2+3cj^2+4dj^2+4ej^2\ri.\cr
&+4f j^2+a^2 j+b^2 j+c^2 j+a b j+a c
   j+b c j+2 a d j+2 b d j+2 c d j+2 a e j+2 b e j+2 c e j\cr
&\lf.+2 a f j+2 b f j+2 c f j+a d^2+b e^2+c f^2+a^2 d+b^2 e+c^2 f\ri\}\cr
&+\zeta(2)\ \zeta(3)\ \lf\{-j^3-a j^2-b j^2-c j^2+d^2 j+e^2 j+f^2 j+a d j-b 
d j-c d j-a e j+b e j-c e j\ri.\cr
&\lf.-d ej-a f j-b f j+c f j-d f j-e f j\ri\}\cr
&+\h\ \zeta(5)\ \lf\{4 j^3+4 a j^2+4 b j^2+4 c j^2+d j^2+e j^2+f j^2+2 a^2 j+2 b^2 j
+2 c^2 j-d^2 j-e^2j\ri.\cr
&-f^2 j+2 a b j+2 a c j+2 b c j-3 a d j+6 b d j+6 c d j+6 a e j-3 b e j+6 c e
   j+7 d e j+6 a f j\cr
&\lf.+6 b f j-3 c f j+7 d f j+7 e f j\ri\}+\ldots\ .}}

\appendix{\appD}{Some group theoretical facts}

The field strength $F_{\mu\nu}^a$ is defined as:
\eqn\fieldST{
F_{\mu\nu}^a=\p_\mu A_\nu^a-\p_\nu A_\mu^a+g f^{abc} A^b_\mu A^c_\nu\ ,}
with the structure constants $f^{abc}$ for an $U(N)$ gauge group.
Let us recall some basic facts about Lie Algebras.
The commutation relations of a representation $T^a$ of the Lie Algebra read:
\eqn\lieal{[T^a,T^b]=i\ f^{abc}\ T^c\ ,}
with the structure constants $f^{abc}$.
We impose the standard normalization condition
\eqn\lienorm{\Tr(T^aT^b)=C(r)\ \delta^{ab}\ ,}
with $C(r)$ being a constant for each representation $r$.
Then we have the relation 
\eqn\Relation{f^{abc}=-\fc{i}{C(r)}\ 
\Tr\lf(\ [T^a,T^b]\ T^c\ \ri)\ ,}
which implies, that $f^{abc}$ is totally anti--symmetric.

The adjoint representation $r=G$ is given by the matrices
\eqn\adjoint{
(T^a)_{bc}=-i\ f^{abc}\ ,}
which obviously satisfies \lieal\ and \Relation.
The covariant derivative $D^{\rm adj.}_\mu$ acting on fields in the adjoint representation is
introduced as:
\eqn\COVDEV{
(D_\lambda^{\rm adj.})_{ab}=\p_\lambda\ \delta_{ab}-i\ g A^m_\lambda\ (T^m)_{ab}=
\p_\lambda\ \delta_{ab}-\ g\ f_{mab}\ A^m_\lambda\ .}
Hence, we have:
\eqn\COVDEVV{
D_\lambda\ F_{\mu\nu}^a=\p_\lambda\ F_{\mu\nu}^a-i\ g\ A^m_\lambda\ (T^m)_{ab}
\ F_{\mu\nu}^b\ .}
In addition, we derive:
\eqn\COVADD{\eqalign{
D_\kappa D_\lambda\ F_{\mu\nu}^a&=\p_\kappa\p_\lambda\ F_{\mu\nu}^a+
g\ f^{and}\ A_\kappa^n\ \p_\lambda F_{\mu\nu}^d -
g\ f^{mac}\ \p_\kappa(A_\lambda^m\ F_{\mu\nu}^c)\cr
&+g^2\ f^{amc}\ f^{mnd}\ A_\kappa^n\ A_\lambda^d\ F_{\mu\nu}^c+g^2\ f^{amc}\ f^{cnd}\ 
A_\lambda^m\ A_\kappa^n\ F_{\mu\nu}^d\ .}}
Finally, for $F_{\mu\nu}\equiv T^a F^a_{\mu\nu}$ 
in the adjoint representation, we may write \fieldST\ as:
\eqn\fieldST{
F_{\mu\nu}=\p_\mu A_\nu-\p_\nu A_\mu-i\ g\ [A_\mu,A_\nu]\ ,}
and \COVDEVV\ as:
\eqn\alter{
D_\lambda=\partial_\lambda-i\ g\ [A_\lambda,\ \star\ ]\ .}
Furthermore, \eqq \COVADD\ gives rise to:
\eqn\alteri{\eqalign{
D_\kappa(D_\lambda\ F_{\mu\nu})&=\p_\kappa\p_\lambda\ F_{\mu\nu}-
i\ g\ [A_\kappa,\ \p_\lambda F_{\mu\nu}]-i\ g\ [\p_\kappa A_\lambda,\ F_{\mu\nu}]\cr
&-i\ g\ [A_\lambda,\ \p_\kappa F_{\mu\nu}]-g^2\ [\ A_\kappa,[A_\lambda,\ F_{\mu\nu}]\ ]}}
To this end, we may prove:
\eqn\MAYDER{
[D_\mu,D_\nu]\ F_{\rho\sigma}=-i\ g\ [F_{\mu\nu},F_{\rho\sigma}]\ .}

\listrefs
\end